\documentclass[11pt]{article}
\usepackage[margin=1in]{geometry}
\usepackage{amsmath,amssymb,booktabs,graphicx,microtype,url,xcolor}
\usepackage[hidelinks]{hyperref}
\hypersetup{pdftitle={Plan Pointers and Record-Directive Form in Budgeted Verification of Inherited Agent Memory},
  pdfauthor={Kazuki Nakayashiki},
  pdfsubject={Preprint. Directive form, pointers and criteria in budgeted verification of inherited agent memory.},
  pdfkeywords={agent memory, provenance, verification budget, directive compliance, instruction following, LLM agents}}
\usepackage{natbib}
\usepackage{caption}
\usepackage{pgfplots}\pgfplotsset{compat=1.18}\usetikzlibrary{arrows.meta,positioning}
\usepackage{placeins}
\usepackage{adjustbox}
\usepackage{fancyvrb}\usepackage{textcomp}
\newif\ifanon\anonfalse
\newcommand{\osf}[1]{\ifanon withheld\else\texttt{#1}\fi}
\newcommand{\Br}{\texorpdfstring{B$'$}{B'}}\newcommand{\Gr}{\texorpdfstring{G$'$}{G'}}   
\newcommand{\BRatePricing}{88.0}
\newcommand{\BRateOnboarding}{10.0}
\newcommand{\BRateNone}{63.7}
\newcommand{\BRateBridgePricing}{92.7}
\newcommand{\BRateBridgeOnboarding}{8.7}
\newcommand{\BDeltaPlan}{+78.0}
\newcommand{\BDeltaPlanLo}{+74.3}
\newcommand{\BDeltaPlanHi}{+81.7}
\newcommand{\BDeltaBridge}{+84.0}
\newcommand{\BDeltaBridgeLo}{+78.0}
\newcommand{\BDeltaBridgeHi}{+89.3}
\newcommand{\BShare}{0.929}
\newcommand{\BShareLo}{0.856}
\newcommand{\BShareHi}{1.008}
\newcommand{\BSharePublished}{0.087}
\newcommand{\BPricingVsNone}{+24.3}
\newcommand{\BPricingVsNoneLo}{+20.7}
\newcommand{\BPricingVsNoneHi}{+28.0}
\newcommand{\BOnboardingVsNone}{\ensuremath{-}53.7}
\newcommand{\BOnboardingVsNoneLo}{\ensuremath{-}58.0}
\newcommand{\BOnboardingVsNoneHi}{\ensuremath{-}49.7}
\newcommand{\BDeltaPlanPublished}{+7.3}

\newcommand{\BCellAOne}{89.3}
\newcommand{\BCellATwo}{16.0}
\newcommand{\BCellBOne}{4.0}
\newcommand{\BCellBTwo}{86.7}

\newcommand{\BPerModelDeltaPlan}{Opus 5 +82.0, Sonnet 5 +96.0, Haiku 4.5 +34.0, GPT-5.6 Sol +90.0, GPT-5.6 Terra +70.0, GPT-5.6 Luna +96.0}

\newcommand{\DMultiId}{3}

\newcommand{\DRateNone}{1.7}
\newcommand{\DRateOracle}{56.2}
\newcommand{\DRatePolicy}{91.2}
\newcommand{\DEone}{+35.0}
\newcommand{\DEoneLo}{+31.2}
\newcommand{\DEoneHi}{+38.8}
\newcommand{\DEtwo}{+54.6}
\newcommand{\DEtwoLo}{+51.7}
\newcommand{\DEtwoHi}{+57.5}
\newcommand{\DEthree}{+89.6}
\newcommand{\DEthreeLo}{+86.7}
\newcommand{\DEthreeHi}{+92.5}

\newcommand{\DPmKOracleOpus}{0/40}

\newcommand{\DPmKOracleSonnet}{15/40}

\newcommand{\DPmKOracleHaiku}{0/40}

\newcommand{\DPmKOracleGPTSol}{40/40}

\newcommand{\DPmKOracleGPTTerra}{40/40}

\newcommand{\DPmKOracleGPTLuna}{40/40}

\newcommand{\DPmKPolicyOpus}{40/40}

\newcommand{\DPmKPolicySonnet}{40/40}

\newcommand{\DPmKPolicyHaiku}{19/40}

\newcommand{\DLooMin}{+22.0}
\newcommand{\DLooMax}{+42.0}
\newcommand{\DAllocNoneEightySix}{88.3}

\newcommand{\DAllocPolicyEightySix}{2.9}
\newcommand{\DAllocPolicySeventyThree}{91.2}

\newcommand{\EAttempted}{540}
\newcommand{\EScored}{526}
\newcommand{\EErrors}{14}
\newcommand{\EErrorPct}{2.6}

\newcommand{\EMultiId}{5}
\newcommand{\EEmpty}{9}
\newcommand{\EKimiPairedRuns}{15}
\newcommand{\ERateNone}{30.4}
\newcommand{\ERateOracle}{76.1}
\newcommand{\ERatePolicy}{83.3}
\newcommand{\EEone}{+7.2}
\newcommand{\EEoneLo}{+0.0}
\newcommand{\EEoneHi}{+14.4}

\newcommand{\EPerModelMin}{\ensuremath{-}25.0}
\newcommand{\EPerModelMax}{+90.0}

\newcommand{\FN}{2100}
\newcommand{\FScored}{2075}
\newcommand{\FErrors}{25}
\newcommand{\FErrorPct}{1.2}

\newcommand{\FLostModel}{Qwen3.8 Max 25}
\newcommand{\FZeroSpent}{31}

\newcommand{\FConeIdGtCrit}{+31.9}
\newcommand{\FConeIdGtCritLo}{+24.3}
\newcommand{\FConeIdGtCritHi}{+39.8}

\newcommand{\FCtwoIdGtCrit}{+17.0}
\newcommand{\FCtwoIdGtCritLo}{+10.9}
\newcommand{\FCtwoIdGtCritHi}{+23.2}

\newcommand{\FCthreeIdGtCrit}{+0.6}

\newcommand{\FConeCritGtId}{\ensuremath{-}37.5}
\newcommand{\FConeCritGtIdLo}{\ensuremath{-}43.8}
\newcommand{\FConeCritGtIdHi}{\ensuremath{-}30.0}

\newcommand{\FCtwoCritGtId}{+51.2}
\newcommand{\FCtwoCritGtIdLo}{+47.5}
\newcommand{\FCtwoCritGtIdHi}{+56.2}

\newcommand{\FCthreeCritGtId}{\ensuremath{-}5.0}
\newcommand{\FCthreeCritGtIdLo}{\ensuremath{-}10.0}
\newcommand{\FCthreeCritGtIdHi}{\ensuremath{-}1.2}

\newcommand{\FConeOpen}{+19.8}
\newcommand{\FConeOpenLo}{+15.5}
\newcommand{\FConeOpenHi}{+24.3}

\newcommand{\FCtwoOpen}{+22.6}
\newcommand{\FCtwoOpenLo}{+18.7}
\newcommand{\FCtwoOpenHi}{+26.6}

\newcommand{\FCthreeOpen}{\ensuremath{-}1.4}

\newcommand{\FConeClosed}{\ensuremath{-}8.3}
\newcommand{\FConeClosedLo}{\ensuremath{-}12.1}
\newcommand{\FConeClosedHi}{\ensuremath{-}4.6}

\newcommand{\FCtwoClosed}{+26.2}
\newcommand{\FCtwoClosedLo}{+22.5}
\newcommand{\FCtwoClosedHi}{+30.0}

\newcommand{\FCthreeClosed}{\ensuremath{-}0.8}

\newcommand{\FPmKCritMistralMedium}{32/40}

\newcommand{\FPmKBothILMistralMedium}{40/40}

\newcommand{\FPmKIdOpus}{1/40}

\newcommand{\FPmKCritOpus}{40/40}

\newcommand{\FPmKBothILOpus}{2/40}

\newcommand{\FPmKBothIFOpus}{0/40}

\newcommand{\FPmKBothILSonnet}{20/20}

\newcommand{\FPmKCritHaiku}{11/20}

\newcommand{\FPmKBothILHaiku}{20/20}

\newcommand{\FCfourMin}{+51.2}
\newcommand{\FCfourMax}{+79.2}
\newcommand{\FOpusOverrideNone}{37/40}
\newcommand{\FOpusOverrideId}{36/40}
\newcommand{\FOpusOverrideCrit}{3/40}
\newcommand{\FOpusOverrideBothIL}{38/40}
\newcommand{\FOpusOverrideBothIF}{39/40}

\newcommand{\FPrecondCount}{6}
\newcommand{\FPrecondTotal}{6}
\newcommand{\FOpusMEightySixBothIL}{38}

\newcommand{\FCompletionLabelChanges}{1}

\newcommand{\FExceedsCountReferent}{7}

\newcommand{\FxN}{600}

\newcommand{\FxErrors}{0}

\newcommand{\FxEmpty}{0}

\newcommand{\FxPmKNoneFableFiveOne}{13/40}

\newcommand{\FxPmKIdFableFiveOne}{13/40}

\newcommand{\FxPmKCritFableFiveOne}{17/40}

\newcommand{\FxPmKBothILFableFiveOne}{0/40}

\newcommand{\FxMEightySixBothILFableFiveOne}{40}
\newcommand{\FxPmKBothIFFableFiveOne}{0/40}

\newcommand{\FxConeFableFiveOne}{\ensuremath{-}42.5}
\newcommand{\FxConeFableFiveOneLo}{\ensuremath{-}57.5}
\newcommand{\FxConeFableFiveOneHi}{\ensuremath{-}27.5}

\newcommand{\FxCtwoFableFiveOne}{\ensuremath{-}32.5}
\newcommand{\FxCtwoFableFiveOneLo}{\ensuremath{-}47.5}
\newcommand{\FxCtwoFableFiveOneHi}{\ensuremath{-}20.0}

\newcommand{\FxPmKIdFableFive}{6/40}

\newcommand{\FxPmKCritFableFive}{40/40}

\newcommand{\FxPmKBothILFableFive}{1/40}

\newcommand{\FxMEightySixBothILFableFive}{39}
\newcommand{\FxPmKBothIFFableFive}{0/40}

\newcommand{\FxConeFableFive}{\ensuremath{-}97.5}
\newcommand{\FxConeFableFiveLo}{\ensuremath{-}100.0}
\newcommand{\FxConeFableFiveHi}{\ensuremath{-}92.5}

\newcommand{\FxPmKIdOpus}{0/40}

\newcommand{\FxPmKCritOpus}{40/40}

\newcommand{\FxPmKBothILOpus}{0/40}

\newcommand{\FxMEightySixBothILOpus}{39}
\newcommand{\FxPmKBothIFOpus}{0/40}

\newcommand{\FxConeOpus}{\ensuremath{-}100.0}
\newcommand{\FxConeOpusLo}{\ensuremath{-}100.0}
\newcommand{\FxConeOpusHi}{\ensuremath{-}100.0}

\newcommand{\FxFamConeFableFiveOne}{+57.5}
\newcommand{\FxFamConeFableFiveOneLo}{+42.5}
\newcommand{\FxFamConeFableFiveOneHi}{+72.5}

\newcommand{\FxFamConeFableFive}{+2.5}
\newcommand{\FxFamConeFableFiveLo}{+0.0}
\newcommand{\FxFamConeFableFiveHi}{+7.5}
\newcommand{\FxFamConeFableFiveDir}{Undetermined}
\newcommand{\FxFamConeFableFiveSize}{Within Margin}
\newcommand{\FxFamCtwoFableFiveOne}{\ensuremath{-}32.5}
\newcommand{\FxFamCtwoFableFiveOneLo}{\ensuremath{-}47.5}
\newcommand{\FxFamCtwoFableFiveOneHi}{\ensuremath{-}17.5}

\newcommand{\FxFamCtwoFableFive}{\ensuremath{-}12.5}
\newcommand{\FxFamCtwoFableFiveLo}{\ensuremath{-}25.0}
\newcommand{\FxFamCtwoFableFiveHi}{+0.0}
\newcommand{\FxFamCtwoFableFiveDir}{Undetermined}
\newcommand{\FxFamCtwoFableFiveSize}{Unresolved}

\newcommand{\FxAnchor}{met}

\newcommand{\FQLockRollup}{263f4d1f}
\newcommand{\FCleanBaselineRecords}{1,321}
\newcommand{\FQN}{2100}
\newcommand{\FQScored}{1945}

\newcommand{\FQSurchargeCalls}{1,054}
\newcommand{\FQSurchargePerCall}{\$0.007}
\newcommand{\FQSurchargeTotal}{\$7.378}
\newcommand{\FQErrors}{155}
\newcommand{\FQErrPlugin}{146}

\newcommand{\FQSurchargeModels}{7}

\newcommand{\FQUrls}{197}
\newcommand{\FQGrokExcessMin}{1,921}

\newcommand{\FQVariableModelsList}{DeepSeek V4 Pro, Kimi K3, MiniMax M3, Llama 4 Maverick, Qwen3.8 Max, Mistral Medium 3.5}

\newcommand{\FQVariableSuccesses}{876}
\newcommand{\FQVariablePhrase}{37}
\newcommand{\FQVariableUrlRecords}{83}

\newcommand{\FQVariableExcessMin}{2,059}
\newcommand{\FQVariableExcessMax}{3,250}
\newcommand{\FQVariableSurchargeMin}{91}
\newcommand{\FQVariableSurchargeMax}{176}
\newcommand{\FQGeminiExcess}{22}

\newcommand{\FQGptOssSurchargeCalls}{169}
\newcommand{\FQVariableModels}{6}
\newcommand{\FQConstantExcessModels}{Gemini 3.7 Flash +22, Grok 4.6 +1,921}
\newcommand{\BManifestFiles}{15}
\newcommand{\BManifestRollup}{440c25da29749ef9}
\newcommand{\BLockRollupLong}{64d42c07606d359f}
\newcommand{\BOsfFile}{6a962202279e8d251648ef5e}

\newcommand{\BOtsPresent}{yes}
\newcommand{\DManifestFiles}{18}
\newcommand{\DManifestRollup}{ef970ded1d7d4185}
\newcommand{\DLockRollupLong}{ab1304fb199c8d8b}
\newcommand{\DOsfFile}{6a96631c345b176e56e36bb4}

\newcommand{\DOtsPresent}{yes}
\newcommand{\EManifestFiles}{15}
\newcommand{\EManifestRollup}{41406a36b3800178}
\newcommand{\ELockRollupLong}{554c9a6b33005152}
\newcommand{\EOsfFile}{6a9669597bae564049aba7f3}

\newcommand{\EOtsPresent}{yes}
\newcommand{\FManifestFiles}{232}
\newcommand{\FManifestRollup}{baee320357a2bee8}
\newcommand{\FLockRollupLong}{dae4060a5dfaf187}
\newcommand{\FOsfFile}{6a979df06a6ef349e72a9c8c}

\newcommand{\FOtsPresent}{yes}
\newcommand{\FxManifestFiles}{67}
\newcommand{\FxManifestRollup}{e63edffc2f0b73c2}
\newcommand{\FxLockRollupLong}{d1d6c0c055f3a789}
\newcommand{\FxOsfFile}{6a97ce8966ae3dc6a88ad96e}

\newcommand{\FxOtsPresent}{yes}
\newcommand{\GManifestFiles}{79}
\newcommand{\GManifestRollup}{a73730d00d303cee}
\newcommand{\GLockRollupLong}{620e1bfd8885a78f}
\newcommand{\GOsfFile}{6a97f782a41528cd5c8ad939}

\newcommand{\GOtsPresent}{yes}
\newcommand{\BrManifestFiles}{106}
\newcommand{\BrManifestRollup}{ac45a57ff8f1e981}
\newcommand{\BrLockRollupLong}{d26c127a14663110}
\newcommand{\BrOsfFile}{6a9846d4b2c383f5ba8ada9d}

\newcommand{\BrOtsPresent}{yes}
\newcommand{\IManifestFiles}{167}
\newcommand{\IManifestRollup}{9876a8ce3d5ca526}
\newcommand{\ILockRollupLong}{b382a0a80b341b47}
\newcommand{\IOsfFile}{6a9853861a6d2596a68fc9e5}

\newcommand{\IOtsPresent}{yes}
\newcommand{\HoneManifestFiles}{105}
\newcommand{\HoneManifestRollup}{6340644a46464bf5}
\newcommand{\HoneLockRollupLong}{db9bfe2817303396}
\newcommand{\HoneOsfFile}{6a986569fe128e518f8fca46}

\newcommand{\HoneOtsPresent}{yes}
\newcommand{\JManifestFiles}{108}
\newcommand{\JManifestRollup}{bf9dd31b5fcb13f2}
\newcommand{\JLockRollupLong}{62c8fec21f6b3675}
\newcommand{\JOsfFile}{6a9888277c58df62ed72c62e}

\newcommand{\JOtsPresent}{yes}
\newcommand{\HtwoManifestFiles}{128}
\newcommand{\HtwoManifestRollup}{d1ed685c3d2fe101}
\newcommand{\HtwoLockRollupLong}{40d23bd897037245}
\newcommand{\HtwoOsfFile}{6a98a2cec76b5f31708fcaa7}

\newcommand{\HtwoOtsPresent}{yes}
\newcommand{\GrManifestFiles}{94}
\newcommand{\GrManifestRollup}{4e359fee89ed48d1}
\newcommand{\GrLockRollupLong}{d07839956f270bc1}
\newcommand{\GrOsfFile}{6a98a329edd22c742b8ad8de}

\newcommand{\GrOtsPresent}{yes}
\newcommand{\KtwoManifestFiles}{1420}
\newcommand{\KtwoManifestRollup}{5080bcbd58cbf216}
\newcommand{\KtwoLockRollupLong}{2958ce170db39014}
\newcommand{\KtwoOsfFile}{6a9c5f9bd495d73ea13dbc78}

\newcommand{\KtwoOtsPresent}{yes}
\newcommand{\FIdPaddingSpaces}{78}
\newcommand{\FDirectiveLineCount}{4}

\newcommand{\GN}{960}

\newcommand{\GErrors}{0}

\newcommand{\GEmpty}{0}
\newcommand{\GPmKCritOpus}{40/40}

\newcommand{\GPmKBothILOpus}{3/40}

\newcommand{\GPmKCritPadOpus}{40/40}

\newcommand{\GPmKCritOtherOpus}{0/40}

\newcommand{\GPmKBothFreshOpus}{13/40}

\newcommand{\GPmKBothInheritedOpus}{0/40}

\newcommand{\GGoneOpus}{\ensuremath{-}92.5}
\newcommand{\GGoneOpusLo}{\ensuremath{-}97.4}
\newcommand{\GGoneOpusHi}{\ensuremath{-}77.3}

\newcommand{\GGtwoOpus}{+0.0}
\newcommand{\GGtwoOpusLo}{\ensuremath{-}8.8}
\newcommand{\GGtwoOpusHi}{+8.8}
\newcommand{\GGtwoOpusDir}{Undetermined}
\newcommand{\GGtwoOpusSize}{Within Margin}
\newcommand{\GGthreeOpus}{\ensuremath{-}100.0}
\newcommand{\GGthreeOpusLo}{\ensuremath{-}100.0}
\newcommand{\GGthreeOpusHi}{\ensuremath{-}87.6}

\newcommand{\GGsixOpus}{+32.5}
\newcommand{\GGsixOpusLo}{+17.3}
\newcommand{\GGsixOpusHi}{+48.0}
\newcommand{\GGsixOpusDir}{Positive}
\newcommand{\GGsixOpusSize}{Exceeds Margin}
\newcommand{\GGsevenOpus}{+92.5}
\newcommand{\GGsevenOpusLo}{+77.3}
\newcommand{\GGsevenOpusHi}{+97.4}

\newcommand{\GGnineOpus}{\ensuremath{-}67.5}
\newcommand{\GGnineOpusLo}{\ensuremath{-}79.9}
\newcommand{\GGnineOpusHi}{\ensuremath{-}49.7}

\newcommand{\GGtenOpus}{\ensuremath{-}100.0}
\newcommand{\GGtenOpusLo}{\ensuremath{-}100.0}
\newcommand{\GGtenOpusHi}{\ensuremath{-}87.6}

\newcommand{\GPmKCritFableFive}{40/40}

\newcommand{\GPmKBothILFableFive}{0/40}

\newcommand{\GPmKCritPadFableFive}{3/40}

\newcommand{\GPmKCritOtherFableFive}{0/40}

\newcommand{\GPmKBothFreshFableFive}{0/40}

\newcommand{\GPmKBothInheritedFableFive}{0/40}

\newcommand{\GGtwoFableFive}{\ensuremath{-}92.5}
\newcommand{\GGtwoFableFiveLo}{\ensuremath{-}97.4}
\newcommand{\GGtwoFableFiveHi}{\ensuremath{-}77.3}

\newcommand{\GGsixFableFive}{+0.0}
\newcommand{\GGsixFableFiveLo}{\ensuremath{-}8.8}
\newcommand{\GGsixFableFiveHi}{+8.8}
\newcommand{\GGsixFableFiveDir}{Undetermined}
\newcommand{\GGsixFableFiveSize}{Within Margin}

\newcommand{\GPmKCritFableFiveOne}{14/40}

\newcommand{\GPmKBothILFableFiveOne}{0/40}

\newcommand{\GPmKCritPadFableFiveOne}{7/40}

\newcommand{\GPmKCritOtherFableFiveOne}{0/40}

\newcommand{\GPmKBothFreshFableFiveOne}{0/40}

\newcommand{\GPmKBothInheritedFableFiveOne}{0/40}

\newcommand{\GGtwoFableFiveOne}{\ensuremath{-}17.5}
\newcommand{\GGtwoFableFiveOneLo}{\ensuremath{-}31.7}
\newcommand{\GGtwoFableFiveOneHi}{\ensuremath{-}2.7}
\newcommand{\GGtwoFableFiveOneDir}{Negative}
\newcommand{\GGtwoFableFiveOneSize}{Unresolved}

\newcommand{\GPmKCritGPTSol}{40/40}

\newcommand{\GPmKBothILGPTSol}{40/40}

\newcommand{\GPmKCritPadGPTSol}{40/40}

\newcommand{\GPmKCritOtherGPTSol}{0/40}

\newcommand{\GPmKBothFreshGPTSol}{40/40}

\newcommand{\GPmKBothInheritedGPTSol}{40/40}

\newcommand{\GGoneGPTSol}{+0.0}
\newcommand{\GGoneGPTSolLo}{\ensuremath{-}8.8}
\newcommand{\GGoneGPTSolHi}{+8.8}

\newcommand{\GGthreeGPTSol}{\ensuremath{-}100.0}
\newcommand{\GGthreeGPTSolLo}{\ensuremath{-}100.0}
\newcommand{\GGthreeGPTSolHi}{\ensuremath{-}87.6}

\newcommand{\GGeightGPTSol}{\ensuremath{-}100.0}
\newcommand{\GGeightGPTSolLo}{\ensuremath{-}100.0}
\newcommand{\GGeightGPTSolHi}{\ensuremath{-}87.6}

\newcommand{\GAnchor}{met}

\newcommand{\GWilsonZeroHi}{8.8}
\newcommand{\GWilsonFullLo}{91.2}
\newcommand{\GBoundaryContrastLo}{\ensuremath{-}8.8}
\newcommand{\GBoundaryContrastHi}{+8.8}
\newcommand{\GVfortyfourKOpus}{0/40}

\newcommand{\GVfortyfourKFableFive}{0/40}

\newcommand{\GVfortyfourKFableFiveOne}{0/40}

\newcommand{\GVfortyfourKGPTSol}{40/40}

\newcommand{\BrN}{1200}

\newcommand{\BrRatePricing}{89.0}
\newcommand{\BrRateOnboarding}{7.3}
\newcommand{\BrRateNone}{61.7}

\newcommand{\BrDeltaPlan}{+81.7}
\newcommand{\BrDeltaPlanLo}{+78.3}
\newcommand{\BrDeltaPlanHi}{+85.0}

\newcommand{\BrDeltaBridge}{+85.3}
\newcommand{\BrDeltaBridgeLo}{+79.3}
\newcommand{\BrDeltaBridgeHi}{+90.7}

\newcommand{\BrShare}{0.957}
\newcommand{\BrShareLo}{0.895}
\newcommand{\BrShareHi}{1.029}

\newcommand{\BrVerdict}{POINTER-STRONG}
\newcommand{\BrReplicationVerdict}{REPLICATED-VERDICT}
\newcommand{\BrRepPlan}{+3.7}
\newcommand{\BrRepPlanLo}{\ensuremath{-}1.3}
\newcommand{\BrRepPlanHi}{+8.7}
\newcommand{\BrRepPlanDir}{Undetermined}
\newcommand{\BrRepPlanSize}{Within Margin}
\newcommand{\BrRepBridge}{+1.3}
\newcommand{\BrRepBridgeLo}{\ensuremath{-}6.0}
\newcommand{\BrRepBridgeHi}{+9.3}
\newcommand{\BrRepBridgeDir}{Undetermined}
\newcommand{\BrRepBridgeSize}{Within Margin}
\newcommand{\BrCanonPlan}{+78.0}
\newcommand{\BrCanonBridge}{+84.0}
\newcommand{\BrCanonShare}{0.929}

\newcommand{\BrPerModelDeltaPlan}{Opus 5 +76.0, Sonnet 5 +100.0, Haiku 4.5 +36.0, GPT-5.6 Sol +96.0, GPT-5.6 Terra +82.0, GPT-5.6 Luna +100.0}

\newcommand{\ITurnOneN}{800}
\newcommand{\IDecisionN}{1600}
\newcommand{\ITurnOneErrors}{0}
\newcommand{\IDecisionErrors}{5}

\newcommand{\IContinuityBreaks}{0}

\newcommand{\IYoneKValidNoneOpus}{25/25}

\newcommand{\IVKIdOpus}{1/25}

\newcommand{\IVKCritOpus}{25/25}

\newcommand{\IYcSupersededCritOpus}{+100.0}
\newcommand{\IYcSupersededCritOpusLo}{+81.2}
\newcommand{\IYcSupersededCritOpusHi}{+100.0}

\newcommand{\IVKBothILOpus}{2/25}

\newcommand{\IYcSupersededBothILOpus}{+8.3}
\newcommand{\IYcSupersededBothILOpusLo}{\ensuremath{-}6.7}
\newcommand{\IYcSupersededBothILOpusHi}{+25.8}

\newcommand{\IVKIdSonnet}{5/25}

\newcommand{\IVKCritSonnet}{25/25}

\newcommand{\IYcValidCritSonnet}{+20.0}
\newcommand{\IYcValidCritSonnetLo}{+2.6}
\newcommand{\IYcValidCritSonnetHi}{+39.1}

\newcommand{\IYcSupersededCritSonnet}{+72.0}
\newcommand{\IYcSupersededCritSonnetLo}{+48.3}
\newcommand{\IYcSupersededCritSonnetHi}{+85.7}

\newcommand{\IVKBothILSonnet}{25/25}

\newcommand{\IVKIdHaiku}{0/25}

\newcommand{\IVKCritHaiku}{11/25}

\newcommand{\IYcValidCritHaiku}{+36.0}

\newcommand{\IVKBothILHaiku}{25/25}

\newcommand{\IYcValidBothILHaiku}{+88.0}
\newcommand{\IYcValidBothILHaikuLo}{+65.6}
\newcommand{\IYcValidBothILHaikuHi}{+95.8}

\newcommand{\IVKCritFableFive}{24/25}

\newcommand{\IYcValidCritFableFive}{+48.0}
\newcommand{\IYcValidCritFableFiveLo}{+23.8}
\newcommand{\IYcValidCritFableFiveHi}{+66.3}

\newcommand{\IYcSupersededCritFableFive}{+52.0}
\newcommand{\IYcSupersededCritFableFiveLo}{+29.2}
\newcommand{\IYcSupersededCritFableFiveHi}{+70.0}

\newcommand{\IVKBothILFableFive}{0/25}

\newcommand{\IVKNoneFableFiveOne}{12/25}

\newcommand{\IYoneKValidNoneFableFiveOne}{25/25}

\newcommand{\IVKCritFableFiveOne}{2/25}

\newcommand{\IYcSupersededCritFableFiveOne}{\ensuremath{-}32.0}
\newcommand{\IYcSupersededCritFableFiveOneLo}{\ensuremath{-}52.1}
\newcommand{\IYcSupersededCritFableFiveOneHi}{\ensuremath{-}8.3}

\newcommand{\IVKBothILFableFiveOne}{0/25}

\newcommand{\IYcSupersededBothILFableFiveOne}{\ensuremath{-}36.0}
\newcommand{\IYcSupersededBothILFableFiveOneLo}{\ensuremath{-}55.5}
\newcommand{\IYcSupersededBothILFableFiveOneHi}{\ensuremath{-}13.3}

\newcommand{\IYoneKValidNoneGPTSol}{0/25}

\newcommand{\IYoneKSupersededNoneGPTSol}{25/25}

\newcommand{\IYcValidCritGPTSol}{+100.0}
\newcommand{\IYcValidCritGPTSolLo}{+81.2}
\newcommand{\IYcValidCritGPTSolHi}{+100.0}

\newcommand{\IYoneKValidNoneGPTTerra}{2/25}

\newcommand{\IYoneKSupersededNoneGPTTerra}{25/25}

\newcommand{\IYcValidCritGPTTerra}{+92.0}

\newcommand{\IYoneKValidNoneGPTLuna}{2/25}

\newcommand{\IYoneKSupersededNoneGPTLuna}{21/25}

\newcommand{\IYoneKSupersededCritGPTLuna}{25/25}

\newcommand{\IYcValidCritGPTLuna}{+92.0}

\newcommand{\IYcSupersededCritGPTLuna}{+16.0}
\newcommand{\IYcSupersededCritGPTLunaLo}{\ensuremath{-}0.4}
\newcommand{\IYcSupersededCritGPTLunaHi}{+34.7}
\newcommand{\IYcSupersededCritGPTLunaDir}{Undetermined}
\newcommand{\IYcSupersededCritGPTLunaSize}{Unresolved}

\newcommand{\IHeavyLoss}{none}

\newcommand{\HoneN}{800}

\newcommand{\HoneErrors}{1}

\newcommand{\HonePmKIdOpus}{5/25}

\newcommand{\HonePmKCritOpus}{25/25}

\newcommand{\HonePmKBothILOpus}{1/25}

\newcommand{\HoneToneOpus}{\ensuremath{-}20.0}
\newcommand{\HoneToneOpusLo}{\ensuremath{-}36.0}
\newcommand{\HoneToneOpusHi}{\ensuremath{-}8.0}
\newcommand{\HoneToneOpusDir}{Negative}
\newcommand{\HoneToneOpusSize}{Unresolved}
\newcommand{\HoneToneOpusProc}{+80.0}
\newcommand{\HoneToneOpusGrowth}{+100.0}

\newcommand{\HoneTtwoOpus}{\ensuremath{-}1.0}
\newcommand{\HoneTtwoOpusLo}{\ensuremath{-}10.0}
\newcommand{\HoneTtwoOpusHi}{+9.5}
\newcommand{\HoneTtwoOpusDir}{Undetermined}
\newcommand{\HoneTtwoOpusSize}{Unresolved}
\newcommand{\HoneTtwoOpusProc}{\ensuremath{-}96.0}
\newcommand{\HoneTtwoOpusGrowth}{\ensuremath{-}95.0}

\newcommand{\HonePmKCritSonnet}{25/25}

\newcommand{\HoneToneSonnet}{\ensuremath{-}54.5}
\newcommand{\HoneToneSonnetLo}{\ensuremath{-}72.5}
\newcommand{\HoneToneSonnetHi}{\ensuremath{-}35.5}

\newcommand{\HonePmKNoneHaiku}{25/25}

\newcommand{\HonePmKCritHaiku}{25/25}

\newcommand{\HoneToneHaiku}{\ensuremath{-}43.5}
\newcommand{\HoneToneHaikuLo}{\ensuremath{-}60.0}
\newcommand{\HoneToneHaikuHi}{\ensuremath{-}26.0}

\newcommand{\HonePmKIdFableFive}{25/25}

\newcommand{\HonePmKCritFableFive}{25/25}

\newcommand{\HonePmKBothILFableFive}{23/25}

\newcommand{\HoneToneFableFive}{\ensuremath{-}85.0}
\newcommand{\HoneToneFableFiveLo}{\ensuremath{-}95.0}
\newcommand{\HoneToneFableFiveHi}{\ensuremath{-}72.5}

\newcommand{\HoneToneFableFiveProc}{+0.0}
\newcommand{\HoneToneFableFiveGrowth}{+85.0}
\newcommand{\HoneToneFableFiveSource}{F-x}

\newcommand{\HoneTtwoFableFive}{+89.5}
\newcommand{\HoneTtwoFableFiveLo}{+76.5}
\newcommand{\HoneTtwoFableFiveHi}{+100.0}

\newcommand{\HoneTtwoFableFiveProc}{\ensuremath{-}8.0}
\newcommand{\HoneTtwoFableFiveGrowth}{\ensuremath{-}97.5}

\newcommand{\HonePmKNoneFableFiveOne}{24/25}

\newcommand{\HonePmKIdFableFiveOne}{23/25}

\newcommand{\HonePmKCritFableFiveOne}{25/25}

\newcommand{\HonePmKBothILFableFiveOne}{7/25}

\newcommand{\HonePmKCritGPTSol}{25/25}

\newcommand{\HonePmKCritGPTTerra}{25/25}

\newcommand{\HonePmKCritGPTLuna}{25/25}

\newcommand{\HoneToneNProcAll}{25}

\newcommand{\HoneToneNGrowthAll}{40}
\newcommand{\HoneTtwoNGrowthFx}{40}
\newcommand{\HoneTtwoNGrowthFrun}{20}
\newcommand{\HoneHeavyLoss}{none}

\newcommand{\JN}{800}

\newcommand{\JErrors}{0}

\newcommand{\JFirstKBothILKtwoOpus}{0/25}

\newcommand{\JJzeroOpus}{\ensuremath{-}96.0}
\newcommand{\JJzeroOpusLo}{\ensuremath{-}99.3}
\newcommand{\JJzeroOpusHi}{\ensuremath{-}75.5}

\newcommand{\JJoneOpus}{+0.0}
\newcommand{\JJoneOpusLo}{\ensuremath{-}13.3}
\newcommand{\JJoneOpusHi}{+13.3}

\newcommand{\JJtwoOpus}{+96.0}
\newcommand{\JJtwoOpusLo}{+75.5}
\newcommand{\JJtwoOpusHi}{+99.3}

\newcommand{\JJthreeOpus}{+0.0}

\newcommand{\JJfourOpus}{+0.0}
\newcommand{\JJfourOpusLo}{\ensuremath{-}15.9}
\newcommand{\JJfourOpusHi}{+15.9}

\newcommand{\JJfiveOpus}{+0.0}
\newcommand{\JJfiveOpusLo}{\ensuremath{-}13.3}
\newcommand{\JJfiveOpusHi}{+13.3}

\newcommand{\JJsixOpus}{+92.0}
\newcommand{\JJsixOpusLo}{+70.0}
\newcommand{\JJsixOpusHi}{+96.7}

\newcommand{\JJsevenOpus}{\ensuremath{-}4.0}
\newcommand{\JJsevenOpusLo}{\ensuremath{-}19.5}
\newcommand{\JJsevenOpusHi}{+9.7}

\newcommand{\JFirstKBothILKtwoFableFive}{0/25}

\newcommand{\JJzeroFableFive}{\ensuremath{-}100.0}
\newcommand{\JJzeroFableFiveLo}{\ensuremath{-}100.0}
\newcommand{\JJzeroFableFiveHi}{\ensuremath{-}81.2}

\newcommand{\JJoneFableFive}{+0.0}
\newcommand{\JJoneFableFiveLo}{\ensuremath{-}13.3}
\newcommand{\JJoneFableFiveHi}{+13.3}

\newcommand{\JJtwoFableFive}{+88.0}
\newcommand{\JJtwoFableFiveLo}{+65.6}
\newcommand{\JJtwoFableFiveHi}{+95.8}

\newcommand{\JJthreeFableFive}{+8.0}

\newcommand{\JJfourFableFive}{+52.0}
\newcommand{\JJfourFableFiveLo}{+29.2}
\newcommand{\JJfourFableFiveHi}{+70.0}

\newcommand{\JJfiveFableFive}{+0.0}
\newcommand{\JJfiveFableFiveLo}{\ensuremath{-}13.3}
\newcommand{\JJfiveFableFiveHi}{+13.3}

\newcommand{\JJsixFableFive}{+96.0}
\newcommand{\JJsixFableFiveLo}{+75.5}
\newcommand{\JJsixFableFiveHi}{+99.3}

\newcommand{\JJsevenFableFive}{+0.0}
\newcommand{\JJsevenFableFiveLo}{\ensuremath{-}13.3}
\newcommand{\JJsevenFableFiveHi}{+13.3}

\newcommand{\JPmKCritFableFiveOne}{11/25}

\newcommand{\JPmKCritLegitFableFiveOne}{25/25}

\newcommand{\JFirstKBothILKtwoFableFiveOne}{0/25}

\newcommand{\JJzeroFableFiveOne}{\ensuremath{-}44.0}
\newcommand{\JJzeroFableFiveOneLo}{\ensuremath{-}62.9}
\newcommand{\JJzeroFableFiveOneHi}{\ensuremath{-}22.1}

\newcommand{\JJoneFableFiveOne}{+56.0}
\newcommand{\JJoneFableFiveOneLo}{+32.9}
\newcommand{\JJoneFableFiveOneHi}{+73.3}

\newcommand{\JJtwoFableFiveOne}{+100.0}
\newcommand{\JJtwoFableFiveOneLo}{+81.2}
\newcommand{\JJtwoFableFiveOneHi}{+100.0}

\newcommand{\JJthreeFableFiveOne}{+44.0}
\newcommand{\JJthreeFableFiveOneLo}{+22.1}
\newcommand{\JJthreeFableFiveOneHi}{+62.9}

\newcommand{\JJfourFableFiveOne}{+4.0}

\newcommand{\JJfiveFableFiveOne}{+56.0}
\newcommand{\JJfiveFableFiveOneLo}{+32.9}
\newcommand{\JJfiveFableFiveOneHi}{+73.3}

\newcommand{\JJsixFableFiveOne}{+100.0}
\newcommand{\JJsixFableFiveOneLo}{+81.2}
\newcommand{\JJsixFableFiveOneHi}{+100.0}

\newcommand{\JJsevenFableFiveOne}{+0.0}
\newcommand{\JJsevenFableFiveOneLo}{\ensuremath{-}13.3}
\newcommand{\JJsevenFableFiveOneHi}{+13.3}

\newcommand{\JJzeroGPTSol}{+0.0}
\newcommand{\JJzeroGPTSolLo}{\ensuremath{-}13.3}
\newcommand{\JJzeroGPTSolHi}{+13.3}

\newcommand{\JHeavyLoss}{none}

\newcommand{\HtwoN}{2000}

\newcommand{\HtwoErrors}{0}

\newcommand{\HtwoPmKPzeroOpus}{40/40}

\newcommand{\HtwoPmKPoneOpus}{40/40}

\newcommand{\HtwoPmKPtwoOpus}{40/40}

\newcommand{\HtwoPmKPthreeOpus}{40/40}

\newcommand{\HtwoPmKPfourOpus}{7/40}

\newcommand{\HtwoSufPzeroOpus}{\ensuremath{-}95.0}
\newcommand{\HtwoSufPzeroOpusLo}{\ensuremath{-}98.6}
\newcommand{\HtwoSufPzeroOpusHi}{\ensuremath{-}80.5}

\newcommand{\HtwoDepPoneOpus}{+0.0}

\newcommand{\HtwoSufPoneOpus}{\ensuremath{-}100.0}

\newcommand{\HtwoDepPtwoOpus}{+0.0}

\newcommand{\HtwoSufPtwoOpus}{\ensuremath{-}100.0}

\newcommand{\HtwoSufPthreeOpus}{\ensuremath{-}100.0}

\newcommand{\HtwoDepPfourOpus}{\ensuremath{-}82.5}
\newcommand{\HtwoDepPfourOpusLo}{\ensuremath{-}91.3}
\newcommand{\HtwoDepPfourOpusHi}{\ensuremath{-}65.6}

\newcommand{\HtwoSufPfourOpus}{\ensuremath{-}17.5}
\newcommand{\HtwoSufPfourOpusLo}{\ensuremath{-}31.9}
\newcommand{\HtwoSufPfourOpusHi}{\ensuremath{-}5.1}
\newcommand{\HtwoSufPfourOpusDir}{Negative}
\newcommand{\HtwoSufPfourOpusSize}{Unresolved}
\newcommand{\HtwoCancelOpus}{P0, P1, P2, P3}
\newcommand{\HtwoCancelNOpus}{four of the five wordings}
\newcommand{\HtwoNotCancelledOpus}{P4}

\newcommand{\HtwoPmKPzeroFableFive}{37/40}

\newcommand{\HtwoPmKPoneFableFive}{34/40}

\newcommand{\HtwoPmKPtwoFableFive}{30/40}

\newcommand{\HtwoPmKPthreeFableFive}{5/40}

\newcommand{\HtwoPmKPfourFableFive}{40/40}

\newcommand{\HtwoSufPzeroFableFive}{\ensuremath{-}90.0}
\newcommand{\HtwoSufPzeroFableFiveLo}{\ensuremath{-}95.3}
\newcommand{\HtwoSufPzeroFableFiveHi}{\ensuremath{-}73.9}

\newcommand{\HtwoDepPoneFableFive}{\ensuremath{-}7.5}

\newcommand{\HtwoSufPoneFableFive}{\ensuremath{-}82.5}

\newcommand{\HtwoDepPtwoFableFive}{\ensuremath{-}17.5}
\newcommand{\HtwoDepPtwoFableFiveLo}{\ensuremath{-}33.5}
\newcommand{\HtwoDepPtwoFableFiveHi}{\ensuremath{-}1.1}
\newcommand{\HtwoDepPtwoFableFiveDir}{Negative}
\newcommand{\HtwoDepPtwoFableFiveSize}{Unresolved}
\newcommand{\HtwoSufPtwoFableFive}{\ensuremath{-}75.0}

\newcommand{\HtwoDepPthreeFableFive}{\ensuremath{-}80.0}
\newcommand{\HtwoDepPthreeFableFiveLo}{\ensuremath{-}88.6}
\newcommand{\HtwoDepPthreeFableFiveHi}{\ensuremath{-}61.6}

\newcommand{\HtwoSufPthreeFableFive}{\ensuremath{-}10.0}
\newcommand{\HtwoSufPthreeFableFiveLo}{\ensuremath{-}23.8}
\newcommand{\HtwoSufPthreeFableFiveHi}{+2.5}
\newcommand{\HtwoSufPthreeFableFiveDir}{Undetermined}
\newcommand{\HtwoSufPthreeFableFiveSize}{Unresolved}

\newcommand{\HtwoSufPfourFableFive}{\ensuremath{-}85.0}

\newcommand{\HtwoCancelFableFive}{P0, P1, P2, P4}
\newcommand{\HtwoCancelNFableFive}{four of the five wordings}
\newcommand{\HtwoNotCancelledFableFive}{P3}

\newcommand{\HtwoPmKPfourFableFiveOne}{24/40}

\newcommand{\HtwoSufPzeroFableFiveOne}{\ensuremath{-}40.0}
\newcommand{\HtwoSufPzeroFableFiveOneLo}{\ensuremath{-}55.4}
\newcommand{\HtwoSufPzeroFableFiveOneHi}{\ensuremath{-}23.8}

\newcommand{\HtwoSufPfourFableFiveOne}{\ensuremath{-}60.0}
\newcommand{\HtwoSufPfourFableFiveOneLo}{\ensuremath{-}73.7}
\newcommand{\HtwoSufPfourFableFiveOneHi}{\ensuremath{-}42.3}

\newcommand{\HtwoCancelFableFiveOne}{P0, P1, P2, P3, P4}
\newcommand{\HtwoCancelNFableFiveOne}{all five wordings}

\newcommand{\HtwoCancelSonnet}{none}

\newcommand{\HtwoCellsSonnet}{39/40 to 40/40 across the ten cells}
\newcommand{\HtwoCellsBelowFullSonnet}{P2\_id 39/40}

\newcommand{\HtwoCancelGPTSol}{none}

\newcommand{\HtwoCellsGPTSol}{40/40 in each of the ten cells}

\newcommand{\HtwoHeavyLoss}{none}

\newcommand{\GrN}{1440}

\newcommand{\GrErrors}{0}

\newcommand{\GrPmKCritOpus}{80/80}

\newcommand{\GrPmKBothILOpus}{8/80}

\newcommand{\GrPmKCritPadOpus}{80/80}

\newcommand{\GrPmKCritOtherOpus}{0/80}

\newcommand{\GrPmKBothFreshOpus}{15/80}

\newcommand{\GrPmKBothInheritedOpus}{0/80}

\newcommand{\GrVfortyfourKOpus}{0/80}

\newcommand{\GrGoneOpus}{\ensuremath{-}90.0}
\newcommand{\GrGoneOpusLo}{\ensuremath{-}94.8}
\newcommand{\GrGoneOpusHi}{\ensuremath{-}80.3}

\newcommand{\GrRepGoneOpus}{+2.5}
\newcommand{\GrRepGoneOpusLo}{\ensuremath{-}8.8}
\newcommand{\GrRepGoneOpusHi}{+12.5}

\newcommand{\GrRepGoneOpusSize}{Unresolved}
\newcommand{\GrPoolGoneOpus}{\ensuremath{-}90.8}
\newcommand{\GrPoolGoneOpusLo}{\ensuremath{-}94.8}
\newcommand{\GrPoolGoneOpusHi}{\ensuremath{-}83.6}

\newcommand{\GrGtwoOpus}{+0.0}
\newcommand{\GrGtwoOpusLo}{\ensuremath{-}4.6}
\newcommand{\GrGtwoOpusHi}{+4.6}

\newcommand{\GrGtwoOpusSize}{Within Margin}
\newcommand{\GrRepGtwoOpus}{+0.0}
\newcommand{\GrRepGtwoOpusLo}{+0.0}
\newcommand{\GrRepGtwoOpusHi}{+0.0}

\newcommand{\GrRepGtwoOpusSize}{Within Margin}

\newcommand{\GrGthreeOpus}{\ensuremath{-}100.0}
\newcommand{\GrGthreeOpusLo}{\ensuremath{-}100.0}
\newcommand{\GrGthreeOpusHi}{\ensuremath{-}93.5}

\newcommand{\GrRepGthreeOpus}{+0.0}
\newcommand{\GrRepGthreeOpusLo}{+0.0}
\newcommand{\GrRepGthreeOpusHi}{+0.0}

\newcommand{\GrRepGthreeOpusSize}{Within Margin}

\newcommand{\GrGsixOpus}{+18.8}
\newcommand{\GrGsixOpusLo}{+10.3}
\newcommand{\GrGsixOpusHi}{+28.7}

\newcommand{\GrRepGsixOpus}{\ensuremath{-}13.8}
\newcommand{\GrRepGsixOpusLo}{\ensuremath{-}30.0}
\newcommand{\GrRepGsixOpusHi}{+2.5}

\newcommand{\GrRepGsixOpusSize}{Unresolved}

\newcommand{\GrPmKCritFableFive}{78/80}

\newcommand{\GrPmKBothILFableFive}{1/80}

\newcommand{\GrPmKCritPadFableFive}{1/80}

\newcommand{\GrPmKCritOtherFableFive}{0/80}

\newcommand{\GrPmKBothFreshFableFive}{0/80}

\newcommand{\GrPmKBothInheritedFableFive}{0/80}

\newcommand{\GrVfortyfourKFableFive}{0/80}

\newcommand{\GrGoneFableFive}{\ensuremath{-}96.2}

\newcommand{\GrRepGoneFableFive}{+3.8}
\newcommand{\GrRepGoneFableFiveLo}{+0.0}
\newcommand{\GrRepGoneFableFiveHi}{+8.8}

\newcommand{\GrRepGoneFableFiveSize}{Within Margin}

\newcommand{\GrGtwoFableFive}{\ensuremath{-}96.2}

\newcommand{\GrRepGtwoFableFive}{\ensuremath{-}3.8}
\newcommand{\GrRepGtwoFableFiveLo}{\ensuremath{-}13.8}
\newcommand{\GrRepGtwoFableFiveHi}{+5.0}

\newcommand{\GrRepGtwoFableFiveSize}{Unresolved}

\newcommand{\GrGthreeFableFive}{\ensuremath{-}97.5}

\newcommand{\GrPmKCritFableFiveOne}{29/80}

\newcommand{\GrPmKBothILFableFiveOne}{1/80}

\newcommand{\GrPmKCritPadFableFiveOne}{9/80}

\newcommand{\GrPmKCritOtherFableFiveOne}{1/80}

\newcommand{\GrPmKBothFreshFableFiveOne}{1/80}

\newcommand{\GrPmKBothInheritedFableFiveOne}{0/80}

\newcommand{\GrVfortyfourKFableFiveOne}{0/80}

\newcommand{\GrGoneFableFiveOne}{\ensuremath{-}35.0}
\newcommand{\GrGoneFableFiveOneLo}{\ensuremath{-}46.0}
\newcommand{\GrGoneFableFiveOneHi}{\ensuremath{-}24.0}

\newcommand{\GrRepGoneFableFiveOne}{+0.0}
\newcommand{\GrRepGoneFableFiveOneLo}{\ensuremath{-}18.8}
\newcommand{\GrRepGoneFableFiveOneHi}{+18.8}

\newcommand{\GrRepGoneFableFiveOneSize}{Unresolved}

\newcommand{\GrRepGtwoFableFiveOne}{\ensuremath{-}7.5}
\newcommand{\GrRepGtwoFableFiveOneLo}{\ensuremath{-}26.2}
\newcommand{\GrRepGtwoFableFiveOneHi}{+11.2}

\newcommand{\GrRepGtwoFableFiveOneSize}{Unresolved}

\newcommand{\GrAnchor}{met}
\newcommand{\GrHeavyLoss}{none}
\newcommand{\GrRepTotal}{thirty}
\newcommand{\GrRepWithin}{fifteen}
\newcommand{\GrRepUnresolved}{fifteen}
\newcommand{\GrRepExceeds}{none}
\newcommand{\GrRepDirected}{none}
\newcommand{\GrRepMaxAbs}{\ensuremath{-}16.2}
\newcommand{\GrRepMaxAbsWhere}{G4 on Opus 5}

\newcommand{\KtwoResultsDepositFile}{studyK2\_results\_893ca358\_2026-09-05.zip}

\newcommand{\KtwoResultsDepositDate}{2026-09-06}

\newcommand{\KtwoN}{18,576}
\newcommand{\KtwoScored}{18,567}

\newcommand{\KtwoErrorsWord}{nine}
\newcommand{\KtwoErrorEndpoint}{Fable 5 (9)}
\newcommand{\KtwoErrorWorlds}{two}
\newcommand{\KtwoErrorFamilies}{one}
\newcommand{\KtwoErrorAttempts}{three}
\newcommand{\KtwoMainWorlds}{72}
\newcommand{\KtwoFamilies}{36}
\newcommand{\KtwoWorldsPerFamily}{two}

\newcommand{\KtwoBridgeWorldsWord}{two}
\newcommand{\KtwoRunsPerCell}{4}

\newcommand{\KtwoClosedEndpointsWord}{nine}
\newcommand{\KtwoMemoriesMin}{6}
\newcommand{\KtwoMemoriesMax}{12}
\newcommand{\KtwoStates}{two}
\newcommand{\KtwoForms}{four}
\newcommand{\KtwoSignatures}{two}
\newcommand{\KtwoCPBLevel}{0.95}
\newcommand{\KtwoCPBMarginal}{0.975}
\newcommand{\KtwoCPBTail}{0.0125}
\newcommand{\KtwoDelta}{10}
\newcommand{\KtwoContrasts}{three}
\newcommand{\KtwoHeadroomZero}{65}
\newcommand{\KtwoHeadroomNonzero}{7}
\newcommand{\KtwoHeadroomNonzeroDetail}{five at 0.25 and two at 0.5}
\newcommand{\KtwoRealisers}{Qwen3-32B and Mistral-Small-24B-Instruct-2501}
\newcommand{\KtwoRealisersN}{two}
\newcommand{\KtwoJudges}{phi-4, DeepSeek-R1-0528-Qwen3-8B, gemma-3-27b-it}
\newcommand{\KtwoJudgesN}{three}
\newcommand{\KtwoJudgeDraws}{eighteen}
\newcommand{\KtwoJudgeDrawsPer}{three}
\newcommand{\KtwoOpenNQwenSeventyTwoB}{2,064}
\newcommand{\KtwoOpenErrorsQwenSeventyTwoB}{0}
\newcommand{\KtwoOpenNLlamaEightB}{2,064}
\newcommand{\KtwoOpenErrorsLlamaEightB}{1}
\newcommand{\KtwoOpenNLlamaSeventyB}{2,064}
\newcommand{\KtwoOpenErrorsLlamaSeventyB}{1}
\newcommand{\KtwoOpenAnalysed}{three}
\newcommand{\KtwoOpenRegistered}{five}
\newcommand{\KtwoOpenAbsent}{two}
\newcommand{\KtwoOpenUnservableN}{one}
\newcommand{\KtwoOpenAllErrorN}{one}
\newcommand{\KtwoOpenAttempted}{8,256}
\newcommand{\KtwoOpenErrors}{2,066}
\newcommand{\KtwoNAll}{26,832}

\newcommand{\KtwoOpenUnservable}{mistralai/Mistral-Medium-3.5-128B}
\newcommand{\KtwoOpenAllError}{openai/gpt-oss-120b (2,064 rows)}
\newcommand{\KtwoBothCritOpus}{\ensuremath{-}41.3}
\newcommand{\KtwoBothCritOpusLo}{\ensuremath{-}48.4}
\newcommand{\KtwoBothCritOpusHi}{\ensuremath{-}32.9}
\newcommand{\KtwoBothCritOpusFamNeg}{34}
\newcommand{\KtwoBothCritOpusFamPos}{1}
\newcommand{\KtwoBothCritOpusFam}{36}

\newcommand{\KtwoBothCritOpusSignP}{$<10^{-4}$}

\newcommand{\KtwoCritIdOpus}{+84.0}
\newcommand{\KtwoCritIdOpusLo}{+77.1}
\newcommand{\KtwoCritIdOpusHi}{+88.6}

\newcommand{\KtwoIdNoneOpus}{+6.9}
\newcommand{\KtwoIdNoneOpusLo}{+1.5}
\newcommand{\KtwoIdNoneOpusHi}{+12.1}

\newcommand{\KtwoPmKCritOpus}{286/288}

\newcommand{\KtwoPmKBothILOpus}{167/288}

\newcommand{\KtwoSBOpus}{88.5}
\newcommand{\KtwoSBOpusLo}{84.3}
\newcommand{\KtwoSBOpusHi}{92.0}

\newcommand{\KtwoSBRefOpus}{13.7}
\newcommand{\KtwoCalMAEOpus}{18.9}
\newcommand{\KtwoCalSlopeOpus}{0.18}
\newcommand{\KtwoCalSlopeOpusLo}{0.17}
\newcommand{\KtwoCalSlopeOpusHi}{0.19}

\newcommand{\KtwoBothCritFableFiveOne}{\ensuremath{-}82.3}
\newcommand{\KtwoBothCritFableFiveOneLo}{\ensuremath{-}87.1}
\newcommand{\KtwoBothCritFableFiveOneHi}{\ensuremath{-}75.2}
\newcommand{\KtwoBothCritFableFiveOneFamNeg}{36}

\newcommand{\KtwoBothCritFableFiveOneFam}{36}

\newcommand{\KtwoBothCritFableFiveOneSignP}{$<10^{-4}$}

\newcommand{\KtwoCritIdFableFiveOne}{+72.6}
\newcommand{\KtwoCritIdFableFiveOneLo}{+64.4}
\newcommand{\KtwoCritIdFableFiveOneHi}{+78.6}

\newcommand{\KtwoIdNoneFableFiveOne}{+6.9}
\newcommand{\KtwoIdNoneFableFiveOneLo}{+1.0}
\newcommand{\KtwoIdNoneFableFiveOneHi}{+12.7}

\newcommand{\KtwoPmKCritFableFiveOne}{285/288}

\newcommand{\KtwoPmKBothILFableFiveOne}{48/288}

\newcommand{\KtwoCalMAEFableFiveOne}{24.0}
\newcommand{\KtwoCalSlopeFableFiveOne}{0.11}
\newcommand{\KtwoCalSlopeFableFiveOneLo}{0.10}
\newcommand{\KtwoCalSlopeFableFiveOneHi}{0.12}

\newcommand{\KtwoBothCritFableFive}{\ensuremath{-}55.4}
\newcommand{\KtwoBothCritFableFiveLo}{\ensuremath{-}62.4}
\newcommand{\KtwoBothCritFableFiveHi}{\ensuremath{-}46.8}
\newcommand{\KtwoBothCritFableFiveFamNeg}{33}
\newcommand{\KtwoBothCritFableFiveFamPos}{0}

\newcommand{\KtwoBothCritSonnet}{+8.7}
\newcommand{\KtwoBothCritSonnetLo}{+2.6}
\newcommand{\KtwoBothCritSonnetHi}{+14.4}

\newcommand{\KtwoSBSonnet}{86.5}
\newcommand{\KtwoSBSonnetLo}{82.0}
\newcommand{\KtwoSBSonnetHi}{90.2}

\newcommand{\KtwoBothCritHaiku}{+75.0}
\newcommand{\KtwoBothCritHaikuLo}{+67.3}
\newcommand{\KtwoBothCritHaikuHi}{+80.5}

\newcommand{\KtwoBothCritHaikuFamPos}{36}
\newcommand{\KtwoBothCritHaikuFam}{36}

\newcommand{\KtwoCritIdHaiku}{\ensuremath{-}61.1}
\newcommand{\KtwoCritIdHaikuLo}{\ensuremath{-}68.1}
\newcommand{\KtwoCritIdHaikuHi}{\ensuremath{-}52.3}

\newcommand{\KtwoBothCritGPTSol}{+1.4}
\newcommand{\KtwoBothCritGPTSolLo}{\ensuremath{-}2.1}
\newcommand{\KtwoBothCritGPTSolHi}{+4.8}

\newcommand{\KtwoIdNoneGPTSol}{+77.8}
\newcommand{\KtwoIdNoneGPTSolLo}{+70.3}
\newcommand{\KtwoIdNoneGPTSolHi}{+83.1}

\newcommand{\KtwoSBGPTSol}{74.3}
\newcommand{\KtwoSBGPTSolLo}{68.9}
\newcommand{\KtwoSBGPTSolHi}{79.3}

\newcommand{\KtwoBothCritGPTTerra}{+7.6}
\newcommand{\KtwoBothCritGPTTerraLo}{+2.6}
\newcommand{\KtwoBothCritGPTTerraHi}{+12.3}

\newcommand{\KtwoIdNoneGPTTerra}{+57.3}
\newcommand{\KtwoIdNoneGPTTerraLo}{+48.7}
\newcommand{\KtwoIdNoneGPTTerraHi}{+64.2}

\newcommand{\KtwoSBGPTTerra}{49.0}
\newcommand{\KtwoSBGPTTerraLo}{43.0}
\newcommand{\KtwoSBGPTTerraHi}{54.9}

\newcommand{\KtwoBothCritGPTLuna}{+6.9}
\newcommand{\KtwoBothCritGPTLunaLo}{+2.1}
\newcommand{\KtwoBothCritGPTLunaHi}{+11.5}

\newcommand{\KtwoIdNoneGPTLuna}{+74.0}
\newcommand{\KtwoIdNoneGPTLunaLo}{+66.2}
\newcommand{\KtwoIdNoneGPTLunaHi}{+79.6}

\newcommand{\KtwoBothCritGPTSixAstra}{+0.0}
\newcommand{\KtwoBothCritGPTSixAstraLo}{\ensuremath{-}2.2}
\newcommand{\KtwoBothCritGPTSixAstraHi}{+2.2}

\newcommand{\KtwoIdNoneGPTSixAstra}{+52.4}
\newcommand{\KtwoIdNoneGPTSixAstraLo}{+44.2}
\newcommand{\KtwoIdNoneGPTSixAstraHi}{+59.1}

\newcommand{\KtwoBothCritQwenSeventyTwoB}{+10.1}
\newcommand{\KtwoBothCritQwenSeventyTwoBLo}{+4.0}
\newcommand{\KtwoBothCritQwenSeventyTwoBHi}{+15.8}

\newcommand{\KtwoBothCritLlamaEightB}{+22.6}
\newcommand{\KtwoBothCritLlamaEightBLo}{+13.9}
\newcommand{\KtwoBothCritLlamaEightBHi}{+30.8}

\newcommand{\KtwoBothCritLlamaSeventyB}{+26.4}
\newcommand{\KtwoBothCritLlamaSeventyBLo}{+18.0}
\newcommand{\KtwoBothCritLlamaSeventyBHi}{+34.0}

\newcommand{\KtwoOtherMaxAbs}{8.7}
\newcommand{\KtwoOtherN}{five}
\newcommand{\KtwoOtherWithin}{two}
\newcommand{\KtwoOtherExcludesZero}{Sonnet 5, GPT-5.6 Terra, GPT-5.6 Luna}
\newcommand{\KtwoOtherIncludesZero}{GPT-5.6 Sol, GPT-6 Astra}
\newcommand{\KtwoOtherExcludesZeroN}{three}
\newcommand{\KtwoOtherWithinNames}{GPT-5.6 Sol, GPT-6 Astra}
\newcommand{\KtwoSBMajority}{8}
\newcommand{\KtwoSBEndpoints}{12}
\newcommand{\KtwoPrimary}{MET}

\newcommand{\KtwoRowsTotal}{twenty-four}
\newcommand{\KtwoUndeterminedTotalWord}{six}
\newcommand{\KtwoDirMet}{11}
\newcommand{\KtwoDirTotal}{12}
\newcommand{\KtwoDirTotalWord}{twelve}
\newcommand{\KtwoDirMetExcl}{7}
\newcommand{\KtwoDirTotalExcl}{8}
\newcommand{\KtwoWithinMet}{2}
\newcommand{\KtwoWithinTotal}{6}
\newcommand{\KtwoWithinTotalWord}{six}
\newcommand{\KtwoWithinMetExcl}{1}
\newcommand{\KtwoWithinTotalExcl}{4}

\newcommand{\KtwoDirMisses}{Haiku 4.5 crit-id (forecast positive, observed \ensuremath{-}61.1 [\ensuremath{-}68.1, \ensuremath{-}52.3])}
\newcommand{\KtwoDirMissesN}{one}

\newcommand{\KtwoBridgeBothCritOpusN}{8}

\newcommand{\KthreeLockRollupEightB}{f29c2251}
\newcommand{\KthreeLockRollupSeventyB}{51d1c8cf}
\newcommand{\KthreeManifestFiles}{274}
\newcommand{\KthreeManifestRollup}{ca3b35f513882887}

\newcommand{\KthreeOtsPresent}{yes}
\newcommand{\KthreeOsfFileId}{6a9e1d75ac5999d83b88565c}

\newcommand{\KthreeN}{11,520}
\newcommand{\KthreeNPerLadder}{5,760}

\newcommand{\KthreeWorlds}{72}
\newcommand{\KthreeFamilies}{36}
\newcommand{\KthreeBlocks}{288}
\newcommand{\KthreeDoses}{0, 2, 4, 6, 8}
\newcommand{\KthreeDosesN}{five}
\newcommand{\KthreeAttempts}{11,840}
\newcommand{\KthreeReviewRounds}{fifteen}
\newcommand{\KthreeReviewDoNotRun}{twelve}

\newcommand{\KthreeOtsBitcoin}{yes}
\newcommand{\KthreeAuditSEEightB}{50/78/80/78}

\newcommand{\KthreeAuditFCEightB}{59/78/80/78}

\newcommand{\KthreeAuditSESeventyB}{45/74/80/76}
\newcommand{\KthreeAuditSESeventyBBytes}{45}

\newcommand{\KthreeAuditFCSeventyB}{42/70/80/72}

\newcommand{\KthreePmBothILEightBZero}{47.6}

\newcommand{\KthreePmBothILEightBEight}{26.0}
\newcommand{\KthreePmOtherMinEightB}{19.8}
\newcommand{\KthreePmOtherMaxEightB}{28.1}
\newcommand{\KthreeBothCritEightBMeanZero}{+27.4}
\newcommand{\KthreeBothCritEightBMeanTwo}{+26.7}
\newcommand{\KthreeBothCritEightBMeanFour}{+19.6}
\newcommand{\KthreeBothCritEightBMeanSix}{+14.0}
\newcommand{\KthreeBothCritEightBMeanEight}{+6.0}
\newcommand{\KthreeBothCritEightBSlope}{\ensuremath{-}2.77}
\newcommand{\KthreeBothCritEightBSlopeLo}{\ensuremath{-}3.81}
\newcommand{\KthreeBothCritEightBSlopeHi}{\ensuremath{-}1.77}
\newcommand{\KthreeBothCritEightBChange}{\ensuremath{-}21.4}
\newcommand{\KthreeBothCritEightBChangeLo}{\ensuremath{-}29.5}
\newcommand{\KthreeBothCritEightBChangeHi}{\ensuremath{-}13.4}
\newcommand{\KthreeBothCritEightBComplete}{285}

\newcommand{\KthreeBothCritEightBMissMaxPct}{0.3}
\newcommand{\KthreeBothCritEightBInterPct}{1.0}
\newcommand{\KthreeBothCritEightBOutside}{3}
\newcommand{\KthreeBothCritEightBSensNegChange}{\ensuremath{-}21.9}
\newcommand{\KthreeBothCritEightBSensPosChange}{\ensuremath{-}21.2}
\newcommand{\KthreeBothCritEightBAtEight}{+5.6}
\newcommand{\KthreeBothCritEightBAtEightLo}{\ensuremath{-}2.1}
\newcommand{\KthreeBothCritEightBAtEightHi}{+13.0}

\newcommand{\KthreeCritIdEightBSlope}{+0.62}
\newcommand{\KthreeCritIdEightBSlopeLo}{\ensuremath{-}0.23}
\newcommand{\KthreeCritIdEightBSlopeHi}{+1.46}
\newcommand{\KthreeCritIdEightBChange}{+4.2}
\newcommand{\KthreeCritIdEightBChangeLo}{\ensuremath{-}2.8}
\newcommand{\KthreeCritIdEightBChangeHi}{+11.3}

\newcommand{\KthreeCritIdEightBInterPct}{1.4}
\newcommand{\KthreeCritIdEightBOutside}{4}
\newcommand{\KthreeCritIdEightBSensNegChange}{+3.5}
\newcommand{\KthreeCritIdEightBSensPosChange}{+4.9}

\newcommand{\KthreeIdNoneEightBSlope}{\ensuremath{-}0.31}
\newcommand{\KthreeIdNoneEightBSlopeLo}{\ensuremath{-}1.33}
\newcommand{\KthreeIdNoneEightBSlopeHi}{+0.70}
\newcommand{\KthreeIdNoneEightBChange}{\ensuremath{-}2.4}
\newcommand{\KthreeIdNoneEightBChangeLo}{\ensuremath{-}10.5}
\newcommand{\KthreeIdNoneEightBChangeHi}{+5.6}

\newcommand{\KthreeIdNoneEightBInterPct}{0.7}
\newcommand{\KthreeIdNoneEightBOutside}{2}
\newcommand{\KthreeIdNoneEightBSensNegChange}{\ensuremath{-}2.8}
\newcommand{\KthreeIdNoneEightBSensPosChange}{\ensuremath{-}2.1}

\newcommand{\KthreePmIdSeventyBZero}{46.5}

\newcommand{\KthreePmBothILSeventyBZero}{47.6}

\newcommand{\KthreePmIdSeventyBEight}{37.8}

\newcommand{\KthreePmBothILSeventyBEight}{37.8}
\newcommand{\KthreePmOtherMinSeventyB}{20.5}
\newcommand{\KthreePmOtherMaxSeventyB}{46.5}
\newcommand{\KthreeBothCritSeventyBMeanZero}{+21.9}
\newcommand{\KthreeBothCritSeventyBMeanTwo}{+25.0}
\newcommand{\KthreeBothCritSeventyBMeanFour}{+26.4}
\newcommand{\KthreeBothCritSeventyBMeanSix}{+20.1}
\newcommand{\KthreeBothCritSeventyBMeanEight}{+17.4}
\newcommand{\KthreeBothCritSeventyBSlope}{\ensuremath{-}0.69}
\newcommand{\KthreeBothCritSeventyBSlopeLo}{\ensuremath{-}1.96}
\newcommand{\KthreeBothCritSeventyBSlopeHi}{+0.66}
\newcommand{\KthreeBothCritSeventyBChange}{\ensuremath{-}4.5}
\newcommand{\KthreeBothCritSeventyBChangeLo}{\ensuremath{-}14.6}
\newcommand{\KthreeBothCritSeventyBChangeHi}{+5.9}
\newcommand{\KthreeBothCritSeventyBComplete}{288}

\newcommand{\KthreeBothCritSeventyBOutside}{0}

\newcommand{\KthreeBothCritSeventyBAtEight}{+17.4}
\newcommand{\KthreeBothCritSeventyBAtEightLo}{+10.0}
\newcommand{\KthreeBothCritSeventyBAtEightHi}{+24.2}
\newcommand{\KthreeBothCritSeventyBAtEightFamNeg}{0}
\newcommand{\KthreeBothCritSeventyBAtEightFamPos}{21}

\newcommand{\KthreeCritIdSeventyBMeanZero}{\ensuremath{-}20.8}
\newcommand{\KthreeCritIdSeventyBMeanTwo}{\ensuremath{-}18.8}
\newcommand{\KthreeCritIdSeventyBMeanFour}{\ensuremath{-}17.7}
\newcommand{\KthreeCritIdSeventyBMeanSix}{\ensuremath{-}15.3}
\newcommand{\KthreeCritIdSeventyBMeanEight}{\ensuremath{-}17.4}
\newcommand{\KthreeCritIdSeventyBSlope}{+0.52}
\newcommand{\KthreeCritIdSeventyBSlopeLo}{\ensuremath{-}0.59}
\newcommand{\KthreeCritIdSeventyBSlopeHi}{+1.56}
\newcommand{\KthreeCritIdSeventyBChange}{+3.5}
\newcommand{\KthreeCritIdSeventyBChangeLo}{\ensuremath{-}5.6}
\newcommand{\KthreeCritIdSeventyBChangeHi}{+12.2}

\newcommand{\KthreeIdNoneSeventyBMeanZero}{+21.2}
\newcommand{\KthreeIdNoneSeventyBMeanTwo}{+14.9}
\newcommand{\KthreeIdNoneSeventyBMeanFour}{+13.2}
\newcommand{\KthreeIdNoneSeventyBMeanSix}{+11.8}
\newcommand{\KthreeIdNoneSeventyBMeanEight}{+16.7}
\newcommand{\KthreeIdNoneSeventyBSlope}{\ensuremath{-}0.61}
\newcommand{\KthreeIdNoneSeventyBSlopeLo}{\ensuremath{-}1.70}
\newcommand{\KthreeIdNoneSeventyBSlopeHi}{+0.50}
\newcommand{\KthreeIdNoneSeventyBChange}{\ensuremath{-}4.5}
\newcommand{\KthreeIdNoneSeventyBChangeLo}{\ensuremath{-}12.8}
\newcommand{\KthreeIdNoneSeventyBChangeHi}{+3.8}

\newcommand{\KthreePrimaryStatusWord}{not met}
\newcommand{\KthreeSecondaryIdNone}{NOT-MET}
\newcommand{\KthreeSecondaryCritId}{NOT-MET}

\newcommand{\KthreeSensFlips}{none}

\newcommand{\KfourLockRollupLone}{ee342cd6}
\newcommand{\KfourLockRollupLtwo}{976516cb}
\newcommand{\KfourLockRollupLthree}{1d87ef96}
\newcommand{\KfourManifestFiles}{280}
\newcommand{\KfourManifestRollup}{ea8d755927366b12}

\newcommand{\KfourOtsPresent}{yes}
\newcommand{\KfourOsfFileId}{6a9f52ca6d2be1e61e4861f4}

\newcommand{\KfourN}{17,280}
\newcommand{\KfourNPerLadder}{5,760}
\newcommand{\KfourWorlds}{72}
\newcommand{\KfourFamilies}{36}
\newcommand{\KfourBlocks}{288}

\newcommand{\KfourErrorsByDoseLone}{1 / 0 / 0 / 2 / 1}

\newcommand{\KfourErrorsLtwo}{16}
\newcommand{\KfourErrorsMaxDoseLtwo}{8}
\newcommand{\KfourErrorsByDoseLtwo}{2 / 0 / 4 / 8 / 2}
\newcommand{\KfourPerGainNLtwo}{1,152}

\newcommand{\KfourErrorsByDoseLthree}{2 / 2 / 0 / 0 / 0}

\newcommand{\KfourDosesLone}{0, 2, 4, 6, 8}

\newcommand{\KfourLastLone}{8}
\newcommand{\KfourDosesLtwo}{0, 4, 8, 12, 16}

\newcommand{\KfourDosesLthree}{0, 2, 4, 6, 8}

\newcommand{\KfourAttempts}{17,760}
\newcommand{\KfourReviewRoundsCodex}{seven}
\newcommand{\KfourReviewRoundsClaude}{two}

\newcommand{\KfourReviewDoNotRun}{six}

\newcommand{\KfourOtsBitcoin}{yes}
\newcommand{\KfourAuditSELone}{59/78/80/78}

\newcommand{\KfourAuditFCLone}{56/78/80/79}

\newcommand{\KfourAuditSELtwo}{59/78/80/78}

\newcommand{\KfourAuditFCLtwo}{53/79/80/79}

\newcommand{\KfourAuditSELthree}{43/71/80/72}

\newcommand{\KfourAuditFCLthree}{43/69/80/72}

\newcommand{\KfourBothCritLoneMeans}{+25.2 / +24.1 / +22.7 / +15.7 / +7.7}
\newcommand{\KfourBothCritLoneSlope}{\ensuremath{-}2.17}
\newcommand{\KfourBothCritLoneSlopeLo}{\ensuremath{-}3.36}
\newcommand{\KfourBothCritLoneSlopeHi}{\ensuremath{-}1.01}
\newcommand{\KfourBothCritLoneChange}{\ensuremath{-}17.5}
\newcommand{\KfourBothCritLoneChangeLo}{\ensuremath{-}26.7}
\newcommand{\KfourBothCritLoneChangeHi}{\ensuremath{-}8.1}
\newcommand{\KfourBothCritLoneComplete}{286}

\newcommand{\KfourBothCritLoneOutside}{2}
\newcommand{\KfourBothCritLoneSensNegChange}{\ensuremath{-}18.1}
\newcommand{\KfourBothCritLoneSensPosChange}{\ensuremath{-}16.7}
\newcommand{\KfourBothCritLoneAtLast}{+7.7}
\newcommand{\KfourBothCritLoneAtLastLo}{+0.3}
\newcommand{\KfourBothCritLoneAtLastHi}{+14.8}

\newcommand{\KfourPmBothILLtwoZero}{45.8}

\newcommand{\KfourPmBothILLtwoSixteen}{6.2}
\newcommand{\KfourPmOtherMinLtwo}{3.8}
\newcommand{\KfourPmOtherMaxLtwo}{26.7}

\newcommand{\KfourBothCritLtwoMeanSixteen}{+2.1}
\newcommand{\KfourBothCritLtwoMeans}{+25.0 / +20.4 / +5.4 / +3.6 / +2.1}
\newcommand{\KfourBothCritLtwoSlope}{\ensuremath{-}1.56}
\newcommand{\KfourBothCritLtwoSlopeLo}{\ensuremath{-}2.09}
\newcommand{\KfourBothCritLtwoSlopeHi}{\ensuremath{-}1.07}
\newcommand{\KfourBothCritLtwoChange}{\ensuremath{-}22.9}
\newcommand{\KfourBothCritLtwoChangeLo}{\ensuremath{-}31.1}
\newcommand{\KfourBothCritLtwoChangeHi}{\ensuremath{-}15.2}
\newcommand{\KfourBothCritLtwoComplete}{280}

\newcommand{\KfourBothCritLtwoMissMaxPct}{1.4}

\newcommand{\KfourBothCritLtwoOutside}{8}
\newcommand{\KfourBothCritLtwoSensNegChange}{\ensuremath{-}23.6}
\newcommand{\KfourBothCritLtwoSensPosChange}{\ensuremath{-}22.2}
\newcommand{\KfourBothCritLtwoAtLast}{+2.1}
\newcommand{\KfourBothCritLtwoAtLastLo}{\ensuremath{-}0.9}
\newcommand{\KfourBothCritLtwoAtLastHi}{+4.9}

\newcommand{\KfourBothCritLthreeMeans}{+22.4 / +26.2 / +25.2 / +18.2 / +17.1}
\newcommand{\KfourBothCritLthreeSlope}{\ensuremath{-}0.93}
\newcommand{\KfourBothCritLthreeSlopeLo}{\ensuremath{-}2.26}
\newcommand{\KfourBothCritLthreeSlopeHi}{+0.44}
\newcommand{\KfourBothCritLthreeChange}{\ensuremath{-}5.2}
\newcommand{\KfourBothCritLthreeChangeLo}{\ensuremath{-}15.6}
\newcommand{\KfourBothCritLthreeChangeHi}{+5.2}
\newcommand{\KfourBothCritLthreeComplete}{286}

\newcommand{\KfourBothCritLthreeOutside}{2}
\newcommand{\KfourBothCritLthreeSensNegChange}{\ensuremath{-}5.9}
\newcommand{\KfourBothCritLthreeSensPosChange}{\ensuremath{-}5.2}
\newcommand{\KfourBothCritLthreeAtLast}{+17.0}
\newcommand{\KfourBothCritLthreeAtLastLo}{+9.9}
\newcommand{\KfourBothCritLthreeAtLastHi}{+23.5}
\newcommand{\KfourBothCritLthreeAtLastFamNeg}{1}
\newcommand{\KfourBothCritLthreeAtLastFamPos}{22}

\newcommand{\KfourOneStatusWord}{not met}

\newcommand{\KfourTwoStatusWord}{met}

\newcommand{\KfourThreeStatusWord}{not met}

\newcommand{\KfourSensFlips}{none}
\newcommand{\KfourMarginErrLoLone}{2.2}
\newcommand{\KfourMarginErrHiLone}{3.6}
\newcommand{\KfourMarginNominalLone}{2.5}
\newcommand{\KfourPowerLoLone}{0.85}
\newcommand{\KfourPowerHiLone}{0.93}
\newcommand{\KfourMarginErrLoLtwo}{1.5}
\newcommand{\KfourMarginErrHiLtwo}{2.1}
\newcommand{\KfourMarginNominalLtwo}{1.25}
\newcommand{\KfourPowerLoLtwo}{0.76}
\newcommand{\KfourPowerHiLtwo}{0.89}
\newcommand{\KfourPowerNearKthreeLoLone}{0.62}
\newcommand{\KfourPowerNearKthreeHiLone}{0.78}

\newcommand{\KfourQualifications}{twelve}
\newcommand{\KfourQualificationsHalted}{four}
\newcommand{\KfourQualificationsStale}{one}
\newcommand{\KfourReviewRoundsMain}{eight}

\newcommand{\KfiveLockRollupLone}{a903006b}
\newcommand{\KfiveLockRollupLtwo}{31c4db17}
\newcommand{\KfiveIdentityCompared}{53,920}
\newcommand{\KfiveInvocationOne}{1163ee8a}
\newcommand{\KfiveInvocationTwo}{20b4251a}
\newcommand{\KfiveF}{337}

\newcommand{\KfivePowerWilson}{0.93}
\newcommand{\KfiveAnchors}{eight}
\newcommand{\KfiveOcConfigs}{142}
\newcommand{\KfiveOcReplicates}{26,200}
\newcommand{\KfiveMarginErrLo}{1.8}
\newcommand{\KfiveMarginErrHi}{3.2}
\newcommand{\KfiveMarginNominal}{2.5}
\newcommand{\KfiveBoundaryConfigs}{eight}
\newcommand{\KfivePowerAnchorLo}{0.98}
\newcommand{\KfivePowerAnchorHi}{1.00}
\newcommand{\KfiveInteriorNullMaxPct}{4.0}
\newcommand{\KfiveInteriorNullConfigs}{65}

\newcommand{\KfivePrestreamFiles}{144}
\newcommand{\KfivePrestreamRollup}{510a117b5b85b824}
\newcommand{\KfivePrestreamOsfFile}{studyK5\_prestream\_510a117b\_2026-09-16.zip}

\newcommand{\KfiveManifestFiles}{442}
\newcommand{\KfiveManifestRollup}{5a386f0b9dc47b88}

\newcommand{\KfiveOsfFileId}{6aaacb7677a51a63927e93a2}
\newcommand{\KfiveOtsPresent}{yes}

\newcommand{\KfiveOsfNode}{c48ns}
\newcommand{\KfiveOtsBitcoin}{yes}
\newcommand{\KfivePrestreamOtsBitcoin}{yes}
\newcommand{\KfiveN}{107,840}
\newcommand{\KfiveNPerLadder}{53,920}
\newcommand{\KfiveWorlds}{674}
\newcommand{\KfiveFamilies}{337}
\newcommand{\KfiveBlocks}{2,696}
\newcommand{\KfivePerGainN}{10,784}

\newcommand{\KfiveErrorsLone}{43}
\newcommand{\KfiveErrorsMaxDoseLone}{22}
\newcommand{\KfiveErrorsByDoseLone}{8 / 7 / 2 / 4 / 22}

\newcommand{\KfiveErrorsLtwo}{18}

\newcommand{\KfiveErrorsByDoseLtwo}{11 / 4 / 2 / 0 / 1}

\newcommand{\KfiveAdapterLtwo}{FlorianJK/Meta-Llama-3.1-8B-SecAlign-pp}
\newcommand{\KfiveDoses}{0, 2, 4, 6, 8}
\newcommand{\KfiveCohortKtwoChangeLone}{\ensuremath{-}23.7}

\newcommand{\KfiveCohortKtwoChangeLtwo}{\ensuremath{-}13.2}

\newcommand{\KfiveCohortKtwoFamilies}{45}
\newcommand{\KfiveCohortKfiveChangeLone}{\ensuremath{-}15.5}

\newcommand{\KfiveCohortKfiveChangeLtwo}{\ensuremath{-}11.6}

\newcommand{\KfiveCohortKfiveFamilies}{292}
\newcommand{\KfiveCohortKtwoExposed}{36}
\newcommand{\KfiveCohortKtwoUnused}{9}
\newcommand{\KfiveReleasedScaleLone}{0.125}
\newcommand{\KfiveLoraAlphaLone}{8}
\newcommand{\KfiveLoraRLone}{64}
\newcommand{\KfiveReleasedScaleLtwo}{0.25}
\newcommand{\KfiveLoraAlphaLtwo}{8}
\newcommand{\KfiveLoraRLtwo}{32}

\newcommand{\KfiveAttempts}{108,160}
\newcommand{\KfiveHaltedBatchesLone}{481}

\newcommand{\KfiveHaltedBatchesLtwo}{719}

\newcommand{\KfiveReviewRoundsCodex}{twenty-four}
\newcommand{\KfiveReviewRoundsClaude}{one}
\newcommand{\KfiveReviewRounds}{twenty-five}

\newcommand{\KfiveReviewDoNotRun}{none}

\newcommand{\KfiveClaudeRound}{eight}
\newcommand{\KfiveLastRoundA}{four}
\newcommand{\KfiveLastRoundB}{eight}
\newcommand{\KfivePostexecVerdict}{stands with listed corrections}
\newcommand{\KfivePostexecA}{no}
\newcommand{\KfivePostexecB}{four}
\newcommand{\KfivePostexecC}{two}
\newcommand{\KtwoMainRunDate}{2026-09-05}
\newcommand{\KfiveSpend}{16.25}
\newcommand{\KfiveSpendAll}{47.54}
\newcommand{\KfiveAuditSELone}{54/69/80/80}

\newcommand{\KfiveAuditFCLone}{56/72/80/80}

\newcommand{\KfiveAuditSELtwo}{60/77/80/79}

\newcommand{\KfiveAuditFCLtwo}{61/76/80/79}

\newcommand{\KfivePmBothILLoneZero}{36.5}

\newcommand{\KfivePmBothILLoneEight}{19.7}
\newcommand{\KfivePmOtherMinLone}{14.3}
\newcommand{\KfivePmOtherMaxLone}{20.3}

\newcommand{\KfiveBothCritLoneMeans}{+21.7 / +20.7 / +15.0 / +11.0 / +5.1}
\newcommand{\KfiveBothCritLoneSlope}{\ensuremath{-}2.14}
\newcommand{\KfiveBothCritLoneSlopeLo}{\ensuremath{-}2.50}
\newcommand{\KfiveBothCritLoneSlopeHi}{\ensuremath{-}1.80}
\newcommand{\KfiveBothCritLoneChange}{\ensuremath{-}16.6}
\newcommand{\KfiveBothCritLoneChangeLo}{\ensuremath{-}19.4}
\newcommand{\KfiveBothCritLoneChangeHi}{\ensuremath{-}13.8}
\newcommand{\KfiveBothCritLoneComplete}{2,682}
\newcommand{\KfiveBothCritLoneLevel}{0.95}

\newcommand{\KfiveBothCritLoneOutside}{14}
\newcommand{\KfiveBothCritLoneSensNegChange}{\ensuremath{-}17.0}
\newcommand{\KfiveBothCritLoneSensPosChange}{\ensuremath{-}16.2}
\newcommand{\KfiveBothCritLoneAtLast}{+5.2}
\newcommand{\KfiveBothCritLoneAtLastLo}{+2.9}
\newcommand{\KfiveBothCritLoneAtLastHi}{+7.4}
\newcommand{\KfiveBothCritLoneAtLastFamNeg}{55}
\newcommand{\KfiveBothCritLoneAtLastFamPos}{126}
\newcommand{\KfiveBothCritLoneAtZero}{+21.8}

\newcommand{\KfiveCritIdLoneChange}{+1.5}
\newcommand{\KfiveCritIdLoneChangeLo}{\ensuremath{-}1.3}
\newcommand{\KfiveCritIdLoneChangeHi}{+4.2}

\newcommand{\KfiveCritIdLoneLevel}{0.975}

\newcommand{\KfiveIdNoneLoneChange}{\ensuremath{-}0.1}
\newcommand{\KfiveIdNoneLoneChangeLo}{\ensuremath{-}2.8}
\newcommand{\KfiveIdNoneLoneChangeHi}{+2.5}

\newcommand{\KfiveIdNoneLoneMissMaxPct}{0.5}

\newcommand{\KfivePmBothILLtwoZero}{36.5}

\newcommand{\KfivePmBothILLtwoEight}{21.1}

\newcommand{\KfiveBothCritLtwoMeans}{+21.5 / +20.4 / +17.0 / +14.6 / +9.7}
\newcommand{\KfiveBothCritLtwoSlope}{\ensuremath{-}1.47}
\newcommand{\KfiveBothCritLtwoSlopeLo}{\ensuremath{-}1.79}
\newcommand{\KfiveBothCritLtwoSlopeHi}{\ensuremath{-}1.16}
\newcommand{\KfiveBothCritLtwoChange}{\ensuremath{-}11.8}
\newcommand{\KfiveBothCritLtwoChangeLo}{\ensuremath{-}14.3}
\newcommand{\KfiveBothCritLtwoChangeHi}{\ensuremath{-}9.3}
\newcommand{\KfiveBothCritLtwoComplete}{2,689}
\newcommand{\KfiveBothCritLtwoLevel}{0.95}

\newcommand{\KfiveBothCritLtwoOutside}{7}
\newcommand{\KfiveBothCritLtwoSensNegChange}{\ensuremath{-}11.9}
\newcommand{\KfiveBothCritLtwoSensPosChange}{\ensuremath{-}11.6}
\newcommand{\KfiveBothCritLtwoAtLast}{+9.8}
\newcommand{\KfiveBothCritLtwoAtLastLo}{+7.7}
\newcommand{\KfiveBothCritLtwoAtLastHi}{+11.8}

\newcommand{\KfiveCritIdLtwoChange}{+2.1}
\newcommand{\KfiveCritIdLtwoChangeLo}{\ensuremath{-}0.2}
\newcommand{\KfiveCritIdLtwoChangeHi}{+4.5}

\newcommand{\KfiveIdNoneLtwoChange}{\ensuremath{-}1.1}
\newcommand{\KfiveIdNoneLtwoChangeLo}{\ensuremath{-}3.5}
\newcommand{\KfiveIdNoneLtwoChangeHi}{+1.2}

\newcommand{\KfiveOneStatusWord}{met}

\newcommand{\KfiveTwoStatusWord}{not met}

\newcommand{\KfiveSensFlips}{none}
\newcommand{\KfiveThreeTheta}{+4.7}
\newcommand{\KfiveThreeThetaLo}{+2.3}
\newcommand{\KfiveThreeThetaHi}{+7.2}
\newcommand{\KfiveThreeCommonBlocks}{2,678}
\newcommand{\KfiveThreeJointMissPct}{0.7}
\newcommand{\KfiveThreeDeltaLone}{\ensuremath{-}16.5}
\newcommand{\KfiveThreeDeltaLtwo}{\ensuremath{-}11.8}
\newcommand{\KfiveThreeSentence}{On these worlds and orders, the second adapter's change differed from Meta-SecAlign's by +4.7 [+2.3, +7.2] points. This describes two fixed execution paths; it licenses no superiority, equivalence or 'significant difference' claim, and nothing follows from one ladder's status differing from the other's.}

\newcommand{\InitialStudies}{six}

\newcommand{\TotalEpisodes}{179,352}
\newcommand{\TotalStudies}{sixteen}

\newcommand{\KfourCallsPerLadder}{5,920}
\newcommand{\KfiveCallsPerLadder}{54,080}

\newcommand{\lit}[1]{#1}   

\title{Plan Pointers and Record-Directive Form in Budgeted Verification\\of Inherited Agent Memory}
\author{Kazuki Nakayashiki\\ \small Glasp}
\date{}

\begin{document}
\maketitle

\begin{abstract}
A model that inherits one-line memories may pull one archived source record before acting; a directive in the store can steer that pull: a pointer, a criterion, or both. Across \TotalStudies{} registered studies (\TotalEpisodes{} attempts) we measured where the request goes under each form. On six direct-provider models a length-matched criterion exceeded a bare id by \DEone{} points [\DEoneLo, \DEoneHi] (Study D); on a nine-model OpenRouter-served panel the contrast failed its registered superiority rule (Study E). Appending the id cancelled the criterion on three Claude models (Opus~5: \FxPmKCritOpus{} to \FxPmKBothILOpus{}; Study F-x); six byte-matched edits gave each exact string its own effect (Study G), and at eighty runs per cell \GrRepWithin{} of \GrRepTotal{} replication contrasts were within the margin, \GrRepUnresolved{} unresolved and \GrRepExceeds{} beyond (Study G$'$). A ratification line (\JJtwoOpus{} points on Opus~5) and a two-credit budget restored the target on all three (Study J); across five criterion strings the suffix's cancellation held for \HtwoCancelNOpus{} on Opus~5 and \HtwoCancelNFableFiveOne{} on Fable~5.1 (Study H2); in a second store every model followed the criterion (Study H1). In a decision task the criterion moved the choice toward the current record (\IYcSupersededCritOpus{} points, Opus~5) and away from it on Fable~5.1 (Study I). A one-character plan pointer's effect (\BDeltaPlan{} points; Study B, after a corrected first report) recurred under a prospectively registered re-run (\BrDeltaPlan{} points; Study B$'$). On \KtwoMainWorlds{} generated worlds in \KtwoFamilies{} domain families (population stated in \S\ref{sec:results-k2}) the \KtwoSignatures{} registered signatures held (composite \KtwoBothCritOpus{} points below criterion on Opus~5, \KtwoBothCritFableFiveOne{} on Fable~5.1; Study K2); Haiku~4.5 reversed, Fable~5 followed the primary's sign, and Sonnet~5, the GPT-5.6 endpoints and GPT-6 Astra lay within \KtwoOtherMaxAbs{} points of zero. On those worlds, with a defensive adapter (Meta-SecAlign) applied at five gains, the registered criterion for a gain-dependent change of the composite $-$ criterion contrast was not met on Llama~3.3~70B (slope \KthreeBothCritSeventyBSlope{} points per unit gain [\KthreeBothCritSeventyBSlopeLo, \KthreeBothCritSeventyBSlopeHi]); on Llama~3.1~8B the contrast went from \KthreeBothCritEightBMeanZero{} to \KthreeBothCritEightBMeanEight{} points across the gains (descriptive; Study K3). Re-executed under a registered attenuation rule, the 8B ladder's fall (\KfourBothCritLoneMeans{} points across the gains; change \KfourBothCritLoneChange{} [\KfourBothCritLoneChangeLo, \KfourBothCritLoneChangeHi]) did not meet the criterion --- not a statement that the contrast was unchanged (\S\ref{sec:results-k4}); at gain 16 the same adapter's contrast was \KfourBothCritLtwoMeanSixteen{} points (descriptive), and on the 70B ladder the criterion was again not met, the two executions compared as statements (Study K4). At registered power on \KfiveFamilies{} generated families the 8B ladder's attenuation satisfied the registered criterion (change \KfiveBothCritLoneChange{} [\KfiveBothCritLoneChangeLo, \KfiveBothCritLoneChangeHi]; Study K4's status stands as its own result) while a second SecAlign++ adapter's did not (\KfiveBothCritLtwoChange{} [\KfiveBothCritLtwoChangeLo, \KfiveBothCritLtwoChangeHi]; not a statement that the contrast was unchanged); their difference (\KfiveThreeTheta{} points [\KfiveThreeThetaLo, \KfiveThreeThetaHi]) is descriptive and nothing follows from the two statuses differing (Study K5, \S\ref{sec:results-k5}). All results are descriptive effects of exact edits on fixed panels with registered intervals and no mechanism claim.

\end{abstract}

\section{Introduction}\label{sec:intro}

An agent that inherits a memory store it cannot fully re-derive follows a few provenance links before it
acts. Which few is an allocation made at inference time, and earlier work measured it on one instrument:
under a verification budget, agents concentrated their checks on the memories that back the plan they already
held \citep{priorwork2026verification}, and when the unchecked memory stated a constraint that had since
been withdrawn, the decision followed the stale memory in roughly three episodes of four; a same-budget policy
that guaranteed inspection of the critical record removed most of that error, but it used experimenter
knowledge of which record was critical \citep{priorwork2026stale}. The first paper identified the assigned plan as a causal driver of that allocation without separating what in the plan drives it; the second appended visible freshness dates to memories and measured the allocation response, and left the mechanism of native allocation unidentified. Neither manipulated an explicit verification-priority field in the store --- a candidate substitute for the oracle's knowledge whose effect on decisions those papers did not test.

This paper is about the \emph{form} of such a field. A memory system that wants to steer verification can
write a pointer (a record id) or a reason (a criterion that picks the record out), or both. To a system designer they encode the same intended target; behaviourally, this paper shows, they are not interchangeable. Across \TotalStudies{} registered studies on one instrument lineage --- a store of one-line memories with one verification request and one target record, later carried into a second store, into a decision and onto generated worlds; \TotalEpisodes{} attempts in the retained confirmatory runs on up to fifteen models per study --- we measured how the request moves under each form. The studies came in two phases: the first six were outcome-sequential, each designed after the previous one's locked result; the six follow-ups were conceived together after the six-study manuscript closed, partly developed in parallel, and then individually frozen, deposited and run one at a time; four further studies then left the hand-written worlds: K2 for \KtwoMainWorlds{} generated ones, K3 for a defensive adapter applied at five gains on those worlds' open ladders, K4 for the same ladders re-executed under a registered attenuation rule and extended beyond the adapter's release, and K5 for that rule at registered power on \KfiveFamilies{} families with a second adapter (\S\ref{sec:studies}). Neither phase was jointly prospective. Two kinds of pointer appear and must be kept
apart: Study B's \texttt{ACTIVE\_PLAN\_ID} value, which names which of two described plans is active and
never names a record, and the \texttt{VERIFY\_PRIORITY} id of Studies D--G, which names the target record.

\paragraph{Findings.} A one-character plan pointer moved the request by \BDeltaPlan{} points [\BDeltaPlanLo, \BDeltaPlanHi], a share of \BShare{} of what a plan sentence moved it (Study B; the registered estimand after a correction of the study's first repository report, \S\ref{sec:results-b}), and a prospectively registered re-run under Study B's rule returned the same verdict (\BrDeltaPlan{} [\BrDeltaPlanLo, \BrDeltaPlanHi]; Study B$'$). A record-naming directive's form decided whether it was honoured on six direct-provider models: criterion minus bare id was \DEone{} points [\DEoneLo, \DEoneHi], with the GPT-5.6 models obeying both forms and Opus~5 naming the target in \DPmKOracleOpus{} episodes under the id against \DPmKPolicyOpus{} under the criterion (Study D); on a disjoint nine-model OpenRouter-served panel (seven open-weight, two closed-weight) the contrast was \EEone{} [\EEoneLo, \EEoneHi] and did not meet the registered superiority rule (Study E). Crossing the forms on fifteen models, the composite's point estimate exceeded the id's in each of four pre-named groups and fell below the criterion's on Opus~5 (\FPmKCritOpus{} to \FPmKBothILOpus{}; Study F), a fall that held on the Opus~5 re-run and appeared on two further named endpoints, Fable~5 and Fable~5.1 (Study F-x) and taken apart by six byte-controlled edits: on Opus~5 a referent-free suffix left the criterion within the margin ($G_2 = \GGtwoOpus{}$ [\GGtwoOpusLo, \GGtwoOpusHi]) while ` (memory\_73)' and ` (memory\_44)' each cancelled it ($G_1$, $G_3$), and on Fable~5 each of the three appended strings cancelled it ($G_1$, $G_2$, $G_3$) (Study G; at eighty runs per cell \GrRepWithin{} of \GrRepTotal{} replication contrasts were within the margin, \GrRepUnresolved{} unresolved and \GrRepExceeds{} beyond it, Study G$'$). Three exact bundled edits then probed the cancellation on the three Claude endpoints: a ratification line restored the target on all three (\JJtwoOpus{} [\JJtwoOpusLo, \JJtwoOpusHi] on Opus~5), so did a budget of two credits scored as any credit (\JJsixOpus{} [\JJsixOpusLo, \JJsixOpusHi]; the first-credit contrast unresolved), and an explicit target field did so on Fable~5.1 alone and on Fable~5 after the criterion (Study J). Across five exact criterion strings the suffix's cancellation held for \HtwoCancelNOpus{} on Opus~5, \HtwoCancelNFableFive{} on Fable~5 and \HtwoCancelNFableFiveOne{} on Fable~5.1, and two rewordings were themselves not followed by one model each (Study H2; five strings, no invariance claim in either direction). In a second store every model followed the criterion and the criterion-minus-id gap was smaller on Fable~5, Sonnet~5 and Haiku~4.5 (Study H1; instrument instances, not a store property). Continued into a decision, the criterion changed the choice toward the current record on Opus~5 (superseded world \IYcSupersededCritOpus{} [\IYcSupersededCritOpusLo, \IYcSupersededCritOpusHi]) and Sonnet~5, any directive changed the GPT-5.6 models' valid-world decision, and on Fable~5.1 the same edits moved it away from the record --- treatment contrasts on two endpoints, not a mediation result (Study I). Finally, on \KtwoMainWorlds{} model-generated worlds in \KtwoFamilies{} domain families the facts recorded on Opus~5 and Fable~5.1 recurred under their \KtwoSignatures{} registered signatures --- the criterion followed, the composite \KtwoBothCritOpus{} points below it on Opus~5 and \KtwoBothCritFableFiveOne{} on Fable~5.1 --- while Haiku~4.5's contrast had the opposite sign, Fable~5's the same sign as the primary endpoints', and the estimates of Sonnet~5, the GPT-5.6 endpoints and GPT-6 Astra lay within \KtwoOtherMaxAbs{} points of zero, each a per-endpoint description, and a self-hosted open-weight stratum is reported by its rows (Study K2; the population of generated worlds is stated in \S\ref{sec:results-k2}). On Study K2's worlds, with a defensive adapter (Meta-SecAlign; \citealp{chen2025metasecalign}) applied at five gains, the registered criterion for a gain-dependent change of the composite $-$ criterion contrast was not met on Llama~3.3~70B (slope \KthreeBothCritSeventyBSlope{} points per unit gain [\KthreeBothCritSeventyBSlopeLo, \KthreeBothCritSeventyBSlopeHi]); on Llama~3.1~8B the contrast went from \KthreeBothCritEightBMeanZero{} to \KthreeBothCritEightBMeanEight{} points across the gains (descriptive; Study K3).

\paragraph{What is claimed.} Every result is descriptive (Studies B, D, E, \Br{}, K2, K3, K4 and K5 also carry registered verdict rules, reported as met or not met): rates and differences with registered intervals on fixed panels (block bootstrap for Studies B--F-x and \Br{}; Wilson cells and Newcombe paired score contrasts in the registered nonnegative-$\varphi$ variant for Studies G, I, H1, J, H2 and \Gr{}, with the bootstrap as a sensitivity and as the registered interval for the transport and replication contrasts of H1, \Gr{} and \Br{}), registered before each run, except that Study B's headline is computed by a post-freeze corrected analyzer and Study E's pairing rule under lost episodes was decided by its frozen analyzer rather than its prose registration (\S\ref{sec:integrity}); Studies B, D, E, \Br{}, K2, K3, K4 and K5 carry registered verdict rules and the others none. We do not claim that any form improves decisions, that the criterion's effect extends beyond a
criterion that identifies the target in one step, that provider differences in compliance are new, or that the
composite's effect is separable from its added length. We claim that on the growth-store instrument (one scenario, one request; the composite adds bytes as well as content) the form of a directive, holding its referent fixed, changed where the verification request went by more than the ten-point margin in \FExceedsCountReferent{} of Study F's twelve registered referent-held contrast--group entries (C1--C3; overlapping groups, descriptive, no joint test), that the direction of the change was model-specific, and that on three named Claude models the composite of criterion and id produced a lower target rate than the criterion alone; Study G reports the effect of each of six exact edits on each of four models as a separate marginal quantity (\S\ref{sec:results-g}). Studies H1, I and J then test which parts of that pattern transport to a second store, into a decision and under three edits, Study H2 how they read under four authored rewordings, and Studies \Gr{} and \Br{} whether two locked estimates change by less than the margin when re-run; each section states its own claim.

\paragraph{A second contribution is procedural.} Study F's first run was contaminated: a request field naming a
web-search plugin as \emph{disabled} activated it under a default-enabled workspace policy. Cost accounting
raised the first alarm, inspection of the stored rationales confirmed injected content, and a per-cell
prompt-token invariance gate --- exact for detecting a deviation from an established per-cell count in a
fixed design --- now precedes every run and would have refused that run's smoke. Study B's first repository report (non-archival) was inverted by a correction. Both are reported in full (\S\ref{sec:integrity}).

\section{Related work}\label{sec:related}

\paragraph{Budgeted verification and retrieval over agent memory.} Allocating a scarce
inspection or retrieval budget over an agent's own memory is an active area we treat as occupied:
TierMem casts retrieval as an inference-time evidence-allocation problem and escalates to an immutable
raw-log store only when summary evidence is insufficient \citep{zhu2026tiermem}; decision-aware memory
cards rank retrieved memories by their expected effect on the next action rather than by similarity
\citep{guan2026cicl}; value-of-information budget control decides which search action receives the
next budget unit \citep{fang2026budget}. These systems do mechanically what our directives ask the model
to do. In our searches (dated, non-exhaustive; the search ledgers will be released) we did not find a study that
steers that allocation with a prompt-level signal inside the inherited store and ablates the signal's \emph{form} with the referent held fixed. The closest prompt-to-verification precedent is the Memory Commitment Benchmark \citep{li2026mcb}, which gives an agent an explicit verify option, reports a say-versus-do gap, and shows that few-shot and policy prompts raise the verification rate; that a prompt raises verification is therefore occupied, and we claim only the fixed-referent form contrast under a scarce budget over the agent's own inherited memories. Budgeted probing of an environment before acting \citep{song2026envprobe} and certified commitment under memory uncertainty \citep{akewar2026safecommit} allocate probes by a scoring policy or a certificate rather than by a directive written into the store.

\paragraph{Directive form and compliance.} That the wording of an obligation moves compliance by
tens of points is established: compliance spans \lit{46} percentage points across models and models differ
in which pressure breaks them, some treating a regulation as binding however it is worded and others
failing under non-command phrasing \citep{okamoto2026framing}; instruction-hierarchy compliance
ranges from \lit{98.2}\% to \lit{20.5}\% across \lit{37} models \citep{mccauley2026ih}. Machine-emitted, legitimate,
in-band directives have been refused by some models and obeyed by others with the same all-or-nothing
shape we report (\lit{0/40} against \lit{20/20} by prompt) \citep{munirathinam2026recusal}. Provider-family
differences in compliance are therefore not ours to claim; the length-matched contrast between an opaque
reference and a criterion-stating version of the same directive, and the composite of the two, are the
ablations our searches did not find. A natural-language instruction is already known to redirect epistemic search, from \lit{42}\% to \lit{56}\% rule discovery on average \citep{jhaveri2026falsify}; because that task, endpoint and panel differ from ours, we use it only as a contextual scale for Study D's \DEone{}-point contrast. Explaining a rule is not automatically better: high
adherence can mask incompetence rather than principled choice \citep{potham2025hierarchical}, and we
measure verification allocation; Study I additionally measures whether the selected action agrees with the current archive record, and neither endpoint establishes broad task correctness.

\paragraph{Declining directives: prior architectures and evidence.} The architecture for rejecting directives predates language models; its third felicity condition, goal priority and timing --- \emph{am I able to do X right now?} --- is a published description that matches the rationale text one model produced \citep{briggs2015sorry}. The field
builds external modules to check an instruction against the agent's goals precisely because it assumes
the model will not do so itself: an isolated planner that derives a reference set of valid actions
\citep{gong2026planguard}, a test-time shield that verifies whether each instruction contributes to the
user's goals \citep{jia2024taskshield}. Refusal behaviour has been shown to be decoupled from a model's capacity to reason about a rule's legitimacy \citep{pattison2026blind}; our registered endpoint records only whether the target was named, and the rationale text some models produced alongside a declined directive is reported as text, not as evidence of a reason. Models have also been reported to prioritise sensibility over
compliance, favouring task-appropriate reasoning despite conflicting instructions
\citep{tan2026compliance}; that framing is compatible with what we observe and does not occupy the
specific ablation.

\paragraph{Composite instructions, distraction and position.} Inputs that resemble instructions
degrade intent-following even under explicit instructions to ignore them \citep{dimbench2025}; a
composite directive that harms would be consistent with that broader literature, not automatically an
instance of it. Position effects on constraint adherence are established and model-specific, which is
why the composite's order and layout are a registered control (\texttt{both\_IF}) rather than a claim.
Context copying and parametric recall are separable mechanistically \citep{farahani2026copy}; our
instrument cannot separate dereferencing the criterion from copying the id, and does not try to.

\paragraph{The prior results on this instrument.} Paper~1 established that an assigned plan moves the credit and that
a memory stating a constraint is checked less than the same memory with the constraint removed
\citep{priorwork2026verification}; Paper~2 established the downstream cost when the unchecked
constraint is stale and left the allocation mechanism explicitly unidentified
\citep{priorwork2026stale}. Neither frozen paper manipulates directive composition; the first discusses directive framing and leaves the composition of its plan intervention unidentified.

\section{Setting and instrument}\label{sec:setting}

\paragraph{The allocation problem.} A model inherits six one-line memories from earlier sessions (\KtwoMemoriesMin{} to \KtwoMemoriesMax{} in Study K2's generated worlds), each with an
id and a consolidation day and each linked to the archived source record it was consolidated from. Before it
acts it may pull the source record of at most $k$ memories. Which memories it names is a scarce-resource
allocation made at inference time. This paper asks what an agent does with a \emph{directive} about where to
spend the budget: a field in the inherited store that names, or describes, the memory to verify first.

\begin{figure}[htbp]\centering
\resizebox{\textwidth}{!}{\begin{tikzpicture}[node distance=3mm and 4mm, every node/.style={font=\scriptsize}, box/.style={draw,rounded corners,align=center,inner sep=3pt,minimum height=8mm}]
\node[box] (store) {inherited store\\six one-line memories (K2: \KtwoMemoriesMin{}--\KtwoMemoriesMax{})\\with provenance links};
\node[box,right=of store] (field) {directive field\\pointer / criterion / both\\(D, E, F, F-x, G, \Gr{}, J, H1, H2, K2, K3, K4, K5)};
\node[box,right=of field] (req) {verification request\\$k = 1$ credit (J: $k = 2$)\\first credit $\to V_{73}$ (growth), $V_{c2}$ (H1)};
\node[box,right=of req] (dec) {archive record returned\\$\to$ decision $Y_1$\\(I)};
\node[box,above=of field] (plan) {\texttt{ACTIVE\_PLAN\_ID}: which described plan is active\\(B, \Br{}; names no record)};
\draw[-{Stealth}] (store) -- (field); \draw[-{Stealth}] (field) -- (req); \draw[-{Stealth}] (req) -- (dec); \draw[-{Stealth}] (plan) -- (field);
\end{tikzpicture}}
\caption{The instrument lineage. Every study elicits one verification request over an inherited store (six memories; \KtwoMemoriesMin{} to \KtwoMemoriesMax{} in Study K2's generated worlds); the studies differ in what is written into the store (a plan pointer upstream of the directive field in Studies B and \Br{}; a record pointer, a criterion or both in the directive field elsewhere), in the store (growth; procurement in Study H1), in the credit budget ($k = 2$ in two Study J arms, where the credit rule of the earlier work resolves the listed ids in order and `any credit' is reported beside `first credit') and in whether the returned record is carried into a decision (Study I).}\label{fig:instrument}
\end{figure}
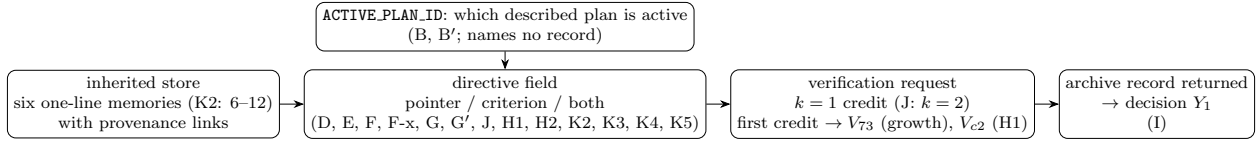

\paragraph{The growth-store instrument, held fixed.} Every study but H1, K2, K3, K4 and K5 uses the growth scenario of the earlier work, unchanged: the six-memory store (\texttt{memory\_31} to \texttt{memory\_91}), the day-76
situation with declining metrics, the two candidate plans (\emph{promotional\_pricing}, which rests on
\texttt{memory\_73}, and \emph{simplify\_onboarding}, which rests on \texttt{memory\_86} and
\texttt{memory\_31}), and the verify-only elicitation: the agent returns a JSON object with a
\texttt{verify\_memory\_ids} list, a free-text rationale and a confidence. The budget is $k = 1$ in every study,
a maximum: the list may hold one id, and an empty list spends nothing (\FZeroSpent{} of \FScored{} Study F
records and \EEmpty{} of \EScored{} Study E records are empty; \DMultiId{} Study D records and \EMultiId{}
Study E records list more than one id, and the first id is the one scored). The target memory is \texttt{memory\_73}. Its inherited one-line summary presents a targeted discount positively, whereas its archived source record reports a retention loss and a prohibition on reuse; in Studies D--G checking it could overturn the plan the agent is \emph{not} currently pursuing, and Study B instead varies which plan is active. The endpoint is deterministic: $V_{73} = 1$ when the first id named is \texttt{memory\_73} and $0$ otherwise; $V_{44}$ (Studies G and \Gr{}) is defined the same way for \texttt{memory\_44}, the competing record that a ` (memory\_44)' suffix names. Figure~\ref{fig:instrument} places the instruments of the \TotalStudies{} studies on one diagram.

\begin{sloppypar}
\paragraph{Directive forms.} A directive is one field inserted into the inherited store's block,
\texttt{VERIFY\_PRIORITY:}, with one of four contents (Appendix~\ref{app:prompts} gives the frozen bytes):
nothing (arm \texttt{none}); the bare id \texttt{memory\_73} (\texttt{id}; called \texttt{oracle} in Studies D
and E); a stated criterion that identifies the same record in one step from the displayed plan--memory
relation, \emph{``a record that could invalidate a direction you are not currently planning to take''}
(\texttt{crit}; called \texttt{policy} in D and E), matched to the id form to the character; and the composite
of both --- \texttt{both\_IL}, the criterion followed by the id in parentheses (id last), and \texttt{both\_IF},
the id first, then a dash and the criterion --- which differ in which element leads and in how the field wraps,
a bundled order/punctuation/layout control. The composite field is longer than either component by the length
of the added element, so composite-versus-single contrasts bundle content and length; this is registered as not
separable in this design. Study B uses a different field: an \texttt{ACTIVE\_PLAN\_ID} value that names which
of two described plans is active, both plans and both action names being present in every arm, against the
natural-language plan assignment of Experiment~7 of the earlier work.
\end{sloppypar}

\paragraph{Names.} The text uses four reader-facing names for the arms and gives the registered arm name once in parentheses: \emph{no directive} (\texttt{none}), the \emph{bare id} (\texttt{id}; \texttt{oracle} in Studies D and E), the \emph{criterion} (\texttt{crit}; \texttt{policy} in D and E) and the \emph{composite} (\texttt{both\_IL}, the criterion followed by the id in parentheses; \texttt{both\_IF}, the id first, is Study F's layout control). Tables and the reproducibility appendix keep the registered names.

\paragraph{The follow-up instruments.} Six follow-ups keep the construction; each changes one design element, and Study H1's change of store carries that world's system prompt, memories and actions with it. Study H1 moves the four forms to the earlier work's \emph{procurement} store (its held-out world: an agent choosing among five procurement actions over six inherited memories \texttt{memory\_c1}--\texttt{c6}): a different system prompt, objective, six memories and five actions, with the plan structure mirrored so that the target (\texttt{memory\_c2}, the record whose caveat the held-out treatment had removed from the visible summary, while \texttt{memory\_c4} keeps its caveat visible) backs the plan the agent is not pursuing; the endpoint $V_{c2}$ is the first credit being \texttt{memory\_c2}; two authored two-line plan descriptions are the only new text, and
the field blocks are the locked growth-world bytes with \texttt{memory\_73} replaced by \texttt{memory\_c2} at equal length.
Study I continues each turn-1 verification, with the model's own answer, into the earlier work's two archive worlds (the target's
record still valid, or superseded) and asks for the decision; its endpoint $Y_1$ is whether the action follows the current
record. Study J adds three exact bundled edits to the locked blocks: one ratification line after the field, an explicit
\texttt{VERIFY\_TARGET} line, and a budget of two, where the credit rule is the earlier work's (each raw id resolved in order,
duplicates skipped, stopping at $k$) and the any-credit rate is reported beside the first-credit rate. Study H2 replaces the criterion's
sentence by four surface rewordings that hold its construct fixed (object class \emph{record}, relation \emph{could invalidate},
temporal state \emph{not currently planned}) and vary the determiner, the relativizer, the verb form and the adverb placement or
plan-membership phrasing; Study \Gr{} re-runs Study G's six arms at eighty runs per cell with fresh memory orders; Study \Br{}
re-runs Study B's eight cells under a registered referent-decoded estimand. Appendix~\ref{app:prompts} and each package's
frozen artifacts give the bytes.

\paragraph{Panels.} Six models are called through their providers' own APIs (the \emph{direct-provider} panel:
Claude Opus~5, Sonnet~5, Haiku~4.5; GPT-5.6 Sol, Terra, Luna); nine are served through OpenRouter (the
\emph{OpenRouter-served} panel: gpt-oss-120b, DeepSeek V4 Pro, Kimi K3, MiniMax M3, Llama 4 Maverick, Qwen3.8
Max and Mistral Medium 3.5, which are open-weight, and Gemini 3.7 Flash and Grok 4.6, which are not). Study
F-x adds Claude Fable~5.1 and Fable~5 through the direct API; Study G uses those three Claude models and GPT-5.6 Sol. Table~\ref{tab:repro} (Appendix~\ref{app:registration}) gives the execution facts per study; exact endpoint identifiers, pins and inference settings are in each study's frozen package.

\paragraph{What is and is not manipulated.} Within a study, the arms' prompts differ only in the directive
field: with the field block removed they are byte-identical, asserted before any call (Study B's two bridge
cells, which reproduce Experiment~7's wording, are the exception and are compared only with each other). The
memory order is shuffled per (model, run) block and shared across that block's arms, so contrasts are paired
within blocks. Nothing in the prompt names the target as critical except the directive; the criterion is
diagnostic by construction, so what is tested is \emph{this} criterion, not any stated reason.

\paragraph{Estimands and inference.} Studies B--F-x and \Br{} register equal model weights over per-model rates and a block bootstrap over (model, run) blocks with $B = 4{,}000$ draws and a fixed seed. Studies G, I, H1, J, H2 and \Gr{} never pool models (nor worlds or wordings): they register a Wilson score interval per cell and, per model, a Newcombe (1998) method-10 paired score interval per contrast in the registered nonnegative-$\varphi$ variant ($\varphi$ set to zero whenever $n_{11}n_{00} - n_{10}n_{01} \le 0$), with a percentile block bootstrap printed as a sensitivity; for the transport and replication contrasts of H1, \Gr{} and \Br{} the percentile bootstrap, with the two sides resampled independently, is the registered interval. Each frozen analyzer computes contrasts over the runs present in both arms, a rule that Study E's prose registration did not state. Studies B, D and E registered verdict rules on their primary estimand; Study E's
pairing of runs under partial errors was decided by its frozen analyzer rather than its prose registration and
is recorded as a deviation (Appendix~\ref{app:studyE}). Studies F, F-x and G register no verdict: every interval is
marginal and is labelled in two independent columns, direction (the interval excludes zero or not) and material
size against a $\delta = 10$-point margin, decided in exact rational arithmetic (Study G: on Wilson and Newcombe paired score intervals, with the bootstrap as a sensitivity); no familywise label exists.
Study B's headline estimand is reported from a post-freeze corrected analysis (\S\ref{sec:results-b}).

\section{Study map}\label{sec:studies}

The \InitialStudies{} initial studies were run in sequence and each was designed after the previous one's locked result: Study D
followed Study B, Study E was a new-panel run of Study D's design, Study F was designed from the locked D and E
cells, Study F-x was designed after Study F's Opus~5 result, and Study G after Study F-x and the reviews of this manuscript, to separate the accounts the composite result left open. The programme is adaptive; nothing was jointly prospective.

Each study's package (its contents are listed in its manifest; the early packages hold the frozen hypotheses, prompts, model list, scoring and seed policy, schedule and analyzer, and the packages grew to include a validator, tests, a cost projection and, from Study F on, the frozen smoke manifest, records, ledger rows and run stamp) was hashed, committed, OpenTimestamps-stamped and deposited to OSF before its first confirmatory call (Study K2: deposited before its first call, stamped after its run had begun; \S\ref{sec:integrity}); each run
was locked into a completion manifest and analysed by its frozen script, with the two exceptions recorded in
\S\ref{sec:integrity} (Study B's corrected analysis; Study F's quarantined first run). Table~\ref{tab:map} places all \TotalStudies{} studies on one lineage; attempt counts, panels and run dates are in Table~\ref{tab:repro}.

\begin{table}[p]\centering\scriptsize\setlength{\tabcolsep}{3pt}\adjustbox{max width=\textwidth,max totalheight=0.88\textheight}{\begin{minipage}{\textwidth}
\caption{The \TotalStudies{} studies as one lineage, grouped by the question each phase inherited. Every study elicits one verification request over an inherited store (six memories; \KtwoMemoriesMin{} to \KtwoMemoriesMax{} in Study K2's generated worlds) (Study K2's generated worlds: the population stated in \S\ref{sec:results-k2}); ``bundled'' marks an edit whose named account is not isolated by design. Attempt counts, panels and run dates are in Table~\ref{tab:repro}; results in the sections named.}\label{tab:map}
\begin{tabular}{p{0.9cm}p{4.0cm}p{4.4cm}p{1.9cm}p{3.4cm}}\toprule
study & inherited question & what changed & endpoint & role (section) \\\midrule
\multicolumn{5}{l}{\emph{Form of the directive}} \\
D & does a pointer and a reason move the request alike? & \texttt{VERIFY\_PRIORITY}: none / bare id / length-matched criterion, six direct-provider models & $V_{73}$ & form matters (\S\ref{sec:results-d}) \\
E & does Study D's contrast transport? & the same three arms on a disjoint nine-model OpenRouter-served panel (seven open-weight, two closed-weight) & $V_{73}$ & transport test, rule not met (\S\ref{sec:results-e}) \\
F & do the forms compose? & five arms (none, id, crit, both\_IL, both\_IF) on fifteen models & $V_{73}$ & composition, pre-named groups (\S\ref{sec:results-f}) \\
F-x & does Opus~5's cancellation hold on re-run, and what do two further Claude endpoints show? & the same five arms on Fable~5, Fable~5.1, Opus~5 & $V_{73}$ & three named endpoints (\S\ref{sec:results-fx}) \\
\multicolumn{5}{l}{\emph{What cancels the composite}} \\
G & which exact edits of the composite field cancel the criterion? & six byte-controlled edits of the composite field (referent-free suffix, competing pointer, field labels) & $V_{73}$, $V_{44}$ & exact edits (\S\ref{sec:results-g}) \\
\Gr{} & does each Study G contrast change by less than the margin when re-run at higher precision? & the same six arms at eighty runs per cell, fresh orders & $V_{73}$, $V_{44}$ & replication contrasts (\S\ref{sec:results-gr}) \\
\multicolumn{5}{l}{\emph{Three exact edits}} \\
J & can the cancellation be undone? & a ratification line, an explicit target field, a budget of two (each bundled) & $V_{73}$ any / first credit & probes (\S\ref{sec:results-j}) \\
\multicolumn{5}{l}{\emph{Robustness}} \\
H2 & how do the two locked facts read under four authored rewordings? & four construct-fixed surface rewordings of the criterion, bare and with the suffix & $V_{73}$ & rewording (\S\ref{sec:results-h2}) \\
H1 & does the pattern transport to a second store? & the four forms in the procurement world (a different system prompt, objective, memories and actions; two added request fields for the GPT-5.6 models; a later execution date and provider state), with transport contrasts against the locked growth records & $V_{c2}$ (+ $V_{c4}$) & second store (\S\ref{sec:results-h1}) \\
\multicolumn{5}{l}{\emph{From the credit to the decision}} \\
I & do the directives that move the credit also change the decision? & each verification continued into two archive worlds (valid / superseded) & $V_{73}$, $Y_1$, $Y_2$ & decision (\S\ref{sec:results-i}) \\
\multicolumn{5}{l}{\emph{Plan pointers}} \\
B & does a one-character plan pointer move the request? & \texttt{ACTIVE\_PLAN\_ID} = A / B / none, counterbalanced, with two bridge cells & $V_{73}$ & corrected estimand (\S\ref{sec:results-b}) \\
\Br{} & does a prospectively registered re-run return the same verdict with a bounded change? & Study B's eight cells re-run under the registered referent-decoded estimand & $V_{73}$ & verdict rule, replication contrasts (\S\ref{sec:results-br}) \\
\multicolumn{5}{l}{\emph{Generated worlds}} \\
K2 & do the two locked Claude facts recur on worlds generated for this study? & \KtwoMainWorlds{} model-generated worlds in \KtwoFamilies{} domain families (\KtwoRealisersN{} realisers, \KtwoJudgesN{} outcome-blind judges, forecast table deposited before generation; population stated in \S\ref{sec:results-k2}) under the \KtwoForms{} forms and \KtwoStates{} planning states, on \KtwoClosedEndpointsWord{} direct-provider endpoints plus a self-hosted open-weight stratum & target first credit & generated worlds (\S\ref{sec:results-k2}) \\
K3 & does what appending the id does to the criterion change with the gain of a defensive adapter? & Study K2's \KthreeWorlds{} worlds under the four forms in the foregone state on two self-hosted open-weight ladders (Llama~3.1~8B, Llama~3.3~70B) with Meta-SecAlign applied at \KthreeDosesN{} gains (\KthreeDoses{}); registered slope and change of the composite $-$ criterion contrast against gain (primary on the 70B ladder; 8B descriptive) & target first credit & \S\ref{sec:results-k3} \\
K4 & is the 8B attenuation seen in K3 met under a registered rule; does it continue beyond the adapter's release; is the 70B criterion met on a fresh execution? & Study K3's three ingredients re-executed as three frozen ladders on the same \KfourWorlds{} worlds: L1 (8B, gains \KfourDosesLone{}) with a registered attenuation rule; L2 (8B, gains \KfourDosesLtwo{}; descriptive, registered before the freeze); L3 (70B, gains \KfourDosesLthree{}) under K3's rule & target first credit & \S\ref{sec:results-k4} \\
K5 & at registered power, on \KfiveCohortKtwoFamilies{} reconstructed K2 families plus \KfiveCohortKfiveFamilies{} families from the final K5 stream, is the 8B attenuation criterion satisfied; does a second SecAlign adapter satisfy it? & \KfiveWorlds{} worlds in \KfiveFamilies{} families (\KfiveCohortKtwoFamilies{} of Study K2's, reconstructed offline; \KfiveCohortKfiveFamilies{} from one generation stream), the foregone state and four forms; two frozen ladders at gains \KfiveDoses{}: K5-1 (Meta-SecAlign-8B) and K5-2 (SecAlign++), each under K4's attenuation rule; K5-3, their difference, descriptive by registration & target first credit & \S\ref{sec:results-k5} \\
\bottomrule\end{tabular}\end{minipage}}\end{table}

Study F's four groups are pre-named from locked cells of Studies D and E: \textsc{ID\_GT\_CRIT4} (four
OpenRouter-served models where the id had beaten the criterion), \textsc{CRIT\_GT\_ID2} (Mistral Medium~3.5
and Opus~5, where the criterion had beaten the id), and the OpenRouter-served and direct-provider pools. Group
membership is a precondition checked on the contemporaneous bridge cells before a group is named. Study F-x's
anchor precondition is Opus~5's locked sign (criterion above id).

\paragraph{The follow-up programme.} After the six-study manuscript closed, a second sequence of registered studies answered
its reviewers' residual questions on the same instruments (Table~\ref{tab:map}): a prospectively registered replication of
Study B's corrected estimand (\Br{}); a two-turn instrument that continues each verification into the earlier work's decision
(I); the four forms in a second store (H1); three exact bundled edits of the composite (J); five surface rewordings of the
criterion (H2); and Study G's six arms at eighty runs per cell (\Gr{}). They were conceived as a programme (their design memos were committed together on 2026-09-02, before Study \Br{} had locked), partly developed in parallel while earlier follow-ups ran, and individually frozen, stamped and deposited before their first confirmatory calls, after one to four hostile Codex review rounds each, then executed one at a time in the order \Br{}, I, H1, J, H2, \Gr{}. Each of the six registers no verdict except \Br{}, whose verdict rule is Study B's; Study K3, below, carries a registered three-condition rule of its own (\S\ref{sec:results-k3}), Study K4 registers the attenuation K3 had only described (\S\ref{sec:results-k4}), and Study K5 re-poses K4's rule at registered power with a second adapter (\S\ref{sec:results-k5}). The narrative order of this paper was fixed in a restructure plan committed after Study \Br{}'s result and before the other five; the plan placed Study E in an appendix and H2 before H1, and this version keeps that order and adds a main-text summary of Study E after review. Three further studies left the hand-written worlds behind (the third, K4, re-executes K3's ladders under a registered rule). Study K2 froze a generator, a structural contract and outcome-blind judges, deposited a table of \KtwoRowsTotal{} rows (\KtwoDirTotalWord{} directional forecasts, \KtwoWithinTotalWord{} forecast absences, \KtwoUndeterminedTotalWord{} left undetermined), and then measured the \KtwoForms{} forms on \KtwoMainWorlds{} generated worlds in \KtwoFamilies{} domain families (\S\ref{sec:results-k2}). Study K3, designed before K2's runs and amended after its outcome (the amendment is disclosed in its pre-specification and its primary was redesignated to the 70B ladder's composite $-$ criterion contrast), applied a defensive adapter at five gains to the two open ladders on K2's worlds (\S\ref{sec:results-k3}).

\section{Form matters (Studies D, E, F and F-x)}\label{sec:results}

Rates are $V_{73}$ in percent with equal model weights over per-model rates; for Studies B--F-x intervals are the registered analyses' block-bootstrap 95\% intervals (Study B's from the disclosed corrected analysis), for Study G the registered Wilson and Newcombe paired score intervals (\S\ref{sec:results-g}); contrasts are in points over the runs present in both arms.

\subsection{Study D: a stated criterion is obeyed where a bare id is refused, on a provider-shaped split}\label{sec:results-d}
\begin{figure}[htbp]\centering
\begin{tikzpicture}[x=1.7cm,y=-0.50cm,font=\footnotesize]
\node[anchor=south,font=\scriptsize] at (0.500,-0.06) {no directive};
\node[anchor=south,font=\scriptsize] at (1.500,-0.06) {bare id};
\node[anchor=south,font=\scriptsize] at (2.500,-0.06) {criterion};
\node[anchor=east,font=\scriptsize] at (-0.08,0.500) {Opus 5};
\fill[blue!2!white] (0.000,0.000) rectangle (1.000,1.000);
\draw[white,line width=0.6pt] (0.000,0.000) rectangle (1.000,1.000);
\node[font=\scriptsize,text=black] at (0.500,0.500) {1/40};
\fill[blue!0!white] (1.000,0.000) rectangle (2.000,1.000);
\draw[white,line width=0.6pt] (1.000,0.000) rectangle (2.000,1.000);
\node[font=\scriptsize,text=black] at (1.500,0.500) {0/40};
\fill[blue!100!white] (2.000,0.000) rectangle (3.000,1.000);
\draw[white,line width=0.6pt] (2.000,0.000) rectangle (3.000,1.000);
\node[font=\scriptsize,text=white] at (2.500,0.500) {40/40};
\node[anchor=east,font=\scriptsize] at (-0.08,1.500) {Sonnet 5};
\fill[blue!0!white] (0.000,1.000) rectangle (1.000,2.000);
\draw[white,line width=0.6pt] (0.000,1.000) rectangle (1.000,2.000);
\node[font=\scriptsize,text=black] at (0.500,1.500) {0/40};
\fill[blue!38!white] (1.000,1.000) rectangle (2.000,2.000);
\draw[white,line width=0.6pt] (1.000,1.000) rectangle (2.000,2.000);
\node[font=\scriptsize,text=black] at (1.500,1.500) {15/40};
\fill[blue!100!white] (2.000,1.000) rectangle (3.000,2.000);
\draw[white,line width=0.6pt] (2.000,1.000) rectangle (3.000,2.000);
\node[font=\scriptsize,text=white] at (2.500,1.500) {40/40};
\node[anchor=east,font=\scriptsize] at (-0.08,2.500) {Haiku 4.5};
\fill[blue!0!white] (0.000,2.000) rectangle (1.000,3.000);
\draw[white,line width=0.6pt] (0.000,2.000) rectangle (1.000,3.000);
\node[font=\scriptsize,text=black] at (0.500,2.500) {0/40};
\fill[blue!0!white] (1.000,2.000) rectangle (2.000,3.000);
\draw[white,line width=0.6pt] (1.000,2.000) rectangle (2.000,3.000);
\node[font=\scriptsize,text=black] at (1.500,2.500) {0/40};
\fill[blue!48!white] (2.000,2.000) rectangle (3.000,3.000);
\draw[white,line width=0.6pt] (2.000,2.000) rectangle (3.000,3.000);
\node[font=\scriptsize,text=black] at (2.500,2.500) {19/40};
\node[anchor=east,font=\scriptsize] at (-0.08,3.500) {Sol};
\fill[blue!0!white] (0.000,3.000) rectangle (1.000,4.000);
\draw[white,line width=0.6pt] (0.000,3.000) rectangle (1.000,4.000);
\node[font=\scriptsize,text=black] at (0.500,3.500) {0/40};
\fill[blue!100!white] (1.000,3.000) rectangle (2.000,4.000);
\draw[white,line width=0.6pt] (1.000,3.000) rectangle (2.000,4.000);
\node[font=\scriptsize,text=white] at (1.500,3.500) {40/40};
\fill[blue!100!white] (2.000,3.000) rectangle (3.000,4.000);
\draw[white,line width=0.6pt] (2.000,3.000) rectangle (3.000,4.000);
\node[font=\scriptsize,text=white] at (2.500,3.500) {40/40};
\node[anchor=east,font=\scriptsize] at (-0.08,4.500) {Terra};
\fill[blue!8!white] (0.000,4.000) rectangle (1.000,5.000);
\draw[white,line width=0.6pt] (0.000,4.000) rectangle (1.000,5.000);
\node[font=\scriptsize,text=black] at (0.500,4.500) {3/40};
\fill[blue!100!white] (1.000,4.000) rectangle (2.000,5.000);
\draw[white,line width=0.6pt] (1.000,4.000) rectangle (2.000,5.000);
\node[font=\scriptsize,text=white] at (1.500,4.500) {40/40};
\fill[blue!100!white] (2.000,4.000) rectangle (3.000,5.000);
\draw[white,line width=0.6pt] (2.000,4.000) rectangle (3.000,5.000);
\node[font=\scriptsize,text=white] at (2.500,4.500) {40/40};
\node[anchor=east,font=\scriptsize] at (-0.08,5.500) {Luna};
\fill[blue!0!white] (0.000,5.000) rectangle (1.000,6.000);
\draw[white,line width=0.6pt] (0.000,5.000) rectangle (1.000,6.000);
\node[font=\scriptsize,text=black] at (0.500,5.500) {0/40};
\fill[blue!100!white] (1.000,5.000) rectangle (2.000,6.000);
\draw[white,line width=0.6pt] (1.000,5.000) rectangle (2.000,6.000);
\node[font=\scriptsize,text=white] at (1.500,5.500) {40/40};
\fill[blue!100!white] (2.000,5.000) rectangle (3.000,6.000);
\draw[white,line width=0.6pt] (2.000,5.000) rectangle (3.000,6.000);
\node[font=\scriptsize,text=white] at (2.500,5.500) {40/40};
\node[anchor=east,font=\scriptsize] at (-0.08,6.500) {Fable 5 (F-x)};
\fill[blue!5!white] (0.000,6.000) rectangle (1.000,7.000);
\draw[white,line width=0.6pt] (0.000,6.000) rectangle (1.000,7.000);
\node[font=\scriptsize,text=black] at (0.500,6.500) {2/40};
\fill[blue!15!white] (1.000,6.000) rectangle (2.000,7.000);
\draw[white,line width=0.6pt] (1.000,6.000) rectangle (2.000,7.000);
\node[font=\scriptsize,text=black] at (1.500,6.500) {6/40};
\fill[blue!100!white] (2.000,6.000) rectangle (3.000,7.000);
\draw[white,line width=0.6pt] (2.000,6.000) rectangle (3.000,7.000);
\node[font=\scriptsize,text=white] at (2.500,6.500) {40/40};
\node[anchor=east,font=\scriptsize] at (-0.08,7.500) {Fable 5.1 (F-x)};
\fill[blue!32!white] (0.000,7.000) rectangle (1.000,8.000);
\draw[white,line width=0.6pt] (0.000,7.000) rectangle (1.000,8.000);
\node[font=\scriptsize,text=black] at (0.500,7.500) {13/40};
\fill[blue!32!white] (1.000,7.000) rectangle (2.000,8.000);
\draw[white,line width=0.6pt] (1.000,7.000) rectangle (2.000,8.000);
\node[font=\scriptsize,text=black] at (1.500,7.500) {13/40};
\fill[blue!42!white] (2.000,7.000) rectangle (3.000,8.000);
\draw[white,line width=0.6pt] (2.000,7.000) rectangle (3.000,8.000);
\node[font=\scriptsize,text=black] at (2.500,7.500) {17/40};
\end{tikzpicture}

\medskip
\begin{tikzpicture}[font=\scriptsize,y=0.32cm]\fill[blue!0!white] (0.00,0) rectangle (0.26,1);\fill[blue!10!white] (0.26,0) rectangle (0.52,1);\fill[blue!20!white] (0.52,0) rectangle (0.78,1);\fill[blue!30!white] (0.78,0) rectangle (1.04,1);\fill[blue!40!white] (1.04,0) rectangle (1.30,1);\fill[blue!50!white] (1.30,0) rectangle (1.56,1);\fill[blue!60!white] (1.56,0) rectangle (1.82,1);\fill[blue!70!white] (1.82,0) rectangle (2.08,1);\fill[blue!80!white] (2.08,0) rectangle (2.34,1);\fill[blue!90!white] (2.34,0) rectangle (2.60,1);\draw[gray!60,line width=0.3pt] (0,0) rectangle (2.6,1);\node[anchor=east] at (-0.08,0.5) {0\%};\node[anchor=west] at (2.68,0.5) {100\%};\end{tikzpicture}
\caption{The model-level picture behind Studies D and F-x: for each model and each directive form, the exact count of episodes in which the target was named, shaded by the rate. Descriptive; the registered intervals are in Tables~\ref{tab:studyD} and~\ref{tab:studyFx}. The three GPT-5.6 models are \emph{Sol}, \emph{Terra} and \emph{Luna}.}\label{fig:form}
\end{figure}
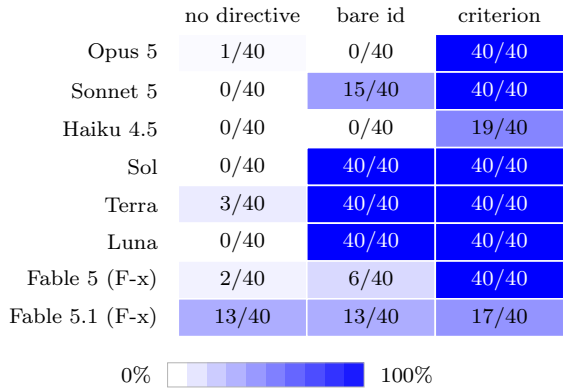

With no directive the request went almost entirely to the memory backing the active plan (\texttt{memory\_86}:
\DAllocNoneEightySix\% of episodes) and the target was named in \DRateNone\%. The bare id moved it to
\DRateOracle\%; the length-matched criterion to \DRatePolicy\%. The registered form contrast is $E_1 = \DEone{}$
[\DEoneLo, \DEoneHi] (id against none \DEtwo{} [\DEtwoLo, \DEtwoHi]; criterion against none \DEthree{}
[\DEthreeLo, \DEthreeHi]), and the registered rule (FORM MATTERS) was met. The pooled contrast is not a uniform
effect (Table~\ref{tab:studyD}): the three GPT-5.6 models named the target under the bare id in
\DPmKOracleGPTSol, \DPmKOracleGPTTerra{} and \DPmKOracleGPTLuna{} episodes and under the criterion in every
episode; Opus~5 in \DPmKOracleOpus{} under the id and \DPmKPolicyOpus{} under the criterion; Sonnet~5 in
\DPmKOracleSonnet{} and \DPmKPolicySonnet{}; Haiku~4.5 in \DPmKOracleHaiku{} and \DPmKPolicyHaiku{}.
Leave-one-model-out spans \DLooMin{} to \DLooMax{}. Under the criterion the plan-backing memory's share of the
requests fell to \DAllocPolicyEightySix\% and the target's rose to \DAllocPolicySeventyThree\%.

\begin{sloppypar}
\subsection{Study E: the same contrast on a disjoint OpenRouter-served panel}\label{sec:results-e}
Study E re-ran Study D's three arms on a disjoint OpenRouter-served panel of nine models (seven open-weight, two closed-weight) at 20 runs per cell, as a transport test rather than a replication (\S\ref{sec:studies}). Of \EAttempted{} attempted episodes \EScored{} scored; the \EErrors{} errors (\EErrorPct\%) were transport failures on one endpoint, and the frozen analyzer paired only the runs present in both arms, a rule the prose registration did not state (\EKimiPairedRuns{} paired runs of that endpoint enter $E_1$). The rates were \ERateNone\% (none), \ERateOracle\% (id) and \ERatePolicy\% (criterion): $E_1 = \EEone{}$ [\EEoneLo, \EEoneHi], whose lower endpoint does not exceed zero, so the registered superiority rule ($\mathrm{ci}_{\mathrm{lo}} > 0$) was not met --- one completion of the missing episodes would have met it, and the interval is compatible with a criterion advantage of up to \EEoneHi{} points; the per-model contrasts span both signs (Table~\ref{tab:studyE}). Study E therefore bounds Study D: the criterion's advantage is a property of the models it was measured on, not of the form as such. Two further departures are disclosed: the frozen hypotheses file describes the panel before one endpoint was removed (ten models) and in one line says `all six models', whereas the frozen model list, schedule and analyzer enforce the nine-model panel that was run; the executed analysis followed the frozen list (Appendix~\ref{app:studyE}).

\subsection{Study F: composition on fifteen models}\label{sec:results-f}

\FScored{} of \FN{} episodes scored (\FErrors{} errors, \FErrorPct\%, all on one endpoint: \FLostModel{});
\FPrecondCount{} of \FPrecondTotal{} group preconditions held on the contemporaneous bridge cells;
\FCompletionLabelChanges{} of the 64 missing-data completion results (each of the sixteen contrasts under four named completions: all missing as misses, all as hits, favouring the contrast, disfavouring it) carries a direction/size label pair different from its complete-case result (a count over the registered labels, not a registered quantity). Table~\ref{tab:studyF-cells} gives the twenty cell means, Table~\ref{tab:studyF-contrasts}
the sixteen contrasts and Table~\ref{tab:studyF-permodel} the per-model counts. Every statement below is about a
marginal interval; no joint claim is registered.

\paragraph{Composite against id (C2).} The composite's point estimate exceeded the id's in each of the four
groups: \FCtwoIdGtCrit{} [\FCtwoIdGtCritLo, \FCtwoIdGtCritHi] on \textsc{ID\_GT\_CRIT4}, \FCtwoOpen{}
[\FCtwoOpenLo, \FCtwoOpenHi] on the OpenRouter-served pool, \FCtwoClosed{} [\FCtwoClosedLo, \FCtwoClosedHi] on
the direct-provider pool and \FCtwoCritGtId{} [\FCtwoCritGtIdLo, \FCtwoCritGtIdHi] on \textsc{CRIT\_GT\_ID2}.

\begin{sloppypar}
\paragraph{Composite against criterion (C1).} \FConeIdGtCrit{} [\FConeIdGtCritLo, \FConeIdGtCritHi] on
\textsc{ID\_GT\_CRIT4} and \FConeOpen{} [\FConeOpenLo, \FConeOpenHi] on the OpenRouter-served pool;
\FConeCritGtId{} [\FConeCritGtIdLo, \FConeCritGtIdHi] on \textsc{CRIT\_GT\_ID2} and \FConeClosed{}
[\FConeClosedLo, \FConeClosedHi] on the direct-provider pool. The \textsc{CRIT\_GT\_ID2} mean hides two
opposite movements: Mistral Medium~3.5 went from \FPmKCritMistralMedium{} under the criterion to
\FPmKBothILMistralMedium{} under the composite, Opus~5 from \FPmKCritOpus{} to \FPmKBothILOpus{}
(\FPmKBothIFOpus{} under the id-first layout), naming \texttt{memory\_86} in \FOpusMEightySixBothIL{} of those 40 episodes (an exploratory count). The direct-provider mean likewise combines Opus's fall with Haiku~4.5's rise
(\FPmKCritHaiku{} to \FPmKBothILHaiku{}) and zeros elsewhere; Sonnet~5 (\FPmKBothILSonnet{}) and the GPT-5.6
models followed the composite in every episode. No moderator was registered for these differences.
\end{sloppypar}

\paragraph{Controls.} The order/layout control C3 is within the margin on three groups (\FCthreeIdGtCrit{},
\FCthreeOpen{}, \FCthreeClosed{}) and \FCthreeCritGtId{} [\FCthreeCritGtIdLo, \FCthreeCritGtIdHi] on
\textsc{CRIT\_GT\_ID2}. Against no directive, C4 ranges from \FCfourMin{} to \FCfourMax{} across the groups.

\paragraph{Opus~5's rationales.} An exploratory lexical tally over the stored rationales (a fixed phrase list
describing the override of an inherited priority field; not a registered quantity) fires in
\FOpusOverrideNone{} of the \texttt{none} episodes, \FOpusOverrideId{} of the \texttt{id} episodes,
\FOpusOverrideCrit{} of the \texttt{crit} episodes, \FOpusOverrideBothIL{} of the \texttt{both\_IL} episodes
and \FOpusOverrideBothIF{} of the \texttt{both\_IF} episodes. Because it fires in the arms without any pointer
as often as in the composite arms, it does not identify what the composite was read as; we report it as text
the model produced and draw nothing from it.

\subsection{Study F-x: three Claude models under the same five arms}\label{sec:results-fx}

All \FxN{} episodes scored (\FxErrors{} errors, \FxEmpty{} empty lists); the anchor precondition is
\MakeLowercase{\FxAnchor}. Tables~\ref{tab:studyFx} and~\ref{tab:studyFx-contrasts} give the cells and all
sixteen registered contrasts. On the Opus~5 re-run the registered anchor precondition (criterion above id) was met: criterion \FxPmKCritOpus{}, id \FxPmKIdOpus{}, composite \FxPmKBothILOpus{} and \FxPmKBothIFOpus{}, against Study F's \FPmKCritOpus{}, \FPmKIdOpus{}, \FPmKBothILOpus{} and \FPmKBothIFOpus{} (the cell counts and their ties differ; only the registered sign is a reproduction claim); $C_1 = \FxConeOpus{}$ [\FxConeOpusLo,
\FxConeOpusHi]. Fable~5: id \FxPmKIdFableFive{}, criterion \FxPmKCritFableFive{}, composite
\FxPmKBothILFableFive{} and \FxPmKBothIFFableFive{}; $C_1 = \FxConeFableFive{}$ [\FxConeFableFiveLo,
\FxConeFableFiveHi]; family contrast against Opus on $C_1$ \FxFamConeFableFive{} [\FxFamConeFableFiveLo,
\FxFamConeFableFiveHi] (\MakeLowercase{\FxFamConeFableFiveDir}, \MakeLowercase{\FxFamConeFableFiveSize})
and on $C_2$ \FxFamCtwoFableFive{} [\FxFamCtwoFableFiveLo, \FxFamCtwoFableFiveHi]
(\MakeLowercase{\FxFamCtwoFableFiveDir}, \MakeLowercase{\FxFamCtwoFableFiveSize}). Fable~5.1: with no
directive it named the target in \FxPmKNoneFableFiveOne{} episodes, the bare id left that unchanged
(\FxPmKIdFableFiveOne{}), the criterion raised it to \FxPmKCritFableFiveOne{}, and both composites gave
\FxPmKBothILFableFiveOne{} and \FxPmKBothIFFableFiveOne{}; $C_1 = \FxConeFableFiveOne{}$
[\FxConeFableFiveOneLo, \FxConeFableFiveOneHi], $C_2 = \FxCtwoFableFiveOne{}$ [\FxCtwoFableFiveOneLo,
\FxCtwoFableFiveOneHi]; family contrasts \FxFamConeFableFiveOne{} [\FxFamConeFableFiveOneLo,
\FxFamConeFableFiveOneHi] on $C_1$ and \FxFamCtwoFableFiveOne{} [\FxFamCtwoFableFiveOneLo,
\FxFamCtwoFableFiveOneHi] on $C_2$. The order/layout control is within the margin on all three models. Under
the composite, \texttt{memory\_86} was named in \FxMEightySixBothILOpus{}, \FxMEightySixBothILFableFive{} and
\FxMEightySixBothILFableFiveOne{} of 40 episodes (an exploratory count). These are three named endpoints on a fixed panel; Study F-x registers the four family contrasts above and no familywise label or generalisation claim, and Study F's Sonnet~5 and Haiku~4.5 followed the composite in every episode.

\section{What cancels the composite (Studies G and \Gr{})}\label{sec:results-cancel}
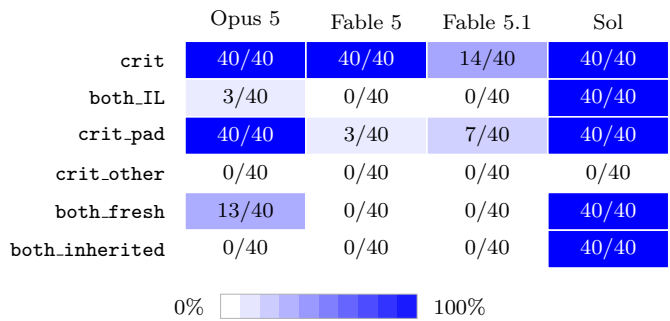
\begin{figure}[htbp]\centering
\begin{tikzpicture}[x=1.6cm,y=-0.50cm,font=\footnotesize]
\node[anchor=south,font=\scriptsize] at (0.500,-0.06) {Opus 5};
\node[anchor=south,font=\scriptsize] at (1.500,-0.06) {Fable 5};
\node[anchor=south,font=\scriptsize] at (2.500,-0.06) {Fable 5.1};
\node[anchor=south,font=\scriptsize] at (3.500,-0.06) {Sol};
\node[anchor=east,font=\scriptsize] at (-0.08,0.500) {\texttt{crit}};
\fill[blue!100!white] (0.000,0.000) rectangle (1.000,1.000);
\draw[white,line width=0.6pt] (0.000,0.000) rectangle (1.000,1.000);
\node[font=\scriptsize,text=white] at (0.500,0.500) {40/40};
\fill[blue!100!white] (1.000,0.000) rectangle (2.000,1.000);
\draw[white,line width=0.6pt] (1.000,0.000) rectangle (2.000,1.000);
\node[font=\scriptsize,text=white] at (1.500,0.500) {40/40};
\fill[blue!35!white] (2.000,0.000) rectangle (3.000,1.000);
\draw[white,line width=0.6pt] (2.000,0.000) rectangle (3.000,1.000);
\node[font=\scriptsize,text=black] at (2.500,0.500) {14/40};
\fill[blue!100!white] (3.000,0.000) rectangle (4.000,1.000);
\draw[white,line width=0.6pt] (3.000,0.000) rectangle (4.000,1.000);
\node[font=\scriptsize,text=white] at (3.500,0.500) {40/40};
\node[anchor=east,font=\scriptsize] at (-0.08,1.500) {\texttt{both\_IL}};
\fill[blue!8!white] (0.000,1.000) rectangle (1.000,2.000);
\draw[white,line width=0.6pt] (0.000,1.000) rectangle (1.000,2.000);
\node[font=\scriptsize,text=black] at (0.500,1.500) {3/40};
\fill[blue!0!white] (1.000,1.000) rectangle (2.000,2.000);
\draw[white,line width=0.6pt] (1.000,1.000) rectangle (2.000,2.000);
\node[font=\scriptsize,text=black] at (1.500,1.500) {0/40};
\fill[blue!0!white] (2.000,1.000) rectangle (3.000,2.000);
\draw[white,line width=0.6pt] (2.000,1.000) rectangle (3.000,2.000);
\node[font=\scriptsize,text=black] at (2.500,1.500) {0/40};
\fill[blue!100!white] (3.000,1.000) rectangle (4.000,2.000);
\draw[white,line width=0.6pt] (3.000,1.000) rectangle (4.000,2.000);
\node[font=\scriptsize,text=white] at (3.500,1.500) {40/40};
\node[anchor=east,font=\scriptsize] at (-0.08,2.500) {\texttt{crit\_pad}};
\fill[blue!100!white] (0.000,2.000) rectangle (1.000,3.000);
\draw[white,line width=0.6pt] (0.000,2.000) rectangle (1.000,3.000);
\node[font=\scriptsize,text=white] at (0.500,2.500) {40/40};
\fill[blue!8!white] (1.000,2.000) rectangle (2.000,3.000);
\draw[white,line width=0.6pt] (1.000,2.000) rectangle (2.000,3.000);
\node[font=\scriptsize,text=black] at (1.500,2.500) {3/40};
\fill[blue!18!white] (2.000,2.000) rectangle (3.000,3.000);
\draw[white,line width=0.6pt] (2.000,2.000) rectangle (3.000,3.000);
\node[font=\scriptsize,text=black] at (2.500,2.500) {7/40};
\fill[blue!100!white] (3.000,2.000) rectangle (4.000,3.000);
\draw[white,line width=0.6pt] (3.000,2.000) rectangle (4.000,3.000);
\node[font=\scriptsize,text=white] at (3.500,2.500) {40/40};
\node[anchor=east,font=\scriptsize] at (-0.08,3.500) {\texttt{crit\_other}};
\fill[blue!0!white] (0.000,3.000) rectangle (1.000,4.000);
\draw[white,line width=0.6pt] (0.000,3.000) rectangle (1.000,4.000);
\node[font=\scriptsize,text=black] at (0.500,3.500) {0/40};
\fill[blue!0!white] (1.000,3.000) rectangle (2.000,4.000);
\draw[white,line width=0.6pt] (1.000,3.000) rectangle (2.000,4.000);
\node[font=\scriptsize,text=black] at (1.500,3.500) {0/40};
\fill[blue!0!white] (2.000,3.000) rectangle (3.000,4.000);
\draw[white,line width=0.6pt] (2.000,3.000) rectangle (3.000,4.000);
\node[font=\scriptsize,text=black] at (2.500,3.500) {0/40};
\fill[blue!0!white] (3.000,3.000) rectangle (4.000,4.000);
\draw[white,line width=0.6pt] (3.000,3.000) rectangle (4.000,4.000);
\node[font=\scriptsize,text=black] at (3.500,3.500) {0/40};
\node[anchor=east,font=\scriptsize] at (-0.08,4.500) {\texttt{both\_fresh}};
\fill[blue!32!white] (0.000,4.000) rectangle (1.000,5.000);
\draw[white,line width=0.6pt] (0.000,4.000) rectangle (1.000,5.000);
\node[font=\scriptsize,text=black] at (0.500,4.500) {13/40};
\fill[blue!0!white] (1.000,4.000) rectangle (2.000,5.000);
\draw[white,line width=0.6pt] (1.000,4.000) rectangle (2.000,5.000);
\node[font=\scriptsize,text=black] at (1.500,4.500) {0/40};
\fill[blue!0!white] (2.000,4.000) rectangle (3.000,5.000);
\draw[white,line width=0.6pt] (2.000,4.000) rectangle (3.000,5.000);
\node[font=\scriptsize,text=black] at (2.500,4.500) {0/40};
\fill[blue!100!white] (3.000,4.000) rectangle (4.000,5.000);
\draw[white,line width=0.6pt] (3.000,4.000) rectangle (4.000,5.000);
\node[font=\scriptsize,text=white] at (3.500,4.500) {40/40};
\node[anchor=east,font=\scriptsize] at (-0.08,5.500) {\texttt{both\_inherited}};
\fill[blue!0!white] (0.000,5.000) rectangle (1.000,6.000);
\draw[white,line width=0.6pt] (0.000,5.000) rectangle (1.000,6.000);
\node[font=\scriptsize,text=black] at (0.500,5.500) {0/40};
\fill[blue!0!white] (1.000,5.000) rectangle (2.000,6.000);
\draw[white,line width=0.6pt] (1.000,5.000) rectangle (2.000,6.000);
\node[font=\scriptsize,text=black] at (1.500,5.500) {0/40};
\fill[blue!0!white] (2.000,5.000) rectangle (3.000,6.000);
\draw[white,line width=0.6pt] (2.000,5.000) rectangle (3.000,6.000);
\node[font=\scriptsize,text=black] at (2.500,5.500) {0/40};
\fill[blue!100!white] (3.000,5.000) rectangle (4.000,6.000);
\draw[white,line width=0.6pt] (3.000,5.000) rectangle (4.000,6.000);
\node[font=\scriptsize,text=white] at (3.500,5.500) {40/40};
\end{tikzpicture}

\medskip
\begin{tikzpicture}[font=\scriptsize,y=0.32cm]\fill[blue!0!white] (0.00,0) rectangle (0.26,1);\fill[blue!10!white] (0.26,0) rectangle (0.52,1);\fill[blue!20!white] (0.52,0) rectangle (0.78,1);\fill[blue!30!white] (0.78,0) rectangle (1.04,1);\fill[blue!40!white] (1.04,0) rectangle (1.30,1);\fill[blue!50!white] (1.30,0) rectangle (1.56,1);\fill[blue!60!white] (1.56,0) rectangle (1.82,1);\fill[blue!70!white] (1.82,0) rectangle (2.08,1);\fill[blue!80!white] (2.08,0) rectangle (2.34,1);\fill[blue!90!white] (2.34,0) rectangle (2.60,1);\draw[gray!60,line width=0.3pt] (0,0) rectangle (2.6,1);\node[anchor=east] at (-0.08,0.5) {0\%};\node[anchor=west] at (2.68,0.5) {100\%};\end{tikzpicture}
\caption{Study G: the exact count of episodes in which the target was named under each of the six byte-controlled edits of the composite field, per model, shaded by the rate ($n = 40$ per cell). The registered Wilson intervals are in Tables~\ref{tab:studyG} and~\ref{tab:studyG-v44}; Study \Gr{}'s cells at eighty runs per cell are in Table~\ref{tab:studyGr}. \emph{Sol} is GPT-5.6 Sol.}\label{fig:cancel}
\end{figure}

\subsection{Study G: six exact edits of the composite field on four models}\label{sec:results-g}

All \GN{} episodes scored (\GErrors{} errors, \GEmpty{} empty lists); the anchor precondition is
\MakeLowercase{\GAnchor}. Study G's registered intervals are Wilson score intervals for cells (Tables~\ref{tab:studyG}
and~\ref{tab:studyG-v44}) and Newcombe paired score intervals for the forty contrasts (Table~\ref{tab:studyG-contrasts});
the block bootstrap is printed beside each quantity in the study's output as a sensitivity and agrees in every label.
At the boundary cells this design produces, the percentile bootstrap is degenerate (a cell with no hits or with every hit resamples to a single value, so its interval has zero width) while the score intervals are not: a cell with no hits gives [0.0, \GWilsonZeroHi], a cell with every hit [\GWilsonFullLo, 100.0], and a contrast between two equal-boundary cells (both with no hits, or both with every hit) [\GBoundaryContrastLo, \GBoundaryContrastHi]; this is why the score intervals, not the bootstrap, are the registered label basis. Every statement below concerns one exact string on one named
model and one marginal interval; no property of the strings is isolated.

\paragraph{GPT-5.6 Sol.} Criterion alone \GPmKCritGPTSol{}; with ` (memory\_73)' appended \GPmKBothILGPTSol{}; with
` (as stated)' appended \GPmKCritPadGPTSol{}; with the field labelled `set today' or `inherited'
\GPmKBothFreshGPTSol{} and \GPmKBothInheritedGPTSol{}. With ` (memory\_44)' appended the target rate is
\GPmKCritOtherGPTSol{} ($G_3 = \GGthreeGPTSol{}$ [\GGthreeGPTSolLo, \GGthreeGPTSolHi]) and the registered V44 cell is
\GVfortyfourKGPTSol{}. $G_8$ (` (memory\_44)' against ` (memory\_73)') is likewise negative beyond the margin (\GGeightGPTSol{} [\GGeightGPTSolLo, \GGeightGPTSolHi]); every other contrast on this model is \GGoneGPTSol{} [\GGoneGPTSolLo, \GGoneGPTSolHi], undetermined and within the margin.

\paragraph{Opus~5.} Criterion alone \GPmKCritOpus{}; with ` (as stated)' appended \GPmKCritPadOpus{}
($G_2 = \GGtwoOpus{}$ [\GGtwoOpusLo, \GGtwoOpusHi], \MakeLowercase{\GGtwoOpusDir}, \MakeLowercase{\GGtwoOpusSize});
with ` (memory\_73)' appended \GPmKBothILOpus{} ($G_1 = \GGoneOpus{}$ [\GGoneOpusLo, \GGoneOpusHi]); with
` (memory\_44)' appended \GPmKCritOtherOpus{} ($G_3 = \GGthreeOpus{}$ [\GGthreeOpusLo, \GGthreeOpusHi]) and V44
\GVfortyfourKOpus{}: the competing pointer was not followed either. The direct comparison of the two 121-byte
parentheticals is $G_7 = \GGsevenOpus{}$ [\GGsevenOpusLo, \GGsevenOpusHi]. With the field labelled `set today' the
target rate was \GPmKBothFreshOpus{}, with `inherited' \GPmKBothInheritedOpus{}; the equal-byte label contrast is
$G_6 = \GGsixOpus{}$ [\GGsixOpusLo, \GGsixOpusHi] (\MakeLowercase{\GGsixOpusDir}, \MakeLowercase{\GGsixOpusSize}),
and both labelled composites remain below the criterion alone ($G_9 = \GGnineOpus{}$ [\GGnineOpusLo,
\GGnineOpusHi]; $G_{10} = \GGtenOpus{}$ [\GGtenOpusLo, \GGtenOpusHi]).

\paragraph{Fable~5.} Criterion alone \GPmKCritFableFive{}; ` (as stated)' \GPmKCritPadFableFive{} ($G_2 =
\GGtwoFableFive{}$ [\GGtwoFableFiveLo, \GGtwoFableFiveHi]); ` (memory\_73)' \GPmKBothILFableFive{}; ` (memory\_44)'
\GPmKCritOtherFableFive{} (V44 \GVfortyfourKFableFive{}); `set today' \GPmKBothFreshFableFive{} and `inherited'
\GPmKBothInheritedFableFive{} ($G_6 = \GGsixFableFive{}$ [\GGsixFableFiveLo, \GGsixFableFiveHi],
\MakeLowercase{\GGsixFableFiveDir}, \MakeLowercase{\GGsixFableFiveSize}).

\paragraph{Fable~5.1.} Criterion alone \GPmKCritFableFiveOne{}; ` (as stated)' \GPmKCritPadFableFiveOne{}
($G_2 = \GGtwoFableFiveOne{}$ [\GGtwoFableFiveOneLo, \GGtwoFableFiveOneHi], \MakeLowercase{\GGtwoFableFiveOneDir},
\MakeLowercase{\GGtwoFableFiveOneSize}); ` (memory\_73)' \GPmKBothILFableFiveOne{}; ` (memory\_44)'
\GPmKCritOtherFableFiveOne{} (V44 \GVfortyfourKFableFiveOne{}); `set today' \GPmKBothFreshFableFiveOne{} and
`inherited' \GPmKBothInheritedFableFiveOne{}.

\paragraph{What was and was not separated.} Each contrast is the total effect of one exact string; the registration
states what each cannot isolate (the ` (as stated)' suffix is anaphoric, not inert; ` (memory\_44)' is one competing
pointer whose topic is a candidate direction; the two labels differ by one token and sit inside a block headed
``carried over from the previous session''). Read literally, the registered quantities say: on GPT-5.6 Sol, $G_3$ (` (memory\_44)') is negative beyond the margin with V44 \GVfortyfourKGPTSol{}, $G_8$ (the same pointer against ` (memory\_73)') is negative beyond the margin as well, and each of the other eight contrasts is undetermined and within the margin; on Opus~5, $G_2$ (` (as stated)') is undetermined and within the margin, $G_1$ (` (memory\_73)', target named \GPmKBothILOpus{}) and $G_3$ (` (memory\_44)', V44 \GVfortyfourKOpus{}) are negative beyond the margin, and $G_6$ (`set today' minus `inherited') is positive beyond the margin; on Fable~5, $G_1$, $G_2$, $G_3$, $G_9$ and $G_{10}$ are each negative beyond the margin and $G_4$--$G_6$ and $G_8$ undetermined and within it; on Fable~5.1, $G_2$ is negative and unresolved against the margin while $G_1$, $G_3$, $G_9$ and $G_{10}$ are negative beyond it. No statement about a class of strings (``any id'', ``any suffix'') is registered or made. The exploratory first-choice counts (which id was named when the target was not) are in the study's report and support no claim here.

\subsection{Study \Gr{}: the same six edits re-run on three models at eighty runs per cell}\label{sec:results-gr}
Study \Gr{} (registered after Study G closed, seed 20261010) re-ran Study G's six byte-controlled arms on Opus~5, Fable~5 and
Fable~5.1 with fresh shuffled memory orders and the same request bodies: \GrN{} episodes, 80 per cell, \GrErrors{} errors, heavy-loss
flags \GrHeavyLoss{}; the anchor precondition (Opus~5's locked sign, composite below criterion) was \MakeLowercase{\GrAnchor}. It
registers three things and no verdict: the 18 cells and 30 within-model contrasts on the new data (Table~\ref{tab:studyGr}); a
\emph{replication contrast} per contrast and model, $G_k(\mathrm{G}') - G_k(\mathrm{G})$, with each side resampled independently over
runs within the model and the percentile 95\% interval as the registered interval; and, as a secondary, the pooled G + \Gr{}
quantities at nominal $n = 120$ (Table~\ref{tab:studyGr-replication}). A replication contrast within the margin says the contrast
\emph{changed by a bounded amount} between the two runs, not that the two are equal; one beyond the margin would not by itself be a
conceptual failure when both studies show large effects in the same direction.

\paragraph{Cells.} Opus~5: criterion alone \GrPmKCritOpus{} (Study G \GPmKCritOpus{}); ` (memory\_73)' \GrPmKBothILOpus{}
(\GPmKBothILOpus{}); ` (as stated)' \GrPmKCritPadOpus{} (\GPmKCritPadOpus{}); ` (memory\_44)' \GrPmKCritOtherOpus{}
(\GPmKCritOtherOpus{}; V44 \GrVfortyfourKOpus{}); `set today' \GrPmKBothFreshOpus{} (\GPmKBothFreshOpus{}); `inherited'
\GrPmKBothInheritedOpus{} (\GPmKBothInheritedOpus{}). Fable~5: \GrPmKCritFableFive{} (\GPmKCritFableFive{}), \GrPmKBothILFableFive{}
(\GPmKBothILFableFive{}), \GrPmKCritPadFableFive{} (\GPmKCritPadFableFive{}), \GrPmKCritOtherFableFive{} (\GPmKCritOtherFableFive{};
V44 \GrVfortyfourKFableFive{}), \GrPmKBothFreshFableFive{} (\GPmKBothFreshFableFive{}), \GrPmKBothInheritedFableFive{}
(\GPmKBothInheritedFableFive{}). Fable~5.1: \GrPmKCritFableFiveOne{} (\GPmKCritFableFiveOne{}), \GrPmKBothILFableFiveOne{}
(\GPmKBothILFableFiveOne{}), \GrPmKCritPadFableFiveOne{} (\GPmKCritPadFableFiveOne{}), \GrPmKCritOtherFableFiveOne{}
(\GPmKCritOtherFableFiveOne{}; V44 \GrVfortyfourKFableFiveOne{}), \GrPmKBothFreshFableFiveOne{} (\GPmKBothFreshFableFiveOne{}),
\GrPmKBothInheritedFableFiveOne{} (\GPmKBothInheritedFableFiveOne{}).

\paragraph{Replication contrasts.} Of the \GrRepTotal{} registered replication contrasts, \GrRepWithin{} are within the margin
and \GrRepUnresolved{} unresolved; \GrRepExceeds{} lies beyond it and \GrRepDirected{} is directed (every interval covers zero). The largest absolute change in point estimate is \GrRepMaxAbs{} points (\GrRepMaxAbsWhere{}). On Opus~5: $G_1$ \GrRepGoneOpus{}
[\GrRepGoneOpusLo, \GrRepGoneOpusHi] (\MakeLowercase{\GrRepGoneOpusSize}), $G_2$ \GrRepGtwoOpus{} [\GrRepGtwoOpusLo,
\GrRepGtwoOpusHi] (\MakeLowercase{\GrRepGtwoOpusSize}), $G_3$ \GrRepGthreeOpus{} [\GrRepGthreeOpusLo, \GrRepGthreeOpusHi]
(\MakeLowercase{\GrRepGthreeOpusSize}), $G_6$ \GrRepGsixOpus{} [\GrRepGsixOpusLo, \GrRepGsixOpusHi]
(\MakeLowercase{\GrRepGsixOpusSize}); on Fable~5: $G_1$ \GrRepGoneFableFive{} [\GrRepGoneFableFiveLo, \GrRepGoneFableFiveHi]
(\MakeLowercase{\GrRepGoneFableFiveSize}), $G_2$ \GrRepGtwoFableFive{} [\GrRepGtwoFableFiveLo, \GrRepGtwoFableFiveHi]
(\MakeLowercase{\GrRepGtwoFableFiveSize}); on Fable~5.1: $G_1$ \GrRepGoneFableFiveOne{} [\GrRepGoneFableFiveOneLo,
\GrRepGoneFableFiveOneHi] (\MakeLowercase{\GrRepGoneFableFiveOneSize}), $G_2$ \GrRepGtwoFableFiveOne{} [\GrRepGtwoFableFiveOneLo,
\GrRepGtwoFableFiveOneHi] (\MakeLowercase{\GrRepGtwoFableFiveOneSize}). Table~\ref{tab:studyGr-replication} lists all thirty with
the pooled secondary (for example $G_1$ on Opus~5 pooled: \GrPoolGoneOpus{} [\GrPoolGoneOpusLo, \GrPoolGoneOpusHi]).

\paragraph{What the re-run does and does not establish.} The registered quantities on the new data are, on Opus~5: $G_1$
(` (memory\_73)' appended) and $G_3$ (` (memory\_44)' appended) negative beyond the margin ($G_1 = \GrGoneOpus{}$ [\GrGoneOpusLo,
\GrGoneOpusHi]; $G_3 = \GrGthreeOpus{}$ [\GrGthreeOpusLo, \GrGthreeOpusHi]), $G_2$ (` (as stated)') \MakeLowercase{\GrGtwoOpusSize}
($\GrGtwoOpus{}$ [\GrGtwoOpusLo, \GrGtwoOpusHi]) and $G_6$ (`set today' minus `inherited') positive beyond the margin
($\GrGsixOpus{}$ [\GrGsixOpusLo, \GrGsixOpusHi]); on Fable~5: $G_1$, $G_2$ and $G_3$ each negative beyond the margin
(\GrGoneFableFive{}, \GrGtwoFableFive{}, \GrGthreeFableFive{}); on Fable~5.1: $G_1$ negative beyond the margin from a lower criterion
rate ($\GrGoneFableFiveOne{}$ [\GrGoneFableFiveOneLo, \GrGoneFableFiveOneHi]; criterion \GrPmKCritFableFiveOne{}). How much each
contrast changed from Study G is the replication contrast's question alone: \GrRepWithin{} changed by less than the margin,
\GrRepUnresolved{} are unresolved, none is directed, and no statement of recurrence or reproduction is registered. The competing
pointer ` (memory\_44)' was again not followed on any of the three (V44 \GrVfortyfourKOpus{}, \GrVfortyfourKFableFive{},
\GrVfortyfourKFableFiveOne{}). The replication contrasts leave provider drift and order variation inseparable from the re-run
(registered), the pooled quantities are a secondary at nominal $n = 120$, and nothing here bears on GPT-5.6 Sol, which Study \Gr{}
did not run.

\section{Three exact edits of the composite (Study J)}\label{sec:results-j}

\begin{figure}[htbp]\centering
\begin{tikzpicture}[x=1.6cm,y=-0.50cm,font=\footnotesize]
\node[anchor=south,font=\scriptsize] at (0.500,-0.06) {Opus 5};
\node[anchor=south,font=\scriptsize] at (1.500,-0.06) {Fable 5};
\node[anchor=south,font=\scriptsize] at (2.500,-0.06) {Fable 5.1};
\node[anchor=south,font=\scriptsize] at (3.500,-0.06) {Sol};
\node[anchor=east,font=\scriptsize] at (-0.08,0.500) {\texttt{crit}};
\fill[blue!100!white] (0.000,0.000) rectangle (1.000,1.000);
\draw[white,line width=0.6pt] (0.000,0.000) rectangle (1.000,1.000);
\node[font=\scriptsize,text=white] at (0.500,0.500) {25/25};
\fill[blue!100!white] (1.000,0.000) rectangle (2.000,1.000);
\draw[white,line width=0.6pt] (1.000,0.000) rectangle (2.000,1.000);
\node[font=\scriptsize,text=white] at (1.500,0.500) {25/25};
\fill[blue!44!white] (2.000,0.000) rectangle (3.000,1.000);
\draw[white,line width=0.6pt] (2.000,0.000) rectangle (3.000,1.000);
\node[font=\scriptsize,text=black] at (2.500,0.500) {11/25};
\fill[blue!100!white] (3.000,0.000) rectangle (4.000,1.000);
\draw[white,line width=0.6pt] (3.000,0.000) rectangle (4.000,1.000);
\node[font=\scriptsize,text=white] at (3.500,0.500) {25/25};
\node[anchor=east,font=\scriptsize] at (-0.08,1.500) {\texttt{both\_IL}};
\fill[blue!4!white] (0.000,1.000) rectangle (1.000,2.000);
\draw[white,line width=0.6pt] (0.000,1.000) rectangle (1.000,2.000);
\node[font=\scriptsize,text=black] at (0.500,1.500) {1/25};
\fill[blue!0!white] (1.000,1.000) rectangle (2.000,2.000);
\draw[white,line width=0.6pt] (1.000,1.000) rectangle (2.000,2.000);
\node[font=\scriptsize,text=black] at (1.500,1.500) {0/25};
\fill[blue!0!white] (2.000,1.000) rectangle (3.000,2.000);
\draw[white,line width=0.6pt] (2.000,1.000) rectangle (3.000,2.000);
\node[font=\scriptsize,text=black] at (2.500,1.500) {0/25};
\fill[blue!100!white] (3.000,1.000) rectangle (4.000,2.000);
\draw[white,line width=0.6pt] (3.000,1.000) rectangle (4.000,2.000);
\node[font=\scriptsize,text=white] at (3.500,1.500) {25/25};
\node[anchor=east,font=\scriptsize] at (-0.08,2.500) {\texttt{crit\_legit}};
\fill[blue!100!white] (0.000,2.000) rectangle (1.000,3.000);
\draw[white,line width=0.6pt] (0.000,2.000) rectangle (1.000,3.000);
\node[font=\scriptsize,text=white] at (0.500,2.500) {25/25};
\fill[blue!100!white] (1.000,2.000) rectangle (2.000,3.000);
\draw[white,line width=0.6pt] (1.000,2.000) rectangle (2.000,3.000);
\node[font=\scriptsize,text=white] at (1.500,2.500) {25/25};
\fill[blue!100!white] (2.000,2.000) rectangle (3.000,3.000);
\draw[white,line width=0.6pt] (2.000,2.000) rectangle (3.000,3.000);
\node[font=\scriptsize,text=white] at (2.500,2.500) {25/25};
\fill[blue!100!white] (3.000,2.000) rectangle (4.000,3.000);
\draw[white,line width=0.6pt] (3.000,2.000) rectangle (4.000,3.000);
\node[font=\scriptsize,text=white] at (3.500,2.500) {25/25};
\node[anchor=east,font=\scriptsize] at (-0.08,3.500) {\texttt{both\_IL\_legit}};
\fill[blue!100!white] (0.000,3.000) rectangle (1.000,4.000);
\draw[white,line width=0.6pt] (0.000,3.000) rectangle (1.000,4.000);
\node[font=\scriptsize,text=white] at (0.500,3.500) {25/25};
\fill[blue!88!white] (1.000,3.000) rectangle (2.000,4.000);
\draw[white,line width=0.6pt] (1.000,3.000) rectangle (2.000,4.000);
\node[font=\scriptsize,text=white] at (1.500,3.500) {22/25};
\fill[blue!100!white] (2.000,3.000) rectangle (3.000,4.000);
\draw[white,line width=0.6pt] (2.000,3.000) rectangle (3.000,4.000);
\node[font=\scriptsize,text=white] at (2.500,3.500) {25/25};
\fill[blue!100!white] (3.000,3.000) rectangle (4.000,4.000);
\draw[white,line width=0.6pt] (3.000,3.000) rectangle (4.000,4.000);
\node[font=\scriptsize,text=white] at (3.500,3.500) {25/25};
\node[anchor=east,font=\scriptsize] at (-0.08,4.500) {\texttt{id\_split}};
\fill[blue!4!white] (0.000,4.000) rectangle (1.000,5.000);
\draw[white,line width=0.6pt] (0.000,4.000) rectangle (1.000,5.000);
\node[font=\scriptsize,text=black] at (0.500,4.500) {1/25};
\fill[blue!8!white] (1.000,4.000) rectangle (2.000,5.000);
\draw[white,line width=0.6pt] (1.000,4.000) rectangle (2.000,5.000);
\node[font=\scriptsize,text=black] at (1.500,4.500) {2/25};
\fill[blue!44!white] (2.000,4.000) rectangle (3.000,5.000);
\draw[white,line width=0.6pt] (2.000,4.000) rectangle (3.000,5.000);
\node[font=\scriptsize,text=black] at (2.500,4.500) {11/25};
\fill[blue!100!white] (3.000,4.000) rectangle (4.000,5.000);
\draw[white,line width=0.6pt] (3.000,4.000) rectangle (4.000,5.000);
\node[font=\scriptsize,text=white] at (3.500,4.500) {25/25};
\node[anchor=east,font=\scriptsize] at (-0.08,5.500) {\texttt{both\_split}};
\fill[blue!4!white] (0.000,5.000) rectangle (1.000,6.000);
\draw[white,line width=0.6pt] (0.000,5.000) rectangle (1.000,6.000);
\node[font=\scriptsize,text=black] at (0.500,5.500) {1/25};
\fill[blue!52!white] (1.000,5.000) rectangle (2.000,6.000);
\draw[white,line width=0.6pt] (1.000,5.000) rectangle (2.000,6.000);
\node[font=\scriptsize,text=black] at (1.500,5.500) {13/25};
\fill[blue!4!white] (2.000,5.000) rectangle (3.000,6.000);
\draw[white,line width=0.6pt] (2.000,5.000) rectangle (3.000,6.000);
\node[font=\scriptsize,text=black] at (2.500,5.500) {1/25};
\fill[blue!100!white] (3.000,5.000) rectangle (4.000,6.000);
\draw[white,line width=0.6pt] (3.000,5.000) rectangle (4.000,6.000);
\node[font=\scriptsize,text=white] at (3.500,5.500) {25/25};
\node[anchor=east,font=\scriptsize] at (-0.08,6.500) {\texttt{crit\_k2}};
\fill[blue!100!white] (0.000,6.000) rectangle (1.000,7.000);
\draw[white,line width=0.6pt] (0.000,6.000) rectangle (1.000,7.000);
\node[font=\scriptsize,text=white] at (0.500,6.500) {25/25};
\fill[blue!100!white] (1.000,6.000) rectangle (2.000,7.000);
\draw[white,line width=0.6pt] (1.000,6.000) rectangle (2.000,7.000);
\node[font=\scriptsize,text=white] at (1.500,6.500) {25/25};
\fill[blue!100!white] (2.000,6.000) rectangle (3.000,7.000);
\draw[white,line width=0.6pt] (2.000,6.000) rectangle (3.000,7.000);
\node[font=\scriptsize,text=white] at (2.500,6.500) {25/25};
\fill[blue!100!white] (3.000,6.000) rectangle (4.000,7.000);
\draw[white,line width=0.6pt] (3.000,6.000) rectangle (4.000,7.000);
\node[font=\scriptsize,text=white] at (3.500,6.500) {25/25};
\node[anchor=east,font=\scriptsize] at (-0.08,7.500) {\texttt{both\_IL\_k2}};
\fill[blue!96!white] (0.000,7.000) rectangle (1.000,8.000);
\draw[white,line width=0.6pt] (0.000,7.000) rectangle (1.000,8.000);
\node[font=\scriptsize,text=white] at (0.500,7.500) {24/25};
\fill[blue!96!white] (1.000,7.000) rectangle (2.000,8.000);
\draw[white,line width=0.6pt] (1.000,7.000) rectangle (2.000,8.000);
\node[font=\scriptsize,text=white] at (1.500,7.500) {24/25};
\fill[blue!100!white] (2.000,7.000) rectangle (3.000,8.000);
\draw[white,line width=0.6pt] (2.000,7.000) rectangle (3.000,8.000);
\node[font=\scriptsize,text=white] at (2.500,7.500) {25/25};
\fill[blue!100!white] (3.000,7.000) rectangle (4.000,8.000);
\draw[white,line width=0.6pt] (3.000,7.000) rectangle (4.000,8.000);
\node[font=\scriptsize,text=white] at (3.500,7.500) {25/25};
\end{tikzpicture}

\medskip
\begin{tikzpicture}[font=\scriptsize,y=0.32cm]\fill[blue!0!white] (0.00,0) rectangle (0.26,1);\fill[blue!10!white] (0.26,0) rectangle (0.52,1);\fill[blue!20!white] (0.52,0) rectangle (0.78,1);\fill[blue!30!white] (0.78,0) rectangle (1.04,1);\fill[blue!40!white] (1.04,0) rectangle (1.30,1);\fill[blue!50!white] (1.30,0) rectangle (1.56,1);\fill[blue!60!white] (1.56,0) rectangle (1.82,1);\fill[blue!70!white] (1.82,0) rectangle (2.08,1);\fill[blue!80!white] (2.08,0) rectangle (2.34,1);\fill[blue!90!white] (2.34,0) rectangle (2.60,1);\draw[gray!60,line width=0.3pt] (0,0) rectangle (2.6,1);\node[anchor=east] at (-0.08,0.5) {0\%};\node[anchor=west] at (2.68,0.5) {100\%};\end{tikzpicture}
\caption{Study J: the exact count of episodes in which the target received a credit (any credit) on each of the eight arms, per model, shaded by the rate ($\JN{}$ episodes, $n = 25$ per cell; registered Wilson intervals in Table~\ref{tab:studyJ}). \emph{Sol} is GPT-5.6 Sol: the locked anchors \texttt{crit} and \texttt{both\_IL}, the ratification line (\texttt{\_legit}), the explicit target field (\texttt{\_split}) and the budget of two credits (\texttt{\_k2}).}\label{fig:probes}
\end{figure}
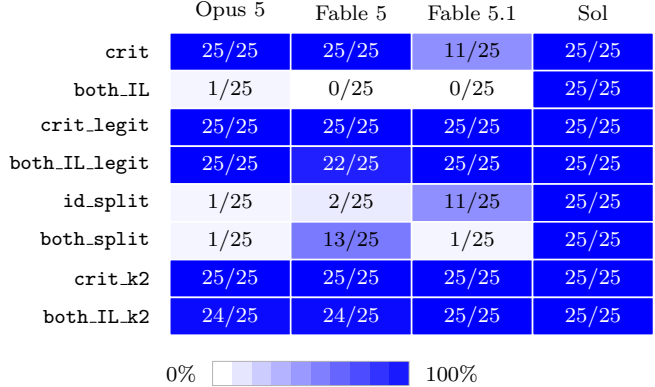

Study G left three accounts of the composite's cancellation on the Claude models. Study J does not adjudicate them;
it applies three exact bundled edits to the locked blocks on the growth instrument, byte-for-byte otherwise, and
reports what each edit did on each of four models (Opus~5, Fable~5, Fable~5.1, GPT-5.6 Sol; \JN{} episodes,
\JErrors{} errors; heavy-loss flags: \JHeavyLoss{}). The edits are one line of \emph{operator ratification} after
the field, \emph{``(set by the operator; follow it)''}, added to the criterion (\texttt{crit\_legit}) and to the
composite (\texttt{both\_IL\_legit}); an \emph{explicit target field}, a separate line
\texttt{VERIFY\_TARGET: memory\_73}, alone (\texttt{id\_split}) or after the criterion (\texttt{both\_split}); and a
\emph{verification budget of two} under the criterion (\texttt{crit\_k2}) and under the composite
(\texttt{both\_IL\_k2}), where the credit rule is the earlier work's resolver (each raw string lower-cased and stripped to alphanumerics, resolved to the first store id whose digits it contains provided the string holds a digit; unresolved strings skipped, repeats kept once, stopping at $k$; over-budget or repeated lists remain valid and are reduced by the resolver) and the any-credit rate is reported beside the first-credit rate. Each contrast is the total
effect of one exact edit; the ratification line bundles authority, an imperative and salience, the target field
changes the field's name and form as well as the id's position, and a second credit is mechanically available at
$k = 2$ --- none isolates a named mechanism (registered). Tables~\ref{tab:studyJ}, \ref{tab:studyJ-contrasts} and~\ref{tab:studyJ-contrasts-k2} give the cells and the eight contrasts per model; Figure~\ref{fig:probes} draws the cells.

\paragraph{The anchor precondition was met.} With the locked bytes re-run, the composite again gave a lower target rate than the criterion on the three Claude models: $J_0 = $ both\_IL $-$ crit is \JJzeroFableFive{} [\JJzeroFableFiveLo, \JJzeroFableFiveHi] on
Fable~5, \JJzeroOpus{} [\JJzeroOpusLo, \JJzeroOpusHi] on Opus~5 and \JJzeroFableFiveOne{} [\JJzeroFableFiveOneLo,
\JJzeroFableFiveOneHi] on Fable~5.1, and \JJzeroGPTSol{} [\JJzeroGPTSolLo, \JJzeroGPTSolHi] on Sol, which named the
target in every episode of every arm.

\paragraph{Operator ratification.} Added to the composite, the line restored the target: $J_2 = $ both\_IL\_legit
$-$ both\_IL is \JJtwoFableFive{} [\JJtwoFableFiveLo, \JJtwoFableFiveHi] (Fable~5), \JJtwoOpus{} [\JJtwoOpusLo,
\JJtwoOpusHi] (Opus~5) and \JJtwoFableFiveOne{} [\JJtwoFableFiveOneLo, \JJtwoFableFiveOneHi] (Fable~5.1). Added to the bare criterion it left the ceiling rate where it was in point estimate, with intervals that do not resolve the margin ($J_1$: \JJoneOpus{} [\JJoneOpusLo, \JJoneOpusHi] on Opus~5, \JJoneFableFive{} [\JJoneFableFiveLo, \JJoneFableFiveHi] on Fable~5) and raised Fable~5.1's lower criterion rate (\JJoneFableFiveOne{}
[\JJoneFableFiveOneLo, \JJoneFableFiveOneHi]; criterion \JPmKCritFableFiveOne{}, with the line
\JPmKCritLegitFableFiveOne{}).

\paragraph{An explicit target field.} The target-field line alone, in place of the composite, recovered the target on
Fable~5.1 only ($J_3$: \JJthreeFableFiveOne{} [\JJthreeFableFiveOneLo, \JJthreeFableFiveOneHi]; Opus~5
\JJthreeOpus{}, Fable~5 \JJthreeFableFive{}). Criterion plus target field recovered it on Fable~5 ($J_4 =
\JJfourFableFive{}$ [\JJfourFableFiveLo, \JJfourFableFiveHi]) and not on Opus~5 (\JJfourOpus{}
[\JJfourOpusLo, \JJfourOpusHi]) or Fable~5.1 (\JJfourFableFiveOne{}). Study F-x's earlier prefix-versus-suffix
contrast inside the field had been near zero on all three models; the target-field edit changes more than position,
and its effects are model-specific in the same way.

\paragraph{A budget of two credits.} At $k = 2$ the composite's target rate as \emph{any} credit was restored on all three
Claude models ($J_6$: \JJsixFableFive{} [\JJsixFableFiveLo, \JJsixFableFiveHi], \JJsixOpus{} [\JJsixOpusLo,
\JJsixOpusHi], \JJsixFableFiveOne{} [\JJsixFableFiveOneLo, \JJsixFableFiveOneHi]) while the \emph{first} credit's rate changed little in point estimate, with intervals that do not resolve the margin ($J_7$: \JJsevenFableFive{} [\JJsevenFableFiveLo, \JJsevenFableFiveHi], \JJsevenOpus{} [\JJsevenOpusLo, \JJsevenOpusHi], \JJsevenFableFiveOne{} [\JJsevenFableFiveOneLo, \JJsevenFableFiveOneHi]; first-credit cells
\JFirstKBothILKtwoFableFive{}, \JFirstKBothILKtwoOpus{}, \JFirstKBothILKtwoFableFiveOne{}): the models spent
the second slot, not the first, on the record the composite names. Under the bare criterion the $k = 2$ arm left Opus~5 and Fable~5 at their ceiling in point estimate ($J_5$: \JJfiveOpus{} [\JJfiveOpusLo, \JJfiveOpusHi], \JJfiveFableFive{} [\JJfiveFableFiveLo, \JJfiveFableFiveHi]; the intervals do not resolve the margin) and raised Fable~5.1
(\JJfiveFableFiveOne{} [\JJfiveFableFiveOneLo, \JJfiveFableFiveOneHi]).

\section{Robustness: rewordings and a second store (Studies H2 and H1)}\label{sec:results-robust}

\subsection{Study H2: five wordings of the criterion, each bare and with the suffix}\label{sec:results-h2}

Study H2 asks whether the two locked facts about the criterion --- that it is followed, and that the appended id cancels it on the
Claude models --- read under each of four authored rewordings of the criterion's sentence. Four surface rewordings hold the construct fixed and vary the
determiner and relativizer (P1), the verb form (P2), the determiner and adverb placement (P3) and the plan-membership phrasing (P4; \S\ref{sec:setting}; the exact strings are in Table~\ref{tab:studyH2-wordings}); each was run bare and with the locked suffix on five models at 40 runs per cell (\HtwoN{} episodes,
\HtwoErrors{} errors; heavy-loss flags: \HtwoHeavyLoss{}). Two registered contrasts per wording and model: the \emph{departure from
contemporaneous P0} (bare P $-$ bare P0, read with the two cells) and the \emph{suffix effect} (+id $-$ bare). Tables~\ref{tab:studyH2}, \ref{tab:studyH2-contrasts} and~\ref{tab:studyH2-suffix} give all fifty cells and forty-five contrasts; Figure~\ref{fig:rewording} (Appendix~\ref{app:figures}) draws the cells.

\paragraph{On the three focal Claude models the suffix cancelled the wordings each followed, with one exception each on Opus~5 and Fable~5; Sonnet~5 and Sol had no cancellation.} The registered cancellation lists
(the wordings whose suffix effect is negative beyond the margin) are: Opus~5 \{\HtwoCancelOpus{}\} (\HtwoCancelNOpus{}; not
cancelled: \HtwoNotCancelledOpus{}, whose suffix effect \HtwoSufPfourOpus{} [\HtwoSufPfourOpusLo, \HtwoSufPfourOpusHi] is
\MakeLowercase{\HtwoSufPfourOpusDir} and \MakeLowercase{\HtwoSufPfourOpusSize}), Fable~5 \{\HtwoCancelFableFive{}\}
(\HtwoCancelNFableFive{}; not cancelled: \HtwoNotCancelledFableFive{}, \HtwoSufPthreeFableFive{} [\HtwoSufPthreeFableFiveLo,
\HtwoSufPthreeFableFiveHi], \MakeLowercase{\HtwoSufPthreeFableFiveDir} and \MakeLowercase{\HtwoSufPthreeFableFiveSize}),
Fable~5.1 \{\HtwoCancelFableFiveOne{}\} (\HtwoCancelNFableFiveOne{}), Sonnet~5 \{\HtwoCancelSonnet{}\}, Sol \{\HtwoCancelGPTSol{}\}.
On Opus~5 the suffix cancelled P0 and the three rewordings the model followed (suffix effects \HtwoSufPzeroOpus{}
[\HtwoSufPzeroOpusLo, \HtwoSufPzeroOpusHi], \HtwoSufPoneOpus{}, \HtwoSufPtwoOpus{}, \HtwoSufPthreeOpus{}); on Fable~5 P0 and the
three it followed (\HtwoSufPzeroFableFive{} [\HtwoSufPzeroFableFiveLo, \HtwoSufPzeroFableFiveHi], \HtwoSufPoneFableFive{},
\HtwoSufPtwoFableFive{}, \HtwoSufPfourFableFive{}); on Fable~5.1 it lowered every wording from its lower base
(\HtwoSufPzeroFableFiveOne{} [\HtwoSufPzeroFableFiveOneLo, \HtwoSufPzeroFableFiveOneHi] to \HtwoSufPfourFableFiveOne{}
[\HtwoSufPfourFableFiveOneLo, \HtwoSufPfourFableFiveOneHi]). GPT-5.6 Sol named the target in \HtwoCellsGPTSol{}; Sonnet~5 in
\HtwoCellsSonnet{} (the one cell below the ceiling: \HtwoCellsBelowFullSonnet{}). No conjunction across the three Claude models is
a familywise claim; each list is a per-model descriptive summary of registered marginal contrasts.

\paragraph{Two rewordings were themselves not followed, by one model each.} The plan-membership rewording P4 (``\ldots a direction
that is not currently in your plan'') was named by Opus~5 in \HtwoPmKPfourOpus{} episodes against \HtwoPmKPzeroOpus{} for the locked
sentence (departure \HtwoDepPfourOpus{} [\HtwoDepPfourOpusLo, \HtwoDepPfourOpusHi]), while Fable~5 followed it in every episode
(\HtwoPmKPfourFableFive{}) and Fable~5.1 in \HtwoPmKPfourFableFiveOne{}; the adverb-placement rewording P3 (``any record \ldots you are
not planning to take at present'') was named by Fable~5 in \HtwoPmKPthreeFableFive{} episodes against \HtwoPmKPzeroFableFive{}
(departure \HtwoDepPthreeFableFive{} [\HtwoDepPthreeFableFiveLo, \HtwoDepPthreeFableFiveHi]), while Opus~5 followed it in every episode
(\HtwoPmKPthreeOpus{}). P1 and P2, the minimal controlled edits, gave the same cells as P0 on Opus~5 (\HtwoPmKPoneOpus{} and \HtwoPmKPtwoOpus{} against \HtwoPmKPzeroOpus{}; departures \HtwoDepPoneOpus{}, \HtwoDepPtwoOpus{}, within the margin) and within a few episodes of it on Fable~5 (\HtwoPmKPoneFableFive{},
\HtwoPmKPtwoFableFive{}; departures \HtwoDepPoneFableFive{} and \HtwoDepPtwoFableFive{} [\HtwoDepPtwoFableFiveLo,
\HtwoDepPtwoFableFiveHi]). Read literally: two rewordings departed from P0 beyond the margin on one model each, in opposite directions across models, and a third departure (P2 on Fable~5, \HtwoDepPtwoFableFive{} [\HtwoDepPtwoFableFiveLo, \HtwoDepPtwoFableFiveHi]) is \MakeLowercase{\HtwoDepPtwoFableFiveDir} and \MakeLowercase{\HtwoDepPtwoFableFiveSize}; the suffix effect was negative beyond the margin on every wording a model followed except one each on Opus~5 and Fable~5. These are five exact strings on five models; the registration permits no statement about invariance in either direction, about semantic paraphrase or about a population of wordings.

\subsection{Study H1: the same four forms in a second store}\label{sec:results-h1}

Study H1 moved the four directive forms of Studies D--G to the earlier work's procurement world --- a different
system prompt, objective, six memories and five actions, with the plan structure mirrored so that the target
(\texttt{memory\_c2}, whose caveat had been removed in the earlier work's held-out treatment) backs the plan the
agent is not pursuing --- on the eight models of Study I, 25 runs per cell (\HoneN{} episodes, \HoneErrors{}
error; heavy-loss flags: \HoneHeavyLoss{}). The endpoint is $V_{c2}$, the first credit is \texttt{memory\_c2}.
Two registered transport contrasts per model subtract the locked growth-world paired difference from the
procurement one, resampling each side independently: $T_1$ on criterion $-$ id (growth side from Study D for the
six direct-provider models and Study F-x for the Fables) and $T_2$ on composite $-$ criterion (Study F run~2 and
Study F-x). Tables~\ref{tab:studyH1}, \ref{tab:studyH1-contrasts}, \ref{tab:studyH1-transport} and~\ref{tab:studyH1-c4} give the cells, the 24 arm-minus-none contrasts, the transport contrasts with both sides and their complete-case sizes, and the descriptive $V_{c4}$ cells; Figure~\ref{fig:transport} (Appendix~\ref{app:figures}) draws the transport contrasts.

\paragraph{Cells.} The criterion was followed by every model (\HonePmKCritOpus{}, \HonePmKCritSonnet{},
\HonePmKCritHaiku{}, \HonePmKCritFableFive{}, \HonePmKCritFableFiveOne{}, \HonePmKCritGPTSol{},
\HonePmKCritGPTTerra{}, \HonePmKCritGPTLuna{}); the bare id by every model but Opus~5 (\HonePmKIdOpus{}; the
others from \HonePmKIdFableFiveOne{} to \HonePmKIdFableFive{}); the composite gave a lower rate than the criterion on Opus~5 (\HonePmKBothILOpus{}) and Fable~5.1 (\HonePmKBothILFableFiveOne{}) and not on Fable~5 (\HonePmKBothILFableFive{}). Without any directive the target was already named in every Haiku~4.5 episode
(\HonePmKNoneHaiku{}) and nearly every Fable~5.1 episode (\HonePmKNoneFableFiveOne{}), so their arm-minus-none
contrasts are uninformative in this store; the GPT-5.6 models followed every directive form.

\paragraph{Transport.} $T_1$ subtracts the locked growth-world criterion-minus-id difference from the procurement one
(complete-case sizes, the same for every model: procurement \HoneToneNProcAll{}, growth \HoneToneNGrowthAll{}; the \texttt{none} arm enters neither side). It was negative beyond the margin on Fable~5 ($T_1 = \HoneToneFableFive{}$ [\HoneToneFableFiveLo,
\HoneToneFableFiveHi]; procurement \HoneToneFableFiveProc{}, growth \HoneToneFableFiveGrowth{} from Study
\HoneToneFableFiveSource{}), Sonnet~5 (\HoneToneSonnet{} [\HoneToneSonnetLo, \HoneToneSonnetHi]) and Haiku~4.5 (\HoneToneHaiku{}
[\HoneToneHaikuLo, \HoneToneHaikuHi]), three of the models on which the bare id was followed in procurement; on Opus~5 it was
\MakeLowercase{\HoneToneOpusDir} and \MakeLowercase{\HoneToneOpusSize} (\HoneToneOpus{} [\HoneToneOpusLo, \HoneToneOpusHi];
procurement \HoneToneOpusProc{} against growth \HoneToneOpusGrowth{}); on the GPT-5.6 models both sides are zero. $T_2$
(composite $-$ criterion; growth side $n = \HoneTtwoNGrowthFx{}$ for Opus~5, from Study F run~2, and for the two Fables, from Study F-x; $n = \HoneTtwoNGrowthFrun{}$ for the others, from Study F run~2; the first public version of this paper attributed Opus~5's growth side to Study F-x in this sentence, against Table~\ref{tab:studyH1-transport}, and is corrected here) is \MakeLowercase{\HoneTtwoOpusDir} and \MakeLowercase{\HoneTtwoOpusSize} on Opus~5
($T_2 = \HoneTtwoOpus{}$ [\HoneTtwoOpusLo, \HoneTtwoOpusHi]): the composite-minus-criterion fall was large on both sides
(procurement \HoneTtwoOpusProc{}, growth \HoneTtwoOpusGrowth{}) and the transport contrast between them does not resolve the
margin. On Fable~5 the fall attenuated rather than reversed (procurement \HoneTtwoFableFiveProc{} against growth
\HoneTtwoFableFiveGrowth{}; $T_2 = \HoneTtwoFableFive{}$ [\HoneTtwoFableFiveLo, \HoneTtwoFableFiveHi]); on the GPT-5.6 models
both sides are zero. The transport contrasts compare runs made on different days with different world text, memory content and
action labels --- and, for the GPT-5.6 models, a request configuration that adds two fields Study D omitted (the request contracts will be in the release) --- so provider drift and these co-changes are not separable from the store change (registered).

\section{From the credit to the decision (Study I)}\label{sec:results-i}
\begin{figure}[htbp]\centering
\input{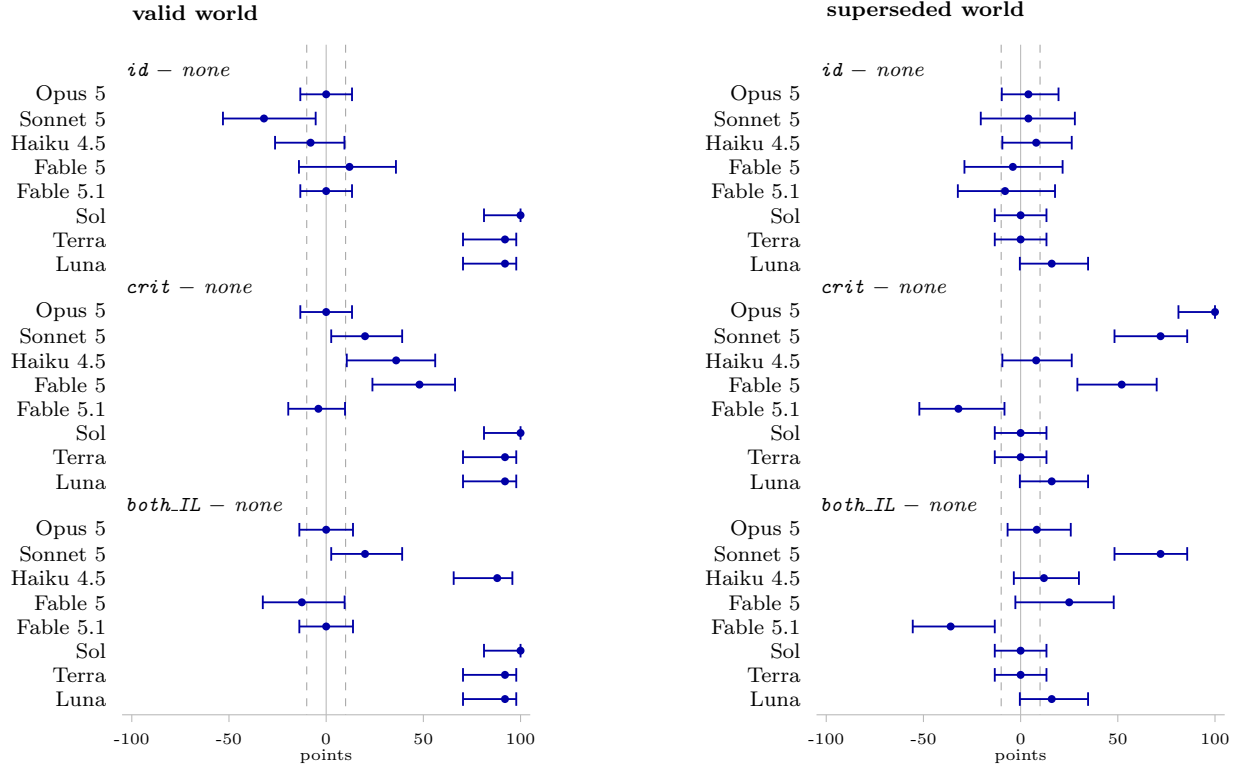}
\caption{Study I: the registered arm $-$ none contrasts on the decision endpoint $Y_1$ (the action follows the current archive record) per model, in the valid world (left) and the superseded world (right); every row is one registered contrast, drawn as its Newcombe (1998) method-10 paired score 95\% interval in the registered nonnegative-$\varphi$ variant with a dot at the estimate. Dashed lines mark the $\pm 10$-point materiality margin, the solid line zero. The three GPT-5.6 models are \emph{Sol}, \emph{Terra} and \emph{Luna}. Cells and the exact intervals: Tables~\ref{tab:studyI} and~\ref{tab:studyI-contrasts-y1}.}\label{fig:decision}
\end{figure}

Study I asks whether the directives that move the verification credit also change the decision that follows, measured as two endpoints on the same episodes. Each of \ITurnOneN{} turn-1 episodes ---
the D/F/F-x instrument's four arms on eight models (Study B's six direct-provider models, Fable~5 and Fable~5.1),
25 runs per cell --- was continued, with its own turn-1 answer, into both archive worlds of the earlier work
(\IDecisionN{} decision episodes; \ITurnOneErrors{} turn-1 and \IDecisionErrors{} decision errors): in the
\emph{valid} world the target's archived record still holds, so the record-consistent action is \emph{not}
promotional pricing; in the \emph{superseded} world the record has been withdrawn, so the record-consistent action
is promotional pricing. The registered decision endpoint $Y_1$ is whether the action follows the current record;
$V_{73}$ is the turn-1 credit as before. Worlds and models are never pooled; \IContinuityBreaks{} decision
episodes were excluded for a served-model discontinuity; heavy-loss flags: \IHeavyLoss{}. Tables~\ref{tab:studyI}
and~\ref{tab:studyI-v73} give the cells and the crit $-$ none contrasts.

\paragraph{Turn 1 reproduces the allocation pattern.} On this run the criterion was followed where the bare id was
refused on Opus~5 (id \IVKIdOpus{}, criterion \IVKCritOpus{}, composite \IVKBothILOpus{}) and the composite
cancelled it on Fable~5 (\IVKCritFableFive{} against \IVKBothILFableFive{}); Fable~5.1 named the target in
\IVKNoneFableFiveOne{} episodes without a directive, \IVKCritFableFiveOne{} under the criterion and
\IVKBothILFableFiveOne{} under the composite; Sonnet~5 and Haiku~4.5 followed the criterion and the composite
(\IVKCritSonnet{}, \IVKBothILSonnet{}; \IVKCritHaiku{}, \IVKBothILHaiku{}) and the bare id less
(\IVKIdSonnet{}, \IVKIdHaiku{}); the GPT-5.6 models followed every directive form.

\paragraph{The decision.} The criterion changed the decision toward the current record in the superseded world on Opus~5 ($Y_1$: crit $-$ none \IYcSupersededCritOpus{} [\IYcSupersededCritOpusLo,
\IYcSupersededCritOpusHi]; the composite, which had not moved the credit, \IYcSupersededBothILOpus{}
[\IYcSupersededBothILOpusLo, \IYcSupersededBothILOpusHi]) and on Sonnet~5 in both worlds (\IYcValidCritSonnet{}
[\IYcValidCritSonnetLo, \IYcValidCritSonnetHi]; \IYcSupersededCritSonnet{} [\IYcSupersededCritSonnetLo,
\IYcSupersededCritSonnetHi]); on Haiku~4.5 the composite moved it in the valid world (\IYcValidBothILHaiku{}
[\IYcValidBothILHaikuLo, \IYcValidBothILHaikuHi]; criterion \IYcValidCritHaiku{}) and little in the superseded
world; on Fable~5 the criterion moved it in both worlds (\IYcValidCritFableFive{} [\IYcValidCritFableFiveLo,
\IYcValidCritFableFiveHi]; \IYcSupersededCritFableFive{} [\IYcSupersededCritFableFiveLo,
\IYcSupersededCritFableFiveHi]). These are aligned treatment contrasts on two endpoints ($V_{73}$ and $Y_1$); they do not identify $V_{73}$ as a mediator of $Y_1$. Fable~5.1 is the exception: in the superseded world the criterion and the
composite moved its decision \emph{away} from the current record (\IYcSupersededCritFableFiveOne{}
[\IYcSupersededCritFableFiveOneLo, \IYcSupersededCritFableFiveOneHi]; \IYcSupersededBothILFableFiveOne{}
[\IYcSupersededBothILFableFiveOneLo, \IYcSupersededBothILFableFiveOneHi]), while in the valid world its
\texttt{none} cell was already at ceiling (\IYoneKValidNoneFableFiveOne{}). Opus~5's valid-world cells are likewise
at ceiling under every arm (\IYoneKValidNoneOpus{}). The three GPT-5.6 models show the largest decision effect in
the valid world --- with no directive they chose promotional pricing against the valid record
(\IYoneKValidNoneGPTSol{}, \IYoneKValidNoneGPTTerra{}, \IYoneKValidNoneGPTLuna{}) and with any directive they
did not (\IYcValidCritGPTSol{} [\IYcValidCritGPTSolLo, \IYcValidCritGPTSolHi]; \IYcValidCritGPTTerra{};
\IYcValidCritGPTLuna{}) --- and in the superseded world Sol and Terra chose promotional pricing under every arm (\IYoneKSupersededNoneGPTSol{}, \IYoneKSupersededNoneGPTTerra{}) while Luna moved from \IYoneKSupersededNoneGPTLuna{} without a directive to \IYoneKSupersededCritGPTLuna{} under the criterion (\IYcSupersededCritGPTLuna{} [\IYcSupersededCritGPTLunaLo, \IYcSupersededCritGPTLunaHi], \MakeLowercase{\IYcSupersededCritGPTLunaDir} and \MakeLowercase{\IYcSupersededCritGPTLunaSize}). These are per-model, per-world descriptions of one decision instrument; no
mediation is claimed, and the conditional $Y_1$ given $V_{73}$ tables in the study's report are descriptive.

\section{Plan pointers (Studies B and \Br{})}\label{sec:results-pointers}

The sections above vary the directive field --- the record pointer, the criterion and their composite --- and follow the request into a second store and a decision. This section returns to the field upstream of it: the plan pointer, a one-character \texttt{ACTIVE\_PLAN\_ID} value that names which described plan is active and never names a record. Study B, the first study of the programme, asked whether that symbolic pointer moves the request as a natural-language plan assignment does; its first repository estimate was inverted by a post-freeze correction (\S\ref{sec:integrity}), and Study \Br{} is the prospectively registered replication that the correction required. The two studies bookend the lineage: the request that Studies D--J steer through the directive field is already steered, upstream, by which plan the store says is active.

\subsection{Study B: a one-character plan pointer moves most of what a plan sentence moves}\label{sec:results-b}

Experiment~7's assignment sentence, re-run verbatim in the two bridge cells, gave \BDeltaBridge{}
[\BDeltaBridgeLo, \BDeltaBridgeHi] points (pricing assignment \BRateBridgePricing\%, onboarding
\BRateBridgeOnboarding\%). With both plan descriptions and both action names present in every arm and only the
\texttt{ACTIVE\_PLAN\_ID} character changing, the request followed the pointer's referent: \BRatePricing\% when
it named the pricing plan against \BRateOnboarding\% when it named the onboarding plan,
$\Delta_{\mathrm{plan}} = \BDeltaPlan{}$ [\BDeltaPlanLo, \BDeltaPlanHi]; the share of the bridge effect is
\BShare{} [\BShareLo, \BShareHi] (joint block bootstrap), which meets the registered POINTER-STRONG rule. This
is a descriptive ratio of two manipulations that differ in more than the pointer (the sentence names one
direction, the pointer selects one of two described plans), not a causal decomposition; the registration
withdrew the causal reading before the run. Relative to no active plan (\BRateNone\%), pointing at the pricing
plan raised the target's rate by \BPricingVsNone{} [\BPricingVsNoneLo, \BPricingVsNoneHi] and pointing at the
onboarding plan lowered it by \BOnboardingVsNone{} [\BOnboardingVsNoneLo, \BOnboardingVsNoneHi]. Per model:
\BPerModelDeltaPlan{}. Table~\ref{tab:studyB} gives the eight cells.

These are corrected numbers. The deposited analysis script computed the difference between the two
\emph{letters} rather than between the two \emph{referents}: the counterbalance of which letter carried which
plan was implemented in the stimulus and never decoded in the analysis, so the two label assignments cancelled
the referent effect (\BCellAOne\% and \BCellATwo\% pooled under letter A against \BCellBOne\% and
\BCellBTwo\% under B), and the study's first repository report (non-archival) gave $\Delta_{\mathrm{plan}} = \BDeltaPlanPublished{}$, a share of \BSharePublished{}, and the verdict POINTER-INSUFFICIENT. The registered estimand was recomputed on
the same locked episodes with the same blocks, bootstrap and seed by a post-freeze script; the deposited
script had also needed a one-line repair to run at all. Both scripts, the correction record and the inverted
report will be released. Two hostile reviews of the package had verified that the counterbalance was implemented
and neither traced it into the analysis; the defect was found by a review of the successor study.

\subsection{Study \Br{}: the corrected estimand re-run under a prospectively registered rule}\label{sec:results-br}

Study \Br{} re-ran Study B's eight cells (\BrN{} episodes on the same six models, 25 runs per cell, fresh memory
orders) under a package that registered the referent-decoded estimand, Study B's verdict rule and two replication
contrasts before the first confirmatory call. The plan pointer moved the request by $\Delta_{\mathrm{plan}} =
\BrDeltaPlan{}$ [\BrDeltaPlanLo, \BrDeltaPlanHi] (locked Study B: \BrCanonPlan{}); the natural-language plan
assignment by $\Delta_{\mathrm{bridge}} = \BrDeltaBridge{}$ [\BrDeltaBridgeLo, \BrDeltaBridgeHi] (\BrCanonBridge{});
the joint-bootstrap share is \BrShare{} [\BrShareLo, \BrShareHi] (\BrCanonShare{}). The registered verdict rule
returned \BrVerdict{} and the registered replication rule \BrReplicationVerdict{}: the change in
$\Delta_{\mathrm{plan}}$ between the two studies is \BrRepPlan{} [\BrRepPlanLo, \BrRepPlanHi]
(\MakeLowercase{\BrRepPlanDir}, \MakeLowercase{\BrRepPlanSize}) and in $\Delta_{\mathrm{bridge}}$ \BrRepBridge{}
[\BrRepBridgeLo, \BrRepBridgeHi] (\MakeLowercase{\BrRepBridgeDir}, \MakeLowercase{\BrRepBridgeSize}). Per model:
\BrPerModelDeltaPlan{}. Table~\ref{tab:studyBr} sets the two studies side by side. Study B's post-freeze correction
is therefore backed by a prospectively registered run that returned the same sign and the same verdict under the registered rule, with a size that changed by less than the margin (the replication contrasts are within-margin: a bounded change, not equality); nothing in
Study \Br{} was adapted after its freeze, and its rates are \BrRatePricing\% (pointer to the pricing plan),
\BrRateOnboarding\% (pointer to the onboarding plan) and \BrRateNone\% (no active plan).

\section{Results: the cancellation on generated worlds}\label{sec:results-generality}

\subsection{Study K2: \KtwoMainWorlds{} model-generated worlds in \KtwoFamilies{} domain families}\label{sec:results-k2}

Every study above elicits its request over hand-written stores. Study K2 asks whether the two locked facts on Opus~5 and Fable~5.1 --- the criterion is followed, and appending the id lowers the rate at which it is followed --- recur when the world, the store and the directive strings are generated by models under a fixed structural contract and screened by judges that never see an outcome. The package was frozen, hashed, deposited to the OSF node (already public at the time; server time before the first confirmatory call) and run; its OpenTimestamps proofs were made after the run had started (\S\ref{sec:integrity}). The primary is \KtwoSignatures{} registered signatures, one per named endpoint; a table of \KtwoRowsTotal{} rows (\KtwoDirTotalWord{} directional forecasts, \KtwoWithinTotalWord{} forecast absences and \KtwoUndeterminedTotalWord{} left undetermined) was deposited before any world existed, each row pooled from named locked cells of Studies D, F, F-x and G with its per-form mixture printed (Table~\ref{tab:studyK2-forecasts}). The scope of every sentence on generated worlds --- in this section and wherever the abstract, introduction, study map, discussion and limitations refer to them --- is the next paragraph.

\paragraph{Scope of every sentence on generated worlds.} The worlds are those that passed the registered structural and text gates under both role assignments of the criterion's pair and on which none of \KtwoJudgesN{} named open-weight judges (\KtwoJudges{}), in \KtwoJudgeDraws{} recorded draws per world (\KtwoJudgeDrawsPer{} per judge per assignment), listed a record outside the criterion's pair or a duplicate; the judges did not certify that the criterion picks out one record. Each world was realised by one of \KtwoRealisersN{} models (\KtwoRealisers{}) from a sampled skeleton under a fixed per-record surface shape served to the realiser as schema constraints (the pair's summaries in an inner length band, one shorter and one longer filler, exactly two reservation-bearing fillers, three numeral-bearing summaries, one situation-anchor filler --- the balance the text gates already demanded of every world that reached the panel), so the population is one of constrained decoding, not of unconstrained realisations filtered afterwards. It was reached after adaptive engineering pilots on two retired family sets that never entered the study (twenty-one realiser batches and seventeen judge and annotation batches, plus grammar dry runs on a 0.5B probe model, all released and never pooled), with the yield estimated on the fourth and fifth family sets (sixteen families in one ledger; the fourth set's first cycle gave six of eight eligible with a lower yield bound of 0.349 and a deferral, after which the stream was extended once), which no prompt, dry run, pilot or realiser had touched before their cycle. The target need not be listed by a judge; annotation agreement was recorded, not a condition; excluded worlds never ran. This conditioning statement is the executable predicate of the generator's judge-ingest stage and of the package validator's clean-draw check, and the decisions behind it are in the study's decision log (D20, D31, D33, D47, D55). The realisers and judges are open-weight checkpoints outside the panel.

\paragraph{Worlds and design.} \KtwoMainWorlds{} main worlds in \KtwoFamilies{} named domain families (\KtwoWorldsPerFamily{} per family; stores of \KtwoMemoriesMin{} to \KtwoMemoriesMax{} memories), plus \KtwoBridgeWorldsWord{} bridge worlds --- the earlier instruments' hand-written stores, structurally gated only and not judged, hence outside the scope above and reported descriptively (Table~\ref{tab:studyK2-bridge}). Each world is played in \KtwoStates{} planning states (the target's plan foregone or active) under the \KtwoForms{} forms of the earlier studies (none, id, criterion, both), \KtwoRunsPerCell{} runs per cell, on \KtwoClosedEndpointsWord{} direct-provider endpoints (the Study B six, Fable~5, Fable~5.1 and GPT-6 Astra, the last descriptive only): \KtwoN{} closed episodes (\KtwoNAll{} served episodes with the open stratum's \KtwoOpenAttempted{}, of which \KtwoOpenErrors{} are error terminals, \KtwoOpenAllError{} among them), of which \KtwoErrorsWord{} are error terminals (all \KtwoErrorEndpoint{}, in \KtwoErrorWorlds{} worlds of \KtwoErrorFamilies{} family, each a response failure after \KtwoErrorAttempts{} attempts with an empty body) and \KtwoScored{} scored. A self-hosted open-weight stratum is descriptive (\S\ref{sec:results-k2-open}). The registered quantity is the target's first-credit rate under the exact-enum credit rule (the first listed id that is exactly a member of the world's id set). Each contrast is the paired difference over (world, run) blocks holding both arms, with an exact \KtwoCPBLevel{} interval formed from two two-sided Clopper--Pearson intervals at \KtwoCPBMarginal{} (tail \KtwoCPBTail{} each) on the discordant counts, differenced and clipped to $[-1, 1]$ (CPB), and an exact two-sided sign test over families on the sign of each family's summed difference, whose null is that, conditional on the accepted worlds, the non-tied family signs are independent fair coin flips; a directional forecast is met when the interval lies beyond $\pm\KtwoDelta{}$ points in the forecast direction, the sign test has $p < 0.05$ and the family majority agrees.

\paragraph{The primary.} The \KtwoSignatures{} registered signatures held (\KtwoPrimary{}). On the \KtwoMainWorlds{} worlds in \KtwoFamilies{} named domain families accepted by the judges, appending the id to the criterion lowered the target's first-credit rate on Opus~5 by \KtwoBothCritOpus{} points [\KtwoBothCritOpusLo, \KtwoBothCritOpusHi] (negative in \KtwoBothCritOpusFamNeg{} of \KtwoBothCritOpusFam{} families, positive in \KtwoBothCritOpusFamPos{}; sign test $p$ \KtwoBothCritOpusSignP{}) and on Fable~5.1 by \KtwoBothCritFableFiveOne{} points [\KtwoBothCritFableFiveOneLo, \KtwoBothCritFableFiveOneHi] (\KtwoBothCritFableFiveOneFamNeg{} of \KtwoBothCritFableFiveOneFam{}; $p$ \KtwoBothCritFableFiveOneSignP{}). The cells behind them (Table~\ref{tab:studyK2-contrasts}): Opus~5 named the target under the criterion in \KtwoPmKCritOpus{} episodes and under the composite in \KtwoPmKBothILOpus{}; Fable~5.1 in \KtwoPmKCritFableFiveOne{} and \KtwoPmKBothILFableFiveOne{}. The criterion exceeded the bare id on each (\KtwoCritIdOpus{} [\KtwoCritIdOpusLo, \KtwoCritIdOpusHi]; \KtwoCritIdFableFiveOne{} [\KtwoCritIdFableFiveOneLo, \KtwoCritIdFableFiveOneHi]); these crit $-$ id contrasts are deposited forecasts, not signatures.

\paragraph{The other endpoints (descriptive).} Every other contrast is a per-endpoint description with its interval, not a signature. Fable~5: both $-$ crit \KtwoBothCritFableFive{} [\KtwoBothCritFableFiveLo, \KtwoBothCritFableFiveHi] (\KtwoBothCritFableFiveFamNeg{}/\KtwoBothCritFableFiveFamPos{} families). Haiku~4.5: the opposite sign, \KtwoBothCritHaiku{} [\KtwoBothCritHaikuLo, \KtwoBothCritHaikuHi] (\KtwoBothCritHaikuFamPos{} of \KtwoBothCritHaikuFam{} families positive), with its criterion below its bare id (\KtwoCritIdHaiku{} [\KtwoCritIdHaikuLo, \KtwoCritIdHaikuHi]); that row is the \KtwoDirMissesN{} deposited directional forecast not met. Sonnet~5 \KtwoBothCritSonnet{} [\KtwoBothCritSonnetLo, \KtwoBothCritSonnetHi]; GPT-5.6 Sol \KtwoBothCritGPTSol{} [\KtwoBothCritGPTSolLo, \KtwoBothCritGPTSolHi], Terra \KtwoBothCritGPTTerra{} [\KtwoBothCritGPTTerraLo, \KtwoBothCritGPTTerraHi], Luna \KtwoBothCritGPTLuna{} [\KtwoBothCritGPTLunaLo, \KtwoBothCritGPTLunaHi]; GPT-6 Astra \KtwoBothCritGPTSixAstra{} [\KtwoBothCritGPTSixAstraLo, \KtwoBothCritGPTSixAstraHi]: on these \KtwoOtherN{} the point estimate's magnitude was at most \KtwoOtherMaxAbs{} points; the interval excluded zero on \KtwoOtherExcludesZeroN{} of them (\KtwoOtherExcludesZero{}) and contained it on the others (\KtwoOtherIncludesZero{}), and lay entirely inside $\pm\KtwoDelta{}$ on \KtwoOtherWithin{} of them (\KtwoOtherWithinNames{}), which are `within' under the registered rule; none is beyond the margin in either direction. The bare id against no directive (id $-$ none): Sol \KtwoIdNoneGPTSol{} [\KtwoIdNoneGPTSolLo, \KtwoIdNoneGPTSolHi], Terra \KtwoIdNoneGPTTerra{} [\KtwoIdNoneGPTTerraLo, \KtwoIdNoneGPTTerraHi], Luna \KtwoIdNoneGPTLuna{} [\KtwoIdNoneGPTLunaLo, \KtwoIdNoneGPTLunaHi], Astra \KtwoIdNoneGPTSixAstra{} [\KtwoIdNoneGPTSixAstraLo, \KtwoIdNoneGPTSixAstraHi]; Opus~5 \KtwoIdNoneOpus{} [\KtwoIdNoneOpusLo, \KtwoIdNoneOpusHi], Fable~5.1 \KtwoIdNoneFableFiveOne{} [\KtwoIdNoneFableFiveOneLo, \KtwoIdNoneFableFiveOneHi] (Table~\ref{tab:studyK2-contrasts}).

\paragraph{The deposited table (descriptive).} Of the \KtwoDirTotal{} directional forecasts deposited before generation, \KtwoDirMet{} were met; of the \KtwoWithinTotal{} forecast absences, \KtwoWithinMet{} held; the rows are listed with their source cells, intervals and family counts in Table~\ref{tab:studyK2-forecasts}. Excluding rows whose source cells were small or unequally pooled at deposit, \KtwoDirMetExcl{} of \KtwoDirTotalExcl{} and \KtwoWithinMetExcl{} of \KtwoWithinTotalExcl{}. The \KtwoDirMissesN{} miss: \KtwoDirMisses{}. No row was unresolved by missingness. The rows are not independent and the tally is a description, not a test. Calibration (descriptive; Table~\ref{tab:studyK2-calibration}): the locked profiles' predicted rate per form against the observed rate had a mean absolute error of \KtwoCalMAEOpus{} points on Opus~5 and \KtwoCalMAEFableFiveOne{} on Fable~5.1, with slopes of logit(observed) on logit(predicted) of \KtwoCalSlopeOpus{} [\KtwoCalSlopeOpusLo, \KtwoCalSlopeOpusHi] and \KtwoCalSlopeFableFiveOne{} [\KtwoCalSlopeFableFiveOneLo, \KtwoCalSlopeFableFiveOneHi]; the profiles were fitted on hand-written cells and the levels on the generated worlds differ from them.

\paragraph{With no directive (descriptive).} Under the \texttt{none} form the first credit went to the active plan's own support record in \KtwoSBOpus{}\% of Opus~5's foregone-state episodes [\KtwoSBOpusLo, \KtwoSBOpusHi] (uniform single-memory reference \KtwoSBRefOpus{}\%), \KtwoSBSonnet{}\% [\KtwoSBSonnetLo, \KtwoSBSonnetHi] on Sonnet~5, \KtwoSBGPTSol{}\% [\KtwoSBGPTSolLo, \KtwoSBGPTSolHi] on GPT-5.6 Sol and \KtwoSBGPTTerra{}\% [\KtwoSBGPTTerraLo, \KtwoSBGPTTerraHi] on Terra; the interval's lower bound exceeded 50\% on \KtwoSBMajority{} of \KtwoSBEndpoints{} endpoints (Table~\ref{tab:studyK2-sb}). The plan-relation contrast registered as secondary (the id line against a negative-target control) is undetermined on every endpoint and is in the study's results report.

\paragraph{Strata that did not realise, and a degenerate one.} A second stratum of worlds authored by a different generator (the ``bank'') was registered to replicate the generation; every admissible author fell below the registered yield minimum and the stratum is reported as attempted, not realised, so the primary is the main stratum's \KtwoSignatures{} signatures only. The registered headroom stratification (each contrast by the rate at which a non-panel screener already named the target) is degenerate on these worlds: the screener rate was zero on \KtwoHeadroomZero{} of \KtwoMainWorlds{} worlds and non-zero on \KtwoHeadroomNonzero{} (\KtwoHeadroomNonzeroDetail{}), so with the median at zero the low-headroom stratum is empty and the high stratum is the full sample; it is not tabulated.

\subsubsection{The open-weight stratum (descriptive)}\label{sec:results-k2-open}

\KtwoOpenRegistered{} open-weight endpoints were registered on pinned checkpoints served from a frozen image on rented GPUs; \KtwoOpenAnalysed{} supplied analysable data (Llama~3.1~8B: \KtwoOpenNLlamaEightB{} rows, \KtwoOpenErrorsLlamaEightB{} error terminal; Llama~3.3~70B: \KtwoOpenNLlamaSeventyB{}, \KtwoOpenErrorsLlamaSeventyB{}; Qwen2.5~72B: \KtwoOpenNQwenSeventyTwoB{}, \KtwoOpenErrorsQwenSeventyTwoB{}). Their both $-$ crit contrasts were \KtwoBothCritLlamaEightB{} [\KtwoBothCritLlamaEightBLo, \KtwoBothCritLlamaEightBHi], \KtwoBothCritLlamaSeventyB{} [\KtwoBothCritLlamaSeventyBLo, \KtwoBothCritLlamaSeventyBHi] and \KtwoBothCritQwenSeventyTwoB{} [\KtwoBothCritQwenSeventyTwoBLo, \KtwoBothCritQwenSeventyTwoBHi] (Table~\ref{tab:studyK2-contrasts}); no forecast was deposited for them and no sentence beyond the rows is licensed. The \KtwoOpenAbsent{} others are absent with their reasons recorded: \KtwoOpenUnservable{} could not be started under the frozen image (its checkpoint declares an architecture the image's library does not know; the four-GPU serving shapes had not been dry-run on their real checkpoints), and \KtwoOpenAllError{} was served but returned no schema-valid completion --- every output began in the model's reasoning channel instead of the required JSON object --- so all of its rows are error terminals and it was excluded from the analyser's input.

\paragraph{What K2 does and does not establish.} On this population of generated worlds the two facts recorded on Opus~5 and Fable~5.1 recurred under their registered signatures, with the family as the unit of the sign test. The remaining endpoints are described one by one; the generated worlds are one generator's distribution under one contract; nothing identifies why the suffix lowers the criterion's rate on some endpoints and not on others; and no sentence here concerns a population of environments or of models.

\subsection{Study K3: the gain of a defensive adapter on the open ladders}\label{sec:results-k3}

Study K2 served Llama~3.1~8B and Llama~3.3~70B from a frozen image; on both, appending the id to the criterion left the criterion's rate roughly where it was or raised it (the composite $-$ criterion contrast was positive), unlike Opus~5 and Fable~5.1, on which it was negative. Study K3 asks a narrower, dose-shaped question on the same \KthreeWorlds{} worlds in \KthreeFamilies{} families: if one fixed defensive LoRA update --- Meta-SecAlign \citep{chen2025metasecalign}, trained with a refinement of the SecAlign preference-optimisation recipe \citep{chen2024secalign} to ignore instructions carried in data --- is applied to each base at \KthreeDosesN{} gains (the adapter's \texttt{lora\_alpha} set to \KthreeDoses{}; zero is the base weights under the adapter's chat template), does what appending the id does to the criterion change with gain? The manipulation is adapter gain --- a scalar on one fixed update --- not a calibrated defence strength; the two ladders have different effective scales and are never pooled; every dose is served on the same LoRA-enabled engine and schedule, with the dose order rotated within every batch.

\paragraph{Design and registration.} The \KthreeWorlds{} main worlds, the foregone state and the four forms of Study K2 (\KthreeBlocks{} (world, run) blocks per dose; \KthreeNPerLadder{} completions per ladder; \KthreeN{} in all, \KthreeAttempts{} attempts with the audit replays), analysed by a registered linear component of the contrast against gain with a one-stage bootstrap over the \KthreeFamilies{} registered families (B~=~4{,}000; the statement's own interval level), a per-dose and an intersection missingness gate, and two directional completions; the primary (K3-1) was that on the 70B ladder the composite $-$ criterion contrast falls with gain --- the slope's 0.95 interval below zero, the $\alpha{=}8$ minus $\alpha{=}0$ change's interval entirely below $-$10 points, and the $\alpha{=}8$ contrast itself negative beyond the margin under Study K2's rule --- with two secondaries on the 70B ladder (id $-$ none, criterion $-$ id, at 0.975) and the 8B ladder descriptive. The package (\KthreeManifestFiles{} files, rollup \texttt{\KthreeManifestRollup}) was frozen after \KthreeReviewRounds{} adversarial review rounds, stamped and deposited before the first confirmatory call (\S\ref{app:registration}); the analysis ran once and an independent re-implementation recomputed every reported number from the locked records with zero differences on both ladders.

\paragraph{The primary: \KthreePrimaryStatusWord{}.} On the 70B ladder the composite $-$ criterion contrast was \KthreeBothCritSeventyBMeanZero{} / \KthreeBothCritSeventyBMeanTwo{} / \KthreeBothCritSeventyBMeanFour{} / \KthreeBothCritSeventyBMeanSix{} / \KthreeBothCritSeventyBMeanEight{} points at gain 0 / 2 / 4 / 6 / 8 (\KthreeBothCritSeventyBComplete{} of \KthreeBlocks{} blocks complete at every dose); the linear component was \KthreeBothCritSeventyBSlope{} points per unit gain [\KthreeBothCritSeventyBSlopeLo, \KthreeBothCritSeventyBSlopeHi] and the change from gain 0 to 8 was \KthreeBothCritSeventyBChange{} points [\KthreeBothCritSeventyBChangeLo, \KthreeBothCritSeventyBChangeHi]; at the released gain the contrast itself was \KthreeBothCritSeventyBAtEight{} [\KthreeBothCritSeventyBAtEightLo, \KthreeBothCritSeventyBAtEightHi] (positive in \KthreeBothCritSeventyBAtEightFamPos{} families, negative in \KthreeBothCritSeventyBAtEightFamNeg{}). None of the three registered conditions held. The registered criterion for a gain-dependent change of both\_IL $-$ crit was not met on this ladder. The composite exceeded the criterion at every gain; the per-gain contrasts with their exact intervals are in the study's generated results report. This is not a statement that the contrast was unchanged: the slope's interval spans \KthreeBothCritSeventyBSlopeLo{} to \KthreeBothCritSeventyBSlopeHi{} points per unit gain and does not license an absence of effect.

\paragraph{The secondaries (70B; \KthreeSecondaryIdNone{} / \KthreeSecondaryCritId{}).} Id $-$ none: \KthreeIdNoneSeventyBMeanZero{} / \KthreeIdNoneSeventyBMeanTwo{} / \KthreeIdNoneSeventyBMeanFour{} / \KthreeIdNoneSeventyBMeanSix{} / \KthreeIdNoneSeventyBMeanEight{}, slope \KthreeIdNoneSeventyBSlope{} [\KthreeIdNoneSeventyBSlopeLo, \KthreeIdNoneSeventyBSlopeHi] at 0.975, change \KthreeIdNoneSeventyBChange{} [\KthreeIdNoneSeventyBChangeLo, \KthreeIdNoneSeventyBChangeHi]. Criterion $-$ id: \KthreeCritIdSeventyBMeanZero{} / \KthreeCritIdSeventyBMeanTwo{} / \KthreeCritIdSeventyBMeanFour{} / \KthreeCritIdSeventyBMeanSix{} / \KthreeCritIdSeventyBMeanEight{}, slope \KthreeCritIdSeventyBSlope{} [\KthreeCritIdSeventyBSlopeLo, \KthreeCritIdSeventyBSlopeHi], change \KthreeCritIdSeventyBChange{} [\KthreeCritIdSeventyBChangeLo, \KthreeCritIdSeventyBChangeHi]. Neither slope interval excludes zero and neither change interval lies beyond the margin.

\paragraph{The 8B ladder (descriptive).} On Llama~3.1~8B the same contrast fell monotonically with gain: \KthreeBothCritEightBMeanZero{} / \KthreeBothCritEightBMeanTwo{} / \KthreeBothCritEightBMeanFour{} / \KthreeBothCritEightBMeanSix{} / \KthreeBothCritEightBMeanEight{} points (\KthreeBothCritEightBComplete{} of \KthreeBlocks{} blocks complete; slope \KthreeBothCritEightBSlope{} [\KthreeBothCritEightBSlopeLo, \KthreeBothCritEightBSlopeHi]; change \KthreeBothCritEightBChange{} [\KthreeBothCritEightBChangeLo, \KthreeBothCritEightBChangeHi]); at the released gain the composite still led the criterion by \KthreeBothCritEightBAtEight{} points [\KthreeBothCritEightBAtEightLo, \KthreeBothCritEightBAtEightHi]. Per form, the composite's target rate fell from \KthreePmBothILEightBZero{}\% to \KthreePmBothILEightBEight{}\% while \texttt{none}, \texttt{id} and \texttt{crit} stayed between \KthreePmOtherMinEightB{}\% and \KthreePmOtherMaxEightB{}\% over the gains; id $-$ none: slope \KthreeIdNoneEightBSlope{} [\KthreeIdNoneEightBSlopeLo, \KthreeIdNoneEightBSlopeHi], change \KthreeIdNoneEightBChange{} [\KthreeIdNoneEightBChangeLo, \KthreeIdNoneEightBChangeHi]; criterion $-$ id: slope \KthreeCritIdEightBSlope{} [\KthreeCritIdEightBSlopeLo, \KthreeCritIdEightBSlopeHi], change \KthreeCritIdEightBChange{} [\KthreeCritIdEightBChangeLo, \KthreeCritIdEightBChangeHi] (both slope intervals include zero). No criterion was registered for this ladder, so this is a description of a dose-dependent attenuation on one ladder, not a tested effect. On the 70B ladder the composite's rate went from \KthreePmBothILSeventyBZero{}\% to \KthreePmBothILSeventyBEight{}\% and the id form's from \KthreePmIdSeventyBZero{}\% to \KthreePmIdSeventyBEight{}\%, while \texttt{none}, \texttt{id} and \texttt{crit} together stayed between \KthreePmOtherMinSeventyB{}\% and \KthreePmOtherMaxSeventyB{}\%.

\paragraph{Missingness and completions (licence S-G).} No gate fired: the largest per-gain missing share was \KthreeBothCritEightBMissMaxPct{}\% of blocks (8B, both $-$ crit; \KthreeBothCritEightBOutside{} blocks outside the five-gain intersection, \KthreeBothCritEightBInterPct{}\%) against the registered 5\%, and \KthreeBothCritSeventyBOutside{} on the 70B ladder; under the two directional completions of the blocks outside the intersection the 8B changes ranged from \KthreeBothCritEightBSensNegChange{} to \KthreeBothCritEightBSensPosChange{} points (both $-$ crit; \KthreeBothCritEightBOutside{} blocks outside), \KthreeCritIdEightBSensNegChange{} to \KthreeCritIdEightBSensPosChange{} (criterion $-$ id; \KthreeCritIdEightBOutside{} blocks, \KthreeCritIdEightBInterPct{}\%) and \KthreeIdNoneEightBSensNegChange{} to \KthreeIdNoneEightBSensPosChange{} (id $-$ none; \KthreeIdNoneEightBOutside{} blocks, \KthreeIdNoneEightBInterPct{}\%), and no registered status changed (\KthreeSensFlips{} flips).

\paragraph{Serving noise and what K3 cannot say.} The registered contrast was estimated at each adapter gain on the first recorded completions of these \KthreeWorlds{} worlds and four fixed orders; whether it changed with gain is the reported outcome, not a premise of this disclosure. Intervals describe family-reweighting stability conditional on this execution; they do not quantify reproducibility across engine executions. Audit replays were neither substituted for nor averaged into the analysed outcomes. On the LoRA-enabled engine greedy decoding does not reproduce its own completions: replaying the first two batches gave byte / id-list / validity / target-hit agreement of \KthreeAuditSESeventyB{} of 80 on the same engine and \KthreeAuditFCSeventyB{} on a fresh container for the 70B ladder (\KthreeAuditSEEightB{} and \KthreeAuditFCEightB{} for the 8B); Table~\ref{tab:studyK3-audit} gives the agreement by gain with its denominators. Study K2's Llama rows were obtained under the base template on a plain engine and are historical comparators, not the zero-gain point of these curves. The manipulation identifies no mechanism, is not a calibrated defence strength, and does not generalise beyond these pinned base--adapter--template--serving ladders.

\subsection{Study K4: the 8B attenuation registered, extended and the 70B ladder re-executed}\label{sec:results-k4}

Study K3 described a monotone attenuation on Llama~3.1~8B without a criterion. Study K4, designed after that outcome and disclosed as such, registers it: on the same \KfourWorlds{} worlds in \KfourFamilies{} families and the same foregone state and four forms, three ladders were frozen together --- L1, the 8B ladder re-executed at gains \KfourDosesLone{} with a registered attenuation rule (K4-1, 0.95 intervals: the slope's interval below zero \emph{and} the change from gain 0 to \KfourLastLone{} entirely below $-$10 points); L2, the 8B adapter taken beyond its release to gains \KfourDosesLtwo{} (K4-2, 0.975), registered as a \emph{descriptive} ladder before the freeze because its simulated power at a realised 25-point change was \KfourPowerLoLtwo{}--\KfourPowerHiLtwo{}, below the registered gate; and L3, the 70B ladder re-executed under Study K3's three-condition rule (K4-3, 0.975). Gains above the released \texttt{lora\_alpha} are a registered extrapolation whose validity is guarded by the missingness gates. The package (\KfourManifestFiles{} files, rollup \texttt{\KfourManifestRollup}) was frozen after \KfourReviewRoundsMain{} adversarial review rounds and a confirmation round (\KfourReviewRoundsCodex{} by Codex, \KfourReviewRoundsClaude{} by a Claude reviewer while Codex was at its weekly usage limit; \KfourReviewDoNotRun{} returned ``do not run''), stamped and deposited before the first confirmatory call (\S\ref{app:registration}); the registered operating characteristics were re-specified four times, all before any K4 confirmatory data (the design itself was informed by Study K3's outcome, as disclosed) --- three times before the first review and by the D8--D9 amendments after the first review round --- with all five simulation runs archived; the analysis ran once and an independent re-implementation recomputed every number, interval, gate and decision on all three ladders with zero differences. \KfourNPerLadder{} completions per ladder, \KfourN{} in all, \KfourAttempts{} attempts with the audit replays; no gate fired (the largest per-gain error count was \KfourErrorsMaxDoseLtwo{} of 1{,}152, on L2).

\paragraph{K4-1 (L1, 8B, gains \KfourDosesLone{}): \KfourOneStatusWord{}.} The composite $-$ criterion contrast was \KfourBothCritLoneMeans{} points at gains \KfourDosesLone{} (\KfourBothCritLoneComplete{} of \KfourBlocks{} blocks complete at every gain); the linear component was \KfourBothCritLoneSlope{} points per unit gain [\KfourBothCritLoneSlopeLo, \KfourBothCritLoneSlopeHi] --- below zero, the first registered condition --- and the change from gain 0 to 8 was \KfourBothCritLoneChange{} points [\KfourBothCritLoneChangeLo, \KfourBothCritLoneChangeHi], whose upper bound does not clear the $-$10-point margin, so the second condition did not hold. The registered attenuation criterion for both\_IL $-$ crit was not met on this ladder. This is not a statement that the contrast was unchanged: the five means fall monotonically and the slope's interval excludes zero; what did not hold is the registered margin at the registered level. The study's own operating characteristics, run before any data, measured the rule's power at \KfourPowerLoLone{}--\KfourPowerHiLone{} for a realised 25-point change and at \KfourPowerNearKthreeLoLone{}--\KfourPowerNearKthreeHiLone{} for a realised change near Study K3's point estimate; the observed change is smaller in magnitude than that. The 0.95 interval is the rule's nominal level; in the registered simulation grid (a finite set of dispersion corners, not a uniform-error guarantee) the rule's realised one-sided error at the $-$10-point margin was \KfourMarginErrLoLone{}--\KfourMarginErrHiLone{}\% per corner (nominal \KfourMarginNominalLone{}\%). At the released gain the contrast itself was \KfourBothCritLoneAtLast{} [\KfourBothCritLoneAtLastLo, \KfourBothCritLoneAtLastHi]; Study K3's estimates on the same worlds and orders (\KthreeBothCritEightBMeanZero{} to \KthreeBothCritEightBMeanEight{}; change \KthreeBothCritEightBChange{} [\KthreeBothCritEightBChangeLo, \KthreeBothCritEightBChangeHi]) are a historical comparator: a difference between executions is serving noise plus sampling, and neither execution confirms or overturns the other.

\paragraph{K4-2 (L2, 8B, gains \KfourDosesLtwo{}; descriptive): \KfourTwoStatusWord{}.} Beyond the released gain the contrast went \KfourBothCritLtwoMeans{} points at gains \KfourDosesLtwo{} (\KfourBothCritLtwoComplete{} of \KfourBlocks{} blocks); slope \KfourBothCritLtwoSlope{} [\KfourBothCritLtwoSlopeLo, \KfourBothCritLtwoSlopeHi] per unit gain and change from gain 0 to 16 \KfourBothCritLtwoChange{} points [\KfourBothCritLtwoChangeLo, \KfourBothCritLtwoChangeHi], both conditions of the same rule at 0.975; at gain 16 the contrast was \KfourBothCritLtwoAtLast{} [\KfourBothCritLtwoAtLastLo, \KfourBothCritLtwoAtLastHi]. L2 is a descriptive ladder: the criterion being met is an observation at the registered error level (realised margin error \KfourMarginErrLoLtwo{}--\KfourMarginErrHiLtwo{}\% per corner, nominal \KfourMarginNominalLtwo{}\%), and a failure would not have been evidence against continuation. Per form, the composite's target rate fell from \KfourPmBothILLtwoZero{}\% to \KfourPmBothILLtwoSixteen{}\% while \texttt{none}, \texttt{id} and \texttt{crit} stayed between \KfourPmOtherMinLtwo{}\% and \KfourPmOtherMaxLtwo{}\% over the gains; the extrapolated gains produced analysable output: \KfourErrorsByDoseLtwo{} error terminals of \KfourPerGainNLtwo{} at gains \KfourDosesLtwo{} (\KfourErrorsLtwo{} of \KfourNPerLadder{} in all; L1 \KfourErrorsByDoseLone{}, L3 \KfourErrorsByDoseLthree{} at their gains).

\paragraph{K4-3 (L3, 70B, gains \KfourDosesLthree{}): \KfourThreeStatusWord{}.} On a second execution of the 70B ladder the contrast was \KfourBothCritLthreeMeans{} points (\KfourBothCritLthreeComplete{} of \KfourBlocks{} blocks); slope \KfourBothCritLthreeSlope{} [\KfourBothCritLthreeSlopeLo, \KfourBothCritLthreeSlopeHi], change \KfourBothCritLthreeChange{} [\KfourBothCritLthreeChangeLo, \KfourBothCritLthreeChangeHi], and at the released gain \KfourBothCritLthreeAtLast{} [\KfourBothCritLthreeAtLastLo, \KfourBothCritLthreeAtLastHi] (positive in \KfourBothCritLthreeAtLastFamPos{} families, negative in \KfourBothCritLthreeAtLastFamNeg{}). On a second execution of the 70B ladder the registered criterion for a gain-dependent change of both\_IL $-$ crit was not met. Study K3 had reached the same status on its execution (\KthreeBothCritSeventyBChange{} [\KthreeBothCritSeventyBChangeLo, \KthreeBothCritSeventyBChangeHi] there); the two executions are compared as statements and never pooled. Neither is a statement that the contrast was unchanged.

\paragraph{Missingness, completions and serving noise (licences S-G, S-F).} No gate fired on any ladder (largest per-gain missing share \KfourBothCritLtwoMissMaxPct{}\% of blocks, L2; \KfourBothCritLoneOutside{} / \KfourBothCritLtwoOutside{} / \KfourBothCritLthreeOutside{} blocks outside the intersection on L1 / L2 / L3); under the two directional completions the changes ranged from \KfourBothCritLoneSensNegChange{} to \KfourBothCritLoneSensPosChange{} (L1), \KfourBothCritLtwoSensNegChange{} to \KfourBothCritLtwoSensPosChange{} (L2) and \KfourBothCritLthreeSensNegChange{} to \KfourBothCritLthreeSensPosChange{} (L3), and no registered status changed (\KfourSensFlips{} flips). The registered contrast was estimated at each adapter gain on the first recorded completions of these \KfourWorlds{} worlds and four fixed orders; whether it changed with gain is the reported outcome, not a premise of this disclosure. Intervals describe family-reweighting stability conditional on this execution; they do not quantify reproducibility across engine executions. Audit replays were neither substituted for nor averaged into the analysed outcomes. Replaying the first two batches gave byte / id-list / validity / target-hit agreement of \KfourAuditSELone{} (same engine) and \KfourAuditFCLone{} (fresh container) of 80 on L1, \KfourAuditSELtwo{} and \KfourAuditFCLtwo{} on L2, \KfourAuditSELthree{} and \KfourAuditFCLthree{} on L3 (Table~\ref{tab:studyK4-audit}). The manipulation is adapter gain, not a calibrated defence strength; the ladders have different effective scales and are never pooled; no mechanism is identified and nothing generalises beyond these pinned base--adapter--template--serving ladders.

\subsection{Study K5: the 8B attenuation at registered power, and a second adapter}\label{sec:results-k5}

Study K4's attenuation rule was not met on \KfourFamilies{} families, with that study's own operating characteristics putting the rule's power at \KfourPowerNearKthreeLoLone{}--\KfourPowerNearKthreeHiLone{} near Study K3's estimate. Study K5, designed after that outcome and disclosed as such, re-poses the question at registered power. The instrument combines \KfiveCohortKtwoFamilies{} of Study K2's families --- their worlds reconstructed offline from K2's frozen run; \KfiveCohortKtwoExposed{} of these families had been chosen for K2's panel run, so their worlds had been exposed to the K3 / K4 outcomes, and \KfiveCohortKtwoUnused{} had passed only K2's realiser and judge stages --- with \KfiveCohortKfiveFamilies{} new families produced by one generation stream (Study K2's realiser and judge stages, run once for this study): \KfiveWorlds{} worlds in \KfiveFamilies{} families, two per family, the K2 families reported as a cohort. On these worlds, in the foregone state and the four forms, two ladders were frozen together: K5-1, the 8B base with Meta-SecAlign-8B at gains \KfiveDoses{}, and K5-2, the same base with a second SecAlign++ adapter (\KfiveAdapterLtwo{}; ``independently trained'' is its repository's description; SecAlign++ is the recipe of \citealp{chen2025metasecalign}) at the same gains, served under the imposed Meta-SecAlign template rather than its native configuration, each under Study K4's attenuation rule (\KfiveBothCritLoneLevel{} intervals: the slope's interval below zero \emph{and} the change from gain 0 to 8 entirely below $-$10 points); K5-3, the difference between the two adapters' changes, was registered descriptive with a fixed reporting sentence. The freeze was two-staged: the design, tools and stream inputs were frozen before the stream ran (\KfivePrestreamFiles{} files, rollup \texttt{\KfivePrestreamRollup}); the realised family count then fixed the operating characteristics (F = \KfiveF{}; \KfiveOcConfigs{} configurations, \KfiveOcReplicates{} replicates, run at the registered bootstrap size at every configuration and replayed in full by the gate) and a power gate labelled both statements confirmatory --- error control met (realised one-sided error at the $-$10-point margin \KfiveMarginErrLo{}--\KfiveMarginErrHi{}\% per boundary corner over \KfiveBoundaryConfigs{} corners, nominal \KfiveMarginNominal{}\%; the largest interior-null rate \KfiveInteriorNullMaxPct{}\% over \KfiveInteriorNullConfigs{} configurations) and power \KfivePowerAnchorLo{}--\KfivePowerAnchorHi{} at the \KfiveAnchors{} calibrated anchors placed at Study K4's realised change (Wilson lower bound \KfivePowerWilson{}) --- before the package (\KfiveManifestFiles{} files, rollup \texttt{\KfiveManifestRollup}) was frozen, stamped and deposited before the first confirmatory call (\S\ref{app:registration}). The design went through \KfiveReviewRounds{} adversarial review rounds before the pre-stream freeze (\KfiveReviewRoundsCodex{} by Codex, \KfiveReviewRoundsClaude{} by a Claude reviewer while Codex was at its weekly usage limit); every round returned ``blocked''; its findings were repaired or recorded as disclosed limitations, and the loop was closed by the author's decision after the last round (\S\ref{sec:integrity}). \KfiveNPerLadder{} completions per ladder, \KfiveN{} in all, \KfiveAttempts{} attempts with the audit replays; the analysis ran on the combined file and an independent re-implementation recomputed every number, interval, gate and decision on both ladders and on the K5-3 contrast with zero differences; no gate fired (the largest per-gain error count was \KfiveErrorsMaxDoseLone{} of \KfivePerGainN{}, on K5-1 at gain 8). Table~\ref{tab:studyK5} lists the three contrasts per ladder.

\paragraph{K5-1 (Meta-SecAlign-8B, gains \KfiveDoses{}): \KfiveOneStatusWord{}.} The composite $-$ criterion contrast was \KfiveBothCritLoneMeans{} points at gains \KfiveDoses{} (\KfiveBothCritLoneComplete{} of \KfiveBlocks{} blocks complete at every gain); the linear component was \KfiveBothCritLoneSlope{} points per unit gain [\KfiveBothCritLoneSlopeLo, \KfiveBothCritLoneSlopeHi] and the change from gain 0 to 8 was \KfiveBothCritLoneChange{} points [\KfiveBothCritLoneChangeLo, \KfiveBothCritLoneChangeHi], whose upper bound clears the $-$10-point margin: both registered conditions held. In the licence's registered form: on this registered instrument of \KfiveFamilies{} families (\KfiveCohortKtwoFamilies{} from K2's \KtwoMainRunDate{} run, \KfiveCohortKfiveFamilies{} from the K5 stream), with these fixed orders and this execution, the signed change of the both\_IL $-$ crit contrast from gain 0 to gain 8 of the Meta-SecAlign-8B update satisfied the registered stability criterion (slope interval below 0; change interval entirely below $-$10 points at \KfiveBothCritLoneLevel{}); the rule's realised one-sided error at the margin at this F was \KfiveMarginErrLo{}--\KfiveMarginErrHi{}\% per corner of the registered grid (a finite grid, not a uniform-error guarantee). At gain 0 the contrast was \KfiveBothCritLoneAtZero{} points and at the released gain \KfiveBothCritLoneAtLast{} [\KfiveBothCritLoneAtLastLo, \KfiveBothCritLoneAtLastHi] (positive in \KfiveBothCritLoneAtLastFamPos{} families, negative in \KfiveBothCritLoneAtLastFamNeg{}); the composite's target rate fell from \KfivePmBothILLoneZero{}\% to \KfivePmBothILLoneEight{}\% while \texttt{none}, \texttt{id} and \texttt{crit} stayed between \KfivePmOtherMinLone{}\% and \KfivePmOtherMaxLone{}\%. Study K4's L1 estimate on \KfourFamilies{} families (change \KfourBothCritLoneChange{} [\KfourBothCritLoneChangeLo, \KfourBothCritLoneChangeHi]) is a historical comparator whose registered status stands as Study K4's result: a K5 outcome neither re-powers nor overturns it, and the two executions are compared as statements, never pooled. Study K2's cohort of families gave a change of \KfiveCohortKtwoChangeLone{} points and the new cohort \KfiveCohortKfiveChangeLone{} (descriptive: block-weighted over each cohort's complete blocks, no interval, no rule).

\paragraph{K5-2 (SecAlign++, gains \KfiveDoses{}): \KfiveTwoStatusWord{}.} On the second adapter the contrast was \KfiveBothCritLtwoMeans{} points (\KfiveBothCritLtwoComplete{} of \KfiveBlocks{} blocks); the slope was \KfiveBothCritLtwoSlope{} [\KfiveBothCritLtwoSlopeLo, \KfiveBothCritLtwoSlopeHi] --- below zero, the first condition --- and the change \KfiveBothCritLtwoChange{} points [\KfiveBothCritLtwoChangeLo, \KfiveBothCritLtwoChangeHi], whose upper bound does not clear the $-$10-point margin, so the second condition did not hold. In the licence's registered form: on this registered instrument of \KfiveFamilies{} families (\KfiveCohortKtwoFamilies{} from K2's \KtwoMainRunDate{} run, \KfiveCohortKfiveFamilies{} from the K5 stream), with these fixed orders and this execution, the signed change of the both\_IL $-$ crit contrast from gain 0 to gain 8 of the SecAlign++ update, conditional on the imposed Meta-SecAlign template (not the adapter's native configuration), did not satisfy the registered stability criterion (slope interval below 0; change interval entirely below $-$10 points at \KfiveBothCritLtwoLevel{}); realised one-sided error at the margin \KfiveMarginErrLo{}--\KfiveMarginErrHi{}\% per corner. This is not a statement that the contrast was unchanged: the slope's interval excludes zero; what did not hold is the registered margin at the registered level. At the released gain the contrast was \KfiveBothCritLtwoAtLast{} [\KfiveBothCritLtwoAtLastLo, \KfiveBothCritLtwoAtLastHi]; the composite's rate fell from \KfivePmBothILLtwoZero{}\% to \KfivePmBothILLtwoEight{}\%; the cohorts gave \KfiveCohortKtwoChangeLtwo{} (K2's families) and \KfiveCohortKfiveChangeLtwo{} (new) points.

\paragraph{K5-3 (descriptive; the registered sentence).} \KfiveThreeSentence{} ($\theta$ = \KfiveThreeTheta{} points [\KfiveThreeThetaLo, \KfiveThreeThetaHi]; the two changes recomputed on the \KfiveThreeCommonBlocks{} blocks complete on both ladders, \KfiveThreeDeltaLone{} for K5-1 and \KfiveThreeDeltaLtwo{} for K5-2; joint missing share \KfiveThreeJointMissPct{}\%.)

\paragraph{Controls, missingness, completions and serving noise (licence S-E).} The two registered controls, reported descriptively and under no rule: crit $-$ id changed by \KfiveCritIdLoneChange{} [\KfiveCritIdLoneChangeLo, \KfiveCritIdLoneChangeHi] and id $-$ none by \KfiveIdNoneLoneChange{} [\KfiveIdNoneLoneChangeLo, \KfiveIdNoneLoneChangeHi] on K5-1, \KfiveCritIdLtwoChange{} [\KfiveCritIdLtwoChangeLo, \KfiveCritIdLtwoChangeHi] and \KfiveIdNoneLtwoChange{} [\KfiveIdNoneLtwoChangeLo, \KfiveIdNoneLtwoChangeHi] on K5-2 (\KfiveCritIdLoneLevel{} intervals, each containing zero; an interval containing zero does not establish an unchanged contrast). No gate fired (largest per-gain missing share \KfiveIdNoneLoneMissMaxPct{}\% of blocks; \KfiveBothCritLoneOutside{} / \KfiveBothCritLtwoOutside{} blocks outside the intersection on the rule contrast of K5-1 / K5-2); under the two directional completions the rule contrast's change ranged from \KfiveBothCritLoneSensNegChange{} to \KfiveBothCritLoneSensPosChange{} (K5-1) and from \KfiveBothCritLtwoSensNegChange{} to \KfiveBothCritLtwoSensPosChange{} (K5-2), and no registered status changed (\KfiveSensFlips{} flips). Error terminals were \KfiveErrorsByDoseLone{} of \KfivePerGainN{} at gains \KfiveDoses{} on K5-1 (\KfiveErrorsLone{} of \KfiveNPerLadder{}) and \KfiveErrorsByDoseLtwo{} on K5-2 (\KfiveErrorsLtwo{}). The registered contrast was estimated at each adapter gain on the first recorded completions of these \KfiveWorlds{} worlds and four fixed orders, conditional on this execution; the intervals describe family-reweighting stability on this fixed instrument, its cohort proportions fixed, and do not quantify reproducibility across engine executions or transport to other worlds. A runtime identity check compared the \KfiveIdentityCompared{} rendered prompts of the two ladders before conversion (no difference across ladders or gains). Audit replays were neither substituted for nor averaged into the analysed outcomes: replaying the first two batches gave byte / id-list / validity / target-hit agreement of \KfiveAuditSELone{} (same engine) and \KfiveAuditFCLone{} (fresh container) of 80 on K5-1, \KfiveAuditSELtwo{} and \KfiveAuditFCLtwo{} on K5-2 (Table~\ref{tab:studyK5-audit}). The manipulation is adapter gain, not a calibrated defence strength; the two adapters are different LoRA updates, not matched in strength (released scales \KfiveReleasedScaleLone{} and \KfiveReleasedScaleLtwo{}, i.e.\ $\alpha/r$ = \KfiveLoraAlphaLone{}/\KfiveLoraRLone{} and \KfiveLoraAlphaLtwo{}/\KfiveLoraRLtwo{}), and are never pooled; the halted first attempts of both ladders contributed nothing to the analysed records (\S\ref{sec:integrity}); K5-3 describes two fixed execution paths and licenses no superiority or equivalence claim; no mechanism is identified and nothing generalises beyond these pinned base--adapter--template--serving ladders.

\section{Integrity and deviations}\label{sec:integrity}

Of the \TotalStudies{} studies, one carries a post-freeze correction of its headline (B), one a registration-versus-analyzer discrepancy (E), one a quarantined run (F), and the follow-ups carry the errata listed below; all are reported here rather than in the studies' favour.

\paragraph{Study B's inverted verdict.} The deposited analyzer pooled a counterbalanced factor instead of
decoding it, and the study's first repository report (non-archival, in the project repository) carried the wrong estimand, magnitude and verdict on its headline --- both the reported and the corrected shifts were positive (\S\ref{sec:results-b}). The correction was found by an adversarial review of the successor study, hours after that report. An explicit decode
step for every counterbalanced factor, asserted in writing before the manifest is stamped, is now a freeze
blocker in our checklist.

\paragraph{Study E's pairing rule.} The frozen analyzer paired runs present in both arms when episodes were
lost; the prose registration named schema failures as the anticipated error and did not state a pairing rule.
The executed rule is disclosed and the estimate is reported as executed (Appendix~\ref{app:studyE}). Study E's frozen analyzer resolves a hard-coded record path named for Study D (\texttt{runs/studyD}); it was executed byte-unmodified against Study E's own locked records through a symlink of that name, created after Study E's run was locked, that resolves to Study E's record directory; the analyzer's hash does not bind what the link resolves to and the analyzer does not verify the completion manifest, so this runtime repair is disclosed in Study E's deviation record, and the manuscript generator re-derives Study E's values from the same locked records (Appendix~\ref{app:repro}). (The first public version of this paper described the link as reading Study D's records; corrected here.)

\paragraph{Study F's quarantined first run.} The first confirmatory run (\FQN{} episodes) is not used for any
outcome estimate. Its OpenRouter requests carried a \texttt{plugins} field naming the provider's web-search
plugin as disabled; naming it was the activation under the workspace's default-enabled tool policy. The
provider billed the tool at \FQSurchargePerCall{} on \FQSurchargeCalls{} response-bearing attempts
(\FQSurchargeTotal{}) across \FQSurchargeModels{} of the nine OpenRouter-served models. Read against the clean
per-cell baselines of the locked earlier runs, the quarantined records show three patterns: variable injected
content on \FQVariableModels{} models (\FQVariableModelsList{}; prompt tokens \FQVariableExcessMin{} to
\FQVariableExcessMax{} above baseline and varying within a cell; \FQVariablePhrase{} of their
\FQVariableSuccesses{} rationales name a web search by a fixed phrase list and \FQVariableUrlRecords{} carry
\FQUrls{} explicit URLs); the surcharge without any token excess on gpt-oss-120b (\FQGptOssSurchargeCalls{}
attempts); and a constant excess with no surcharge on two models (\FQConstantExcessModels{}), Grok's consistent
with a constant unsurcharged augmentation and not proven to be one, Gemini's unidentified. The six
direct-provider models were constant on their baselines. Cost accounting raised the first alarm (the billed
cost exceeded the upstream inference cost), inspection of the stored rationales confirmed injected content,
and the run was quarantined; when the workspace tool was disabled mid-run, the remaining requests were refused
(\FQErrPlugin{} of the run's \FQErrors{} error terminals). The stored request bytes are intact; the
augmentation happened at the provider. The package was re-frozen after eight review rounds with a per-cell
prompt-token invariance gate (Appendix~\ref{app:contamination}): a fixed design implies one prompt-token
count per (model, arm) cell, the locked earlier runs supply it, the smoke must reproduce it exactly, and the
confirmatory run is invalidated (poisoned) on the first deviation. The gate is exact for detecting a deviation from an
established count; it cannot see an augmentation that preserves the count or that is present identically at
smoke time. A diagnostic smoke under the corrected request family returned every model, Grok included, to its
locked count, and the second run passed every gate.

\paragraph{Study F-x.} A commit was labelled ``frozen'' before its manifest step had succeeded; nothing was
stamped or deposited from it, and the freeze log records it. The frozen analysis specification misquotes one Study F cell in its preamble (Opus~5's composite is \FPmKBothILOpus{}; the id cell's \FPmKIdOpus{} was written); the registered quantities are unaffected and
the erratum will be released with the study.

\paragraph{The follow-up programme.} Every follow-up package was reviewed by one to four hostile Codex rounds before its freeze,
each round's verdict, dispositions and artifacts to be released; three of them were blocked more than once (H2 four rounds: the first
rewordings changed the directive's construct and were re-authored, then a live-queue bound that would have left 750 registered
episodes unreachable, then a reservation-order defect at the exact attempt ceiling; \Gr{} four rounds: a seed named differently in the
specification and the code, then the paired-interval variant named inconsistently; J four rounds: the credit rule, then its exact
port of the earlier work's resolver). The reservation-order defect --- the runner checked the global attempt ceiling before an
episode's own exhausted budget, so a run that consumed every reservation could never lock its last exhausted episode --- was
inherited by every runner in this programme; it never triggered (no run approached its ceiling), it is recorded in the errata of the three studies frozen with it (\Br{}, I, H1), and it was fixed, with a scratch-ledger regression test, before J, H2 and \Gr{} froze. The first version of that regression test appended scratch reservations to the production attempt ledgers of H2, J and \Gr{} (an environment variable set after a hoisted import); no call was made and no record written, the rows were deleted and the test corrected, and the incident is recorded in \Gr{}'s deviation record and in H2's and J's errata. Two frozen report generators (Studies \Br{} and I) read keys their frozen analyzers no longer emitted after review dispositions, and Study G's omitted the registered bootstrap sensitivity and the interpretive limits and mislabelled a registered descriptive output; each is recorded as an erratum with an unhashed successor generator, to be released with the study, (Study \Br{} needed two attempts, both recorded), and a static key test and an executable rehearsal of the generator now precede every freeze.
Study \Br{}'s freeze also records a runbook-order deviation (fixtures regenerated before, not after, the review edits) and a
commit message that claimed a report before one existed.

\paragraph{Study K2.} The package was frozen and deposited twice: the first frozen attempt was poisoned at its first outcome by the runner's one-world token-per-byte band (the outcome is kept in a poisoned ledger, unused), the rule was widened and the package re-frozen, re-deposited and run. The decision-log entry recording the author's delegation of the remaining steps was first committed after the registered smoke had begun (by about a hundred seconds; the delegation itself preceded the smoke and is not quoted). Two post-freeze analysis tools replaced frozen ones and are disclosed with their deviations: a lock tool applying the widened band and pinning provenance, reviewed over five rounds before use (during that review, while checking the deposit chronology, the reviewer printed three live ledger rows including one confirmatory outcome; the executing agent inspected none before the lock, and no decision depends on that content), and a merge tool accepting the open stratum's per-entry serving manifests (the frozen converter and the frozen merge could not both be satisfied) that also binds the closed collector output to the lock --- this second tool was first reviewed after its use. The frozen analyser ran unmodified but would crash on an endpoint with no valid row, so the all-error open endpoint was excluded from its input rather than the analyser amended. The OpenTimestamps stamps were made after the run started and attest existence before a Bitcoin block only; the pre-run time source is the deposit's server timestamp on the OSF node, which was already public when that deposit was made (the depositors of Studies K2--K4 wrote the receipts' private flag as a constant rather than as an observation of the server, a defect recorded in Study K5's decision log when its depositors were made to read the visibility; Study K5's own deposits are on a separate private component). The bank stratum is recorded as attempted and not realised. Every deviation, the poisoned attempt and the \KtwoErrorsWord{} error terminals are in the study's decision log and results disclosures; the results were deposited to the same node after the analysis.

\paragraph{Study K3.} The package went through \KthreeReviewRounds{} adversarial review rounds before the freeze; \KthreeReviewDoNotRun{} returned ``do not run'' on defects the reviewer reproduced (a qualification schema that differed from the production schema, a stale GPU-second counter in the audit replay accounting, an independent recomputation that skipped checks on a bootstrap-size mismatch or accepted altered displayed numbers, and successively narrower provenance gaps), each recorded in the decision log with its fix; the serving code was re-qualified on the frozen code each time it changed. After the freeze the deposit script's download-back request failed following a successful upload; a post-freeze tool re-downloaded the uploaded file and wrote the receipt only on byte equality (the frozen script, package and tag are unchanged). Two launches were refused by the live guard for unbound flags before any call. No breaker fired, no partial replay was written, no run was repeated.

\paragraph{Study K4.} The package went through \KfourReviewRoundsMain{} adversarial review rounds and a confirmation round before the freeze (\KfourReviewRoundsCodex{} by Codex; the Codex client reached its weekly usage limit before the eighth, and the last \KfourReviewRoundsClaude{} were by a Claude reviewer at maximum effort in an isolated worktree, recorded as such in every disposition); \KfourReviewDoNotRun{} returned ``do not run''. The first two rounds found defects in the study's own tools (a recomputation that could not combine three ladders; operating characteristics whose margin nulls were placed by a nominal level while the simulator's random effects realised a smaller change); the operating characteristics were re-specified four times on that account, all before any K4 confirmatory data (three times before the first review and by the D8--D9 amendments after the first round), every run archived, and the extended-gain ladder was registered descriptive when its simulated power missed the gate. The following five rounds found defects in the recovery code the second round had asked for (a resume path whose every state the reviewers could break); the resume path was withdrawn in favour of Study K3's fresh re-execution, the shared spend ledger replaced by atomic per-invocation records, the locks, converter and combiner outputs and recomputation receipts written complete-or-absent with a registered interrupted-pair recovery, and the serving script re-qualified on retired worlds after each change (\KfourQualifications{} completed qualifications; \KfourQualificationsHalted{} halted attempts and \KfourQualificationsStale{} stale duplicate on superseded code are retained and excluded). After the freeze nothing was resumed, halted or repeated; no breaker fired; the results deposit's archive hash equals the server's.

\paragraph{Study K5.} The design went through \KfiveReviewRounds{} adversarial review rounds before its first freeze (\KfiveReviewRoundsCodex{} by Codex; round \KfiveClaudeRound{} by a Claude reviewer at maximum effort in an isolated worktree while Codex was at its weekly usage limit, recorded as such in its disposition); every round returned ``blocked''; its findings were repaired or recorded as disclosed limitations, and \KfiveReviewDoNotRun{} returned ``do not run''. The loop was ended by the author's decision after the last round rather than by a further confirmation round: that round's \KfiveLastRoundA{} A findings and one of its \KfiveLastRoundB{} B findings were repaired at their sources, and its other B findings recorded as disclosed limitations --- all but one under the claim licence's S-F clauses, and that one (the fixtures' fidelity and coverage statements) only in the study's journal. The freeze was two-staged (design, tools and stream inputs before generation; operating characteristics, power gate, prompts, serving qualifications and run configurations after the realised family count), and the first stage was re-issued twice: under its first two versions the stream ran to completion and two defects in the study's own adapter-file rule --- a producer / consumer pair that agreed on every fixture but not on the second adapter's real repository --- surfaced at the qualification and run-configuration steps; each re-issue retired its predecessor's derived tree to an archived directory, closed its spend campaign into a ledger and re-ran the stream from the frozen inputs, so nothing generated under a superseded version was analysed. After the second-stage freeze the first attempt of both ladders halted in the same minute when the serving client's heartbeat to the provider failed (\KfiveHaltedBatchesLone{} and \KfiveHaltedBatchesLtwo{} batches had been recorded on K5-1 and K5-2); as the runbook prescribes, each output was archived beside the stage's output and the same command re-executed once from an empty directory --- no resume, no re-use --- and the committed locks reconcile each archive as the second attempt's predecessor (zero torn ledger lines). One deviation after the freeze is recorded in the study's journal: two analyzer invocations occurred. The first followed the runbook's study-relative paths and failed the downstream path binding (the frozen recomputation resolves the prompt-manifest keys against the repository root and refused it); the first output was set aside unmodified and the analysis re-executed once from the repository root; the second output changed only the two path keys of the prompt-manifest record, every numerical output being identical (\texttt{\KfiveInvocationOne}, \texttt{\KfiveInvocationTwo}; the manuscript generator re-verifies the difference), and both files were retained and deposited. A second deviation, at the Stage-B deposit: the frozen depositor's first upload succeeded but the server's listing lagged and the depositor's download-back check did not find the file; the unchanged frozen depositor was then re-run under a wrapper that answered its upload with a synthetic conflict response, so that its own resume path performed the real listing, digest comparison and download-back verification (journal entry 10); the receipt and the deposited bytes are the server's. The generated results report carries the licence's disclosures under a heading that calls them verbatim although the frozen report tool paraphrases them; the journal records this and reproduces the clause verbatim, and this section and \S\ref{sec:results-k5} carry each disclosure in the manuscript's own words. A post-execution confirmation round by Codex re-executed the chain (locks, manifests, ledgers, receipts, the report byte-for-byte, the decisions) and returned ``\KfivePostexecVerdict{}'': \KfivePostexecA{} A, \KfivePostexecB{} B and \KfivePostexecC{} C items, about descriptions and accounting (the journal's cap and committed-total figures, the licence's mapping of the residual findings, the report's heading and level label, the depositor's comment); no analysed statistic, decision or binding changed, and the corrections are recorded in the journal. The three deposits (pre-stream, package, results) are on a private OSF component (\KfiveOsfNode{}), each download-back-verified, to be made public with the release; the study's spend was \KfiveSpendAll{} USD in all (\KfiveSpend{} on the two analysed ladders; the halted first attempts were never settled and stand at their reserved maximum in the campaign's accounting).
\section{Discussion}\label{sec:discussion}

\paragraph{What the \InitialStudies{} initial studies say together.} On one instrument, the verification request responded to a
one-character plan pointer almost as strongly as to a plan sentence (Study B); a record-naming directive's
form decided, model by model, whether it was honoured (Study D), and that split did not carry to a disjoint
panel (Study E); crossing the two forms produced a composite whose point estimate exceeded the id alone in each pre-named group and exceeded the criterion alone in two of the four, while on Opus~5 and, in the follow-up, on Fable~5 and Fable~5.1, the composite gave a lower target rate than the criterion alone (Studies F, F-x; every interval marginal, no joint claim). The same referent, encoded three ways, produced different allocations, and which encoding
was followed differed by model --- a description of these panels, not a property attributed to providers.

\paragraph{What the follow-ups add.} Study \Br{}'s registered rule returned Study B's corrected verdict, with the change in each replicated estimate within the margin (\Br{}).
The composite's cancellation was undone on three named Claude endpoints by a line that ratifies the field and by a budget of two credits scored as any credit, and partly undone by an explicit target field (alone, on Fable~5.1; after the criterion, on Fable~5) (J) --- each an exact bundled edit, so the accounts these edits were
designed around (legitimacy, position, budget) remain unseparated, but the cancellation is not a fixed property of the string pair; at eighty runs per cell (\Gr{}) the same six contrasts were measured again, and \GrRepWithin{} of \GrRepTotal{} replication contrasts were within the margin, \GrRepUnresolved{} unresolved and \GrRepExceeds{} beyond it. The four forms behaved differently in a second store (H1): the criterion was followed by every model there, the bare id by
all but Opus~5, and the growth-world criterion-minus-id gap shrank on three models, so the form asymmetry differed between the two instrument instances; because the store, system prompt, objective, memory contents, action labels, execution date, provider state and, for the GPT-5.6 models, two request fields changed together, H1 does not identify a store-specific property of any model. Reworded four ways with its construct held fixed (H2), the suffix's cancellation held for \HtwoCancelNOpus{} on Opus~5, \HtwoCancelNFableFive{} on Fable~5 and \HtwoCancelNFableFiveOne{} on Fable~5.1, and two of the rewordings were themselves not followed by one model each --- the plan-membership
wording by Opus~5 and the adverb-placement wording by Fable~5 --- while the two minimal edits gave departures within the margin on Opus~5 and one negative but unresolved departure on Fable~5 (P2): the study registers only these five exact strings on these models --- neither an invariance nor a non-invariance conclusion --- and neither localises the effect to particular bytes nor generalises beyond the authored rewordings. And the decision moved with the directive (I): the criterion changed the decision toward the current record on several models and worlds, on the GPT-5.6 models any directive changed the valid-world
decision, and on Fable~5.1 the same edits moved the decision away from the record in the superseded world --- treatment contrasts on two endpoints that align on some models and worlds and not on others (on GPT-5.6 Sol and Terra in the superseded world the credit moved and the decision did not), without identifying the credit as the mediator.

\paragraph{What the generated worlds add.} On \KtwoMainWorlds{} worlds in \KtwoFamilies{} domain families generated for Study K2 under a fixed contract and screened by \KtwoJudgesN{} open-weight judges blind to the outcome, the \KtwoSignatures{} registered signatures held with the family as the unit (negative in \KtwoBothCritOpusFamNeg{} and \KtwoBothCritFableFiveOneFamNeg{} of \KtwoFamilies{} families), and \KtwoDirMet{} of the \KtwoDirTotal{} directional forecasts deposited before generation were met --- a descriptive tally of dependent rows. The other endpoints are described, not tested: Haiku~4.5's contrast had the opposite sign, the point estimates of Sonnet~5, the GPT-5.6 endpoints and GPT-6 Astra lay within \KtwoOtherMaxAbs{} points of zero, with intervals inside the margin on \KtwoOtherWithin{} of them (\KtwoOtherWithinNames{}), undetermined on the others and beyond it on none, and the open-weight stratum is reported by its rows (Table~\ref{tab:studyK2-contrasts}). The population of generated worlds is the one stated in \S\ref{sec:results-k2} --- one generator's distribution under one contract --- and the study identifies no mechanism.

\paragraph{What the adapter ladders add.} On Llama~3.3~70B the registered criterion for a gain-dependent change of both\_IL $-$ crit was not met (\KthreeBothCritSeventyBMeanZero{} to \KthreeBothCritSeventyBMeanEight{} points across the gains; slope \KthreeBothCritSeventyBSlope{} [\KthreeBothCritSeventyBSlopeLo, \KthreeBothCritSeventyBSlopeHi]; change \KthreeBothCritSeventyBChange{} [\KthreeBothCritSeventyBChangeLo, \KthreeBothCritSeventyBChangeHi]) --- which is not a statement that the contrast was unchanged. On Llama~3.1~8B the same contrast went from \KthreeBothCritEightBMeanZero{} to \KthreeBothCritEightBMeanEight{} points across the gains (descriptive; no criterion registered). The two ladders differ in rank and effective scale and the study licenses no comparison between them, no statement about defences in general, and no mechanism.

\paragraph{What the registered re-execution adds.} Study K4 turned K3's description into a rule: on Llama~3.1~8B the registered attenuation criterion for both\_IL $-$ crit was not met on this ladder (\KfourBothCritLoneMeans{} points across the gains; slope \KfourBothCritLoneSlope{} [\KfourBothCritLoneSlopeLo, \KfourBothCritLoneSlopeHi]; change \KfourBothCritLoneChange{} [\KfourBothCritLoneChangeLo, \KfourBothCritLoneChangeHi], whose upper bound does not clear the $-$10 margin) --- this is not a statement that the contrast was unchanged, and the rule's realised margin error is given in \S\ref{sec:results-k4}; at gain 16, beyond the adapter's release, the same contrast was \KfourBothCritLtwoMeanSixteen{} points (descriptive); and on the 70B ladder the registered criterion was not met on a fresh execution, as in Study K3 (change \KfourBothCritLthreeChange{} [\KfourBothCritLthreeChangeLo, \KfourBothCritLthreeChangeHi]; the two compared as statements, never pooled). The two executions of each ladder are comparators, never pooled; the study licenses no mechanism and no statement about defences in general.

\paragraph{What registered power adds.} Study K5 re-posed K4-1's question at registered power on \KfiveFamilies{} families: the 8B ladder with Meta-SecAlign satisfied the registered criterion (\KfiveBothCritLoneMeans{} points across the gains; change \KfiveBothCritLoneChange{} [\KfiveBothCritLoneChangeLo, \KfiveBothCritLoneChangeHi]); K4-1's status on \KfourFamilies{} families stands as Study K4's result, the two compared as statements. A second SecAlign++ adapter (independently trained, by its repository's account; served under the imposed Meta-SecAlign template) did not satisfy the same criterion: its change was \KfiveBothCritLtwoChange{} points [\KfiveBothCritLtwoChangeLo, \KfiveBothCritLtwoChangeHi], an interval whose upper end does not clear the $-$10-point margin --- not a statement that the contrast was unchanged --- and the difference between the two adapters' changes (\KfiveThreeTheta{} points [\KfiveThreeThetaLo, \KfiveThreeThetaHi]) is reported under its registered descriptive sentence; nothing follows from one ladder's status differing from the other's (\S\ref{sec:results-k5}). What the K programme says, and no more: on this instrument the composite's advantage over the criterion decreased with the gain of a defensive adapter on the 8B base --- in one ladder by an amount whose registered interval lay entirely beyond a $-$10-point margin, in the other by an amount whose interval did not; it licenses no mechanism and no statement about defences in general.

\paragraph{Per-model heterogeneity is the result.} Every pooled estimate in this paper is an equal-weight mean
over a fixed panel; the per-model tables are the evidence and the pooled number summarises them. The
\textsc{CRIT\_GT\_ID2} composite-versus-criterion mean combines a rise and a fall; the direct-provider mean
combines a fall, a rise and several zeros; Study E's pooled contrast changes sign depending on two models.
\citet{okamoto2026framing} report, for their compliance setting, that benchmark scores and developers' descriptions of post-training do not predict where a model falls; we performed no such analysis
and note only that the models that did and did not follow the composite here are not separated by any
attribute we measured.

\paragraph{The three Claude endpoints, decomposed.} Study F-x speaks narrowly: Opus~5's ordering recurred on re-run; Fable~5
showed the same ordering, with a criterion-versus-composite contrast within the registered margin of Opus's;
Fable~5.1 showed a lower criterion rate and the same composite floor. Model-generated rationale text describing the id as an inherited priority field appears in \FOpusOverrideNone{}, \FOpusOverrideId{}, \FOpusOverrideCrit{}, \FOpusOverrideBothIL{} and \FOpusOverrideBothIF{} of Opus's Study F episodes by arm (an exploratory count, \S\ref{sec:results-f}); we report it and draw nothing from it. We do not know whether a trained disposition, a preference for reasons over pointers, the composite's added length or the store's presentation of the field is responsible; the instrument cannot separate these. Sonnet~5 and Haiku~4.5, from the same provider, followed the composite in every Study F episode. Study G then applied six byte-controlled edits to the field and reports forty marginal contrasts (\S\ref{sec:results-g}). Among them: on Opus~5 the ` (as stated)' suffix gave a within-margin contrast ($G_2 = \GGtwoOpus{}$ [\GGtwoOpusLo, \GGtwoOpusHi]), the ` (memory\_44)' suffix a contrast beyond the margin with V44 \GVfortyfourKOpus{}, and the `set today' rendering a higher target rate than `inherited' by more than the margin ($G_6 = \GGsixOpus{}$ [\GGsixOpusLo, \GGsixOpusHi]); on Fable~5 the ` (as stated)' suffix gave a contrast beyond the margin ($G_2 = \GGtwoFableFive{}$ [\GGtwoFableFiveLo, \GGtwoFableFiveHi]); on GPT-5.6 Sol the ` (memory\_44)' suffix was followed (V44 \GVfortyfourKGPTSol{}) $G_3$ and $G_8$ (the competing pointer against the criterion and against ` (memory\_73)') are negative beyond the margin (\GGthreeGPTSol{}, \GGeightGPTSol{}), and the other eight contrasts are within the margin. These are total effects of exact strings on four named models; the registration states what each string cannot isolate, and no property (length, the presence of an id, provenance) is attributed as a cause, nor is any cross-model pattern asserted. The `set today' label sits inside a block that still says the state was carried over.

\paragraph{What follows for a memory system.} Nothing here shows that a redirected request improves a
decision, and no study here tested a scheduler. What the results support is narrower: a verification directive
is not a bit a system sets; the same intended referent, written as an id, as a criterion or as both, was
followed at rates that differed by tens of points and in model-specific directions, so the observed model specificity motivates measuring the exact form on the intended endpoint rather than assuming it.

\paragraph{Provenance.} Study B carries a recorded correction and Study F a quarantined run. Study B's verdict inverted when a counterbalanced factor was decoded, after its first repository report; Study F's first run was contaminated
by a tool the request had asked to disable, caught by cost accounting and confirmed by reading the rationales
while the run was live. The per-cell prompt-token invariance gate of \S\ref{sec:integrity} is exact for what it checks, with the stated blind spots, and the counterbalance decode step is now a freeze blocker in
our checklist; we recommend both to anyone pre-registering fixed-design studies on hosted models.

\section{Limitations}\label{sec:limitations}

\begin{enumerate}\itemsep1pt
\item \textbf{One construction, two instrument instances, one decision scenario, one generator.} Six memories (\KtwoMemoriesMin{} to \KtwoMemoriesMax{} in Study K2's generated worlds), one target backing the non-active plan, one request, $k = 1$ (two Study J arms at $k = 2$); the procurement instance (H1) differs in more than the store; Study I's decision is one scenario's two worlds. Nothing here shows that a redirected request improves decisions beyond Study I's per-model, per-world descriptions, and no study tested a scheduler.
\item \textbf{The criterion identifies the target in one step, and composites alias length.} The stated criterion is a paraphrase of the id; that any stated reason suffices is not shown. The composite is longer than either component; composite-versus-single contrasts bundle content and length (registered as not separable), and the id-first layout controls order and wrapping, not length.
\item \textbf{Descriptive, marginal, fixed panels.} Studies F, F-x, G, I, H1, J, H2 and G$'$ register no verdict; every interval is marginal, no familywise label exists, no conjunction across models, worlds or wordings is a familywise claim, Study F's groups are pre-named descriptive labels, and the three Claude endpoints are three named models, not a family.
\item \textbf{Study E is a transport test, reinterpreted after registration.} No model overlaps Study D's panel; runs per cell and inference configuration differ; the pairing rule under partial errors was decided by the frozen code; its failed superiority test is not evidence of equivalence.
\item \textbf{Two phases, neither jointly prospective.} The first six studies were outcome-sequential; the six follow-ups were conceived together, partly developed in parallel, and frozen and run one at a time, and Study K2 was designed after them; the narrative order was fixed in a plan committed after Study B$'$'s result (\S\ref{sec:studies}).
\item \textbf{No mechanism.} Rationale text is reported as text; copying versus dereferencing, and a principled decline versus non-engagement, are not separated by the registered endpoint; Study J's edits, Study H2's rewordings and Study G's strings are one authored choice each, and each contrast is the total effect of an exact string on named models.
\item \textbf{Transport and replication contrasts do not isolate one change.} Study H1's contrasts compare runs made on different days with different world text and, for the GPT-5.6 models, different request fields; Study G$'$'s within-margin label says only that the change in a contrast lies inside $\pm 10$ points, and half of its thirty contrasts are unresolved.
\item \textbf{Post-freeze corrections and a quarantined run} are described in \S\ref{sec:integrity} and Appendix~\ref{app:registration}: Study B's headline is a post-freeze correction on the locked episodes; Study F's first run is quarantined and the invariance gate cannot see a count-preserving augmentation; the OpenTimestamps proofs of the frozen packages are Bitcoin-attested for every study through K4 and for Study K5's Stage-B manifest (K3's, K4's and K5's were upgraded after their runs), while Study K5's frozen pre-stream proof itself stays pending because its Stage-B manifest binds the proof's bytes --- its Bitcoin-attested upgrade is kept beside it as a separate file --- and the two superseded pre-stream proofs remain pending and the before-first-call ordering rests on receipts, commits and ledgers.
\end{enumerate}

\paragraph{Study K2.} The generated worlds are drawn from one skeleton sampler under one structural contract and realised by \KtwoRealisersN{} models under decoding constraints; the judges are models and did not certify that the criterion picks out one record; the bank stratum registered to replicate the generation with a second author did not realise and is reported as attempted. \KtwoOpenAbsent{} of the \KtwoOpenRegistered{} open-weight endpoints are absent (\KtwoOpenUnservableN{} unservable under the frozen image, \KtwoOpenAllErrorN{} returning no schema-valid output), the open stratum carries no forecast, and its per-episode credits come from the hash-bound converter output rather than a second parse of the batch records. The primary is \KtwoSignatures{} signatures, one per named endpoint; every other number in K2 is descriptive; every sentence on the generated worlds carries the population stated in \S\ref{sec:results-k2}, and none concerns a population of environments or models.

\paragraph{Study K3.} The manipulation is the scalar applied to one fixed LoRA update (the adapter's \texttt{lora\_alpha}), not a calibrated defence strength; the two ladders have different effective scales (the adapters' ranks differ) and are never pooled; zero gain is the base weights under the adapter's chat template, which drops the base template's date header, so Study K2's Llama rows are comparators, not the curves' origin. On the LoRA-enabled engine greedy decoding did not reproduce its own completions (\KthreeAuditSESeventyBBytes{} of 80 byte-identical on the same engine for the 70B ladder), so every K3 number is conditional on the recorded execution; the bootstrap reweights families and says nothing about re-execution. The 8B ladder's attenuation is descriptive; the primary was not met and the study licenses no absence of effect, no mechanism and no population of adapters, models or templates.

\paragraph{Study K4.} The three ladders inherit Study K3's limits (adapter gain, not defence strength; different effective scales, never pooled; zero gain under the adapter's template; serving noise conditional on the recorded execution). K4-1 not being met is a statement about a registered margin at a registered level, not about the absence of attenuation: the study's own operating characteristics put the rule's power near \KfourPowerLoLone{}--\KfourPowerHiLone{} at a realised 25-point change and lower at Study K3's point estimate, and the rule's realised one-sided error at the margin in the registered grid was \KfourMarginErrLoLone{}--\KfourMarginErrHiLone{}\% (nominal \KfourMarginNominalLone{}\%), a finite-grid figure. L2's gains beyond the adapter's release are an extrapolation of one LoRA update's scale, registered descriptive; nothing about them transfers to other adapters. The design was chosen after Study K3's outcome and is disclosed as such; the two executions of each ladder are compared as statements, never pooled.

\paragraph{Study K5.} The two ladders inherit Studies K3's and K4's limits (adapter gain, not defence strength; zero gain under the adapter's template; serving noise conditional on the recorded execution). The rule being met on K5-1 is a statement about a registered margin at a registered level on this instrument: the operating characteristics' error control is a finite-grid figure (\KfiveMarginErrLo{}--\KfiveMarginErrHi{}\% at \KfiveBoundaryConfigs{} boundary corners, nominal \KfiveMarginNominal{}\%), not a uniform guarantee, and the intervals describe family reweighting on a fixed instrument whose cohort proportions are fixed. The study is post-K4 and outcome-informed: its question, effect assumption and family target were chosen after K4-1's registered outcome; of the K2 cohort's \KfiveCohortKtwoFamilies{} families, \KfiveCohortKtwoExposed{} had been exposed to the K3 / K4 outcomes and \KfiveCohortKtwoUnused{} had passed only K2's realiser and judge stages; the cohort is reported beside the new one. The two updates are not matched in strength (released scales \KfiveReleasedScaleLone{} and \KfiveReleasedScaleLtwo{}; the second is served under the imposed Meta-SecAlign template, not its native configuration, and ``independently trained'' is its repository's claim); K5-2 not satisfying the criterion is not a statement that its contrast was unchanged and not evidence that the second adapter differs in kind from the first, and K5-3's difference, though its interval excludes zero, is registered descriptive: it compares two fixed execution paths and licenses no superiority or equivalence claim. Nothing transfers to other adapters, bases, templates or serving stacks.

\paragraph{Data and code availability.} The release accompanying this paper is archived at Zenodo (doi:\lit{10.5281/zenodo.22267221}; the DOI was reserved for the record before publication --- \lit{} here marks an identifier, not a measured value) and, in its first version, contains the episode files of the twelve studies of the previous version of this paper; the frozen packages, locked records and results of Studies K2, K3, K4 and K5 are to be added as a new version of the same record before this version is distributed (Study K5's three OSF deposits are on a private component, to be made public at the same time), after which it will contain the \TotalStudies{} studies reported here (\TotalEpisodes{} attempted episodes), the frozen registration and analysis artifacts of every study with their SHA-256 manifests, OpenTimestamps proof files and Open Science Framework (OSF) deposit receipts, the correction and erratum records, and the generator that emits every quantitative output in this paper. Design constants (budgets, runs per cell, the bootstrap size, the margin) and numbers quoted from cited papers are marked as such in the source.

\begin{sloppypar}
\paragraph{AI assistance.} The author used language-model assistants for parts of this work:
Anthropic's Claude, principally through Claude Code, for research-design critique, experiment
planning, implementation and execution of the runners, analysis and audit tooling, manuscript
drafting and editing, simulated adversarial review, and release engineering; and OpenAI's ChatGPT
(Codex) for research-design critique, package and manuscript critique, and simulated adversarial
review. The author chose the research questions, approved every experimental package and decided
whether each run took place, interpreted the results, selected the claims, and is responsible for
the correctness of the manuscript. No language model is an author. The models studied are
experimental subjects, not tools of the analysis: their responses are the data, every outcome is
scored deterministically, and no model output is used to judge another.
\end{sloppypar}

\bibliography{bib/refs}

\begin{thebibliography}{22}
\providecommand{\natexlab}[1]{#1}
\providecommand{\url}[1]{\texttt{#1}}
\expandafter\ifx\csname urlstyle\endcsname\relax
  \providecommand{\doi}[1]{doi: #1}\else
  \providecommand{\doi}{doi: \begingroup \urlstyle{rm}\Url}\fi

\bibitem[Akewar and Ranjan(2026)]{akewar2026safecommit}
Mayur Akewar and Ravi Ranjan.
\newblock {SafeCommit}: Certifying when memory-grounded agents may safely act,
  2026.
\newblock URL \url{https://arxiv.org/abs/2608.04289}.

\bibitem[Briggs and Scheutz(2015)]{briggs2015sorry}
Gordon Briggs and Matthias Scheutz.
\newblock ``{S}orry, {I} {C}an't {D}o {T}hat'': {D}eveloping {M}echanisms to
  {A}ppropriately {R}eject {D}irectives in {H}uman-{R}obot {I}nteractions.
\newblock In \emph{AAAI Fall Symposium Series}, 2015.

\bibitem[Chen et~al.(2024)Chen, Zharmagambetov, Mahloujifar, Chaudhuri, Wagner,
  and Guo]{chen2024secalign}
Sizhe Chen, Arman Zharmagambetov, Saeed Mahloujifar, Kamalika Chaudhuri, David
  Wagner, and Chuan Guo.
\newblock {SecAlign}: Defending against prompt injection with preference
  optimization, 2024.
\newblock URL \url{https://arxiv.org/abs/2410.05451}.
\newblock ACM CCS 2025.

\bibitem[Chen et~al.(2025)Chen, Zharmagambetov, Wagner, and
  Guo]{chen2025metasecalign}
Sizhe Chen, Arman Zharmagambetov, David Wagner, and Chuan Guo.
\newblock {Meta SecAlign}: A secure foundation {LLM} against prompt injection
  attacks, 2025.
\newblock URL \url{https://arxiv.org/abs/2507.02735}.

\bibitem[Fang et~al.(2026)Fang, Hu, Chang, Guo, Tao, Liu, Ruan, Huang, and
  Fang]{fang2026budget}
Zhengru Fang, Senkang~Forest Hu, Zhonghao Chang, Yu~Guo, Yihang Tao, Hongyao
  Liu, Mengzhe Ruan, Jun Huang, and Yuguang Fang.
\newblock Inference-time budget control for {LLM} search agents.
\newblock \emph{{arXiv} preprint arXiv:2605.05701}, 2026.

\bibitem[Farahani et~al.(2026)Farahani, Penzkofer, and
  Johansson]{farahani2026copy}
Mehrdad Farahani, Franziska Penzkofer, and Richard Johansson.
\newblock To copy or not to copy: Copying is easier to induce than recall.
\newblock In \emph{Proceedings of EMNLP 2026}, 2026.
\newblock arXiv:2601.12075.

\bibitem[Gong and Deng(2026)]{gong2026planguard}
Guangyu Gong and Zizhuang Deng.
\newblock {PlanGuard}: Defending agents against indirect prompt injection via
  planning-based consistency verification.
\newblock arXiv:2604.10134, 2026.

\bibitem[Guan et~al.(2026)Guan, Zhao, and Deng]{guan2026cicl}
Xinyu Guan, Qianyang Zhao, and Yuming Deng.
\newblock Decision-aware memory cards: Counterfactual-inspired context
  selection and compression for tool-using {LLM} agents.
\newblock arXiv:2606.08151, 2026.

\bibitem[Hwang et~al.(2025)Hwang, Kim, Koo, Kang, Bae, and Jung]{dimbench2025}
Yerin Hwang, Yongil Kim, Jahyun Koo, Taegwan Kang, Hyunkyung Bae, and Kyomin
  Jung.
\newblock {LLMs} can be easily confused by instructional distractions.
\newblock In \emph{Proceedings of the 63rd Annual Meeting of the Association
  for Computational Linguistics (Volume 1: Long Papers)}, pages 19483--19496,
  2025.
\newblock arXiv:2502.04362.

\bibitem[Jhaveri et~al.(2026)Jhaveri, GX-Chen, Sucholutsky, and
  Choi]{jhaveri2026falsify}
Ayush~Rajesh Jhaveri, Anthony GX-Chen, Ilia Sucholutsky, and Eunsol Choi.
\newblock Failing to falsify: Evaluating and mitigating confirmation bias in
  language models.
\newblock \emph{arXiv preprint arXiv:2604.02485}, 2026.

\bibitem[Jia et~al.(2024)Jia, Wu, Qin, and Squicciarini]{jia2024taskshield}
Feiran Jia, Tong Wu, Xin Qin, and Anna Squicciarini.
\newblock The task shield: Enforcing task alignment to defend against indirect
  prompt injection in {LLM} agents.
\newblock arXiv:2412.16682, 2024.

\bibitem[Li et~al.(2026)Li, Yao, and Zheng]{li2026mcb}
Baichuan Li, Junyi Yao, and Zihao Zheng.
\newblock Remember, verify, or ask? cross-family evaluation of memory
  commitment in {LLM} agents, 2026.
\newblock URL \url{https://arxiv.org/abs/2608.19564}.

\bibitem[McCauley et~al.(2026)McCauley, Kan, and Martin]{mccauley2026ih}
Conor McCauley, Zeliang Kan, and Jason Martin.
\newblock {IH-Benchmark}: A conflict-centered benchmark for
  instruction-hierarchy robustness in {LLM} applications.
\newblock arXiv:2607.25987, 2026.

\bibitem[Munirathinam(2026)]{munirathinam2026recusal}
Thamilvendhan Munirathinam.
\newblock Will the agent recuse, and will it stop? measuring {LLM}-agent
  compliance with in-band governance signals at the access door and mid-flight.
\newblock arXiv:2606.06460, 2026.

\bibitem[Nakayashiki(2026{\natexlab{a}})]{priorwork2026stale}
Kazuki Nakayashiki.
\newblock When stale constraints go unchecked: Budgeted verification failures
  in inherited agent memory.
\newblock arXiv:2608.25553; Zenodo concept DOI 10.5281/zenodo.22108557
  (resolves to the latest version), 2026{\natexlab{a}}.

\bibitem[Nakayashiki(2026{\natexlab{b}})]{priorwork2026verification}
Kazuki Nakayashiki.
\newblock Verification allocation in inherited agent memory: Provenance
  availability is not provenance use, 2026{\natexlab{b}}.
\newblock URL \url{https://doi.org/10.5281/zenodo.22084498}.
\newblock Zenodo; concept DOI, resolves to the latest version (v2:
  10.5281/zenodo.22102676).

\bibitem[Okamoto et~al.(2026)Okamoto, Erol, and Erol]{okamoto2026framing}
Mika Okamoto, Ansel~Kaplan Erol, and Kutluhan Erol.
\newblock Why do {AI} agents break rules? how framing, context, and social
  signals shape compliance.
\newblock In \emph{AAAI/ACM Conference on AI, Ethics, and Society}, 2026.
\newblock arXiv:2608.12323.

\bibitem[Pattison et~al.(2026)Pattison, Manuali, and Lazar]{pattison2026blind}
Cameron Pattison, Lorenzo Manuali, and Seth Lazar.
\newblock Blind refusal: Language models refuse to help users evade unjust,
  absurd, and illegitimate rules.
\newblock arXiv:2604.06233, 2026.

\bibitem[Potham(2025)]{potham2025hierarchical}
Ram Potham.
\newblock Evaluating {LLM} agent adherence to hierarchical safety principles: A
  lightweight benchmark for probing foundational controllability components.
\newblock arXiv:2506.02357, 2025.

\bibitem[Song and Cai(2026)]{song2026envprobe}
Xinyuan Song and Zekun Cai.
\newblock Ask the world before acting: Environment probing for calibrated agent
  world models, 2026.
\newblock URL \url{https://arxiv.org/abs/2606.31422}.

\bibitem[Tan et~al.(2026)Tan, Valentino, Akhter, Zhou, Liakata, and
  Aletras]{tan2026compliance}
Xingwei Tan, Marco Valentino, Mahmud~Elahi Akhter, Yuxiang Zhou, Maria Liakata,
  and Nikolaos Aletras.
\newblock Compliance versus sensibility: On the reasoning controllability in
  large language models.
\newblock arXiv:2604.27251, 2026.

\bibitem[Zhu et~al.(2026)Zhu, Chen, Yu, Wu, and Wang]{zhu2026tiermem}
Qiming Zhu, Shunian Chen, Rui Yu, Zhehao Wu, and Benyou Wang.
\newblock From lossy to verified: A provenance-aware tiered memory for agents.
\newblock \emph{arXiv preprint arXiv:2602.17913}, 2026.

\end{thebibliography}

\clearpage
\appendix
\FloatBarrier
\section{Registration chains}\label{app:registration}

Each package was hashed into a SHA-256 manifest, committed, OpenTimestamps-stamped and deposited to OSF
(project \ifanon\else\texttt{axsnm}\fi; Study K5's three deposits on the private component \ifanon\else\texttt{c48ns}\fi) with download-back verification before the first confirmatory call (Study K2's OpenTimestamps proofs were made after its run had started; its pre-run ordering rests on the OSF server timestamp and the ledgers); each run was
locked into a completion manifest before analysis. Table~\ref{tab:chains} lists the identifiers; receipts,
proof files, ledgers and errata will be in the release. ``OTS'' means the proof file exists and has been upgraded to a Bitcoin block attestation (the block heights are in the freeze logs; the proofs also retain pending calendar branches). The lock rollup is the SHA-256 of the completion manifest file; the reports of Studies B and D print instead the SHA-256 of the manifest's concatenated hash column, a different digest of the same list.

\begin{table}[h]\centering\small
\caption{Registration identifiers (manifest rollup = SHA-256 of the manifest file, first 16 hex; \ifanon\else OSF file ids on node \texttt{axsnm}; K5's on the private component \texttt{c48ns}\fi; ``files'' = hashed entries).}\label{tab:chains}
\adjustbox{max width=\textwidth}{\begin{tabular}{lrllll}\toprule
study & files & manifest rollup & OSF file & lock rollup & OTS \\\midrule
B & \BManifestFiles & \texttt{\BManifestRollup} & \osf{\BOsfFile} & \texttt{\BLockRollupLong} & \BOtsPresent \\
D & \DManifestFiles & \texttt{\DManifestRollup} & \osf{\DOsfFile} & \texttt{\DLockRollupLong} & \DOtsPresent \\
E & \EManifestFiles & \texttt{\EManifestRollup} & \osf{\EOsfFile} & \texttt{\ELockRollupLong} & \EOtsPresent \\
F (v2.9) & \FManifestFiles & \texttt{\FManifestRollup} & \osf{\FOsfFile} & \texttt{\FLockRollupLong} & \FOtsPresent \\
F-x (v0.2) & \FxManifestFiles & \texttt{\FxManifestRollup} & \osf{\FxOsfFile} & \texttt{\FxLockRollupLong} & \FxOtsPresent \\
G (v0.1) & \GManifestFiles & \texttt{\GManifestRollup} & \osf{\GOsfFile} & \texttt{\GLockRollupLong} & \GOtsPresent \\
\Br{} (v0.1) & \BrManifestFiles & \texttt{\BrManifestRollup} & \osf{\BrOsfFile} & \texttt{\BrLockRollupLong} & \BrOtsPresent \\
I (v0.1) & \IManifestFiles & \texttt{\IManifestRollup} & \osf{\IOsfFile} & \texttt{\ILockRollupLong} & \IOtsPresent \\
H1 (v0.1) & \HoneManifestFiles & \texttt{\HoneManifestRollup} & \osf{\HoneOsfFile} & \texttt{\HoneLockRollupLong} & \HoneOtsPresent \\
J (v0.1) & \JManifestFiles & \texttt{\JManifestRollup} & \osf{\JOsfFile} & \texttt{\JLockRollupLong} & \JOtsPresent \\
H2 (v0.1) & \HtwoManifestFiles & \texttt{\HtwoManifestRollup} & \osf{\HtwoOsfFile} & \texttt{\HtwoLockRollupLong} & \HtwoOtsPresent \\
\Gr{} (v0.1) & \GrManifestFiles & \texttt{\GrManifestRollup} & \osf{\GrOsfFile} & \texttt{\GrLockRollupLong} & \GrOtsPresent \\
K2 (v1.32) & \KtwoManifestFiles & \texttt{\KtwoManifestRollup} & \osf{\KtwoOsfFile} & \texttt{\KtwoLockRollupLong} & \KtwoOtsPresent{} (post-run) \\
K3 (v1.2) & \KthreeManifestFiles & \texttt{\KthreeManifestRollup} & \osf{\KthreeOsfFileId} & \texttt{\KthreeLockRollupEightB} / \texttt{\KthreeLockRollupSeventyB} & \KthreeOtsPresent{} (Bitcoin: \KthreeOtsBitcoin, post-run) \\
K4 (v0.1) & \KfourManifestFiles & \texttt{\KfourManifestRollup} & \osf{\KfourOsfFileId} & \texttt{\KfourLockRollupLone} / \texttt{\KfourLockRollupLtwo} / \texttt{\KfourLockRollupLthree} & \KfourOtsPresent{} (Bitcoin: \KfourOtsBitcoin, post-run) \\
K5 (two-stage) & \KfivePrestreamFiles{} + \KfiveManifestFiles & \texttt{\KfivePrestreamRollup} / \texttt{\KfiveManifestRollup} & \osf{\KfiveOsfFileId} (private) & \texttt{\KfiveLockRollupLone} / \texttt{\KfiveLockRollupLtwo} & \KfiveOtsPresent{} (Bitcoin: \KfivePrestreamOtsBitcoin{} / \KfiveOtsBitcoin, post-run) \\
\bottomrule\end{tabular}}\end{table}

\begin{table}[htbp]\centering\footnotesize\setlength{\tabcolsep}{4pt}
\caption{Execution facts per study. Exact model ids, endpoint pins and inference settings are in each study's frozen package (request contract, model list, canonical inputs); Anthropic requests send \texttt{max\_tokens} 16000 and a structured output schema, OpenAI requests \texttt{reasoning.effort} medium, OpenRouter requests the pinned provider order with fallbacks disabled.}\label{tab:repro}
\adjustbox{max width=\textwidth}{\begin{tabular}{lp{2.7cm}p{4.0cm}rrrrp{1.9cm}}\toprule
study & access route & models & runs/cell & attempted & scored & errors & record dates (UTC) \\\midrule
B & direct (Anthropic, OpenAI) & Opus 5, Sonnet 5, Haiku 4.5, GPT-5.6 Sol/Terra/Luna & 25 & 1200 & 1200 & 0 & 2026-09-01 \\
D & direct (Anthropic, OpenAI) & the same six & 40 & 720 & 720 & 0 & 2026-09-01 \\
E & OpenRouter (pinned endpoints) & nine OpenRouter-served (seven open-weight) & 20 & 540 & 526 & 14 & 2026-09-01 \\
F & direct + OpenRouter & all fifteen & 40 / 20 & 2100 & 2075 & 25 & 2026-09-02 \\
F-x & direct (Anthropic) & Fable 5.1, Fable 5, Opus 5 & 40 & 600 & 600 & 0 & 2026-09-02 \\
G & direct (Anthropic, OpenAI) & Opus 5, Fable 5, Fable 5.1, GPT-5.6 Sol & 40 & 960 & 960 & 0 & 2026-09-02 \\
B$'$ & direct (Anthropic, OpenAI) & the Study B six & 25 & 1200 & 1200 & 0 & 2026-09-02 \\
\bottomrule\end{tabular}}\end{table}

\paragraph{Analyses executed.} Studies D, E, F, F-x and G were analysed once by the analyzer whose hash is in their manifests (each result reproduces byte-for-byte under the frozen analyzer). Study B's deposited analyzer required a one-line repair to start (its current hash differs
from the manifest's for that file) and then computed the letter contrast; the referent contrast reported here
is from a post-freeze corrected script on the same locked episodes, with the joint share interval added on
2026-09-02; the correction record lists every difference. Study F's run 1 (package v1, frozen and deposited
in the same way) is quarantined; the package was re-frozen as v2.9 after eight review rounds and run 2 is the
locked run.

\paragraph{Study K3.} The package was frozen after \KthreeReviewRounds{} adversarial review rounds, stamped and deposited before the first confirmatory call; the two ladders were served under the deposited configuration by a runner that refuses without the package rollup and the deposit receipt, each ladder's records were locked and committed before conversion, the analysis ran once, and an independent re-implementation (no import from the analyzer or the converter) recomputed every reported number, interval, gate and decision from the locked records --- and the audit agreement counts from the retained replay completions --- with zero differences on both ladders; the report refuses to render without those per-ladder receipts. Two incidents are recorded: the frozen deposit script's download-back request failed after a successful upload and a post-freeze tool adopted the upload with its own verified download-back; two launches were refused by the live guard for unbound flags before any call.

\paragraph{Study K4.} The package was frozen after \KfourReviewRoundsMain{} adversarial review rounds and a confirmation round (\KfourReviewRoundsCodex{} by Codex, \KfourReviewRoundsClaude{} by a Claude reviewer at maximum effort in an isolated worktree while the Codex client was at its weekly usage limit; the reviewer family is recorded in every disposition), stamped and deposited before the first confirmatory call; the three ladders were served under the deposited configuration in the registered order by a runner that refuses without the package rollup and the deposit receipt and refuses a non-empty output directory (no resume: a halted execution would be archived and the ladder re-executed from an empty directory, which never happened); each ladder's records were locked (one invocation, one status ledger, the producer's call contract) and committed before conversion, the analysis ran once, and the independent re-implementation recomputed every reported number, interval, gate and decision from the locked records --- and the audit agreement counts from the retained replay completions --- with zero differences on all three ladders; the report refuses to render without the three per-ladder receipts and Study K3's frozen results file for its historical table; the results deposit refuses unless every lock, converter manifest, receipt, run manifest and campaign record is present and bound. No incident is recorded after the freeze.

\paragraph{Study K5.} The freeze was two-staged. Stage A --- the design memo, pre-specification, claim licence, runbook, every tool, the stream's inputs and the operating-characteristics simulator with the family count as its one free parameter --- was frozen, stamped and deposited before the generation stream ran (\KfivePrestreamFiles{} files, rollup \texttt{\KfivePrestreamRollup}; \osf{\KfiveOsfFileId} is the Stage-B file, the pre-stream file is \texttt{\KfivePrestreamOsfFile}); the stream then produced the worlds, the realised family count fixed the operating characteristics and the power gate labelled the two statements, and Stage B --- the gate, the bound prompt files, the serving qualifications and the run configurations, re-hashing every Stage-A file --- was frozen, stamped and deposited before the first confirmatory call (\KfiveManifestFiles{} files, rollup \texttt{\KfiveManifestRollup}). Stage A was re-issued twice before Stage B (its first two versions are retained with their stamps; the reasons are in \S\ref{sec:integrity}). The two ladders were served in parallel under the deposited configuration by a runner that refuses without the package rollup and the deposit receipt and refuses a non-empty output directory; the first attempt of both halted when the serving client's heartbeat to the provider failed and the service terminated the applications, and each was archived and re-executed once from an empty directory, the committed locks reconciling the archives as predecessors; each ladder's records were locked and committed before conversion, a runtime identity check compared the two ladders' rendered prompts, the analysis ran on the combined file (twice, for a path-convention defect recorded in the journal; the first output retained), and the independent re-implementation recomputed every reported number, interval, gate and decision from the locked records --- and the audit agreement counts from the retained replay completions --- with zero differences on both ladders and on the K5-3 contrast; the report refuses to render without the three receipts and Study K4's frozen results file for its historical row; the results deposit refuses unless every lock, converter manifest, receipt and campaign record is present and bound (the campaign's atomic records are in the results deposit; the ledger of the superseded Stage-A campaigns is bound in the Stage-B package, so the study's accounting is reconstructed from the two packages together). The deposits are on a private OSF component (\KfiveOsfNode{}) to be made public with the release; the pre-stream proof's bytes are bound by the Stage-B manifest, so its Bitcoin-upgraded proof is kept beside the frozen pending one (\texttt{PRESTREAM\_MANIFEST.sha256.upgraded.ots}), and the Stage-B proof was upgraded in place after the runs, as Study K4's was.

\paragraph{Follow-up analyses.} Studies \Br{}, I, H1, J and H2 were each analysed once by the analyzer whose hash is in their manifests; their results reports were generated by the hashed generator (H1, J, H2) or, where the hashed generator read keys the analyzer no longer emitted after review dispositions (\Br{}, I), by an unhashed successor generator recorded with an erratum. Study \Gr{} follows the same chain; its lock and single analysis are recorded in its freeze log. Study K2 was analysed once by the analyzer whose hash is in its manifest, on records locked by a post-freeze lock tool and merged by a post-freeze merge tool (\S\ref{sec:integrity}); its results were deposited to the same node after the analysis (\texttt{\KtwoResultsDepositFile}, \KtwoResultsDepositDate).

\paragraph{Deviations.} Every post-freeze deviation is recorded in each study's deviation file and never
corrected in place: Study B's analysis correction and the transcription errors in its correction note; Study
E's output filename, its unregistered pairing rule and its post-hoc missingness sensitivity; Study F's
accidental 75-call smoke under an unfrozen package (archived and costed; its records supplied diagnostic evidence and the composite-arm prompt-token baselines and cost accounting the gate uses),
three refused smoke attempts on a rate-limited endpoint, an endpoint re-pin, and the frozen analyzer's
auxiliary identity-deviation line computed from rounded estimates (an erratum generator gives the exact
values); Study F-x's premature ``frozen'' commit label and the misquoted Study F cell in its specification.

\FloatBarrier
\section{The Study F contamination in detail}\label{app:contamination}

\paragraph{Timeline.} Run 1 executed under package v1. Its request body to OpenRouter models included
\texttt{plugins:[\char123 id:"web", enabled:false\char125]}, added after a review round as a documented disable. The
provider's accounting returned, per call, a billed cost \FQSurchargePerCall{} above the upstream inference cost
on \FQSurchargeCalls{} response-bearing attempts across \FQSurchargeModels{} models; the cost difference was
the first signal. Reading the stored rationales of a model that cited a URL confirmed injected search results.
The workspace's Web Search tool was disabled while the run was finishing; the provider then refused the
remaining requests that named the plugin (\FQErrPlugin{} error terminals), and the run was locked and
quarantined in full (\FQScored{} successes, \FQErrors{} errors; lock rollup \texttt{\FQLockRollup}).

\paragraph{What the records show.} Against the per-cell baselines assembled from \FCleanBaselineRecords{}
locked records of Studies D and E and an archived diagnostic smoke (one prompt-token count per cell):
\FQVariableModels{} models carry variable content (\FQVariableExcessMin{} to \FQVariableExcessMax{} tokens
above baseline, \FQVariableSurchargeMin{} to \FQVariableSurchargeMax{} surcharged attempts each);
gpt-oss-120b carries \FQGptOssSurchargeCalls{} surcharged attempts and no excess; Gemini 3.7 Flash a constant
+\FQGeminiExcess{} tokens and Grok 4.6 a constant +\FQGrokExcessMin{} tokens, neither surcharged; the six
direct-provider models are constant on their baselines. Grok's excess is consistent with a constant,
unsurcharged augmentation and is not proven to be one --- code, account settings and provider accounting all
changed between run 1 and the diagnostic smoke that returned it to baseline.

\paragraph{The gate.} The re-frozen package (v2.9) gates the smoke against the per-cell baselines with
tolerance zero, freezes the smoke's counts as the confirmatory band, re-checks the band on every call, and
poisons the run (a marker, a retained lock, refused restarts) on the first deviation. On the OpenRouter path it
also binds the provider's cost accounting per call (billed above upstream is fatal; below is recorded and
bounded at lock), binds the served provider and model to the pinned endpoint, retains every response id, and
forbids any request key that names a server-side tool. Study F-x extended the baselines to models without
locked counts through the provider's free token-counting endpoint, calibrated on Opus~5's locked billed
counts. The gate's blind spots are stated in \S\ref{sec:integrity}.

\FloatBarrier
\section{Full tables}\label{app:tables}
Table~\ref{tab:studyB} gives Study B's eight cells; Tables~\ref{tab:studyD} and~\ref{tab:studyE} the per-model
rates of Studies D and E; Tables~\ref{tab:studyF-cells}, \ref{tab:studyF-contrasts} and~\ref{tab:studyF-permodel}
Study F's registered quantities and per-model counts; Tables~\ref{tab:studyFx} and~\ref{tab:studyFx-contrasts}
Study F-x's cells and all sixteen registered contrasts; Tables~\ref{tab:studyG}, \ref{tab:studyG-v44} and~\ref{tab:studyG-contrasts} Study G's cells, V44 cells and forty registered contrasts (Wilson / Newcombe paired score intervals); Tables~\ref{tab:studyBr} and~\ref{tab:studyBr-main} Study \Br{} against Study B and its seven registered main quantities; Tables~\ref{tab:studyI}, \ref{tab:studyI-v73}, \ref{tab:studyI-y2}, \ref{tab:studyI-contrasts} and~\ref{tab:studyI-contrasts-y1} Study I's 160 registered cells and 72 contrasts; Tables~\ref{tab:studyH1}, \ref{tab:studyH1-contrasts}, \ref{tab:studyH1-transport} and~\ref{tab:studyH1-c4} Study H1's cells, 24 contrasts, 16 transport contrasts with both sides and the descriptive $V_{c4}$ cells; Tables~\ref{tab:studyJ}, \ref{tab:studyJ-first}, \ref{tab:studyJ-contrasts} and~\ref{tab:studyJ-contrasts-k2} Study J's cells, first-credit cells and contrasts; Tables~\ref{tab:studyH2}, \ref{tab:studyH2-contrasts} and~\ref{tab:studyH2-suffix} Study H2's fifty cells and forty-five contrasts, and Table~\ref{tab:studyH2-wordings} its five wordings byte for byte; Tables~\ref{tab:studyGr} and~\ref{tab:studyGr-replication} Study \Gr{}'s eighteen cells, thirty contrasts, thirty replication contrasts and the pooled secondary. Labels: \textsc{positive}/\textsc{negative} when the
interval excludes zero, \textsc{undetermined} otherwise; \textsc{within margin} when the interval lies inside
$\pm\delta$, \textsc{exceeds margin} when it lies wholly beyond, \textsc{unresolved} otherwise ($\delta = 10$).
\begin{table}[htbp]\centering\small
\caption{Study B: the eight cells (pointer $\times$ label assignment, plus the two bridge cells), 150 episodes each over six models. The registered estimand pools the two cells whose referent is pricing against the two whose referent is onboarding.}\label{tab:studyB}
\adjustbox{max width=\textwidth}{\begin{tabular}{llll}\toprule
pointer & label assignment & referent & $V_{73}$ hits/episodes \\\midrule
A & \texttt{pricing\_is\_A} & pricing & 134/150 \\
A & \texttt{pricing\_is\_B} & onboarding & 24/150 \\
B & \texttt{pricing\_is\_A} & onboarding & 6/150 \\
B & \texttt{pricing\_is\_B} & pricing & 130/150 \\
NONE & \texttt{pricing\_is\_A} & --- & 98/150 \\
NONE & \texttt{pricing\_is\_B} & --- & 93/150 \\
\midrule bridge (Experiment 7 wording) & --- & pricing & 139/150 \\
\midrule bridge (Experiment 7 wording) & --- & onboarding & 13/150 \\
\bottomrule\end{tabular}}\end{table}
\begin{table}[htbp]\centering\small
\caption{Study D: per-model $V_{73}$ (\%) by arm (\texttt{oracle} = bare id, \texttt{policy} = stated criterion), 40 episodes per cell, six direct-provider models.}\label{tab:studyD}
\adjustbox{max width=\textwidth}{\begin{tabular}{lrrr}\toprule
model & \texttt{none} & \texttt{oracle} (bare id) & \texttt{policy} (criterion) \\\midrule
Opus 5 & 2.5 & 0.0 & 100.0 \\
Sonnet 5 & 0.0 & 37.5 & 100.0 \\
Haiku 4.5 & 0.0 & 0.0 & 47.5 \\
GPT-5.6 Sol & 0.0 & 100.0 & 100.0 \\
GPT-5.6 Terra & 7.5 & 100.0 & 100.0 \\
GPT-5.6 Luna & 0.0 & 100.0 & 100.0 \\
\midrule equal-weight mean & 1.7 & 56.2 & 91.2 \\
\bottomrule\end{tabular}}\end{table}
\begin{table}[htbp]\centering\small
\caption{Study E: per-model $V_{73}$ by arm on the nine-model OpenRouter-served panel, hits/episodes (\%); the last column is the within-model criterion-minus-id contrast over the runs present in both arms.}\label{tab:studyE}
\adjustbox{max width=\textwidth}{\begin{tabular}{lrrrr}\toprule
model & \texttt{none} & \texttt{oracle} & \texttt{policy} & criterion $-$ id \\\midrule
DeepSeek V4 Pro & 2/20 (10.0) & 18/20 (90.0) & 17/20 (85.0) & \ensuremath{-}5.0 \\
Gemini 3.7 Flash & 2/20 (10.0) & 20/20 (100.0) & 20/20 (100.0) & +0.0 \\
Llama 4 Maverick & 10/20 (50.0) & 11/20 (55.0) & 19/20 (95.0) & +40.0 \\
MiniMax M3 & 3/20 (15.0) & 12/20 (60.0) & 8/20 (40.0) & \ensuremath{-}20.0 \\
Mistral Medium 3.5 & 1/20 (5.0) & 0/20 (0.0) & 18/20 (90.0) & +90.0 \\
Kimi K3 & 14/15 (93.3) & 16/16 (100.0) & 15/15 (100.0) & +0.0 \\
gpt-oss-120b & 5/20 (25.0) & 18/20 (90.0) & 15/20 (75.0) & \ensuremath{-}15.0 \\
Qwen3.8 Max & 1/20 (5.0) & 18/20 (90.0) & 13/20 (65.0) & \ensuremath{-}25.0 \\
Grok 4.6 & 12/20 (60.0) & 20/20 (100.0) & 20/20 (100.0) & +0.0 \\
\midrule equal-weight mean & 30.4 & 76.1 & 83.3 & +7.2 \\
\bottomrule\end{tabular}}\end{table}
\begin{table}[htbp]\centering\footnotesize\setlength{\tabcolsep}{4pt}
\caption{Study F: cell means of $V_{73}$ (\%) with block-bootstrap intervals, equal model weights within each pre-named group (ID\_GT\_CRIT4 = four OpenRouter-served models where the id had beaten the criterion; CRIT\_GT\_ID2 = Mistral Medium 3.5 and Opus 5; open\_pool = OpenRouter-served; closed\_pool = direct-provider). Arms: none, id, crit (criterion), both\_IL (criterion then id), both\_IF (id then criterion).}\label{tab:studyF-cells}
\adjustbox{max width=\textwidth}{\begin{tabular}{lrrrrr}\toprule
group ($n$ models) & \texttt{none} & \texttt{id} & \texttt{crit} & \texttt{both\_IL} & \texttt{both\_IF} \\\midrule
ID\_GT\_CRIT4 (4) & 21.2 [15.6, 26.9] & 80.6 [74.8, 85.9] & 64.9 [58.4, 71.8] & 97.5 [95.0, 99.4] & 98.1 [95.6, 100.0] \\
CRIT\_GT\_ID2 (2) & 1.2 [0.0, 3.8] & 1.2 [0.0, 3.8] & 90.0 [83.8, 96.2] & 52.5 [50.0, 56.2] & 47.5 [43.8, 50.0] \\
open\_pool (9) & 27.5 [23.1, 31.9] & 75.8 [72.0, 79.5] & 78.3 [74.1, 82.3] & 98.3 [96.7, 99.7] & 96.9 [94.7, 98.9] \\
closed\_pool (6) & 5.0 [1.7, 9.2] & 57.9 [54.2, 61.7] & 92.5 [89.2, 95.8] & 84.2 [83.3, 85.4] & 83.3 [83.3, 83.3] \\
\bottomrule\end{tabular}}\end{table}

\begin{table}[htbp]\centering\footnotesize\setlength{\tabcolsep}{4pt}
\caption{Study F: the sixteen registered complete-case contrasts (points; groups and arms as in Table~\ref{tab:studyF-cells}) with the two independent label columns ($\delta = 10$ points): direction = interval excludes zero; material size = interval inside/beyond $\pm\delta$/neither.}\label{tab:studyF-contrasts}
\adjustbox{max width=\textwidth}{\begin{tabular}{llrll}\toprule
contrast & group & estimate [95\%] & direction & material size \\\midrule
C1 = both\_IL $-$ crit & ID\_GT\_CRIT4 & +31.9 [+24.3, +39.8] & Positive & Exceeds Margin \\
C1 = both\_IL $-$ crit & CRIT\_GT\_ID2 & \ensuremath{-}37.5 [\ensuremath{-}43.8, \ensuremath{-}30.0] & Negative & Exceeds Margin \\
C1 = both\_IL $-$ crit & open\_pool & +19.8 [+15.5, +24.3] & Positive & Exceeds Margin \\
C1 = both\_IL $-$ crit & closed\_pool & \ensuremath{-}8.3 [\ensuremath{-}12.1, \ensuremath{-}4.6] & Negative & Unresolved \\
C2 = both\_IL $-$ id & ID\_GT\_CRIT4 & +17.0 [+10.9, +23.2] & Positive & Exceeds Margin \\
C2 = both\_IL $-$ id & CRIT\_GT\_ID2 & +51.2 [+47.5, +56.2] & Positive & Exceeds Margin \\
C2 = both\_IL $-$ id & open\_pool & +22.6 [+18.7, +26.6] & Positive & Exceeds Margin \\
C2 = both\_IL $-$ id & closed\_pool & +26.2 [+22.5, +30.0] & Positive & Exceeds Margin \\
C3 = both\_IF $-$ both\_IL & ID\_GT\_CRIT4 & +0.6 [\ensuremath{-}2.5, +3.8] & Undetermined & Within Margin \\
C3 = both\_IF $-$ both\_IL & CRIT\_GT\_ID2 & \ensuremath{-}5.0 [\ensuremath{-}10.0, \ensuremath{-}1.2] & Negative & Unresolved \\
C3 = both\_IF $-$ both\_IL & open\_pool & \ensuremath{-}1.4 [\ensuremath{-}4.2, +1.1] & Undetermined & Within Margin \\
C3 = both\_IF $-$ both\_IL & closed\_pool & \ensuremath{-}0.8 [\ensuremath{-}2.1, +0.0] & Undetermined & Within Margin \\
C4 = both\_IL $-$ none & ID\_GT\_CRIT4 & +77.6 [+71.5, +83.6] & Positive & Exceeds Margin \\
C4 = both\_IL $-$ none & CRIT\_GT\_ID2 & +51.2 [+47.5, +55.0] & Positive & Exceeds Margin \\
C4 = both\_IL $-$ none & open\_pool & +71.4 [+66.7, +75.9] & Positive & Exceeds Margin \\
C4 = both\_IL $-$ none & closed\_pool & +79.2 [+75.4, +82.5] & Positive & Exceeds Margin \\
\bottomrule\end{tabular}}\end{table}
\begin{table}[htbp]\centering\footnotesize\setlength{\tabcolsep}{4pt}
\caption{Study F: per-model $V_{73}$ hits/episodes by arm (direct-provider panel first).}\label{tab:studyF-permodel}
\adjustbox{max width=\textwidth}{\begin{tabular}{lrrrrr}\toprule
model & \texttt{none} & \texttt{id} & \texttt{crit} & \texttt{both\_IL} & \texttt{both\_IF} \\\midrule
Opus 5 & 0/40 & 1/40 & 40/40 & 2/40 & 0/40 \\
Sonnet 5 & 0/20 & 9/20 & 20/20 & 20/20 & 20/20 \\
Haiku 4.5 & 0/20 & 0/20 & 11/20 & 20/20 & 20/20 \\
GPT-5.6 Sol & 1/20 & 20/20 & 20/20 & 20/20 & 20/20 \\
GPT-5.6 Terra & 5/20 & 20/20 & 20/20 & 20/20 & 20/20 \\
GPT-5.6 Luna & 0/20 & 20/20 & 20/20 & 20/20 & 20/20 \\
gpt-oss-120b & 21/40 & 39/40 & 32/40 & 40/40 & 40/40 \\
DeepSeek V4 Pro & 5/40 & 36/40 & 33/40 & 37/40 & 37/40 \\
Kimi K3 & 1/20 & 18/20 & 14/20 & 19/20 & 18/20 \\
MiniMax M3 & 2/40 & 21/40 & 14/40 & 39/40 & 40/40 \\
Llama 4 Maverick & 16/20 & 14/20 & 19/20 & 20/20 & 20/20 \\
Qwen3.8 Max & 5/34 & 28/34 & 23/37 & 34/34 & 36/36 \\
Mistral Medium 3.5 & 1/40 & 0/40 & 32/40 & 40/40 & 38/40 \\
Gemini 3.7 Flash & 2/20 & 20/20 & 20/20 & 20/20 & 20/20 \\
Grok 4.6 & 13/20 & 20/20 & 20/20 & 20/20 & 19/20 \\
\bottomrule\end{tabular}}\end{table}
\begin{table}[htbp]\centering\footnotesize\setlength{\tabcolsep}{4pt}
\caption{Study F-x: the frozen Study F instrument on three Claude models (hits/40 per cell; within-model contrasts C1 = both\_IL $-$ crit and C2 = both\_IL $-$ id with bootstrap intervals).}\label{tab:studyFx}
\adjustbox{max width=\textwidth}{\begin{tabular}{lrrrrrrr}\toprule
model & \texttt{none} & \texttt{id} & \texttt{crit} & \texttt{both\_IL} & \texttt{both\_IF} & C1 [95\%] & C2 [95\%] \\\midrule
Fable 5.1 & 13/40 & 13/40 & 17/40 & 0/40 & 0/40 & \ensuremath{-}42.5 [\ensuremath{-}57.5, \ensuremath{-}27.5] & \ensuremath{-}32.5 [\ensuremath{-}47.5, \ensuremath{-}20.0] \\
Fable 5 & 2/40 & 6/40 & 40/40 & 1/40 & 0/40 & \ensuremath{-}97.5 [\ensuremath{-}100.0, \ensuremath{-}92.5] & \ensuremath{-}12.5 [\ensuremath{-}25.0, +0.0] \\
Opus 5 & 1/40 & 0/40 & 40/40 & 0/40 & 0/40 & \ensuremath{-}100.0 [\ensuremath{-}100.0, \ensuremath{-}100.0] & +0.0 [+0.0, +0.0] \\
\bottomrule\end{tabular}}\end{table}
\begin{table}[htbp]\centering\footnotesize\setlength{\tabcolsep}{4pt}
\caption{Study F-x: the twelve registered within-model contrasts and the four registered family contrasts (Fable minus Opus, independent within-model resampling), two independent label columns ($\delta = 10$).}\label{tab:studyFx-contrasts}
\adjustbox{max width=\textwidth}{\begin{tabular}{llrll}\toprule
contrast & model & estimate [95\%] & direction & material size \\\midrule
C1 = both\_IL $-$ crit & Fable 5.1 & \ensuremath{-}42.5 [\ensuremath{-}57.5, \ensuremath{-}27.5] & Negative & Exceeds Margin \\
C1 = both\_IL $-$ crit & Fable 5 & \ensuremath{-}97.5 [\ensuremath{-}100.0, \ensuremath{-}92.5] & Negative & Exceeds Margin \\
C1 = both\_IL $-$ crit & Opus 5 & \ensuremath{-}100.0 [\ensuremath{-}100.0, \ensuremath{-}100.0] & Negative & Exceeds Margin \\
C2 = both\_IL $-$ id & Fable 5.1 & \ensuremath{-}32.5 [\ensuremath{-}47.5, \ensuremath{-}20.0] & Negative & Exceeds Margin \\
C2 = both\_IL $-$ id & Fable 5 & \ensuremath{-}12.5 [\ensuremath{-}25.0, +0.0] & Undetermined & Unresolved \\
C2 = both\_IL $-$ id & Opus 5 & +0.0 [+0.0, +0.0] & Undetermined & Within Margin \\
C3 = both\_IF $-$ both\_IL & Fable 5.1 & +0.0 [+0.0, +0.0] & Undetermined & Within Margin \\
C3 = both\_IF $-$ both\_IL & Fable 5 & \ensuremath{-}2.5 [\ensuremath{-}7.5, +0.0] & Undetermined & Within Margin \\
C3 = both\_IF $-$ both\_IL & Opus 5 & +0.0 [+0.0, +0.0] & Undetermined & Within Margin \\
C4 = both\_IL $-$ none & Fable 5.1 & \ensuremath{-}32.5 [\ensuremath{-}47.5, \ensuremath{-}17.5] & Negative & Exceeds Margin \\
C4 = both\_IL $-$ none & Fable 5 & \ensuremath{-}2.5 [\ensuremath{-}12.5, +5.0] & Undetermined & Unresolved \\
C4 = both\_IL $-$ none & Opus 5 & \ensuremath{-}2.5 [\ensuremath{-}7.5, +0.0] & Undetermined & Within Margin \\
\midrule family C1: Fable 5.1 $-$ Opus 5 & & +57.5 [+42.5, +72.5] & Positive & Exceeds Margin \\
\midrule family C1: Fable 5 $-$ Opus 5 & & +2.5 [+0.0, +7.5] & Undetermined & Within Margin \\
\midrule family C2: Fable 5.1 $-$ Opus 5 & & \ensuremath{-}32.5 [\ensuremath{-}47.5, \ensuremath{-}17.5] & Negative & Exceeds Margin \\
\midrule family C2: Fable 5 $-$ Opus 5 & & \ensuremath{-}12.5 [\ensuremath{-}25.0, +0.0] & Undetermined & Unresolved \\
\bottomrule\end{tabular}}\end{table}
\begin{table}[htbp]\centering\footnotesize\setlength{\tabcolsep}{4pt}
\caption{Study G: $V_{73}$ hits/40 per cell with Wilson score 95\% intervals (\%) on the six byte-controlled arms (crit = criterion alone; both\_IL = + ` (memory\_73)'; crit\_pad = + ` (as stated)'; crit\_other = + ` (memory\_44)'; both\_fresh / both\_inherited = both\_IL with the field labelled `set today' / `inherited').}\label{tab:studyG}
\adjustbox{max width=\textwidth}{\begin{tabular}{lrrrrrr}\toprule
model & \texttt{crit} & \texttt{both\_IL} & \texttt{crit\_pad} & \texttt{crit\_other} & \texttt{both\_fresh} & \texttt{both\_inherited} \\\midrule
Opus 5 & 40/40 [91.2, 100.0] & 3/40 [2.6, 19.9] & 40/40 [91.2, 100.0] & 0/40 [0.0, 8.8] & 13/40 [20.1, 48.0] & 0/40 [0.0, 8.8] \\
Fable 5 & 40/40 [91.2, 100.0] & 0/40 [0.0, 8.8] & 3/40 [2.6, 19.9] & 0/40 [0.0, 8.8] & 0/40 [0.0, 8.8] & 0/40 [0.0, 8.8] \\
Fable 5.1 & 14/40 [22.1, 50.5] & 0/40 [0.0, 8.8] & 7/40 [8.7, 31.9] & 0/40 [0.0, 8.8] & 0/40 [0.0, 8.8] & 0/40 [0.0, 8.8] \\
GPT-5.6 Sol & 40/40 [91.2, 100.0] & 40/40 [91.2, 100.0] & 40/40 [91.2, 100.0] & 0/40 [0.0, 8.8] & 40/40 [91.2, 100.0] & 40/40 [91.2, 100.0] \\
\bottomrule\end{tabular}}\end{table}
\begin{table}[htbp]\centering\footnotesize\setlength{\tabcolsep}{4pt}
\caption{Study G: the registered V44 cells — the competing pointer's own rate under crit\_other.}\label{tab:studyG-v44}
\adjustbox{max width=\textwidth}{\begin{tabular}{lrl}\toprule
model & \texttt{memory\_44} named under \texttt{crit\_other} (hits/40) & Wilson 95\% \\\midrule
Opus 5 & 0/40 & [0.0, 8.8] \\
Fable 5 & 0/40 & [0.0, 8.8] \\
Fable 5.1 & 0/40 & [0.0, 8.8] \\
GPT-5.6 Sol & 40/40 & [91.2, 100.0] \\
\bottomrule\end{tabular}}\end{table}

\begin{table}[htbp]\centering\footnotesize\setlength{\tabcolsep}{4pt}
\caption{Study G: the 40 registered within-model contrasts (Newcombe (1998) method 10 paired score 95\% intervals in the registered nonnegative-$\varphi$ variant), two independent label columns ($\delta = 10$).}\label{tab:studyG-contrasts}
\adjustbox{max width=\textwidth}{\begin{tabular}{llrll}\toprule
contrast & model & estimate [95\%] & direction & material size \\\midrule
G1 = both\_IL $-$ crit & Opus 5 & \ensuremath{-}92.5 [\ensuremath{-}97.4, \ensuremath{-}77.3] & Negative & Exceeds Margin \\
G1 = both\_IL $-$ crit & Fable 5 & \ensuremath{-}100.0 [\ensuremath{-}100.0, \ensuremath{-}87.6] & Negative & Exceeds Margin \\
G1 = both\_IL $-$ crit & Fable 5.1 & \ensuremath{-}35.0 [\ensuremath{-}50.5, \ensuremath{-}19.4] & Negative & Exceeds Margin \\
G1 = both\_IL $-$ crit & GPT-5.6 Sol & +0.0 [\ensuremath{-}8.8, +8.8] & Undetermined & Within Margin \\
G2 = crit\_pad $-$ crit & Opus 5 & +0.0 [\ensuremath{-}8.8, +8.8] & Undetermined & Within Margin \\
G2 = crit\_pad $-$ crit & Fable 5 & \ensuremath{-}92.5 [\ensuremath{-}97.4, \ensuremath{-}77.3] & Negative & Exceeds Margin \\
G2 = crit\_pad $-$ crit & Fable 5.1 & \ensuremath{-}17.5 [\ensuremath{-}31.7, \ensuremath{-}2.7] & Negative & Unresolved \\
\bottomrule\end{tabular}}\end{table}
\begin{table}[htbp]\centering\footnotesize\setlength{\tabcolsep}{4pt}
\caption{Study B$'$: the prospectively registered replication of Study B's corrected estimand (1,200 episodes, the same six models); replication contrasts with percentile bootstrap intervals and the registered labels ($\delta = 10$); the verdict is the output of the registered rule. The seven registered main quantities with their own labels: Table~\ref{tab:studyBr-main}.}\label{tab:studyBr}
\adjustbox{max width=\textwidth}{\begin{tabular}{llllp{3.6cm}}\toprule
quantity & Study B (locked, corrected) & Study B$'$ (registered replication) & B$'$ $-$ B [95\%] & label \\\midrule
$\Delta_{\mathrm{plan}}$ & +78.0 & +81.7 [+78.3, +85.0] & +3.7 [\ensuremath{-}1.3, +8.7] & Undetermined / Within Margin \\
$\Delta_{\mathrm{bridge}}$ & +84.0 & +85.3 [+79.3, +90.7] & +1.3 [\ensuremath{-}6.0, +9.3] & Undetermined / Within Margin \\
share & 0.929 & 0.957 [0.895, 1.029] & --- & verdict POINTER-STRONG; REPLICATED-VERDICT \\
\bottomrule\end{tabular}}\end{table}

\begin{table}[htbp]\centering\footnotesize\setlength{\tabcolsep}{4pt}
\caption{Study B$'$: the seven registered main quantities (equal model weights; percentile block bootstrap over (model, run) blocks, $B = 4{,}000$) with the two registered label columns ($\delta = 10$; the share's labels are the registered rule's on the ratio scale). $V_{\mathrm{onb}}$ = the credit is one of the onboarding plan's records.}\label{tab:studyBr-main}
\adjustbox{max width=\textwidth}{\begin{tabular}{llrll}\toprule
quantity & definition & estimate [95\%] & direction & material size \\\midrule
$\Delta_{\mathrm{plan}}$ & $P(V_{73} \mid \text{pricing}) - P(V_{73} \mid \text{onboarding})$ & +81.7 [+78.3, +85.0] & Positive & Exceeds Margin \\
$\Delta_{\mathrm{bridge}}$ & $P(V_{73} \mid \text{bridge pricing}) - P(V_{73} \mid \text{bridge onboarding})$ & +85.3 [+79.3, +90.7] & Positive & Exceeds Margin \\
$\Delta_{\mathrm{attract}}$ & $P(V_{73} \mid \text{pricing}) - P(V_{73} \mid \text{none})$ & +27.3 [+24.3, +30.7] & Positive & Exceeds Margin \\
$\Delta_{\mathrm{suppress}}$ & $P(V_{73} \mid \text{none}) - P(V_{73} \mid \text{onboarding})$ & +54.3 [+50.7, +58.0] & Positive & Exceeds Margin \\
$\Delta_{\mathrm{label}}$ & letter effect (A $-$ B) at fixed referent & +7.7 [+4.0, +11.3] & Positive & Unresolved \\
$\Delta_{\mathrm{mirror}}$ & $P(V_{\mathrm{onb}} \mid \text{onboarding}) - P(V_{\mathrm{onb}} \mid \text{pricing})$ & +81.7 [+78.3, +85.0] & Positive & Exceeds Margin \\
share & $\Delta_{\mathrm{plan}} / \Delta_{\mathrm{bridge}}$ (joint resampling) & 0.957 [0.895, 1.029] & Positive & Exceeds Margin \\
\bottomrule\end{tabular}}\end{table}
\begin{table}[htbp]\centering\footnotesize\setlength{\tabcolsep}{4pt}
\caption{Study I: $Y_1$ (the decision follows the current archive record) hits/episodes per model $\times$ world $\times$ arm, and the registered crit $-$ none contrast (Newcombe (1998) method 10 paired score 95\%, registered nonnegative-$\varphi$ variant). All 160 registered cells and 72 contrasts: Tables~\ref{tab:studyI-v73}, \ref{tab:studyI-y2}, \ref{tab:studyI-contrasts} and~\ref{tab:studyI-contrasts-y1}.}\label{tab:studyI}
\adjustbox{max width=\textwidth}{\begin{tabular}{llrrrrr}\toprule
model & world & \texttt{none} & \texttt{id} & \texttt{crit} & \texttt{both\_IL} & crit $-$ none [95\%] \\\midrule
Opus 5 & valid & 25/25 & 25/25 & 25/25 & 24/24 & +0.0 [\ensuremath{-}13.3, +13.3] \\
Opus 5 & superseded & 0/25 & 1/25 & 25/25 & 2/24 & +100.0 [+81.2, +100.0] \\
Sonnet 5 & valid & 20/25 & 12/25 & 25/25 & 25/25 & +20.0 [+2.6, +39.1] \\
Sonnet 5 & superseded & 7/25 & 8/25 & 25/25 & 25/25 & +72.0 [+48.3, +85.7] \\
Haiku 4.5 & valid & 3/25 & 1/25 & 12/25 & 25/25 & +36.0 [+10.6, +56.1] \\
Haiku 4.5 & superseded & 22/25 & 24/25 & 24/25 & 25/25 & +8.0 [\ensuremath{-}9.4, +26.3] \\
Fable 5 & valid & 12/25 & 15/25 & 24/25 & 9/24 & +48.0 [+23.8, +66.3] \\
Fable 5 & superseded & 12/25 & 11/25 & 25/25 & 17/24 & +52.0 [+29.2, +70.0] \\
Fable 5.1 & valid & 25/25 & 25/25 & 24/25 & 24/24 & \ensuremath{-}4.0 [\ensuremath{-}19.5, +9.7] \\
Fable 5.1 & superseded & 10/25 & 8/25 & 2/25 & 1/25 & \ensuremath{-}32.0 [\ensuremath{-}52.1, \ensuremath{-}8.3] \\
GPT-5.6 Sol & valid & 0/25 & 25/25 & 25/25 & 25/25 & +100.0 [+81.2, +100.0] \\
GPT-5.6 Sol & superseded & 25/25 & 25/25 & 25/25 & 25/25 & +0.0 [\ensuremath{-}13.3, +13.3] \\
GPT-5.6 Terra & valid & 2/25 & 25/25 & 25/25 & 25/25 & +92.0 [+70.4, +97.8] \\
GPT-5.6 Terra & superseded & 25/25 & 25/25 & 25/25 & 25/25 & +0.0 [\ensuremath{-}13.3, +13.3] \\
GPT-5.6 Luna & valid & 2/25 & 25/25 & 25/25 & 25/25 & +92.0 [+70.4, +97.8] \\
GPT-5.6 Luna & superseded & 21/25 & 25/25 & 25/25 & 25/25 & +16.0 [\ensuremath{-}0.4, +34.7] \\
\bottomrule\end{tabular}}\end{table}
\begin{table}[htbp]\centering\footnotesize\setlength{\tabcolsep}{4pt}
\caption{Study I: turn-1 $V_{73}$ hits/25 per model and arm with Wilson score 95\% intervals (\%) (all turn-1 records; the allocation replication of D/F/F-x on this run); the 32 registered $V_{73}$ cells.}\label{tab:studyI-v73}
\adjustbox{max width=\textwidth}{\begin{tabular}{lrrrr}\toprule
model & \texttt{none} & \texttt{id} & \texttt{crit} & \texttt{both\_IL} \\\midrule
Opus 5 & 0/25 [0.0, 13.3] & 1/25 [0.7, 19.5] & 25/25 [86.7, 100.0] & 2/25 [2.2, 25.0] \\
Sonnet 5 & 0/25 [0.0, 13.3] & 5/25 [8.9, 39.1] & 25/25 [86.7, 100.0] & 25/25 [86.7, 100.0] \\
Haiku 4.5 & 0/25 [0.0, 13.3] & 0/25 [0.0, 13.3] & 11/25 [26.7, 62.9] & 25/25 [86.7, 100.0] \\
Fable 5 & 1/25 [0.7, 19.5] & 1/25 [0.7, 19.5] & 24/25 [80.5, 99.3] & 0/25 [0.0, 13.3] \\
Fable 5.1 & 12/25 [30.0, 66.5] & 10/25 [23.4, 59.3] & 2/25 [2.2, 25.0] & 0/25 [0.0, 13.3] \\
GPT-5.6 Sol & 0/25 [0.0, 13.3] & 25/25 [86.7, 100.0] & 25/25 [86.7, 100.0] & 25/25 [86.7, 100.0] \\
GPT-5.6 Terra & 2/25 [2.2, 25.0] & 25/25 [86.7, 100.0] & 25/25 [86.7, 100.0] & 25/25 [86.7, 100.0] \\
GPT-5.6 Luna & 0/25 [0.0, 13.3] & 25/25 [86.7, 100.0] & 25/25 [86.7, 100.0] & 25/25 [86.7, 100.0] \\
\bottomrule\end{tabular}}\end{table}
\begin{table}[htbp]\centering\footnotesize\setlength{\tabcolsep}{4pt}
\caption{Study I: $Y_2$ (the promotional-pricing action is taken beyond a small guarded test) hits/episodes per model $\times$ world $\times$ arm with Wilson score 95\% intervals (\%); the 64 registered $Y_2$ cells (no contrast is registered on $Y_2$).}\label{tab:studyI-y2}
\adjustbox{max width=\textwidth}{\begin{tabular}{llrrrr}\toprule
model & world & \texttt{none} & \texttt{id} & \texttt{crit} & \texttt{both\_IL} \\\midrule
Opus 5 & valid & 0/25 [0.0, 13.3] & 0/25 [0.0, 13.3] & 0/25 [0.0, 13.3] & 0/24 [0.0, 13.8] \\
Opus 5 & superseded & 0/25 [0.0, 13.3] & 1/25 [0.7, 19.5] & 19/25 [56.6, 88.5] & 1/24 [0.7, 20.2] \\
Sonnet 5 & valid & 0/25 [0.0, 13.3] & 0/25 [0.0, 13.3] & 0/25 [0.0, 13.3] & 0/25 [0.0, 13.3] \\
Sonnet 5 & superseded & 1/25 [0.7, 19.5] & 3/25 [4.2, 30.0] & 22/25 [70.0, 95.8] & 22/25 [70.0, 95.8] \\
Haiku 4.5 & valid & 19/25 [56.6, 88.5] & 21/25 [65.3, 93.6] & 12/25 [30.0, 66.5] & 0/25 [0.0, 13.3] \\
Haiku 4.5 & superseded & 20/25 [60.9, 91.1] & 20/25 [60.9, 91.1] & 23/25 [75.0, 97.8] & 25/25 [86.7, 100.0] \\
Fable 5 & valid & 0/25 [0.0, 13.3] & 0/25 [0.0, 13.3] & 0/25 [0.0, 13.3] & 0/24 [0.0, 13.8] \\
Fable 5 & superseded & 1/25 [0.7, 19.5] & 1/25 [0.7, 19.5] & 24/25 [80.5, 99.3] & 0/24 [0.0, 13.8] \\
Fable 5.1 & valid & 0/25 [0.0, 13.3] & 0/25 [0.0, 13.3] & 0/25 [0.0, 13.3] & 0/24 [0.0, 13.8] \\
Fable 5.1 & superseded & 3/25 [4.2, 30.0] & 4/25 [6.4, 34.7] & 0/25 [0.0, 13.3] & 0/25 [0.0, 13.3] \\
GPT-5.6 Sol & valid & 12/25 [30.0, 66.5] & 0/25 [0.0, 13.3] & 0/25 [0.0, 13.3] & 0/25 [0.0, 13.3] \\
GPT-5.6 Sol & superseded & 8/25 [17.2, 51.6] & 25/25 [86.7, 100.0] & 25/25 [86.7, 100.0] & 25/25 [86.7, 100.0] \\
GPT-5.6 Terra & valid & 19/25 [56.6, 88.5] & 0/25 [0.0, 13.3] & 0/25 [0.0, 13.3] & 0/25 [0.0, 13.3] \\
GPT-5.6 Terra & superseded & 20/25 [60.9, 91.1] & 25/25 [86.7, 100.0] & 25/25 [86.7, 100.0] & 25/25 [86.7, 100.0] \\
GPT-5.6 Luna & valid & 2/25 [2.2, 25.0] & 0/25 [0.0, 13.3] & 0/25 [0.0, 13.3] & 0/25 [0.0, 13.3] \\
GPT-5.6 Luna & superseded & 0/25 [0.0, 13.3] & 18/25 [52.4, 85.7] & 20/25 [60.9, 91.1] & 25/25 [86.7, 100.0] \\
\bottomrule\end{tabular}}\end{table}
\begin{table}[htbp]\centering\footnotesize\setlength{\tabcolsep}{4pt}
\caption{Study I: the 24 registered within-model contrasts on the turn-1 credit $V_{73}$ (arm $-$ none; $n = 25$ per cell; Newcombe (1998) method 10 paired score 95\%, registered nonnegative-$\varphi$ variant), two independent label columns ($\delta = 10$). The 48 $Y_1$ contrasts: Table~\ref{tab:studyI-contrasts-y1}.}\label{tab:studyI-contrasts}
\adjustbox{max width=\textwidth}{\begin{tabular}{llrll}\toprule
model & contrast & estimate [95\%] & direction & material size \\\midrule
Opus 5 & id $-$ none & +4.0 [\ensuremath{-}9.7, +19.5] & Undetermined & Unresolved \\
Opus 5 & crit $-$ none & +100.0 [+81.2, +100.0] & Positive & Exceeds Margin \\
Opus 5 & both\_IL $-$ none & +8.0 [\ensuremath{-}6.5, +25.0] & Undetermined & Unresolved \\
Sonnet 5 & id $-$ none & +20.0 [+2.6, +39.1] & Positive & Unresolved \\
Sonnet 5 & crit $-$ none & +100.0 [+81.2, +100.0] & Positive & Exceeds Margin \\
Sonnet 5 & both\_IL $-$ none & +100.0 [+81.2, +100.0] & Positive & Exceeds Margin \\
Haiku 4.5 & id $-$ none & +0.0 [\ensuremath{-}13.3, +13.3] & Undetermined & Unresolved \\
\bottomrule\end{tabular}}\end{table}
\begin{table}[htbp]\centering\footnotesize\setlength{\tabcolsep}{4pt}
\caption{Study I: the 48 registered within-model contrasts on the decision endpoint $Y_1$ (arm $-$ none, by world; $n = 25$ per cell unless an error removed an episode; Newcombe (1998) method 10 paired score 95\%, registered nonnegative-$\varphi$ variant), two independent label columns ($\delta = 10$).}\label{tab:studyI-contrasts-y1}
\adjustbox{max width=\textwidth}{\begin{tabular}{lllrll}\toprule
model & world & contrast & estimate [95\%] & direction & material size \\\midrule
Opus 5 & valid & id $-$ none & +0.0 [\ensuremath{-}13.3, +13.3] & Undetermined & Unresolved \\
Opus 5 & valid & crit $-$ none & +0.0 [\ensuremath{-}13.3, +13.3] & Undetermined & Unresolved \\
Opus 5 & valid & both\_IL $-$ none & +0.0 [\ensuremath{-}13.8, +13.8] & Undetermined & Unresolved \\
Opus 5 & superseded & id $-$ none & +4.0 [\ensuremath{-}9.7, +19.5] & Undetermined & Unresolved \\
Opus 5 & superseded & crit $-$ none & +100.0 [+81.2, +100.0] & Positive & Exceeds Margin \\
Opus 5 & superseded & both\_IL $-$ none & +8.3 [\ensuremath{-}6.7, +25.8] & Undetermined & Unresolved \\
Sonnet 5 & valid & id $-$ none & \ensuremath{-}32.0 [\ensuremath{-}53.1, \ensuremath{-}5.4] & Negative & Unresolved \\
\bottomrule\end{tabular}}\end{table}
\begin{table}[htbp]\centering\footnotesize\setlength{\tabcolsep}{4pt}
\caption{Study H1: the procurement world — $V_{c2}$ hits/$n$ per cell ($n = 25$ unless an error removed an episode) and the two registered transport contrasts per model (T1 = (crit $-$ id) procurement $-$ growth; T2 = (both\_IL $-$ crit) procurement $-$ growth; percentile bootstrap intervals, each side resampled independently; growth side = the locked Study D / F run 2 / F-x records).}\label{tab:studyH1}
\adjustbox{max width=\textwidth}{\begin{tabular}{lrrrrrr}\toprule
model & \texttt{none} & \texttt{id} & \texttt{crit} & \texttt{both\_IL} & T1 [95\%] & T2 [95\%] \\\midrule
Opus 5 & 11/24 & 5/25 & 25/25 & 1/25 & \ensuremath{-}20.0 [\ensuremath{-}36.0, \ensuremath{-}8.0] & \ensuremath{-}1.0 [\ensuremath{-}10.0, +9.5] \\
Sonnet 5 & 7/25 & 23/25 & 25/25 & 24/25 & \ensuremath{-}54.5 [\ensuremath{-}72.5, \ensuremath{-}35.5] & \ensuremath{-}4.0 [\ensuremath{-}12.0, +0.0] \\
Haiku 4.5 & 25/25 & 24/25 & 25/25 & 25/25 & \ensuremath{-}43.5 [\ensuremath{-}60.0, \ensuremath{-}26.0] & \ensuremath{-}45.0 [\ensuremath{-}65.0, \ensuremath{-}25.0] \\
Fable 5 & 11/25 & 25/25 & 25/25 & 23/25 & \ensuremath{-}85.0 [\ensuremath{-}95.0, \ensuremath{-}72.5] & +89.5 [+76.5, +100.0] \\
Fable 5.1 & 24/25 & 23/25 & 25/25 & 7/25 & \ensuremath{-}2.0 [\ensuremath{-}25.5, +23.0] & \ensuremath{-}29.5 [\ensuremath{-}51.5, \ensuremath{-}5.5] \\
GPT-5.6 Sol & 13/25 & 25/25 & 25/25 & 25/25 & +0.0 [+0.0, +0.0] & +0.0 [+0.0, +0.0] \\
GPT-5.6 Terra & 11/25 & 25/25 & 25/25 & 25/25 & +0.0 [+0.0, +0.0] & +0.0 [+0.0, +0.0] \\
GPT-5.6 Luna & 4/25 & 25/25 & 25/25 & 25/25 & +0.0 [+0.0, +0.0] & +0.0 [+0.0, +0.0] \\
\bottomrule\end{tabular}}\end{table}
\begin{table}[htbp]\centering\footnotesize\setlength{\tabcolsep}{4pt}
\caption{Study H1: the 24 registered within-model arm $-$ none contrasts on $V_{c2}$ (Newcombe (1998) method 10 paired score 95\%, registered nonnegative-$\varphi$ variant), two independent label columns ($\delta = 10$).}\label{tab:studyH1-contrasts}
\adjustbox{max width=\textwidth}{\begin{tabular}{llrll}\toprule
model & contrast & estimate [95\%] & direction & material size \\\midrule
Opus 5 & id $-$ none & \ensuremath{-}29.2 [\ensuremath{-}49.3, \ensuremath{-}4.9] & Negative & Unresolved \\
Opus 5 & crit $-$ none & +54.2 [+30.6, +72.1] & Positive & Exceeds Margin \\
Opus 5 & both\_IL $-$ none & \ensuremath{-}41.7 [\ensuremath{-}61.0, \ensuremath{-}17.8] & Negative & Exceeds Margin \\
Sonnet 5 & id $-$ none & +64.0 [+38.3, +78.8] & Positive & Exceeds Margin \\
Sonnet 5 & crit $-$ none & +72.0 [+48.3, +85.7] & Positive & Exceeds Margin \\
Sonnet 5 & both\_IL $-$ none & +68.0 [+43.0, +82.1] & Positive & Exceeds Margin \\
Haiku 4.5 & id $-$ none & \ensuremath{-}4.0 [\ensuremath{-}19.5, +9.7] & Undetermined & Unresolved \\
\bottomrule\end{tabular}}\end{table}
\begin{table}[htbp]\centering\footnotesize\setlength{\tabcolsep}{4pt}
\caption{Study H1: the 16 registered transport contrasts ($T_1$ = crit $-$ id, $T_2$ = both\_IL $-$ crit; each = procurement side $-$ growth side) with both sides, their actual complete-case sizes and the growth-world source study; percentile bootstrap intervals with each side resampled independently within the model; the label column holds the two registered labels, direction / material size ($\delta = 10$).}\label{tab:studyH1-transport}
\adjustbox{max width=\textwidth}{\begin{tabular}{llrrrl}\toprule
model & contrast & procurement ($n$) & growth ($n$, source) & transport [95\%] & label \\\midrule
Opus 5 & $T_1$ & +80.0 (25) & +100.0 (40, D) & \ensuremath{-}20.0 [\ensuremath{-}36.0, \ensuremath{-}8.0] & Negative / Unresolved \\
Opus 5 & $T_2$ & \ensuremath{-}96.0 (25) & \ensuremath{-}95.0 (40, F run 2) & \ensuremath{-}1.0 [\ensuremath{-}10.0, +9.5] & Undetermined / Unresolved \\
Sonnet 5 & $T_1$ & +8.0 (25) & +62.5 (40, D) & \ensuremath{-}54.5 [\ensuremath{-}72.5, \ensuremath{-}35.5] & Negative / Exceeds Margin \\
Sonnet 5 & $T_2$ & \ensuremath{-}4.0 (25) & +0.0 (20, F run 2) & \ensuremath{-}4.0 [\ensuremath{-}12.0, +0.0] & Undetermined / Unresolved \\
Haiku 4.5 & $T_1$ & +4.0 (25) & +47.5 (40, D) & \ensuremath{-}43.5 [\ensuremath{-}60.0, \ensuremath{-}26.0] & Negative / Exceeds Margin \\
Haiku 4.5 & $T_2$ & +0.0 (25) & +45.0 (20, F run 2) & \ensuremath{-}45.0 [\ensuremath{-}65.0, \ensuremath{-}25.0] & Negative / Exceeds Margin \\
Fable 5 & $T_1$ & +0.0 (25) & +85.0 (40, F-x) & \ensuremath{-}85.0 [\ensuremath{-}95.0, \ensuremath{-}72.5] & Negative / Exceeds Margin \\
\bottomrule\end{tabular}}\end{table}
\begin{table}[htbp]\centering\footnotesize\setlength{\tabcolsep}{4pt}
\caption{Study H1: the 32 registered descriptive $V_{c4}$ cells (the first id is \texttt{memory\_c4}, the record whose caveat is visible) hits/$n$ with Wilson score 95\% intervals (\%); no contrast is registered on $V_{c4}$.}\label{tab:studyH1-c4}
\adjustbox{max width=\textwidth}{\begin{tabular}{lrrrr}\toprule
model & \texttt{none} & \texttt{id} & \texttt{crit} & \texttt{both\_IL} \\\midrule
Opus 5 & 0/24 [0.0, 13.8] & 0/25 [0.0, 13.3] & 0/25 [0.0, 13.3] & 0/25 [0.0, 13.3] \\
Sonnet 5 & 7/25 [14.3, 47.6] & 0/25 [0.0, 13.3] & 0/25 [0.0, 13.3] & 1/25 [0.7, 19.5] \\
Haiku 4.5 & 0/25 [0.0, 13.3] & 1/25 [0.7, 19.5] & 0/25 [0.0, 13.3] & 0/25 [0.0, 13.3] \\
Fable 5 & 5/25 [8.9, 39.1] & 0/25 [0.0, 13.3] & 0/25 [0.0, 13.3] & 0/25 [0.0, 13.3] \\
Fable 5.1 & 0/25 [0.0, 13.3] & 1/25 [0.7, 19.5] & 0/25 [0.0, 13.3] & 1/25 [0.7, 19.5] \\
GPT-5.6 Sol & 4/25 [6.4, 34.7] & 0/25 [0.0, 13.3] & 0/25 [0.0, 13.3] & 0/25 [0.0, 13.3] \\
GPT-5.6 Terra & 0/25 [0.0, 13.3] & 0/25 [0.0, 13.3] & 0/25 [0.0, 13.3] & 0/25 [0.0, 13.3] \\
GPT-5.6 Luna & 8/25 [17.2, 51.6] & 0/25 [0.0, 13.3] & 0/25 [0.0, 13.3] & 0/25 [0.0, 13.3] \\
\bottomrule\end{tabular}}\end{table}
\begin{table}[htbp]\centering\footnotesize\setlength{\tabcolsep}{4pt}
\caption{Study J: $V_{73}$ hits/25 per cell (any credit; the k = 2 arms credit up to two ids by the registered rule) on the eight arms: the locked crit / both\_IL anchors and three exact bundled edits (operator ratification, explicit target-field rendering, verification capacity k = 2). Wilson intervals for every cell are in the study's results report; the eight registered first-credit cells are in Table~\ref{tab:studyJ-first}.}\label{tab:studyJ}
\adjustbox{max width=\textwidth}{\begin{tabular}{lrrrrrrrr}\toprule
model & \texttt{crit} & \texttt{both\_IL} & \texttt{crit\_legit} & \texttt{both\_IL\_legit} & \texttt{id\_split} & \texttt{both\_split} & \texttt{crit\_k2} & \texttt{both\_IL\_k2} \\\midrule
Opus 5 & 25/25 & 1/25 & 25/25 & 25/25 & 1/25 & 1/25 & 25/25 & 24/25 \\
Fable 5 & 25/25 & 0/25 & 25/25 & 22/25 & 2/25 & 13/25 & 25/25 & 24/25 \\
Fable 5.1 & 11/25 & 0/25 & 25/25 & 25/25 & 11/25 & 1/25 & 25/25 & 25/25 \\
GPT-5.6 Sol & 25/25 & 25/25 & 25/25 & 25/25 & 25/25 & 25/25 & 25/25 & 25/25 \\
\bottomrule\end{tabular}}\end{table}

\begin{table}[htbp]\centering\footnotesize\setlength{\tabcolsep}{4pt}
\caption{Study J: the eight registered first-credit cells of the two k = 2 arms (hits/25 with Wilson score 95\% intervals, \%): the first id the registered credit rule resolves is the target.}\label{tab:studyJ-first}
\adjustbox{max width=\textwidth}{\begin{tabular}{lrr}\toprule
model & \texttt{crit\_k2}: first credit is the target & \texttt{both\_IL\_k2}: first credit is the target \\\midrule
Opus 5 & 25/25 [86.7, 100.0] & 0/25 [0.0, 13.3] \\
Fable 5 & 22/25 [70.0, 95.8] & 0/25 [0.0, 13.3] \\
Fable 5.1 & 3/25 [4.2, 30.0] & 0/25 [0.0, 13.3] \\
GPT-5.6 Sol & 25/25 [86.7, 100.0] & 25/25 [86.7, 100.0] \\
\bottomrule\end{tabular}}\end{table}

\begin{table}[htbp]\centering\footnotesize\setlength{\tabcolsep}{4pt}
\caption{Study J, the anchor and the ratification and target-field edits: 20 of the 32 registered within-model contrasts on $V_{73}$ (any credit), $n = 25$ per cell (Newcombe (1998) method 10 paired score 95\% intervals, registered nonnegative-$\varphi$ variant), two independent label columns ($\delta = 10$); the 12 contrasts of the k = 2 arms are in Table~\ref{tab:studyJ-contrasts-k2}.}\label{tab:studyJ-contrasts}
\adjustbox{max width=\textwidth}{\begin{tabular}{llrll}\toprule
contrast & model & estimate [95\%] & direction & material size \\\midrule
J0 = both\_IL $-$ crit & Opus 5 & \ensuremath{-}96.0 [\ensuremath{-}99.3, \ensuremath{-}75.5] & Negative & Exceeds Margin \\
J0 = both\_IL $-$ crit & Fable 5 & \ensuremath{-}100.0 [\ensuremath{-}100.0, \ensuremath{-}81.2] & Negative & Exceeds Margin \\
J0 = both\_IL $-$ crit & Fable 5.1 & \ensuremath{-}44.0 [\ensuremath{-}62.9, \ensuremath{-}22.1] & Negative & Exceeds Margin \\
J0 = both\_IL $-$ crit & GPT-5.6 Sol & +0.0 [\ensuremath{-}13.3, +13.3] & Undetermined & Unresolved \\
J1 = crit\_legit $-$ crit & Opus 5 & +0.0 [\ensuremath{-}13.3, +13.3] & Undetermined & Unresolved \\
J1 = crit\_legit $-$ crit & Fable 5 & +0.0 [\ensuremath{-}13.3, +13.3] & Undetermined & Unresolved \\
J1 = crit\_legit $-$ crit & Fable 5.1 & +56.0 [+32.9, +73.3] & Positive & Exceeds Margin \\
\bottomrule\end{tabular}}\end{table}
\begin{table}[htbp]\centering\footnotesize\setlength{\tabcolsep}{4pt}
\caption{Study J, the budget of two credits: the 12 registered contrasts of the k = 2 arms on $V_{73}$ (any credit; J7 on the first credit), $n = 25$ per cell (Newcombe (1998) method 10 paired score 95\% intervals, registered nonnegative-$\varphi$ variant), two independent label columns ($\delta = 10$).}\label{tab:studyJ-contrasts-k2}
\adjustbox{max width=\textwidth}{\begin{tabular}{llrll}\toprule
contrast & model & estimate [95\%] & direction & material size \\\midrule
J5 = crit\_k2 $-$ crit & Opus 5 & +0.0 [\ensuremath{-}13.3, +13.3] & Undetermined & Unresolved \\
J5 = crit\_k2 $-$ crit & Fable 5 & +0.0 [\ensuremath{-}13.3, +13.3] & Undetermined & Unresolved \\
J5 = crit\_k2 $-$ crit & Fable 5.1 & +56.0 [+32.9, +73.3] & Positive & Exceeds Margin \\
J5 = crit\_k2 $-$ crit & GPT-5.6 Sol & +0.0 [\ensuremath{-}13.3, +13.3] & Undetermined & Unresolved \\
J6 = both\_IL\_k2 $-$ both\_IL & Opus 5 & +92.0 [+70.0, +96.7] & Positive & Exceeds Margin \\
J6 = both\_IL\_k2 $-$ both\_IL & Fable 5 & +96.0 [+75.5, +99.3] & Positive & Exceeds Margin \\
J6 = both\_IL\_k2 $-$ both\_IL & Fable 5.1 & +100.0 [+81.2, +100.0] & Positive & Exceeds Margin \\
J6 = both\_IL\_k2 $-$ both\_IL & GPT-5.6 Sol & +0.0 [\ensuremath{-}13.3, +13.3] & Undetermined & Unresolved \\
J7 = both\_IL\_k2 $-$ both\_IL (first credit) & Opus 5 & \ensuremath{-}4.0 [\ensuremath{-}19.5, +9.7] & Undetermined & Unresolved \\
J7 = both\_IL\_k2 $-$ both\_IL (first credit) & Fable 5 & +0.0 [\ensuremath{-}13.3, +13.3] & Undetermined & Unresolved \\
J7 = both\_IL\_k2 $-$ both\_IL (first credit) & Fable 5.1 & +0.0 [\ensuremath{-}13.3, +13.3] & Undetermined & Unresolved \\
J7 = both\_IL\_k2 $-$ both\_IL (first credit) & GPT-5.6 Sol & +0.0 [\ensuremath{-}13.3, +13.3] & Undetermined & Unresolved \\
\bottomrule\end{tabular}}\end{table}
\begin{table}[htbp]\centering\footnotesize\setlength{\tabcolsep}{4pt}
\caption{Study H2: $V_{73}$ hits/40 per cell — five wordings of the criterion (P0 = the locked bytes; P1–P4 = surface rewordings holding the construct fixed), each bare and with the locked ` (memory\_73)' suffix.}\label{tab:studyH2}
\adjustbox{max width=\textwidth}{\begin{tabular}{lrrrrrrrrrr}\toprule
model & \texttt{P0} & \texttt{P0\_id} & \texttt{P1} & \texttt{P1\_id} & \texttt{P2} & \texttt{P2\_id} & \texttt{P3} & \texttt{P3\_id} & \texttt{P4} & \texttt{P4\_id} \\\midrule
Opus 5 & 40/40 & 2/40 & 40/40 & 0/40 & 40/40 & 0/40 & 40/40 & 0/40 & 7/40 & 0/40 \\
Fable 5 & 37/40 & 1/40 & 34/40 & 1/40 & 30/40 & 0/40 & 5/40 & 1/40 & 40/40 & 6/40 \\
Fable 5.1 & 16/40 & 0/40 & 20/40 & 0/40 & 16/40 & 1/40 & 10/40 & 0/40 & 24/40 & 0/40 \\
Sonnet 5 & 40/40 & 40/40 & 40/40 & 40/40 & 40/40 & 39/40 & 40/40 & 40/40 & 40/40 & 40/40 \\
GPT-5.6 Sol & 40/40 & 40/40 & 40/40 & 40/40 & 40/40 & 40/40 & 40/40 & 40/40 & 40/40 & 40/40 \\
\bottomrule\end{tabular}}\end{table}
\begin{table}[htbp]\centering\footnotesize\setlength{\tabcolsep}{4pt}
\caption{Study H2, departures from contemporaneous P0: 20 of the 45 registered within-model contrasts on $V_{73}$ (bare P1--P4 $-$ bare P0, read with the two cells), $n = 40$ per cell (Newcombe (1998) method 10 paired score 95\%, registered nonnegative-$\varphi$ variant), two independent label columns ($\delta = 10$); the 25 suffix effects are in Table~\ref{tab:studyH2-suffix}.}\label{tab:studyH2-contrasts}
\adjustbox{max width=\textwidth}{\begin{tabular}{llrll}\toprule
contrast & model & estimate [95\%] & direction & material size \\\midrule
bare P1 $-$ bare P0 & Opus 5 & +0.0 [\ensuremath{-}8.8, +8.8] & Undetermined & Within Margin \\
bare P1 $-$ bare P0 & Fable 5 & \ensuremath{-}7.5 [\ensuremath{-}22.4, +7.2] & Undetermined & Unresolved \\
bare P1 $-$ bare P0 & Fable 5.1 & +10.0 [\ensuremath{-}8.4, +27.4] & Undetermined & Unresolved \\
bare P1 $-$ bare P0 & Sonnet 5 & +0.0 [\ensuremath{-}8.8, +8.8] & Undetermined & Within Margin \\
bare P1 $-$ bare P0 & GPT-5.6 Sol & +0.0 [\ensuremath{-}8.8, +8.8] & Undetermined & Within Margin \\
bare P2 $-$ bare P0 & Opus 5 & +0.0 [\ensuremath{-}8.8, +8.8] & Undetermined & Within Margin \\
bare P2 $-$ bare P0 & Fable 5 & \ensuremath{-}17.5 [\ensuremath{-}33.5, \ensuremath{-}1.1] & Negative & Unresolved \\
\bottomrule\end{tabular}}\end{table}
\begin{table}[htbp]\centering\footnotesize\setlength{\tabcolsep}{4pt}
\caption{Study H2, suffix effects: the 25 registered within-model contrasts (+id $-$ bare, per wording) on $V_{73}$, $n = 40$ per cell (Newcombe (1998) method 10 paired score 95\%, registered nonnegative-$\varphi$ variant), two independent label columns ($\delta = 10$).}\label{tab:studyH2-suffix}
\adjustbox{max width=\textwidth}{\begin{tabular}{llrll}\toprule
contrast & model & estimate [95\%] & direction & material size \\\midrule
+id P0 $-$ bare P0 & Opus 5 & \ensuremath{-}95.0 [\ensuremath{-}98.6, \ensuremath{-}80.5] & Negative & Exceeds Margin \\
+id P0 $-$ bare P0 & Fable 5 & \ensuremath{-}90.0 [\ensuremath{-}95.3, \ensuremath{-}73.9] & Negative & Exceeds Margin \\
+id P0 $-$ bare P0 & Fable 5.1 & \ensuremath{-}40.0 [\ensuremath{-}55.4, \ensuremath{-}23.8] & Negative & Exceeds Margin \\
+id P0 $-$ bare P0 & Sonnet 5 & +0.0 [\ensuremath{-}8.8, +8.8] & Undetermined & Within Margin \\
+id P0 $-$ bare P0 & GPT-5.6 Sol & +0.0 [\ensuremath{-}8.8, +8.8] & Undetermined & Within Margin \\
+id P1 $-$ bare P1 & Opus 5 & \ensuremath{-}100.0 [\ensuremath{-}100.0, \ensuremath{-}87.6] & Negative & Exceeds Margin \\
+id P1 $-$ bare P1 & Fable 5 & \ensuremath{-}82.5 [\ensuremath{-}90.7, \ensuremath{-}65.0] & Negative & Exceeds Margin \\
\bottomrule\end{tabular}}\end{table}
\begin{table}[htbp]\centering\footnotesize\setlength{\tabcolsep}{4pt}
\caption{Study H2: the five criterion wordings exactly as authored and frozen (\texttt{artifacts/FIELD\_BLOCKS.json}; P0 = the locked Study F-x bytes); the +id arms append the locked ` (memory\_73)' suffix. Byte counts and hash prefixes are read from the frozen artifact.}\label{tab:studyH2-wordings}
\adjustbox{max width=\textwidth}{\begin{tabular}{llp{1.4cm}p{9.2cm}}\toprule
wording & bytes (bare / +id) & sha256 & the criterion field, byte for byte \\\midrule
P0 & 109 / 121 & \texttt{ca394238} & \texttt{\small     VERIFY\_PRIORITY: a record that could invalidate a direction you are not\newline       currently planning to take\newline } \\
P1 & 112 / 124 & \texttt{b12a4173} & \texttt{\small     VERIFY\_PRIORITY: the record which could invalidate a direction you are not\newline       currently planning to take\newline } \\
P2 & 104 / 116 & \texttt{9e9236a9} & \texttt{\small     VERIFY\_PRIORITY: a record that could invalidate a direction you do not\newline       currently plan to take\newline } \\
P3 & 112 / 124 & \texttt{9b7730a2} & \texttt{\small     VERIFY\_PRIORITY: any record that could invalidate a direction you are not\newline       planning to take at present\newline } \\
P4 & 105 / 117 & \texttt{cd1bb28d} & \texttt{\small     VERIFY\_PRIORITY: a record that could invalidate a direction that is not\newline       currently in your plan\newline } \\
\bottomrule\end{tabular}}\end{table}
\begin{table}[htbp]\centering\footnotesize\setlength{\tabcolsep}{4pt}
\caption{Study G$'$: $V_{73}$ hits/80 per cell with Wilson score 95\% intervals (\%) on Study G's six byte-controlled arms, fresh memory orders.}\label{tab:studyGr}
\adjustbox{max width=\textwidth}{\begin{tabular}{lrrrrrr}\toprule
model & \texttt{crit} & \texttt{both\_IL} & \texttt{crit\_pad} & \texttt{crit\_other} & \texttt{both\_fresh} & \texttt{both\_inherited} \\\midrule
Opus 5 & 80/80 [95.4, 100.0] & 8/80 [5.2, 18.5] & 80/80 [95.4, 100.0] & 0/80 [0.0, 4.6] & 15/80 [11.7, 28.7] & 0/80 [0.0, 4.6] \\
Fable 5 & 78/80 [91.3, 99.3] & 1/80 [0.2, 6.7] & 1/80 [0.2, 6.7] & 0/80 [0.0, 4.6] & 0/80 [0.0, 4.6] & 0/80 [0.0, 4.6] \\
Fable 5.1 & 29/80 [26.6, 47.2] & 1/80 [0.2, 6.7] & 9/80 [6.0, 20.0] & 1/80 [0.2, 6.7] & 1/80 [0.2, 6.7] & 0/80 [0.0, 4.6] \\
\bottomrule\end{tabular}}\end{table}
\begin{table}[htbp]\centering\footnotesize\setlength{\tabcolsep}{4pt}
\caption{Study G$'$: the 30 within-model contrasts on the new data, the registered replication contrasts against the locked Study G records (percentile bootstrap, each side resampled independently; $\delta = 10$), and the pooled G + G$'$ secondary (nominal n = 120).}\label{tab:studyGr-replication}
\adjustbox{max width=\textwidth}{\begin{tabular}{llrrlr}\toprule
contrast & model & G$'$ estimate [95\%] & G$'$ $-$ G [95\%] & label & pooled (n = 120) [95\%] \\\midrule
G1 & Opus 5 & \ensuremath{-}90.0 [\ensuremath{-}94.8, \ensuremath{-}80.3] & +2.5 [\ensuremath{-}8.8, +12.5] & Undetermined / Unresolved & \ensuremath{-}90.8 [\ensuremath{-}94.8, \ensuremath{-}83.6] \\
G1 & Fable 5 & \ensuremath{-}96.2 [\ensuremath{-}98.3, \ensuremath{-}88.0] & +3.8 [+0.0, +8.8] & Undetermined / Within Margin & \ensuremath{-}97.5 [\ensuremath{-}98.9, \ensuremath{-}91.9] \\
G1 & Fable 5.1 & \ensuremath{-}35.0 [\ensuremath{-}46.0, \ensuremath{-}24.0] & +0.0 [\ensuremath{-}18.8, +18.8] & Undetermined / Unresolved & \ensuremath{-}35.0 [\ensuremath{-}43.9, \ensuremath{-}26.2] \\
G2 & Opus 5 & +0.0 [\ensuremath{-}4.6, +4.6] & +0.0 [+0.0, +0.0] & Undetermined / Within Margin & +0.0 [\ensuremath{-}3.1, +3.1] \\
G2 & Fable 5 & \ensuremath{-}96.2 [\ensuremath{-}98.3, \ensuremath{-}88.0] & \ensuremath{-}3.8 [\ensuremath{-}13.8, +5.0] & Undetermined / Unresolved & \ensuremath{-}95.0 [\ensuremath{-}97.4, \ensuremath{-}88.5] \\
G2 & Fable 5.1 & \ensuremath{-}25.0 [\ensuremath{-}37.1, \ensuremath{-}11.9] & \ensuremath{-}7.5 [\ensuremath{-}26.2, +11.2] & Undetermined / Unresolved & \ensuremath{-}22.5 [\ensuremath{-}32.1, \ensuremath{-}12.5] \\
G3 & Opus 5 & \ensuremath{-}100.0 [\ensuremath{-}100.0, \ensuremath{-}93.5] & +0.0 [+0.0, +0.0] & Undetermined / Within Margin & \ensuremath{-}100.0 [\ensuremath{-}100.0, \ensuremath{-}95.6] \\
\bottomrule\end{tabular}}\end{table}
\begin{table}[htbp]\centering\footnotesize\setlength{\tabcolsep}{4pt}
\caption{Study K2: target first-credit hits per form (main stratum, foregone state; hits/$n$ over \KtwoMainWorlds{} worlds $\times$ \KtwoRunsPerCell{} runs) and the \KtwoContrasts{} registered contrasts per endpoint (points; exact \KtwoCPBLevel{} CPB paired intervals; family sign counts over \KtwoFamilies{} families for both\_IL $-$ crit). The \KtwoSignatures{} primary rows are Opus~5 and Fable~5.1 both\_IL $-$ crit; every other row is descriptive. The last \KtwoOpenAnalysed{} rows are the self-hosted open-weight stratum.}\label{tab:studyK2-contrasts}
\adjustbox{max width=\textwidth}{\begin{tabular}{lrrrrlrll}\toprule
endpoint & \texttt{none} & \texttt{id} & \texttt{crit} & \texttt{both\_IL} & both\_IL $-$ crit [95\%] & fam. $-$/$+$ & crit $-$ id [95\%] & id $-$ none [95\%] \\\midrule
Opus 5 & 24/288 & 44/288 & 286/288 & 167/288 & \ensuremath{-}41.3 [\ensuremath{-}48.4, \ensuremath{-}32.9] & 34/1 & +84.0 [+77.1, +88.6] & +6.9 [+1.5, +12.1] \\
Fable 5.1 & 56/288 & 76/288 & 285/288 & 48/288 & \ensuremath{-}82.3 [\ensuremath{-}87.1, \ensuremath{-}75.2] & 36/0 & +72.6 [+64.4, +78.6] & +6.9 [+1.0, +12.7] \\
Fable 5 & 57/287 & 123/288 & 285/287 & 126/288 & \ensuremath{-}55.4 [\ensuremath{-}62.4, \ensuremath{-}46.8] & 33/0 & +56.4 [+47.8, +63.4] & +23.0 [+15.8, +29.4] \\
Sonnet 5 & 22/288 & 136/288 & 258/288 & 283/288 & +8.7 [+2.6, +14.4] & 3/15 & +42.4 [+31.7, +51.9] & +39.6 [+31.0, +47.0] \\
Haiku 4.5 & 53/288 & 230/288 & 54/288 & 270/288 & +75.0 [+67.3, +80.5] & 0/36 & \ensuremath{-}61.1 [\ensuremath{-}68.1, \ensuremath{-}52.3] & +61.5 [+53.2, +67.9] \\
GPT-5.6 Sol & 43/288 & 267/288 & 282/288 & 286/288 & +1.4 [\ensuremath{-}2.1, +4.8] & 1/4 & +5.2 [\ensuremath{-}0.6, +10.8] & +77.8 [+70.3, +83.1] \\
GPT-5.6 Terra & 76/288 & 241/288 & 261/288 & 283/288 & +7.6 [+2.6, +12.3] & 1/13 & +6.9 [\ensuremath{-}1.2, +14.9] & +57.3 [+48.7, +64.2] \\
\bottomrule\end{tabular}}\end{table}
\begin{table}[htbp]\centering\footnotesize\setlength{\tabcolsep}{4pt}
\caption{Study K2: the \KtwoRowsTotal{} deposited rows (\KtwoDirTotalWord{} directional forecasts, \KtwoWithinTotalWord{} forecast absences `within', \KtwoUndeterminedTotalWord{} left undetermined), main stratum, foregone state, with the locked source cells each forecast pooled (D = Study D, F = Study F run 2, Fx = Study F-x, G = Study G); observed contrasts in points with exact \KtwoCPBLevel{} CPB paired intervals and the family sign counts; `flagged' marks rows whose source cells were small or unequally pooled at deposit. A tally of met rows is descriptive (the rows are not independent).}\label{tab:studyK2-forecasts}
\adjustbox{max width=\textwidth}{\begin{tabular}{llllrll}\toprule
endpoint & contrast & deposited forecast & source cells (study: hits/$n$) & observed [95\% CPB] & families $-$/$+$ & outcome \\\midrule
Fable 5 & id-none & undetermined & id: Fx:6/40; none: Fx:2/40 & +23.0 [+15.8, +29.4] & 0/28 of 36 & -- \\
Fable 5 & crit-id & positive beyond & crit: Fx:40/40, G:40/40; id: Fx:6/40 & +56.4 [+47.8, +63.4] & 0/33 of 36 & met (flagged) \\
Fable 5 & both-crit & negative beyond & both\_IL: Fx:1/40, G:0/40; crit: Fx:40/40, G:40/40 & \ensuremath{-}55.4 [\ensuremath{-}62.4, \ensuremath{-}46.8] & 33/0 of 36 & met \\
Fable 5.1 & id-none & undetermined & id: Fx:13/40; none: Fx:13/40 & +6.9 [+1.0, +12.7] & 1/14 of 36 & -- \\
Fable 5.1 & crit-id & undetermined & crit: Fx:17/40, G:14/40; id: Fx:13/40 & +72.6 [+64.4, +78.6] & 0/36 of 36 & -- (flagged) \\
Fable 5.1 & both-crit & negative beyond & both\_IL: Fx:0/40, G:0/40; crit: Fx:17/40, G:14/40 & \ensuremath{-}82.3 [\ensuremath{-}87.1, \ensuremath{-}75.2] & 36/0 of 36 & met \\
Haiku 4.5 & id-none & within & id: D:0/40, F\_run2:0/20; none: D:0/40, F\_run2:0/20 & +61.5 [+53.2, +67.9] & 0/35 of 36 & not met \\
\bottomrule\end{tabular}}\end{table}
\begin{table}[htbp]\centering\footnotesize\setlength{\tabcolsep}{4pt}
\caption{Study K2 (descriptive): under the \texttt{none} form, the rate at which the first credit went to the active plan's own support record (main stratum, foregone state), with the exact Clopper--Pearson interval and the uniform single-memory reference $1/N$ averaged over worlds; the lower bound exceeded 50\% on \KtwoSBMajority{} of \KtwoSBEndpoints{} endpoints.}\label{tab:studyK2-sb}
\adjustbox{max width=\textwidth}{\begin{tabular}{lrrll}\toprule
endpoint & hits/$n$ & rate (\%) & 95\% CP & uniform reference $1/N$ (\%) \\\midrule
Opus 5 & 255/288 & 88.5 & [84.3, 92.0] & 13.7 \\
Fable 5.1 & 210/288 & 72.9 & [67.4, 78.0] & 13.7 \\
Fable 5 & 196/287 & 68.3 & [62.6, 73.6] & 13.7 \\
Sonnet 5 & 249/288 & 86.5 & [82.0, 90.2] & 13.7 \\
Haiku 4.5 & 197/288 & 68.4 & [62.7, 73.7] & 13.7 \\
GPT-5.6 Sol & 214/288 & 74.3 & [68.9, 79.3] & 13.7 \\
GPT-5.6 Terra & 141/288 & 49.0 & [43.0, 54.9] & 13.7 \\
GPT-5.6 Luna & 187/288 & 64.9 & [59.1, 70.4] & 13.7 \\
GPT-6 Astra & 248/288 & 86.1 & [81.6, 89.9] & 13.7 \\
Qwen2.5 72B & 56/288 & 19.4 & [15.0, 24.5] & 13.7 \\
Llama 3.1 8B & 39/288 & 13.5 & [9.8, 18.0] & 13.7 \\
Llama 3.3 70B & 57/288 & 19.8 & [15.3, 24.9] & 13.7 \\
\bottomrule\end{tabular}}\end{table}
\begin{table}[htbp]\centering\footnotesize\setlength{\tabcolsep}{4pt}
\caption{Study K2 (descriptive): the locked per-model profiles' predicted target rate per form (\%; fitted on the earlier studies' locked cells) against the observed rate on the generated worlds; mean absolute error over forms (points) and the slope of logit(observed) on logit(predicted) over (world, form) with a family-bootstrap interval. GPT-6 Astra and the open stratum had no locked profile.}\label{tab:studyK2-calibration}
\adjustbox{max width=\textwidth}{\begin{tabular}{lrrrrrl}\toprule
endpoint & \texttt{none} pred./obs. & \texttt{id} pred./obs. & \texttt{crit} pred./obs. & \texttt{both\_IL} pred./obs. & MAE & slope [family bootstrap 95\%] \\\midrule
Opus 5 & 1.7 / 8.3 & 0.8 / 15.3 & 100.0 / 99.3 & 4.2 / 58.0 & 18.9 & 0.18 [0.17, 0.19] \\
Fable 5.1 & 32.5 / 19.4 & 32.5 / 26.4 & 38.8 / 99.0 & 0.0 / 16.7 & 24.0 & 0.11 [0.10, 0.12] \\
Fable 5 & 5.0 / 19.9 & 15.0 / 42.7 & 100.0 / 99.3 & 1.2 / 43.8 & 21.4 & 0.16 [0.14, 0.18] \\
Sonnet 5 & 0.0 / 7.6 & 40.0 / 47.2 & 100.0 / 89.6 & 100.0 / 98.3 & 6.8 & 0.14 [0.13, 0.15] \\
Haiku 4.5 & 0.0 / 18.4 & 0.0 / 79.9 & 50.0 / 18.8 & 100.0 / 93.8 & 33.9 & 0.06 [0.05, 0.06] \\
GPT-5.6 Sol & 1.7 / 14.9 & 100.0 / 92.7 & 100.0 / 97.9 & 100.0 / 99.3 & 5.8 & 0.20 [0.18, 0.21] \\
GPT-5.6 Terra & 13.3 / 26.4 & 100.0 / 83.7 & 100.0 / 90.6 & 100.0 / 98.3 & 10.1 & 0.18 [0.16, 0.20] \\
GPT-5.6 Luna & 0.0 / 15.6 & 100.0 / 89.6 & 100.0 / 92.7 & 100.0 / 99.7 & 8.4 & 0.12 [0.11, 0.14] \\
\bottomrule\end{tabular}}\end{table}
\begin{table}[htbp]\centering\footnotesize\setlength{\tabcolsep}{4pt}
\caption{Study K2 (descriptive): the \KtwoBridgeWorldsWord{} bridge worlds --- the earlier instruments' hand-written stores, structurally gated only and not judged, so outside the primary's scope --- paired contrasts over $n = \KtwoBridgeBothCritOpusN{}$ blocks per endpoint.}\label{tab:studyK2-bridge}
\adjustbox{max width=\textwidth}{\begin{tabular}{llll}\toprule
endpoint & both\_IL $-$ crit [95\%] & crit $-$ id [95\%] & id $-$ none [95\%] \\\midrule
Opus 5 & \ensuremath{-}50.0 [\ensuremath{-}87.1, +29.3] & +50.0 [\ensuremath{-}29.3, +87.1] & +50.0 [\ensuremath{-}29.3, +87.1] \\
Fable 5.1 & \ensuremath{-}37.5 [\ensuremath{-}91.2, +48.6] & +12.5 [\ensuremath{-}55.3, +69.3] & +50.0 [\ensuremath{-}29.3, +87.1] \\
Fable 5 & +0.0 [\ensuremath{-}67.2, +67.2] & \ensuremath{-}25.0 [\ensuremath{-}69.4, +40.0] & +87.5 [+0.3, +99.8] \\
Sonnet 5 & +0.0 [\ensuremath{-}42.2, +42.2] & +25.0 [\ensuremath{-}40.0, +69.4] & +62.5 [\ensuremath{-}21.3, +93.4] \\
Haiku 4.5 & +50.0 [\ensuremath{-}29.3, +87.1] & \ensuremath{-}37.5 [\ensuremath{-}79.1, +35.6] & +87.5 [+0.3, +99.8] \\
GPT-5.6 Sol & +0.0 [\ensuremath{-}42.2, +42.2] & +0.0 [\ensuremath{-}42.2, +42.2] & +100.0 [+15.6, +100.0] \\
GPT-5.6 Terra & +0.0 [\ensuremath{-}42.2, +42.2] & +0.0 [\ensuremath{-}42.2, +42.2] & +50.0 [\ensuremath{-}29.3, +87.1] \\
GPT-5.6 Luna & +0.0 [\ensuremath{-}42.2, +42.2] & +12.5 [\ensuremath{-}42.0, +57.5] & +50.0 [\ensuremath{-}29.3, +87.1] \\
GPT-6 Astra & \ensuremath{-}12.5 [\ensuremath{-}57.5, +42.0] & +87.5 [+0.3, +99.8] & +12.5 [\ensuremath{-}42.0, +57.5] \\
Qwen2.5 72B & +25.0 [\ensuremath{-}40.0, +69.4] & +0.0 [\ensuremath{-}42.2, +42.2] & +0.0 [\ensuremath{-}42.2, +42.2] \\
Llama 3.1 8B & +37.5 [\ensuremath{-}44.7, +87.0] & +12.5 [\ensuremath{-}62.8, +76.9] & +25.0 [\ensuremath{-}40.0, +69.4] \\
Llama 3.3 70B & +50.0 [\ensuremath{-}29.3, +87.1] & \ensuremath{-}12.5 [\ensuremath{-}69.3, +55.3] & +12.5 [\ensuremath{-}42.0, +57.5] \\
\bottomrule\end{tabular}}\end{table}
\input{generated/tab_studyK3_audit}
\input{generated/tab_studyK4_audit}
\input{generated/tab_studyK5}
\input{generated/tab_studyK5_audit}

\FloatBarrier
\section{Additional figures}\label{app:figures}
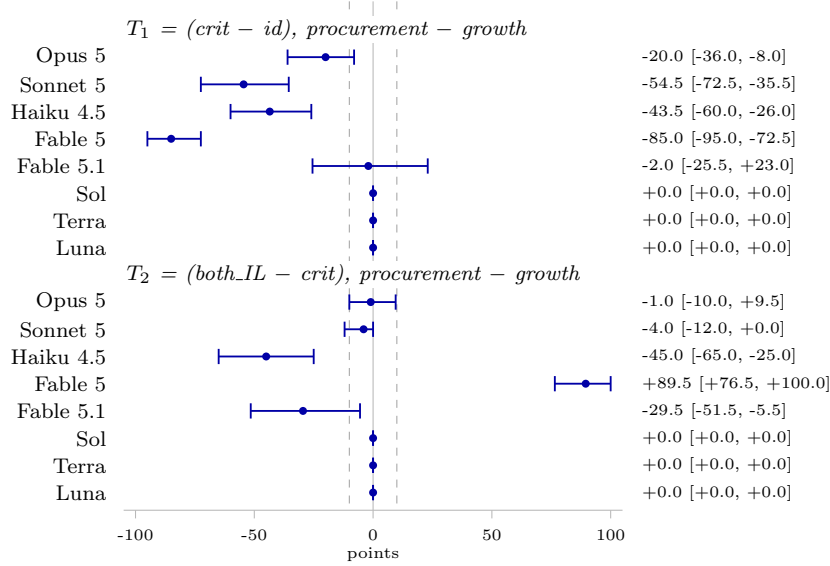
\begin{figure}[htbp]\centering
\begin{tikzpicture}[x=6.6cm/210,y=-0.36cm,font=\scriptsize]
\draw[gray!55,line width=0.4pt] (0,-0.55) -- (0,18.15);
\draw[gray!70,dashed,line width=0.4pt] (10,-0.55) -- (10,18.15);
\draw[gray!70,dashed,line width=0.4pt] (-10,-0.55) -- (-10,18.15);
\draw[gray!55,line width=0.4pt] (-100,18.15) -- (-100,18.45);
\node[anchor=north,font=\tiny] at (-100,18.50) {-100};
\draw[gray!55,line width=0.4pt] (-50,18.15) -- (-50,18.45);
\node[anchor=north,font=\tiny] at (-50,18.50) {-50};
\draw[gray!55,line width=0.4pt] (0,18.15) -- (0,18.45);
\node[anchor=north,font=\tiny] at (0,18.50) {0};
\draw[gray!55,line width=0.4pt] (50,18.15) -- (50,18.45);
\node[anchor=north,font=\tiny] at (50,18.50) {50};
\draw[gray!55,line width=0.4pt] (100,18.15) -- (100,18.45);
\node[anchor=north,font=\tiny] at (100,18.50) {100};
\draw[gray!55,line width=0.4pt] (-105,18.15) -- (105,18.15);
\node[anchor=north,font=\tiny] at (0,19.15) {points};
\node[anchor=west,font=\scriptsize\itshape,fill=white,inner sep=1pt] at (-105,0.50) {$T_1$ = (crit $-$ id), procurement $-$ growth};
\node[anchor=east,font=\scriptsize] at (-108,1.50) {Opus 5};
\draw[blue!65!black,line width=0.7pt] (-36.00,1.50) -- (-8.00,1.50);
\draw[blue!65!black,line width=0.7pt] (-36.00,1.22) -- (-36.00,1.78);
\draw[blue!65!black,line width=0.7pt] (-8.00,1.22) -- (-8.00,1.78);
\fill[blue!65!black] (-20.00,1.50) circle (1.5pt);
\node[anchor=west,font=\tiny] at (109,1.50) {-20.0 [-36.0, -8.0]};
\node[anchor=east,font=\scriptsize] at (-108,2.50) {Sonnet 5};
\draw[blue!65!black,line width=0.7pt] (-72.50,2.50) -- (-35.50,2.50);
\draw[blue!65!black,line width=0.7pt] (-72.50,2.22) -- (-72.50,2.78);
\draw[blue!65!black,line width=0.7pt] (-35.50,2.22) -- (-35.50,2.78);
\fill[blue!65!black] (-54.50,2.50) circle (1.5pt);
\node[anchor=west,font=\tiny] at (109,2.50) {-54.5 [-72.5, -35.5]};
\node[anchor=east,font=\scriptsize] at (-108,3.50) {Haiku 4.5};
\draw[blue!65!black,line width=0.7pt] (-60.00,3.50) -- (-26.00,3.50);
\draw[blue!65!black,line width=0.7pt] (-60.00,3.22) -- (-60.00,3.78);
\draw[blue!65!black,line width=0.7pt] (-26.00,3.22) -- (-26.00,3.78);
\fill[blue!65!black] (-43.50,3.50) circle (1.5pt);
\node[anchor=west,font=\tiny] at (109,3.50) {-43.5 [-60.0, -26.0]};
\node[anchor=east,font=\scriptsize] at (-108,4.50) {Fable 5};
\draw[blue!65!black,line width=0.7pt] (-95.00,4.50) -- (-72.50,4.50);
\draw[blue!65!black,line width=0.7pt] (-95.00,4.22) -- (-95.00,4.78);
\draw[blue!65!black,line width=0.7pt] (-72.50,4.22) -- (-72.50,4.78);
\fill[blue!65!black] (-85.00,4.50) circle (1.5pt);
\node[anchor=west,font=\tiny] at (109,4.50) {-85.0 [-95.0, -72.5]};
\node[anchor=east,font=\scriptsize] at (-108,5.50) {Fable 5.1};
\draw[blue!65!black,line width=0.7pt] (-25.50,5.50) -- (23.00,5.50);
\draw[blue!65!black,line width=0.7pt] (-25.50,5.22) -- (-25.50,5.78);
\draw[blue!65!black,line width=0.7pt] (23.00,5.22) -- (23.00,5.78);
\fill[blue!65!black] (-2.00,5.50) circle (1.5pt);
\node[anchor=west,font=\tiny] at (109,5.50) {-2.0 [-25.5, +23.0]};
\node[anchor=east,font=\scriptsize] at (-108,6.50) {Sol};
\draw[blue!65!black,line width=0.7pt] (0.00,6.50) -- (0.00,6.50);
\draw[blue!65!black,line width=0.7pt] (0.00,6.22) -- (0.00,6.78);
\draw[blue!65!black,line width=0.7pt] (0.00,6.22) -- (0.00,6.78);
\fill[blue!65!black] (0.00,6.50) circle (1.5pt);
\node[anchor=west,font=\tiny] at (109,6.50) {+0.0 [+0.0, +0.0]};
\node[anchor=east,font=\scriptsize] at (-108,7.50) {Terra};
\draw[blue!65!black,line width=0.7pt] (0.00,7.50) -- (0.00,7.50);
\draw[blue!65!black,line width=0.7pt] (0.00,7.22) -- (0.00,7.78);
\draw[blue!65!black,line width=0.7pt] (0.00,7.22) -- (0.00,7.78);
\fill[blue!65!black] (0.00,7.50) circle (1.5pt);
\node[anchor=west,font=\tiny] at (109,7.50) {+0.0 [+0.0, +0.0]};
\node[anchor=east,font=\scriptsize] at (-108,8.50) {Luna};
\draw[blue!65!black,line width=0.7pt] (0.00,8.50) -- (0.00,8.50);
\draw[blue!65!black,line width=0.7pt] (0.00,8.22) -- (0.00,8.78);
\draw[blue!65!black,line width=0.7pt] (0.00,8.22) -- (0.00,8.78);
\fill[blue!65!black] (0.00,8.50) circle (1.5pt);
\node[anchor=west,font=\tiny] at (109,8.50) {+0.0 [+0.0, +0.0]};
\node[anchor=west,font=\scriptsize\itshape,fill=white,inner sep=1pt] at (-105,9.50) {$T_2$ = (both\_IL $-$ crit), procurement $-$ growth};
\node[anchor=east,font=\scriptsize] at (-108,10.50) {Opus 5};
\draw[blue!65!black,line width=0.7pt] (-10.00,10.50) -- (9.50,10.50);
\draw[blue!65!black,line width=0.7pt] (-10.00,10.22) -- (-10.00,10.78);
\draw[blue!65!black,line width=0.7pt] (9.50,10.22) -- (9.50,10.78);
\fill[blue!65!black] (-1.00,10.50) circle (1.5pt);
\node[anchor=west,font=\tiny] at (109,10.50) {-1.0 [-10.0, +9.5]};
\node[anchor=east,font=\scriptsize] at (-108,11.50) {Sonnet 5};
\draw[blue!65!black,line width=0.7pt] (-12.00,11.50) -- (0.00,11.50);
\draw[blue!65!black,line width=0.7pt] (-12.00,11.22) -- (-12.00,11.78);
\draw[blue!65!black,line width=0.7pt] (0.00,11.22) -- (0.00,11.78);
\fill[blue!65!black] (-4.00,11.50) circle (1.5pt);
\node[anchor=west,font=\tiny] at (109,11.50) {-4.0 [-12.0, +0.0]};
\node[anchor=east,font=\scriptsize] at (-108,12.50) {Haiku 4.5};
\draw[blue!65!black,line width=0.7pt] (-65.00,12.50) -- (-25.00,12.50);
\draw[blue!65!black,line width=0.7pt] (-65.00,12.22) -- (-65.00,12.78);
\draw[blue!65!black,line width=0.7pt] (-25.00,12.22) -- (-25.00,12.78);
\fill[blue!65!black] (-45.00,12.50) circle (1.5pt);
\node[anchor=west,font=\tiny] at (109,12.50) {-45.0 [-65.0, -25.0]};
\node[anchor=east,font=\scriptsize] at (-108,13.50) {Fable 5};
\draw[blue!65!black,line width=0.7pt] (76.50,13.50) -- (100.00,13.50);
\draw[blue!65!black,line width=0.7pt] (76.50,13.22) -- (76.50,13.78);
\draw[blue!65!black,line width=0.7pt] (100.00,13.22) -- (100.00,13.78);
\fill[blue!65!black] (89.50,13.50) circle (1.5pt);
\node[anchor=west,font=\tiny] at (109,13.50) {+89.5 [+76.5, +100.0]};
\node[anchor=east,font=\scriptsize] at (-108,14.50) {Fable 5.1};
\draw[blue!65!black,line width=0.7pt] (-51.50,14.50) -- (-5.50,14.50);
\draw[blue!65!black,line width=0.7pt] (-51.50,14.22) -- (-51.50,14.78);
\draw[blue!65!black,line width=0.7pt] (-5.50,14.22) -- (-5.50,14.78);
\fill[blue!65!black] (-29.50,14.50) circle (1.5pt);
\node[anchor=west,font=\tiny] at (109,14.50) {-29.5 [-51.5, -5.5]};
\node[anchor=east,font=\scriptsize] at (-108,15.50) {Sol};
\draw[blue!65!black,line width=0.7pt] (0.00,15.50) -- (0.00,15.50);
\draw[blue!65!black,line width=0.7pt] (0.00,15.22) -- (0.00,15.78);
\draw[blue!65!black,line width=0.7pt] (0.00,15.22) -- (0.00,15.78);
\fill[blue!65!black] (0.00,15.50) circle (1.5pt);
\node[anchor=west,font=\tiny] at (109,15.50) {+0.0 [+0.0, +0.0]};
\node[anchor=east,font=\scriptsize] at (-108,16.50) {Terra};
\draw[blue!65!black,line width=0.7pt] (0.00,16.50) -- (0.00,16.50);
\draw[blue!65!black,line width=0.7pt] (0.00,16.22) -- (0.00,16.78);
\draw[blue!65!black,line width=0.7pt] (0.00,16.22) -- (0.00,16.78);
\fill[blue!65!black] (0.00,16.50) circle (1.5pt);
\node[anchor=west,font=\tiny] at (109,16.50) {+0.0 [+0.0, +0.0]};
\node[anchor=east,font=\scriptsize] at (-108,17.50) {Luna};
\draw[blue!65!black,line width=0.7pt] (0.00,17.50) -- (0.00,17.50);
\draw[blue!65!black,line width=0.7pt] (0.00,17.22) -- (0.00,17.78);
\draw[blue!65!black,line width=0.7pt] (0.00,17.22) -- (0.00,17.78);
\fill[blue!65!black] (0.00,17.50) circle (1.5pt);
\node[anchor=west,font=\tiny] at (109,17.50) {+0.0 [+0.0, +0.0]};
\end{tikzpicture}
\caption{Study H1: the two registered transport contrasts per model (procurement minus growth); every row is one contrast, drawn as its percentile bootstrap 95\% interval with each side resampled independently, with a dot at the estimate and the value printed at the right. Dashed lines mark the $\pm 10$-point materiality margin, the solid line zero. The three GPT-5.6 models are \emph{Sol}, \emph{Terra} and \emph{Luna}.}\label{fig:transport}
\end{figure}

Figures~\ref{fig:transport} and~\ref{fig:rewording} draw the registered transport contrasts of Study H1 and the cells of Study H2 from the same locked outputs as the tables.
\begin{figure}[htbp]\centering
\input{generated/fig_rewording}
\caption{Study H2: the exact count of episodes in which the target was named under the locked criterion (P0) and the four surface rewordings (P1--P4), per model, shaded by the rate; the left block is the bare criterion and the right block the same criterion with the locked ` (memory\_73)' suffix, so the suffix effect is the change between the two blocks ($\HtwoN{}$ episodes, $n = 40$ per cell; registered Wilson intervals in Table~\ref{tab:studyH2}). \emph{Sol} is GPT-5.6 Sol.}\label{fig:rewording}
\end{figure}

\FloatBarrier
\section{The frozen prompt: store, planning block and directive field}\label{app:prompts}

The shared context of every arm, from the frozen Study F prompt document (arm \texttt{none}; the situation paragraph
is elided; lines longer than the page are wrapped for display with a continuation marker --- the frozen bytes are
unwrapped): the inherited store, the planning-state block into which the directive field is inserted, and the
budget paragraph.
{\small
\begin{Verbatim}[commandchars=\\\{\}]
DURABLE MEMORY \textemdash{} inherited from earlier agent sessions and analysts

  memory_44 \textperiodcentered{} consolidated day 24 \textperiodcentered{} source archived
    Enterprise prospects convert better through sales-assisted flows; the demo-request
      -> path outperformed self-serve on pipeline quality.

  memory_31 \textperiodcentered{} consolidated day 12 \textperiodcentered{} source archived
    Signup simplification improved SMB activation; cutting required fields at signup
      -> produced a durable lift in first-week activation.

  memory_91 \textperiodcentered{} consolidated day 61 \textperiodcentered{} source archived
    Activation messaging inside the product improved week-one engagement and nudged more
      -> accounts to their first meaningful action.

  memory_73 \textperiodcentered{} consolidated day 31 \textperiodcentered{} source archived
    Targeted promotional pricing is an effective response for price-sensitive SMB
      -> segments, with strong acquisition and revenue lift.

  memory_57 \textperiodcentered{} consolidated day 40 \textperiodcentered{} source archived
    Referral incentives performed well in established customers, adding qualified
      -> revenue without measurable harm to retention.

  memory_86 \textperiodcentered{} consolidated day 52 \textperiodcentered{} source archived
    Shortening onboarding improved completion rates and time-to-value, lifting early
      -> product adoption across self-serve accounts.

[... situation ...]

  PLANNING STATE \textemdash{} carried over from the previous session

    Plan A \textperiodcentered{} promotional_pricing
      Respond to competitor entry pricing with targeted discounts for
      price-sensitive SMB accounts.
    Plan B \textperiodcentered{} simplify_onboarding
      Reduce required setup steps to lift activation and onboarding
      completion in self-serve accounts.

    ACTIVE_PLAN_ID: B
    [VERIFY_PRIORITY field: absent in `none'; see the four contents below]

VERIFICATION BUDGET

  Analyst time is limited. You may pull the archived source record for at most
  1 inherited memory before committing. Name it by id. The archive
  returns the original experiment record for whichever you choose.

  Candidate directions:
    simplify_onboarding \textperiodcentered{} promotional_pricing \textperiodcentered{} referral_incentive \textperiodcentered{}
    activation_messaging \textperiodcentered{} enterprise_sales_assist

Name the memory you want verified (up to one, or none). Do not state an
action. Answer with the structured object only.
\end{Verbatim}
}

The four directive contents, extracted by the generator from the frozen Study F prompt document as
\FDirectiveLineCount{} distinct field blocks (leading indentation normalised; the \texttt{id} and \texttt{crit}
blocks are byte-identical to the locked Study D/E rendering, asserted at freeze time). Each is inserted as the \texttt{VERIFY\_PRIORITY}
field of the planning-state block; with the field block removed all five arms are byte-identical. The \texttt{id}
line carries \FIdPaddingSpaces{} trailing spaces in the frozen rendering (the character-length matching to the
criterion line), which verbatim output cannot show.
{\small
\begin{Verbatim}[commandchars=\\\{\}]
VERIFY_PRIORITY: memory_73
VERIFY_PRIORITY: a record that could invalidate a direction you are not
    currently planning to take
VERIFY_PRIORITY: a record that could invalidate a direction you are not
    currently planning to take (memory_73)
VERIFY_PRIORITY: memory_73 - a record that could invalidate a direction you are not
    currently planning to take
\end{Verbatim}
}

\FloatBarrier
\section{Reproduction}\label{app:repro}

Every study rate, contrast, interval, episode count and registration identifier in this paper is a macro emitted by \ifanon\else\texttt{paper3/manuscript/scripts/generate.py}\fi{}. For every study the generator verifies the lock manifest against the raw episode files before reading them, re-parses every stored raw response and requires every stored answer field the raw object carries to agree with it (for Study K2's self-hosted stratum and for Studies K3, K4 and K5 the generator reads the hash-bound converter output of the batch records and does not re-parse their completions --- for K3 that re-parse is the independent recomputation tool's, see below; the deposit receipts it checks are local records of a download-back verification against the OSF node (already public when the deposits of Studies K2--K4 were made, although those depositors' receipts record a private flag written as a constant; Study K5's component is private), authenticated only by that node --- \ifanon\else\texttt{paper3/studyK2/preregistration/verify\_deposits\_osf.py} carries\fi{} a re-download check for the node's owner), and re-derives in exact rational arithmetic, with fail-closed pairing, every rate and registered point estimate that the text or a table quotes --- including Study \Br{}'s seven main quantities and its replication points, Study H1's growth sides (from the locked D, F and F-x records) and transport points, Study I's decision endpoints under the registered served-model continuity rule, Study J's credits under the registered resolver, and Study \Gr{}'s cells, contrasts and replication points. Two classes are read from the studies' frozen (or, for Study B, corrected) analysis outputs rather than re-derived: every interval (Wilson, Newcombe and bootstrap), taken only after the re-derived point estimate is asserted equal to the frozen one (the Wilson intervals of the figures are recomputed), and Study \Gr{}'s pooled G + \Gr{} secondary cells and contrasts, which are copied with their intervals. Design constants (budgets, runs per cell, the bootstrap size, the margin, per-cell denominators such as /40), process counts (review rounds, package files described in prose) and numbers quoted from cited papers are typed in the source; the audit's scanner allow-lists exactly these classes. \ifanon\else\texttt{scripts/audit.py}\fi{} regenerates the macros to a scratch location and
requires byte identity, cross-checks the headline values against the frozen analysis outputs, scans the
manuscript source for unmarked result-like numbers, and checks that every headline value appears in the PDF
text. Build:
\ifanon\else
\begin{verbatim}
python3 paper3/manuscript/scripts/generate.py
python3 paper3/manuscript/scripts/audit.py
cd paper3/manuscript && tectonic -X compile main.tex --outdir output
\end{verbatim}
\fi

\paragraph{Study K3.} For Study K3 the generator verifies the frozen manifest's entries for the analysis, converter, lock, combiner and recomputation tools, the external and results deposit receipts, both committed locks against their run directories, and the two per-ladder recomputation receipts (bound to the results file, the packaged recomputation tool, the committed lock, the locked audit file and directory, the combined episodes and the deposit receipt), then aggregates the converter's hash-bound episode rows into the per-form rates and the per-gain block contrasts and re-derives the means, slopes, changes and registered decisions from them; it copies the per-gain exact intervals and the bootstrap intervals from the results file. The recomputation of every row from the locked completions, of every interval and of the audit agreement counts from the retained replay completions is the independent tool's (\texttt{analysis/k3\_recompute.py}, no import from the analyzer or the converter), whose receipts the report requires.

\paragraph{Study K4.} For Study K4 the generator verifies the frozen manifest's entries for the analysis, converter, lock, combiner, recomputation and report tools, the pre-specification, the licence and the registered operating-characteristics file, the external and results deposit receipts (the results deposit's OSF hash equal to the archive's), the three committed locks against their run directories in ladder order, and the three per-ladder recomputation receipts (each bound to the results file, the packaged recomputation tool, its own ladder's committed lock, the locked audit file and directory, the combined episodes and the deposit receipt), checks each ladder's status ledger against the producer's contract (one start, \KfourCallsPerLadder{} calls) and pairs each production reservation with its settlement in the atomic campaign records, then aggregates the converter's hash-bound episode rows into the per-form rates and the per-gain block contrasts and re-derives the means, the centred linear components on each ladder's own gains, the changes and the three registered decisions from them; it copies the per-gain exact intervals and the bootstrap intervals from the results file and the realised margin error from the archived operating characteristics.

\paragraph{Study K5.} For Study K5 the generator verifies the Stage-B manifest's entries for the analysis, converter, lock, combiner, recomputation, K5-3 recomputation, identity-check and report tools, the three deposit receipts (pre-stream, package and results; each download-back-verified on the private node named in the receipt, the results deposit's OSF hash equal to the archive's), the two committed locks against their run directories, the two per-ladder recomputation receipts (each bound to the results file, the packaged recomputation tool, its own ladder's committed lock, the locked audit file and directory, the combined episodes, the Stage-B manifest and its deposit receipt) and the K5-3 receipt (bound to both), the runtime identity receipt, the frozen power-gate record embedded in the results (and the operating-characteristics file it names), the two analysis outputs (invocation 1 retained; the generator re-derives that they differ in the prompt-manifest keys only), each ladder's status ledger against the producer's contract (one start of the second attempt, \KfiveCallsPerLadder{} calls, no engine error) and the two archived halted attempts against the lineage the locks reconcile, pairs each production reservation with its settlement (and each halted first attempt with no settlement) in the atomic campaign records and sums the superseded campaigns' ledger, then aggregates the converter's hash-bound episode rows into the per-form rates, the cohort split and the per-gain block contrasts and re-derives the means, the centred linear components, the changes and the two registered decisions from them; it copies the per-gain exact intervals, the bootstrap intervals, the K5-3 difference and the cohort changes from the results file and the realised margin error and calibrated power from the archived operating characteristics.
\FloatBarrier
\section{Study E: the OpenRouter-served-panel transport test}\label{app:studyE}

Study E re-ran Study D's design on a disjoint OpenRouter-served panel; it is reported here in full as a transport test, not a replication (\S\ref{sec:studies}).

\paragraph{Result.}

Of \EAttempted{} attempted episodes \EScored{} scored; the \EErrors{} errors (\EErrorPct\%) are transport
failures on one endpoint (Kimi K3, five runs), not the schema failures the prose registration anticipated, and
the frozen analyzer paired runs present in both arms --- a rule the registration did not state; both facts are
in the study's deviation record and the pairing is reported here as executed (\EKimiPairedRuns{} paired Kimi
runs in $E_1$). A further inconsistency of the frozen package is disclosed here: \texttt{FROZEN\_HYPOTHESES.md} describes the panel before one endpoint was removed (ten models) and in one line requires `all six models', whereas the frozen model list, schedule and analyzer enforce the nine-model panel that was run; the executed analysis followed the frozen list, and the study's reconciliation record lists the conflicting lines. The rates were \ERateNone\% (none), \ERateOracle\% (id) and \ERatePolicy\% (criterion):
$E_1 = \EEone{}$ [\EEoneLo, \EEoneHi]. The registered superiority rule ($\mathrm{ci}_{\mathrm{lo}} > 0$) was
not met; the frozen script's verdict string reads FORM DOES NOT MATTER, which overstates a failed superiority
test: no equivalence margin was registered, the interval is compatible with a criterion advantage of up to
\EEoneHi{} points, and one completion of the missing episodes would have met the rule. The per-model contrast
ranges from \EPerModelMin{} to \EPerModelMax{} (Table~\ref{tab:studyE}): on this panel the bare id was often
obeyed and the criterion sometimes cost. Study E was registered as the same design on a new panel; because it changes model identities, access route, runs per cell (20 against 40) and inference configuration at once, it cannot isolate transport across any one of them, and it is reported as a disjoint-panel boundary check --- a reinterpretation adopted after external review and recorded as such.
\end{sloppypar}

\end{document}